\documentclass[10pt,a4paper,reqno]{amsart}
\usepackage{acronym}
\usepackage{amsfonts}
\usepackage{amsmath}
\usepackage{amssymb}
\usepackage{amsthm}
\usepackage{bm}
\usepackage{braket}
\usepackage{cite}
\usepackage{color}
\usepackage{geometry}
\usepackage{graphicx}
\usepackage{hyperref}
\usepackage{mathrsfs}
\usepackage{mathtools}
\usepackage{setspace}
\usepackage{epstopdf}

\theoremstyle{definition}

\theoremstyle{definition}

\usepackage{booktabs}
\usepackage{threeparttable}
\usepackage{diagbox}
\usepackage{multirow}
\usepackage{float}
\usepackage{array}

\usepackage{stmaryrd}
\usepackage{subcaption}
\usepackage{tikz}

\usepackage{epsfig}
\usepackage{setspace}
\usepackage{epstopdf}
\usepackage{booktabs}
\usepackage{threeparttable}
\usepackage{diagbox}

\newgeometry{top=2cm,bottom=2cm,outer=1.5cm,inner=1.5cm}
\newcommand{\bse}{\begin{subequations}}
\newcommand{\ese}{\end{subequations}}

\numberwithin{equation}{section}
\DeclareSymbolFont{largesymbols}{OMX}{yhex}{m}{n}
\DeclareMathAccent{\Widehat}{\mathord}{largesymbols}{"62}

\title[Physics-informed neural network methods based on Miura transformations]{Physics-informed neural network methods based on Miura transformations and discovery of new localized wave solutions}
\author{Shuning Lin}
\address[SL]{School of Mathematical Sciences, Shanghai Key Laboratory of Pure Mathematics and Mathematical Practice, and Shanghai Key Laboratory of Trustworthy Computing \\
East China Normal University \\ Shanghai 200241 \\ China}
\author{Yong Chen$^*$}
\address[YC]{School of Mathematical Sciences, Shanghai Key Laboratory of Pure Mathematics and Mathematical Practice, and Shanghai Key Laboratory of Trustworthy Computing \\
East China Normal University \\ Shanghai 200241 \\ China}
\address[YC]{College of Mathematics and Systems Science \\ Shandong University of Science and Technology \\ Qingdao 266590 \\ China}
\email{ychen@sei.ecnu.edu.cn}

\begin{document}

\begin{abstract}
We put forth two physics-informed neural network (PINN) schemes based on Miura transformations and the novelty of this research is the incorporation of Miura transformation constraints into neural networks to solve nonlinear PDEs. The most noteworthy advantage of our method is that we can simply exploit the initial-boundary data of a solution of a certain nonlinear equation to obtain the data-driven solution of another evolution equation with the aid of PINNs and during the process, the Miura transformation plays an indispensable role of a bridge between solutions of two separate equations. It is tailored to the inverse process of the Miura transformation and can overcome the difficulties in solving solutions based on the implicit expression. Moreover, two schemes are applied to perform abundant computational experiments to effectively reproduce dynamic behaviors of solutions for the well-known KdV equation and mKdV equation. Significantly, new data-driven solutions are successfully simulated and one of the most important results is the discovery of a new localized wave solution: kink-bell type solution of the defocusing mKdV equation and it has not been previously observed and reported to our knowledge. It provides a possibility for new types of numerical solutions by fully leveraging the many-to-one relationship between solutions before and after Miura transformations. Performance comparisons in different cases as well as advantages and disadvantages analysis of two schemes are also discussed. On the basis of the performance of two schemes and no free lunch theorem, they both have their own merits and thus more appropriate one should be chosen according to specific cases.


\end{abstract}
\maketitle

\section{Introduction}

The research of the numerical method based on neural networks for solving partial differential equations can be traced back to 1994, which was studied by M. Dissanayake and N. Phan-Thien \cite{Dissanayake1994}. Recently, the physics-informed neural network (PINN) method proposed by Raissi, Perdikaris and Karniadakis \cite{PINN2019} has landmark significance after a series of attempts of solving forward and inverse problems of nonlinear partial differential equations \cite{Raissi1,Raissi2}. It aims to train neural networks (NNs) to solve supervised learning tasks while respecting laws of physics described by nonlinear partial differential equations. This deep learning framework has breathed new life into the area of scientific computing and many efficient and significant variants on the basic structure were devised, such as VPINNs \cite{VPINNs2019}, fPINNs \cite{fPINNs2019}, XPINNs \cite{XPINNs2020}, g-PINNs \cite{gPINNs2020}, B-PINNs \cite{B-PINNs2021}, hp-VPINNs \cite{hp-VPINNs2021}, NSFnets \cite{NSFnets2021} and so on. Furthermore, the development of neural networks is extended from learning mappings between finite-dimensional Euclidean spaces to neural operators that learn mappings between function spaces and two neural operators that have shown early promising results are: the deep operator network (DeepONet) \cite{DeepONet2019} and the Fourier neural operator (FNO) \cite{FNO2020}. The performance of these two neural operators are analyzed and compared both theoretically and computationally and they both exhibit high accuracy for diverse applications \cite{DeepONetFNO2022}. Besides, some efforts to improve the performance of PINNs have been made and have achieved remarkable results. To accelerate the training process and reduce the training cost, Jagtap et al. \cite{Jagtap2020} put forward two approaches of locally adaptive activation functions namely, layer-wise and neuron-wise locally adaptive activation functions. Meanwhile, two new residual-based adaptive sampling methods: residual-based adaptive distribution (RAD) and residual-based adaptive refinement with distribution (RAR-D) are devised to improve the sampling efficiency and the accuracy of PINNs \cite{WuZhu2022}. A meta-learning technique for offline discovery of PINN loss functions, proposed by Psaros et al \cite{Psaros2022}, is also a powerful tool to achieve the significant performance improvement. With continuous research and improvement of this technique, the PINN method has been widely applied to various fields, like flow visualizations \cite{RaissiYazdani2020}, high-speed flows \cite{Mao2020}, heat transfer problems \cite{CaiWang2021} and has shown extraordinary performance. In short, this PINN methodology, as one of the most revolutionary and powerful data-driven approaches, promotes the development of scientific computing and other related fields dramatically.

Integrable systems are a class of nonlinear dynamic systems possessing remarkable properties, such as Lax integrability, Painlev\'{e} integrability, Liouville integrability and so on \cite{Lax1968, Painleve1983, Liouvilie1989}. Numerous exact solutions of integrable equations can be obtained, which are served as abundant training samples for the PINNs. We mainly focus on studying integrable systems via the deep learning algorithm, especially the PINN algorithm, and has made a series of progress. In 2020, Li and Chen (a co-author of this paper) \cite{Li2020, LiChen2020} carried out numerical experiments on second-order and third-order nonlinear evolution equations to simulate localized wave solutions by means of PINNs. This is our first step into the integrable-deep learning, a concept first brought forward by Chen. Then dynamic behaviors of the rogue wave solution for the nonlinear Schr\"{o}dinger equation \cite{Pu2021} and the rogue periodic wave solution for the Chen-Lee-Liu equation \cite{Peng2021} have been reproduced for the first time based on the PINN method. Afterwards, PINNs were applied to high-dimensional integrable systems to solve the ($N$+1)-dimensional initial-boundary value problem with 2$N$+1 hyperplane boundaries and the (2+1)-dimensional resonance rogue solution was successfully simulated \cite{Miao2022}. Furthermore, Peng and Chen utilized the deep learning technique to solve nonlocal integrable systems with the $\mathcal{P}\mathcal{T}$ symmetry term through adding the nonlocal term into the NNs and studied the data-driven soliton solutions and the parameter prediction for the nonlocal Hirota equation \cite{Peng2022}. Beyond that, extensive relevant researches are also carried out have yielded fruitful results \cite{PuLi2021,PuChen2022,Yan2022,LB2022}.

One of the most important features of integrable systems is the possession of infinite conservation laws. Notably, our published work \cite{Lin2022} incorporated explicitly the presence of conservation laws and it was the first time (to our knowledge) that conserved quantities of integrable systems were introduced into neural networks. The most remarkable advantages of this PINN method based on conserved quantities lies in that it can impose physical constraints from a global perspective. Meanwhile, it is tailored to the nature of equations by introducing conserved quantities of nonlinear systems into neural networks, which implies that the underlying information of the given equations is dug out to improve the precision and reliability. In addition, it was applied to simulate the soliton molecule, M-shape double-peak soliton, plateau soliton, interaction solution, etc. Numerical results indicated that this proposed method can obviously improve prediction accuracy and enhance the ability of generalization compared with the original PINN. This methodology provides a promising new direction to devise deep learning algorithms with the advantages of integrable systems.

Utilizing the PINN method to pertinently solve problems arising in the field of integrable systems that cannot be solved by classical methods and further combining deep learning with integrable system theory more tightly and effectively to devise significant integrable-deep learning algorithms are always the goal we pursue. To our knowledge, there is rare correlative study concerning on the consideration of transformations in PINNs. The novelty of this research is the incorporation of Miura transformation constraints into NNs, which is the most important transformation of integrable systems in our view.

In 1883, an interesting property of the sine-Gordon equation is found by German geometer A.V. B\"{a}cklund \cite{LYS1999}. Assume $u$ satisfies the sine-Gordon equation
\begin{align}\label{E1-1}
u_{xt}={\rm{sin}} u,	
\end{align}
and $u^\prime$ is also the solution of \eqref{E1-1} under the transformation
\begin{equation}\label{E1-2}
\begin{split}
\begin{cases}
(\frac{u^\prime+u}{2})_x=a {\rm{sin}} \frac{u^\prime-u}{2},\\
(\frac{u^\prime-u}{2})_t=a {\rm{sin}} \frac{u^\prime+u}{2}.
\end{cases}
\end{split}
\end{equation}
This transformation from one solution of an equation to another is called the B\"{a}cklund transformation \cite{Matveev1991,Rogers2002,Rogers1982,Tenenblat1998,Wu1995}. Its outstanding advantage mainly lies in that new solutions can be derived based on the obtained solutions of the equation. With the development of the theory of solitons, it is found that many well-known nonlinear partial differential equations have B\"{a}cklund transformations, such as the KdV equation and the potential KdV equation \cite{KdVBacklund}, the modified KdV equation \cite{mKdVBacklund} and so on. The B\"{a}cklund transformation mentioned above is the one between the solutions of the same partial differential equation and it can also be defined between different equations, which is called the Miura transformation \cite{Miura1968} (a special B\"{a}cklund transformation).

The Miura transformation is also of great significance in the field of integrable systems. In 1968, by analyzing the similarity between the KdV and mKdV equation, both in form and in possession of many polynomial conservation laws \cite{MiuraGardner1968}, Miura \cite{Miura1968} inferred that their solutions might be intimately related and proposed the best-known Miura transformation between the mKdV equation and the KdV equation. These two nonlinear partial differential equations are both well-known. In 1834, British scientist Russel occasionally observed a kind of water wave (solitary wave) whose shape and speed will not change in the process of traveling but he couldn't develop the appropriate equation. Later the KdV equation that described Russel's solitary waves was first derived by Korteweg and de Vries \cite{KdV1895} in the study of long waves with small amplitude in a relatively shallow channel in 1895. Then in 1965, Zabusky and Kruskal \cite{Zabusky1965} numerically solved the KdV equation by finite difference method (FDM) and discussed special properties of solitary waves. Meanwhile, the concept of soliton was given for the first time. Since then, the theory of modern integrable system has really been revived. This was also the origin of modern soliton theory and was an important step forward in the research of nonlinear systems. The inverse scattering method was proposed by Garder, Greene, Kruskal and Miura \cite{GGKM1967} to derive the analytic solution of the initial value problem of KdV equation in 1967 and was utilized by Ablowitz, Kaup, Newell and Segur \cite{AKNS1973} to obtain soliton solutions of a series of initial value problems for nonlinear evolution equations, such as KdV equation and modified KdV (mKdV) equation in 1973. Since the KdV equation has excellent properties, such as infinite conservation laws \cite{MiuraGardner1968}, it has been widely applied in plasma physics, ocean internal wave and other fields. The mKdV equation, which was derived in the study of anharmonic lattices \cite{Zabusky1967}, can be regarded as the KdV equation with a cubic nonlinearity. Because of its relation to inverse scattering theory, the mKdV equation was also studied extensively \cite{1976Miura,Scott1973}, such as the exact solutions \cite{Wadati1972, Hirota1972, Wadati1973}, the discussion of global well-posedness \cite{Fonsecal1999}, asymptotic stability of solitons \cite{Germain2016} and so on. The importance of Miura transformation is self-evident since a remarkable explicit nonlinear transformation acts as a bridge connecting the solutions of two classical integrable equations in the field of integrable systems. Since then, the research about Miura transformations has been widely carried out. Fordy and Gibbons \cite{Fordy1980} extended the method of factorization of operators, which was used to derived the Miura transformation of the KdV equation, to some third-order scattering operators and derived transformations between several fifth-order nonlinear evolution equations. Miura transformations of many other equations were also discussed, including the Miura transformations between two classes of equations: a family of higher order Korteweg-de Vries (ho-KdV) and an associated family of higher order modified Korteweg-de Vries (ho-mKdV) equations \cite{Chern1979, Chan1988}, between the KP equation and the modified KP (mKP) equation \cite{Konopelchenko1984}, for Sato-type hierarchies \cite{Shaw1993} and for the constrained KP (cKP) and the constrained modified KP (cmKP) hierarchies \cite{Shaw1997}, etc. Moreover, many scholars studied classification problems of Miura transformations  under some restrictions \cite{Xu2006,Cao2011}. The Miura transformation reveals the relationship between the solutions, so it is of profound significance for solving partial differential equations.

For a Miura transformation mapping the solution of a certain partial differential equation to that of another one, the solution $u$ of the latter can be easily obtained based on the solution $v$ of the former, which evokes the question of whether we can solve the former based on the solution $u$ or even the initial-boundary data of the latter. In this paper, two PINN schemes based on Miura transformations are devised and applied to solve nonlinear equations. By taking advantage of PINNs, we can simply use the initial-boundary data of a solution of one nonlinear equation to obtain the data-driven solution of another evolution equation based on the Miura transformation, which has a vital bond role in solutions of two separate equations. The introduction of Miura transformations into the PINNs is the significant difference from the traditional PINNs and its variants mentioned above. Meanwhile, the major superiority of our method lies in that the solution $v$ can be derived based on the implicit expression of $v$ (the Miura transformation). The main difference of two schemes is that only one PINN is constructed to acquire the data-driven solutions (both $\Widehat{u}(x,t)$ and $\Widehat{v}(x,t)$) in Scheme \uppercase\expandafter{\romannumeral 1} while two PINNs are trained to derive $\Widehat{u}(x,t)$ and $\Widehat{v}(x,t)$ respectively in Scheme \uppercase\expandafter{\romannumeral 2}. Considering the richness and diversity of Miura transformations, the proposed method is mainly applied in solving the defocusing and focusing mKdV equations based on the real Miura transformation between the defocusing mKdV equation and the KdV equation, and the complex Miura transformation between the focusing mKdV equation and the KdV equation. Owing to the existence of the many-to-one relationship between solutions before and after Miura transformations, various numerical solutions of the defocusing mKdV equation can be derived by utilizing one set of initial-boundary data of the solution for the KdV equation, which furnishes possibility for the discovery of new localized wave solutions. The proposed method is tailored to the inverse process of the Miura transformation and it can overcome the difficulties in solving solutions based on the implicit expression and give full play to its strength of multiplicity of solutions. It provides basis for the further applications in Miura transformations of other nonlinear evolution equations, which will be continued in the future work.

The outline of this paper is as follows. In Section \ref{Methodology}, we introduce briefly the Miura transformations and put forward two PINN schemes based on Miura transformations. In Section \ref{real Miura}, numerical experiments are carried out with respect to the real Miura transformation between the defocusing mKdV equation and the KdV equation by Scheme \uppercase\expandafter{\romannumeral 1} and new numerical solutions are successfully simulated. We shift our attention to applications of Scheme \uppercase\expandafter{\romannumeral 2} in the complex Miura transformation between the focusing mKdV equation and the KdV equation in Section \ref{complex Miura}. Moreover, supported with mass data, we compare the performance and analyze merits and drawbacks of two schemes in Section \ref{comparison}. Finally, the conclusion and expectation are given in the last section.

\section{Methodology}\label{Methodology}
\subsection{Introduction of Miura transformations}
\quad

As is known to all, Miura transformations play a prominent role in the fields of integrable systems. For the transformation between solutions of different nonlinear partial differential equations, the best-known one is, of course, the map proposed by Miura \cite{Miura1968}
\begin{align}\label{E1-3}
v \mapsto u= v_x-v^2,	
\end{align}
from the mKdV equation
\begin{align}\label{E1-4}
v_t-6v^2 v_x+v_{xxx}=0,		
\end{align}
to the KdV equation
\begin{align}\label{E1-5}
u_t+6u u_x+u_{xxx}=0.
\end{align}
Meanwhile, many other PDEs also have Miura transformations. The Miura transformation has caused a tremendous reaction since it is surprising and rare to find a transformation between two independent nonlinear partial differential equations.

Inspired by the above example, the general form of Miura transformations is given below. Suppose that $v$ satisfies the following partial differential equation
\begin{align}\label{E2-1}
G(v, v_{x}, v_{t}, \cdots, \partial_{x}^{l} v, \cdots, \partial_{t}^{l} v)=0.
\end{align}
If $u=T(v, v_{x}, v_{t}, \cdots)$ is the solution of the nonlinear partial differential equation
\begin{align}\label{E2-2}
F(u, u_{x}, u_{t}, \cdots, \partial_{x}^{k} u, \cdots, \partial_{t}^{k} u)=0,	
\end{align}
the mapping
\begin{align}\label{E2-3}
v \mapsto u=T(v, v_{x}, v_{t}, \cdots),
\end{align}
is called the Miura transformation from \eqref{E2-1} to \eqref{E2-2}, which is a special B\"{a}cklund transformation.

For convenience, we denote \eqref{E2-3} as
\begin{align}\label{E2-4}
M(v, v_{x}, v_{t}, \cdots, u)=0.
\end{align}

\subsection{Physics-informed neural network methods based on Miura transformations}
\quad

In general, it is simpler and more convenient to solve \eqref{E2-2} by Miura transformation than by solving it directly if the solution $v$ of \eqref{E2-1} is given. That's why we study Miura transformations. Conversely, how to acquire the numerical solution $\widehat{v}$ if we solely know the initial-boundary dataset of $u$?

The main purpose of this article is to devise PINN methods based on Miura transformations to obtain the data-driven solution of \eqref{E2-1} by using initial and boundary conditions of the solution of \eqref{E2-2}. In other words, it is tailored to the inverse process of the Miura transformation, which is an implicit expression of $v$.

To this end, we put forth the following two types of schemes:

\textbf{(1) Scheme \uppercase\expandafter{\romannumeral 1}:}

In this scheme, $u$ and $v$ are different outputs of the same physics-informed neural network.

Considering that the depth of neural network depends on the number of weighted layers, we construct a neural network of depth $L$ consisting of one input layer, $L-1$ hidden layers and one output layer. The $l$th ($l=0,1,\cdots,L$) layer has $N_l$ neurons, which represents that it transmits $N_l$-dimensional output vector $\mathbf{x}^l$ to the ($l+1$)th layer as the input data. The connection between layers is achieved by the following affine transformation $\mathcal{A}$ and activation function $\sigma(\cdot)$:
\begin{align}
\mathbf{x}^l=\sigma(\mathcal{A}_l(\mathbf{x}^{l-1}))=\sigma(\mathbf{w}^{l} \mathbf{x}^{l-1}+\mathbf{b}^{l}),	
\end{align}
where $\mathbf{w}^{l}\in \mathbb{R}^{N_{l} \times N_{l-1}}$ and $\mathbf{b}^{l}\in \mathbb{R}^{N_{l}}$ denote the weight matrix and bias vector of the $l$th layer, respectively. Thus, the relation between input $\mathbf{x}^0$ and output $\mathbf{o}(\mathbf{x}^0,\boldsymbol{\Theta})$ is given by
\begin{align}
\mathbf{o}(\mathbf{x}^0,\boldsymbol{\Theta})&=\begin{pmatrix}
u(\mathbf{x}^0,\boldsymbol{\Theta})  \\
v(\mathbf{x}^0,\boldsymbol{\Theta})
\end{pmatrix}\\
&=(\mathcal{A}_L \circ \sigma \circ \mathcal{A}_{L-1} \circ \cdots \circ \sigma \circ \mathcal{A}_1)(\mathbf{x}^0),
\end{align}
and here $\boldsymbol{\Theta}=\left\{\mathbf{w}^{l}, \mathbf{b}^{l}\right\}_{l=1}^{L}$ represents the trainable parameters of PINN.

\begin{figure}[htbp]
\centering
\includegraphics[width=15cm,height=8cm]{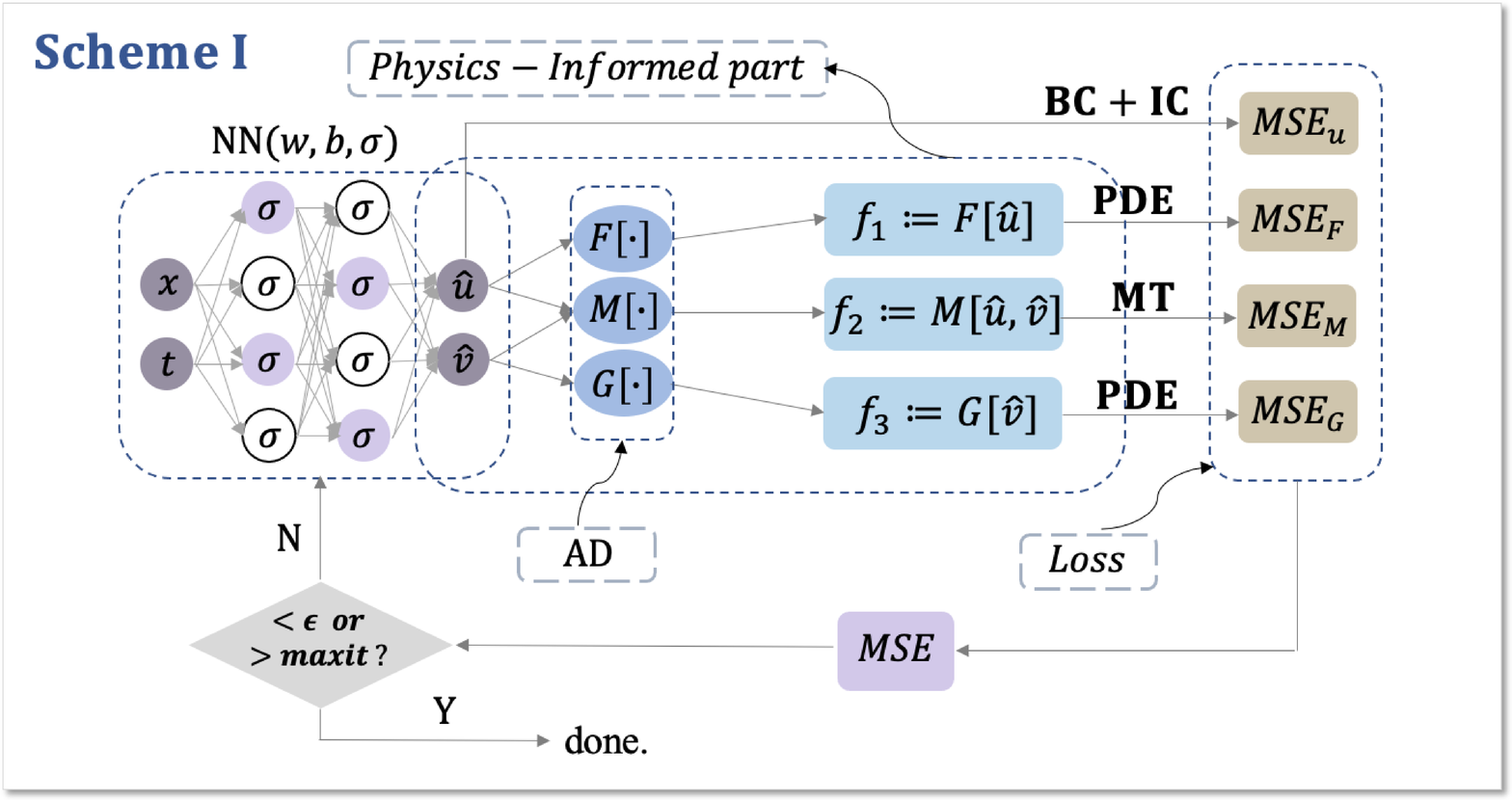}
\caption{(Color online) Schematic diagram of Scheme \uppercase\expandafter{\romannumeral 1}.}
\label{fig2-1}
\end{figure}

Firstly, spatial region $\left[x_{0}, x_{1}\right]$ and time region $\left[t_{0}, t_{1}\right]$ are divided into $N_x$ and $N_t$ discrete equidistance points, respectively. Assume we have the initial-boundary dataset $\{x^i_u,t^i_u,u^i\}^{N_u}_{i=1}$ randomly selected from the above grids ($\mathcal{I} \cup \mathcal{B}, \mathcal{I}=[x_0+j\frac{x_1-x_0}{N_x-1},t_0],(j=0,1,\cdots,N_x-1), \mathcal{B}=[x, t_0+k\frac{t_1-t_0}{N_t-1}],(x=x_0\ {\rm or}\ x_1, k=0,1,\cdots,N_t-1)$) and then obtain a set of collocation points of spatiotemporal region ($\left[x_{0}, x_{1}\right] \times \left[t_{0}, t_{1}\right]$), which is not required to locate at grids and is denoted by $\{x_{f}^i,t_{f}^i\}^{N_{f}}_{i=1}$. Then we construct the mean squared error function as the loss function to measure the difference between the predicted values and the true values of each iteration. The given information is investigated to merge into mean squared errors, including the initial and boundary data, the governing equations as well as the Miura transformation:

\begin{equation}
MSE=MSE_u+MSE_F+MSE_G+MSE_M,
\end{equation}
where
\begin{equation}
MSE_u=\frac{1}{N_u}\sum^{N_u}_{i=1}|\Widehat{u}(x_u^i,t_u^i)-u^i|^2,
\end{equation}

\begin{equation}
MSE_{F}=\frac{1}{N_f}\sum^{N_f}_{i=1}|F(x_{f}^i,t_{f}^i)|^2,
\end{equation}

\begin{equation}
MSE_{G}=\frac{1}{N_f}\sum^{N_f}_{i=1}|G(x_{f}^i,t_{f}^i)|^2,
\end{equation}

\begin{equation}
MSE_{M}=\frac{1}{N_f}\sum^{N_f}_{i=1}|M(x_{f}^i,t_{f}^i)|^2.	
\end{equation}
Here, $G$, $F$ and $M$ are defined in \eqref{E2-1}, \eqref{E2-2} and \eqref{E2-4} separately. Based on MSE criteria, the parameters of neural networks are optimized to approach the initial and boundary training data and satisfy the structure imposed by \eqref{E2-1}-\eqref{E2-3}. Then the numerical solutions $\Widehat{u}(x,t)$ and $\Widehat{v}(x,t)$ can be obtained and their derivatives with respect to time $t$ and space $x$ are derived by automatic differentiation (AD) \cite{AutomaticDifferentiation} to compute $\{F(x_{f}^i,t_{f}^i), G(x_{f}^i,t_{f}^i), M(x_{f}^i,t_{f}^i)\}^{N_{f}}_{i=1}$. Fig. \ref{fig2-1} displays a sketch of Scheme \uppercase\expandafter{\romannumeral 1}, where $maxit$ denotes the maximum number of iterations.

\textbf{(2) Scheme \uppercase\expandafter{\romannumeral 2}:}

 Here, this scheme has two steps.

 \textbf{Step One:}

 Step one is just to solve the initial-boundary value problem of \eqref{E2-2}. Firstly, we construct the following mean squared error function as the loss function to train a neural network by using the initial and boundary data as well as the governing equation:
\begin{equation}
MSE_1=MSE_u+MSE_F,
\end{equation}
where
\begin{equation}
MSE_u=\frac{1}{N_u}\sum^{N_u}_{i=1}|\Widehat{u}(x_u^i,t_u^i)-u^i|^2,
\end{equation}

\begin{equation}
MSE_{F}=\frac{1}{N_f}\sum^{N_f}_{i=1}|F(x_{f}^i,t_{f}^i)|^2,
\end{equation}
and $u$ is the sole output of this physics-informed neural network. Meanwhile, the acquisition method of initial-boundary dataset $\{x^i_u,t^i_u,u^i\}^{N_u}_{i=1}$ and collocation points $\{x_{f}^i,t_{f}^i\}^{N_{f}}_{i=1}$ is same as Scheme \uppercase\expandafter{\romannumeral 1}. Under the principle of minimizing the mean squared error loss, we can acquire the numerical solution $\Widehat{u}(x,t)$ of the given domain and period after parameter optimization.

 \textbf{Step Two:}

Based on the numerical solution $\Widehat{u}(x,t)$ obtained in the first step, a new physics-informed neural network is constructed to acquire the data-driven solution $\Widehat{v}(x,t)$.

We divide spatial region $[x_0,x_1]$ and time region $[t_0,t_1]$ into $N_x$ and  $N_t$ discrete equidistance points, respectively. Then the solution $\Widehat{u}$ is discretized into $N_x \times N_t$ data points. We randomly select a set of collocation points of spatiotemporal region ($\left[x_{0}, x_{1}\right] \times \left[t_{0}, t_{1}\right]$) from $N_x \times N_t$ grid points, denoted by $\{x_{g}^i,t_{g}^i\}^{N_{g}}_{i=1}$ as well as the corresponding numerical solution $\{\Widehat{u}(x_g^i,t_g^i)\}^{N_{g}}_{i=1}$ and even derivatives with respect to time $t$ and space $x$.

\begin{figure}[htbp]
\centering
\includegraphics[width=15cm,height=15cm]{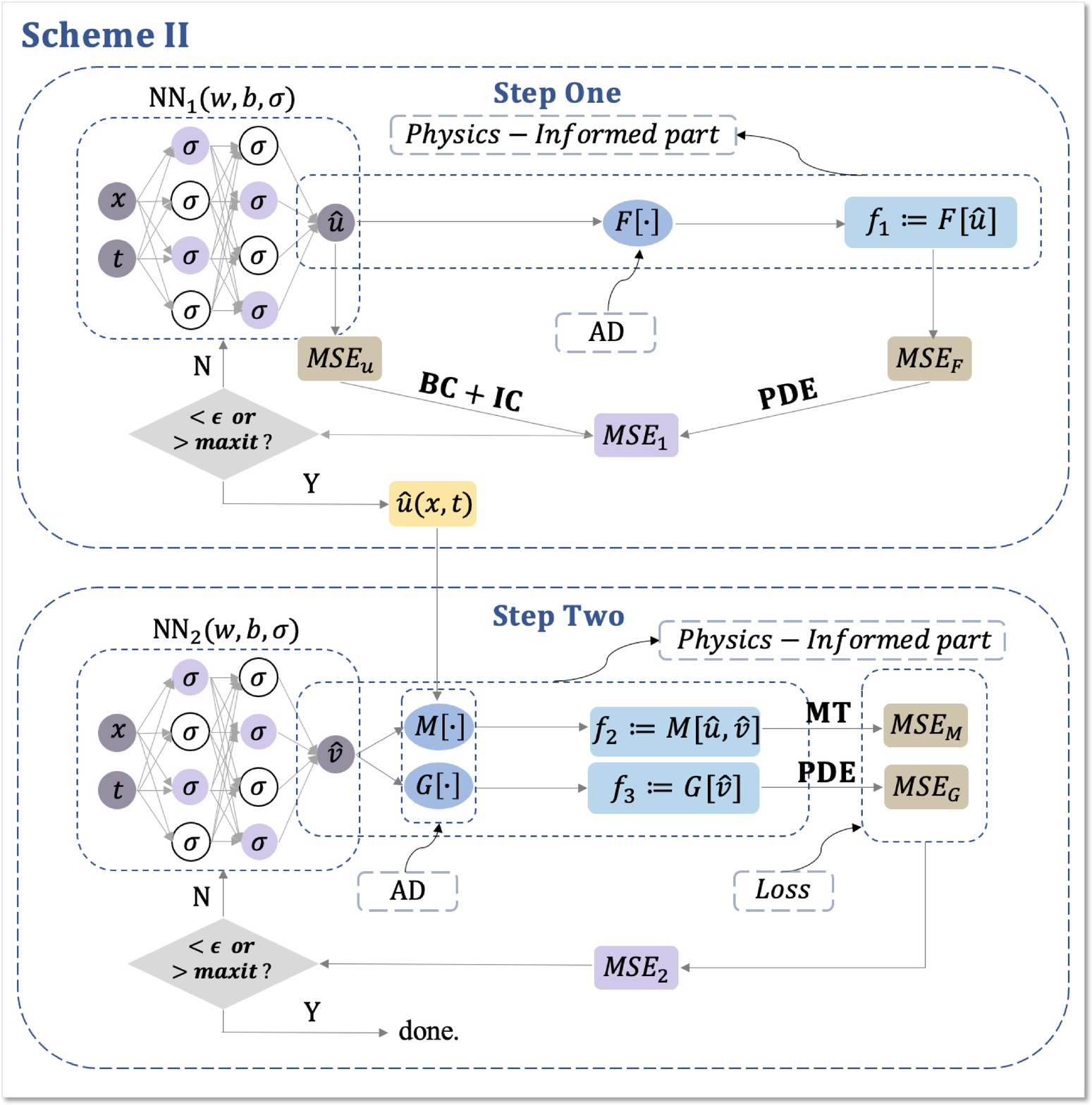}
\caption{(Color online) Schematic diagram of Scheme \uppercase\expandafter{\romannumeral 2}.}
\label{fig2-2}
\end{figure}

Then the mean squared error loss in step two is given by:
\begin{equation}
MSE_2=MSE_G+MSE_M,
\end{equation}
where
\begin{equation}
MSE_{G}=\frac{1}{N_g}\sum^{N_g}_{i=1}|G(x_{g}^i,t_{g}^i)|^2,
\end{equation}

\begin{equation}
MSE_{M}=\frac{1}{N_g}\sum^{N_g}_{i=1}|M(x_g^i,t_g^i)|^2.	
\end{equation}
Here, the information of the governing equation \eqref{E2-1} is merged into $MSE_G$. Meanwhile, $MSE_M$ reflects the constraint of the Miura transformation and the computation of it requires the  numerical solution $\{\Widehat{u}(x_g^i,t_g^i)\}^{N_{g}}_{i=1}$, which implies that gaining data-driven solution $\Widehat{v}(x,t)$ is based on step one. Then the process of Scheme \uppercase\expandafter{\romannumeral 2} is shown schematically in Fig. \ref{fig2-2}.

Since the training data only accounts for a small proportion of total data on grids, the relative $\mathbb{L}_2$ errors of $N_x \times N_t$ data points on grids can be calculated to evaluate the generalization ability of the PINN method based on Miura transformations:
\begin{align}
RE_u=\frac{\sqrt{\sum^{N_x-1}_{j=0} \sum^{N_t-1}_{k=0} |\Widehat{u}(x_0+j\frac{x_1-x_0}{N_x-1},t_0+k\frac{t_1-t_0}{N_t-1})-u^{j, k}|^2}}{\sqrt{\sum^{N_x-1}_{j=0} \sum^{N_t-1}_{k=0} |u^{j, k}|^2}},\\
RE_v=\frac{\sqrt{\sum^{N_x-1}_{j=0} \sum^{N_t-1}_{k=0} |\Widehat{v}(x_0+j\frac{x_1-x_0}{N_x-1},t_0+k\frac{t_1-t_0}{N_t-1})-v^{j, k}|^2}}{\sqrt{\sum^{N_x-1}_{j=0} \sum^{N_t-1}_{k=0} |v^{j, k}|^2}},
\end{align}
where $\Widehat{u}(x_0+j\frac{x_1-x_0}{N_x-1},t_0+k\frac{t_1-t_0}{N_t-1})$ ($\Widehat{v}(x_0+j\frac{x_1-x_0}{N_x-1},t_0+k\frac{t_1-t_0}{N_t-1})$) and $u^{j, k}$ ($v^{j, k}$) represent the predictive value and true value, separately.

\begin{figure}[htbp]
\centering
\includegraphics[width=18cm,height=10cm]{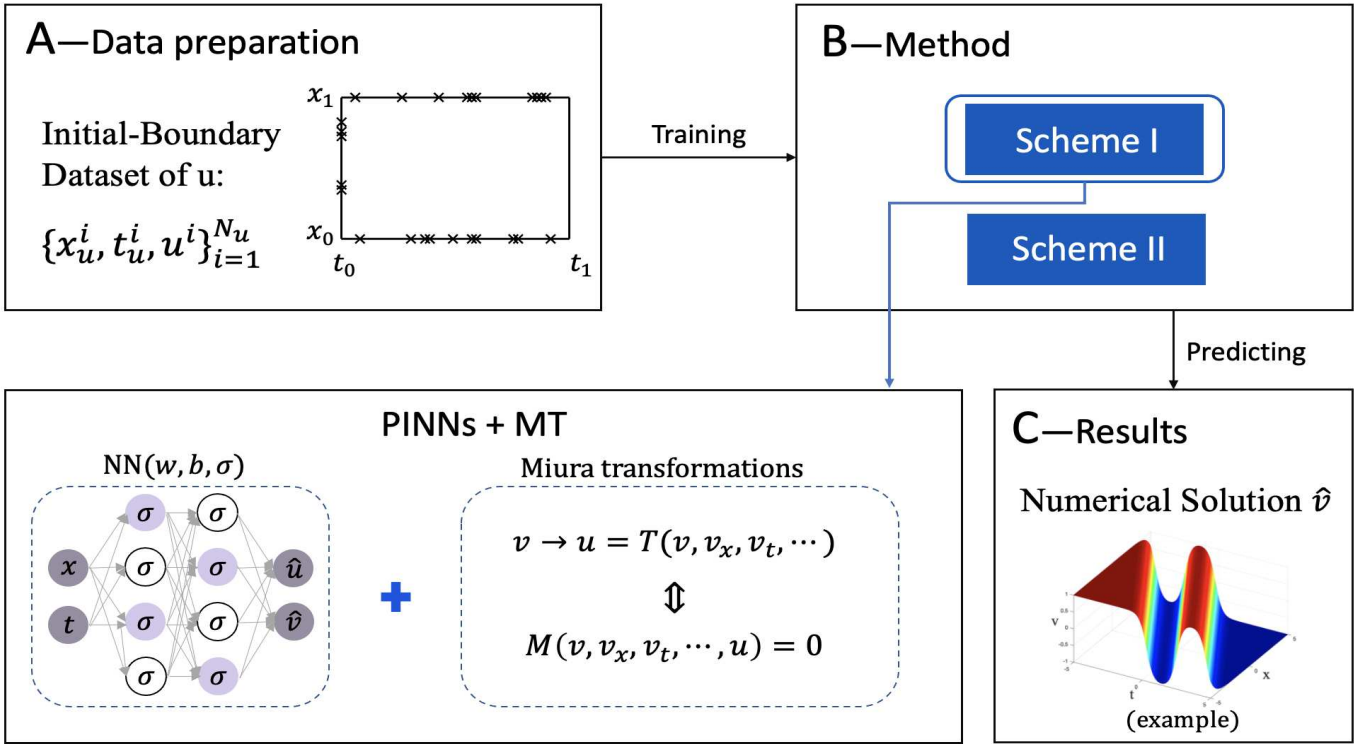}
\caption{(Color online) Illustrations of problem-solving process.}
\label{fig2-3}
\end{figure}

A remarkable advantage of this method is that we can simply use small amounts of initial-boundary data of a solution of a certain nonlinear equation to obtain the data-driven solution of another evolution equation with the aid of PINNs and Miura transformations, which act as a bridge connecting the solutions of two integrable equations during the process. In addition, it is challenging to solve $v$ based on the Miura transformation \eqref{E2-3} since it is an implicit expression of $v$, but our method can overcome this difficulty owing to the superiority of PINNs in solving partial differential equations. Then the process of problem-solving can be roughly summarized as three parts and is shown in Fig. \ref{fig2-3}.

All codes are based upon Python 3.7 and Tensorflow 1.15, and the presented numerical experiments are run on a MacBook Pro computer with 2.3 GHz Intel Core i5 processor and 16-GB memory.

\section{Applications in the real Miura transformation between the defocusing mKdV equation and the KdV equation}\label{real Miura}

In this section, Scheme \uppercase\expandafter{\romannumeral 1} is adopted to carry out numerical experiments with respect to the real Miura transformation between the defocusing mKdV equation and the KdV equation.

\subsection{Case 1}\label{3.1}
\quad

Miura \cite{Miura1968} presented the real Miura transformation in 1968
\begin{equation}\label{E3-1}
u=v_x-v^2,	
\end{equation}
to transform the solution of the defocusing mKdV equation
\begin{equation}\label{E3-2}
v_t-6v^2 v_x+v_{xxx}=0,	
\end{equation}
into that of the KdV equation
\begin{equation}\label{E3-3}
u_t+6u u_x+u_{xxx}=0.	
\end{equation}

Evidently, based on a shock wave solution (kink solution) of the defocusing mKdV equation,
\begin{equation}\label{E3-4}
v=k {\rm{tanh}} (k x+2k^3 t)	,
\end{equation}
the soliton solution of the KdV equation
\begin{equation}\label{E3-5}
u=2 k^2 {\rm{sech}}^2(k x+2k^3 t)-k^2,	
\end{equation}
can be obtained by real Miura transformation \eqref{E3-1}.

Then we consider the KdV equation with the first kind of boundary condition (Dirichlet boundary condition)
\begin{equation}\label{E3-6}
\begin{split}
\begin{cases}
u_t+6u u_x+u_{xxx}=0,x\in[x_0,x_1],t\in[t_0,t_1],\\
u(x,t_0)=u_0(x),\\
u(x_0,t)=a_1(t),\\
u(x_1,t)=a_2(t).
\end{cases}
\end{split}
\end{equation}
After choosing $k=1$ and $[x_0,x_1]\times[t_0,t_1]=[-5,5]\times[-5,5]$ as the training region, the corresponding initial-boundary conditions are given by
\begin{align}
&u_0(x)=2{\rm{sech}}(-10+x)^2-1,\\
&u(-5,t)=2{\rm{sech}}(2t-5)^2-1,\\
&u(5,t)=2{\rm{sech}}(2t+5)^2-1.	
\end{align}
To obtain the training dataset, we divide the spatial region $[x_0,x_1]=[-5,5]$ and time region $[t_0,t_1]=[-5,5]$ into $N_x=513$ and $N_t=201$ discrete equidistance points separately. Thus, the solution $u$ are  discretized into $513 \times 201$ data points in the given spatiotemporal domain. We randomly select $N_u=200$ points ($\{x^i_u,t^i_u,u^i\}^{N_u}_{i=1}$) from the initial-boundary dataset and proceed by sampling $N_f=5000$ collocation points ($\{x_{f}^i,t_{f}^i\}^{N_{f}}_{i=1}$) via the Latin hypercube sampling method \cite{Stein1987}. Wherein, $N_f=5000$ collocation points can be located at any points and don't need to be grid points. In this case, we present the concrete forms of governing equations including the Miura transformation as well as the following partial differential equations
\begin{align}
f_1:&=F=u_t+6u u_x+u_{xxx},\\
f_2:&=M=u-(v_x-v^2),\\
f_3:&=G=v_t-6v^2 v_x+v_{xxx}.
\end{align}
A 5-layer feedforward neural network with 40 neurons per hidden layer is constructed to learn the soliton solution of the KdV equation and more importantly the data-driven solution of the defocusing mKdV equation. In addition, we use the hyperbolic tangent ($tanh$) activation function and initialize weights of the neural network with the Xavier initialization \cite{Glorot2010}. The derivatives of the network $u$ and $v$ with respect to time $t$ and space $x$ are derived by automatic differentiation \cite{AutomaticDifferentiation}.

\begin{figure}[htbp]
\centering
\includegraphics[width=7cm,height=5cm]{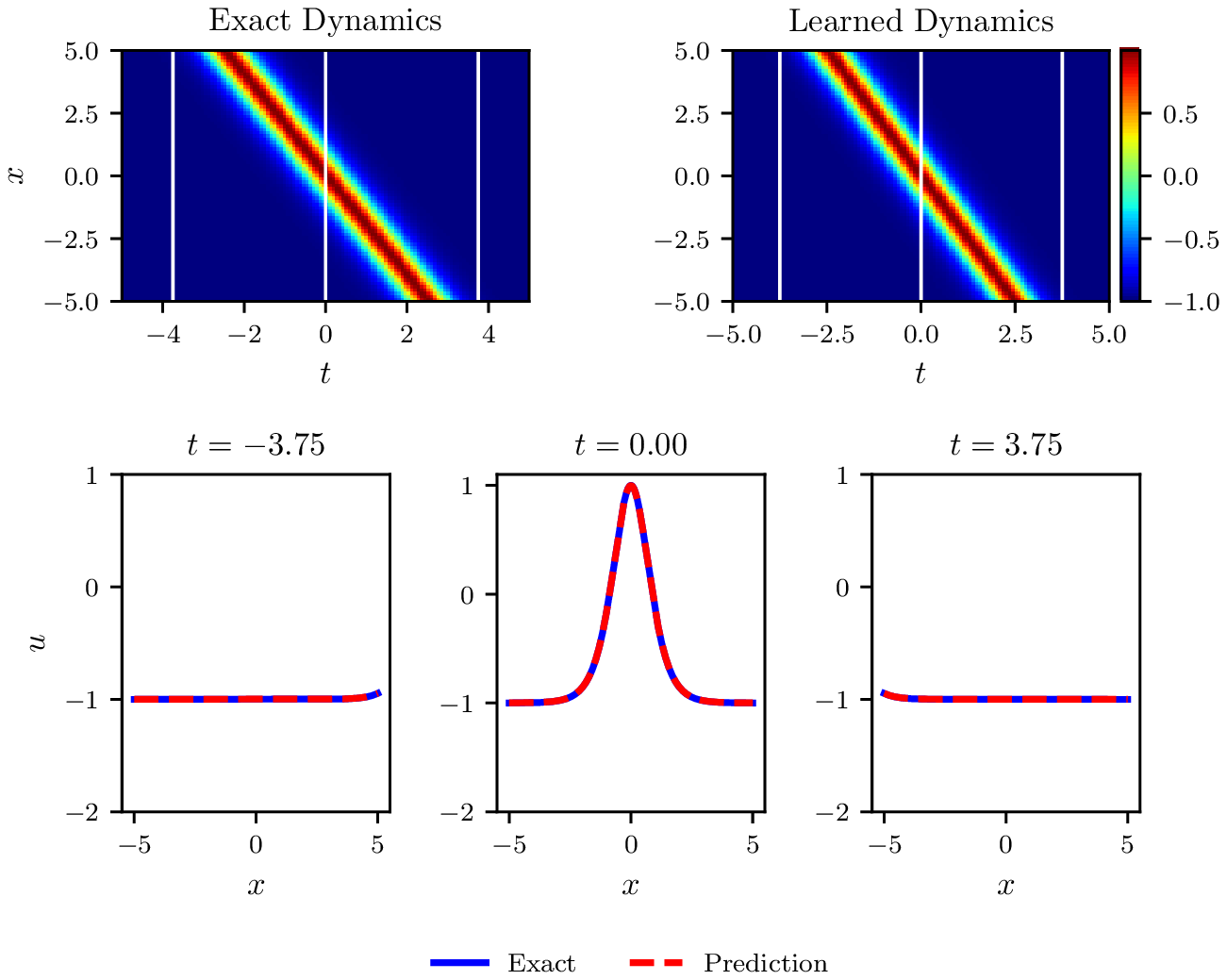}
$a$
\includegraphics[width=7cm,height=5cm]{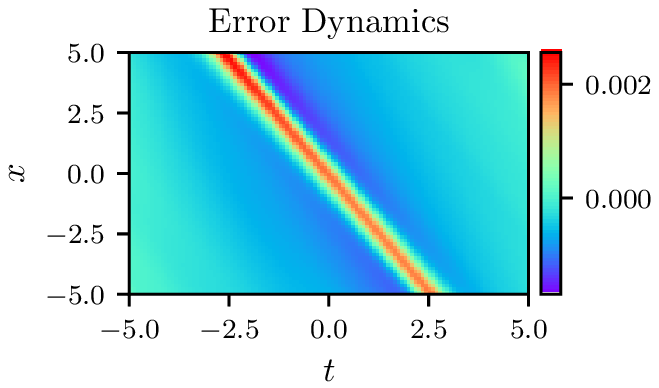}
$b$
\caption{(Color online) One-soliton solution $u(x,t)$ of the KdV equation by Scheme \uppercase\expandafter{\romannumeral 1}: (a) The density diagrams and comparison between the predicted solutions and exact solutions at the three temporal snapshots of $u(x,t)$; (b) The error density diagram of $u(x,t)$.}
\label{fig3-1}
\end{figure}

We adopt the L-BFGS algorithm \cite{L-BFGS} as the optimization algorithm. Our PINN method based on Miura transformations finally succeeds in numerical simulations of the soliton solution \eqref{E3-5} of the KdV equation. After 5267 times iterations in about 503.2289 seconds, the relative $\mathbb{L}_2$ error of $u$ is 7.385367e-04. Fig. \ref{fig3-1} displays the density diagrams of the one-soliton solution, comparison between the predicted solutions and exact solutions as well as the error density diagram of the KdV equation. Besides, the comparisons between exact solutions and predicted solutions at different time points $t=-3.75, 0, 3.75$ are shown in the bottom panel of Fig. \ref{fig3-1} (a). It can be clearly seen that the soliton solution is well simulated.

\begin{figure}[htbp]
\centering
\includegraphics[width=7cm,height=5cm]{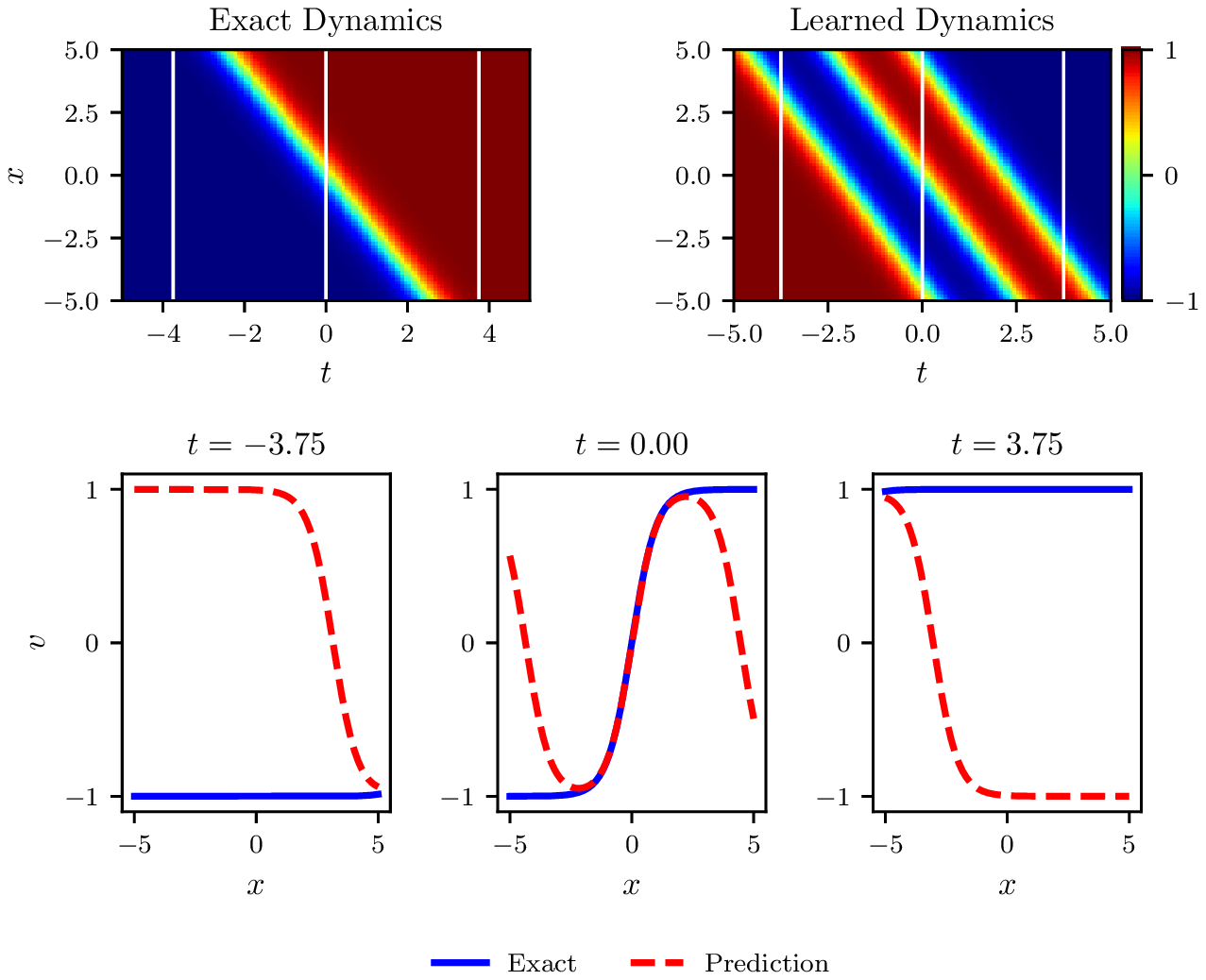}
$a$
\includegraphics[width=7cm,height=5cm]{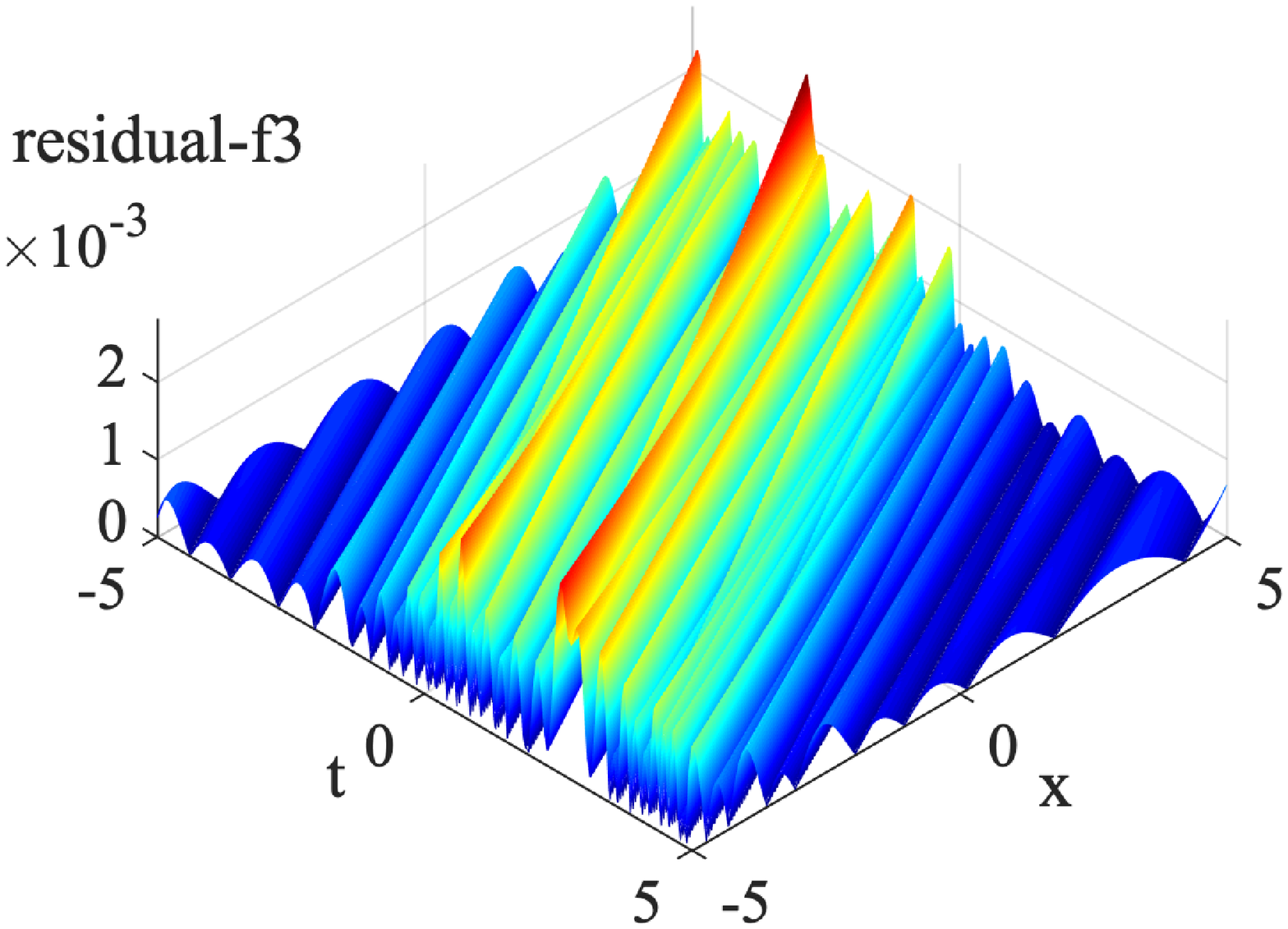}
$b$
\caption{(Color online) Data-driven solution $v(x,t)$ of the defocusing mKdV equation by Scheme \uppercase\expandafter{\romannumeral 1}: (a) The density diagrams and comparison between the predicted solutions and exact solutions at the three temporal snapshots of $v(x,t)$; (b) The three-dimensional plot of the residual corresponding to $f_3$.}
\label{fig3-2}
\end{figure}

Significantly, we obtain a new data-driven solution different from the shock wave solution (kink solution) \eqref{E3-4} of the defocusing mKdV equation. Fig. \ref{fig3-2} (a) shows the density diagrams and comparison between the predicted solutions and exact solutions at the three temporal snapshots of $v(x,t)$ and it is patently obvious that they are completely different. Therefore, we intend to explore if the data-driven one is the numerical solution of the defocusing mKdV equation from the following two aspects:

$\bullet$ \bm{$MSE_G$}

The mean squared error ($MSE_G$)
\begin{equation}
MSE_{G}=\frac{1}{N}\sum^{N}_{i=1}|G(x^i,t^i)|^2, N=N_x \times N_t,
\end{equation}
corresponding to $f_3$
\begin{align}
f_3:=G=v_t-6v^2 v_x+v_{xxx},	
\end{align}
of $N_x \times N_t=513 \times 201$ grid points is calculated to determine whether the numerical solution satisfies the defocusing mKdV equation and it can be used to evaluate the generalization ability of our method since the size of training data is only a small percentage of total data on grids. Finally, the calculated result of $MSE_G$ is 7.247345e-07 and it is small enough.

$\bullet$ \textbf{3D plot of the residuals corresponding to \bm{$f_3$}}

The PDE residuals corresponding to $f_3$, i.e. $|f_3(x^i,t^i)|$ on $N_x \times N_t=513 \times 201$ grid points can also be predicted and its three-dimensional plot is displayed in Fig. \ref{fig3-2} (b). Obviously, the residual is kept small in the order of magnitude of $10^{-3}$ in the given region.

\begin{figure}[htbp]
\centering
\includegraphics[width=7cm,height=5.5cm]{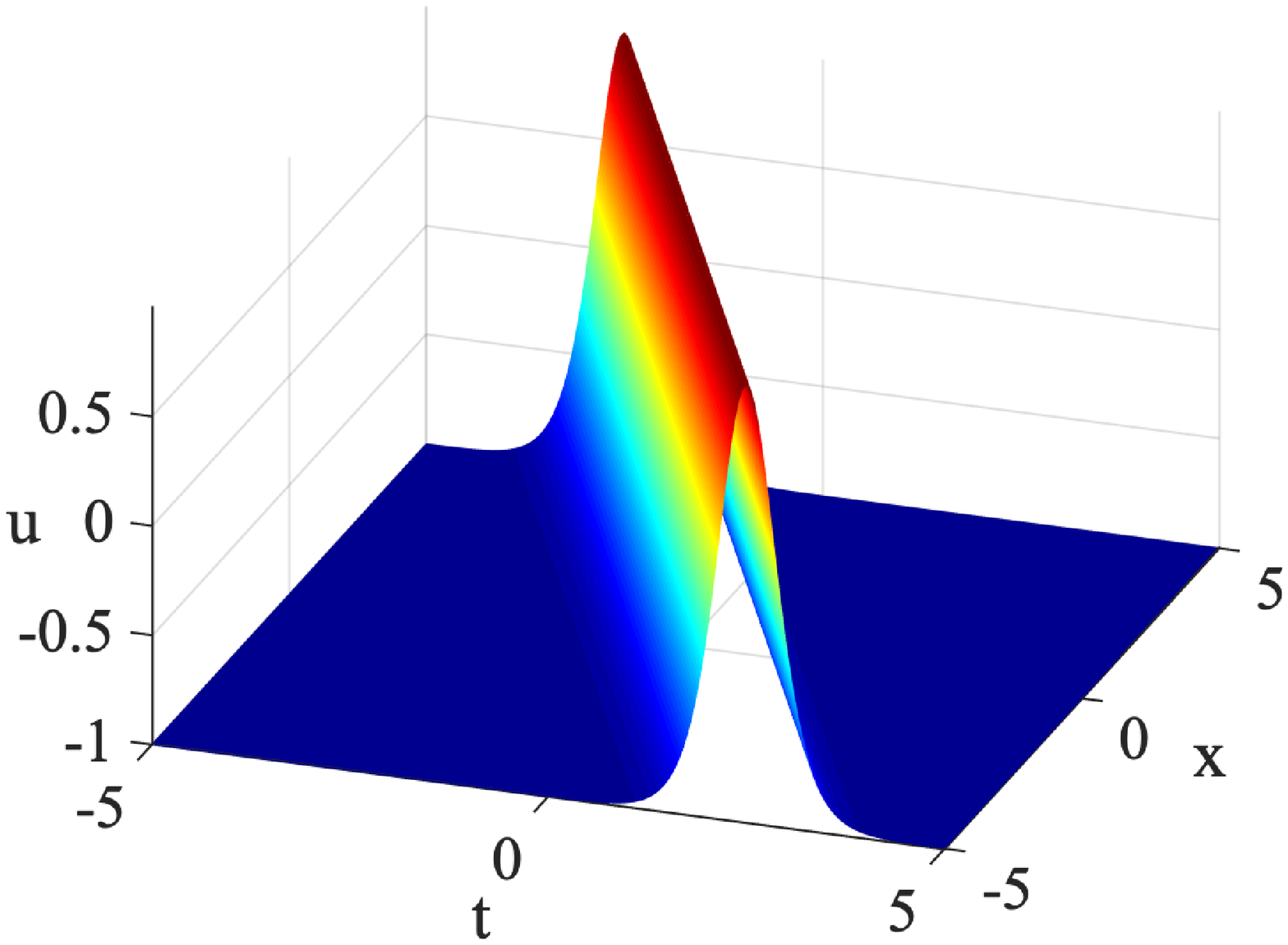}
$a$
\includegraphics[width=7cm,height=5.5cm]{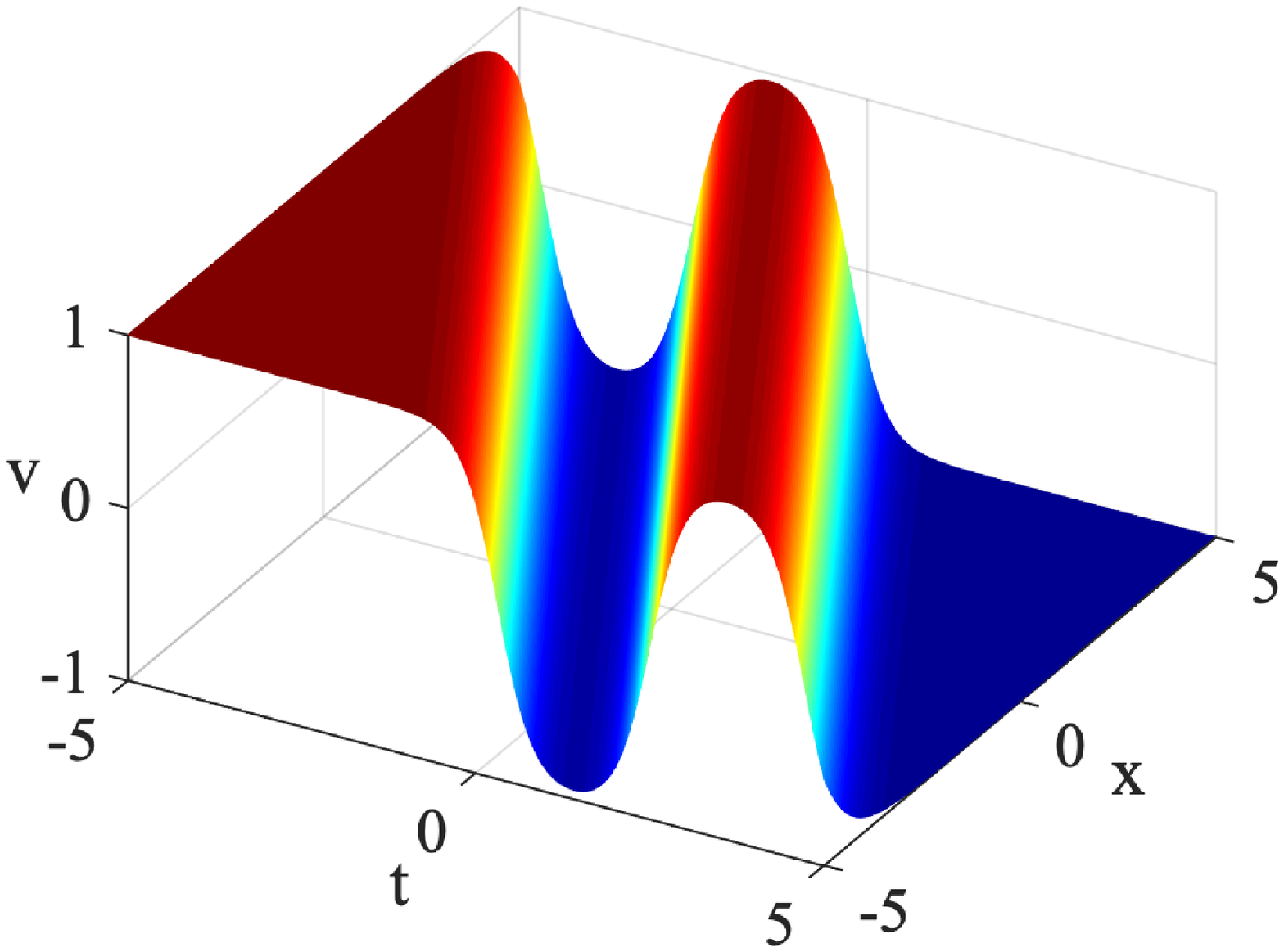}
$b$
\caption{(Color online) Data-driven solutions $u(x,t)$ of the KdV equation and $v(x,t)$ of the defocusing mKdV equation by Scheme \uppercase\expandafter{\romannumeral 1}: (a) The three-dimensional plot of $u(x,t)$; (b) The three-dimensional plot of $v(x,t)$.}
\label{fig3-3}
\end{figure}

Above results fully verify that this data-driven solution is a new numerical solution of the defocusing mKdV equation. Fig. \ref{fig3-3} presents the three-dimensional plots of data-driven solutions.

\begin{figure}[htbp]
\centering
\includegraphics[width=7cm,height=5.5cm]{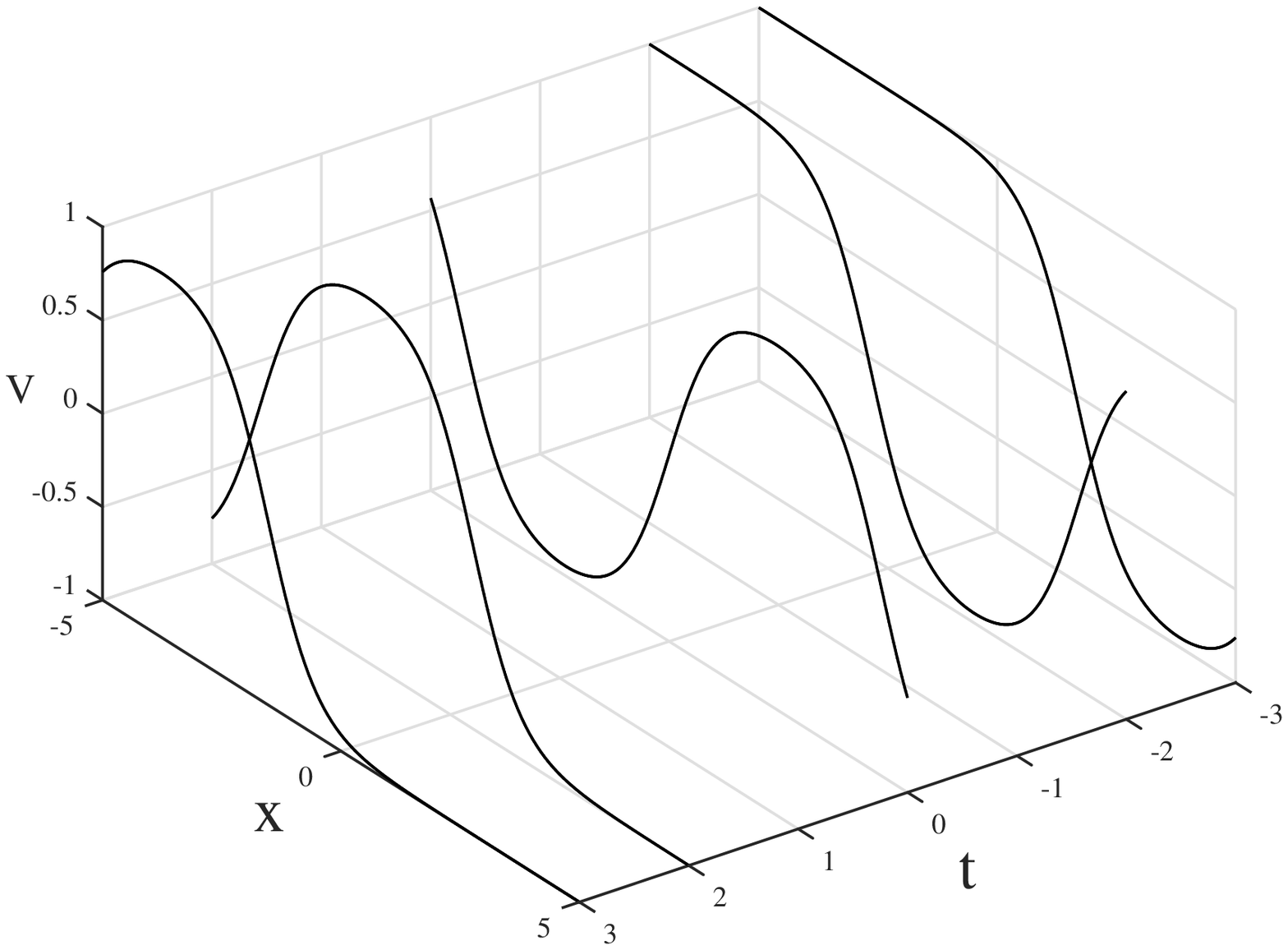}
$a$
\includegraphics[width=7cm,height=5.5cm]{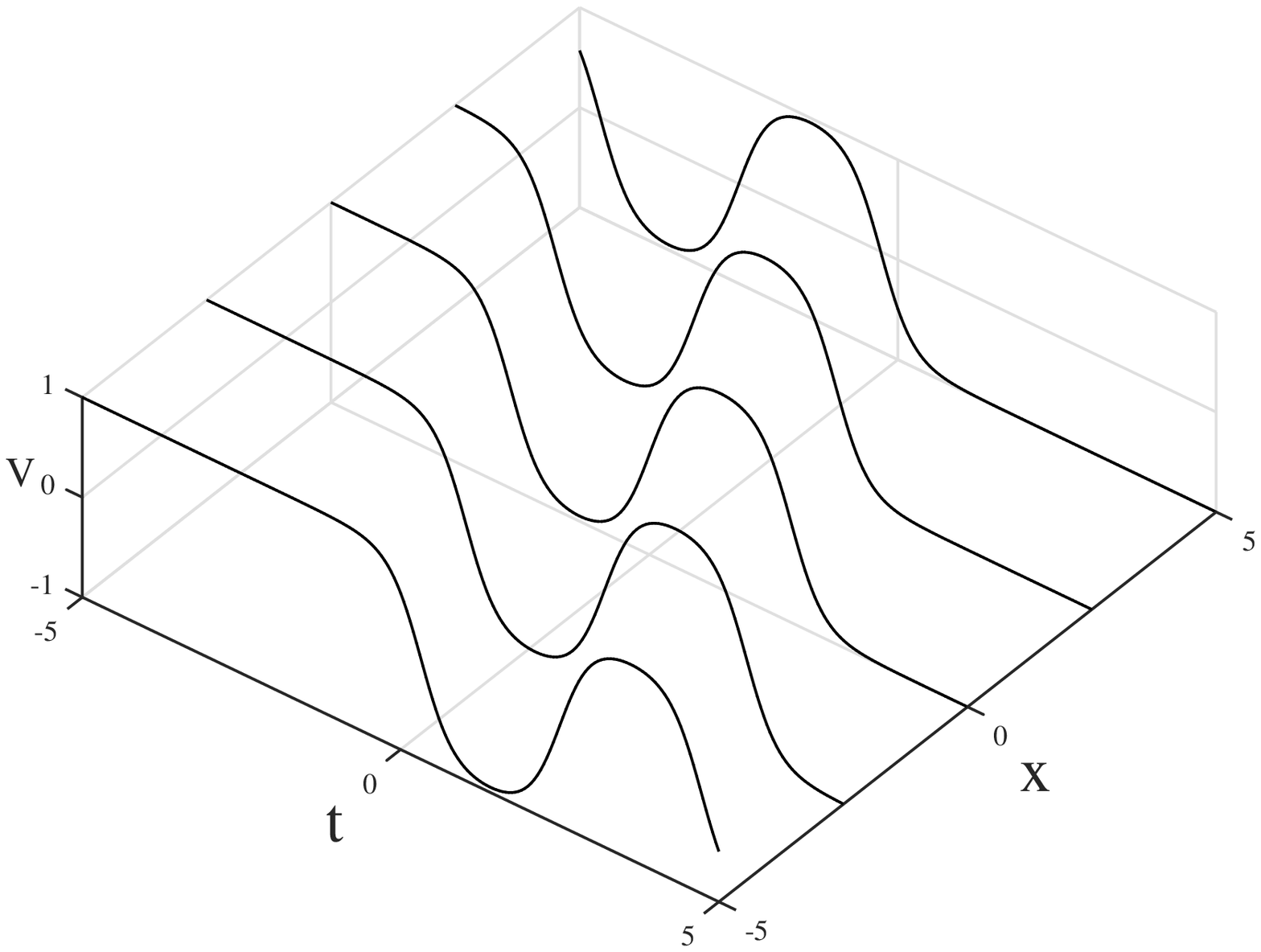}
$b$
\caption{(Color online) Evolution plots of data-driven solution $v(x,t)$: (a) Evolution plots at different time points $t=-3, -2, 0, 2, 3$; (b) Evolution plots at different space points $x=-5, -2.5, 0, 2.5, 5$.}
\label{figxt}
\end{figure}

To our knowledge, the new numerical solution of the defocusing mKdV equation has not been previously observed and reported and we name it 'kink-bell type solution' since it appears that there is a bell curve on the kink curve. From the three-dimensional plot of data-driven $v(x,t)$ in Fig. \ref{fig3-3} (b) and the corresponding evolution plots in Fig. \ref{figxt}, it can be seen that the waveform is almost the same at different space points.

\subsection{Case 2}
\quad

For the mKdV equation in the form
\begin{equation}
v_t+\beta v^2 v_x-\gamma v_{xxx}=0,
\end{equation}
A. Ankiewicz et al.\cite{Ankiewicz2018} studied its first-order rational solution
\begin{equation}\label{rational}
v=\frac{12 \gamma}{3 \gamma-2 \beta (x-\beta t)^2}	-1.
\end{equation}
After setting $\beta=-6, \gamma=-1$, the first-order rational solution of the defocusing mKdV equation can be got.

Similarly, the real Miura transformation \eqref{E3-1} can transform the first-order rational solution of the defocusing mKdV equation
\begin{equation}
v=-\frac{12}{-3+12(x+6t)^2}-1,	
\end{equation}
into the solution of the KdV equation
\begin{equation}
u=\frac{12(24 x+144 t)}{\left(-3+12(x+6 t)^{2}\right)^{2}}-\left(-\frac{12}{-3+12(x+6 t)^{2}}-1\right)^{2}.
\end{equation}

Here we take $[x_0,x_1]=[-10,-5], [t_0,t_1]=[2,5]$, and then the initial-boundary conditions of the KdV equation are obtained
\begin{align}
&u_0(x)=\frac{12(24x+288)}{(-3+12(x+12)^2)^2}-\left(-\frac{12}{-3+12(x+12)^2}-1\right)^2,\\
&u(-10,t)=\frac{12(-240+144t)}{(-3+12(-10+6t)^2)^2}-\left(-\frac{12}{-3+12(-10+6t)^2}-1\right)^2,\\
&u(-5,t)=\frac{12(-120+144t)}{(-3+12(-5+6t)^2)^2}-\left(-\frac{12}{-3+12(-5+6t)^2}-1\right)^2.	
\end{align}
Spatial region $[x_0,x_1]=[-10,-5]$ and time region $[t_0,t_1]=[2,5]$ are divided into $N_x=513$ and $N_t=201$ discrete equidistance points with the aid of MATLAB, respectively. Likewise, the initial-boundary dataset is generated by randomly selecting $N_u=200$ points ($\{x^i_u,t^i_u,u^i\}^{N_u}_{i=1}$) from the original dataset and $N_f=5000$ collocation points ($\{x_{f}^i,t_{f}^i\}^{N_{f}}_{i=1}$) are selected by adopting the same sampling method in Section \ref{3.1}.

Then we aim to obtain the data-driven solutions of two equations above through the use of initial-boundary data of the KdV equation. The loss functions and the governing equations are the same as above as well. Based on the L-BFGS algorithm, a 5-layer feedforward neural network with 40 neurons per hidden layer is established to simulate the solutions of the KdV and mKdV equations after utilizing the same activation function, the method of initializing weights as well as the automatic differentiation technique.

\begin{figure}[htbp]
\centering
\includegraphics[width=7cm,height=5cm]{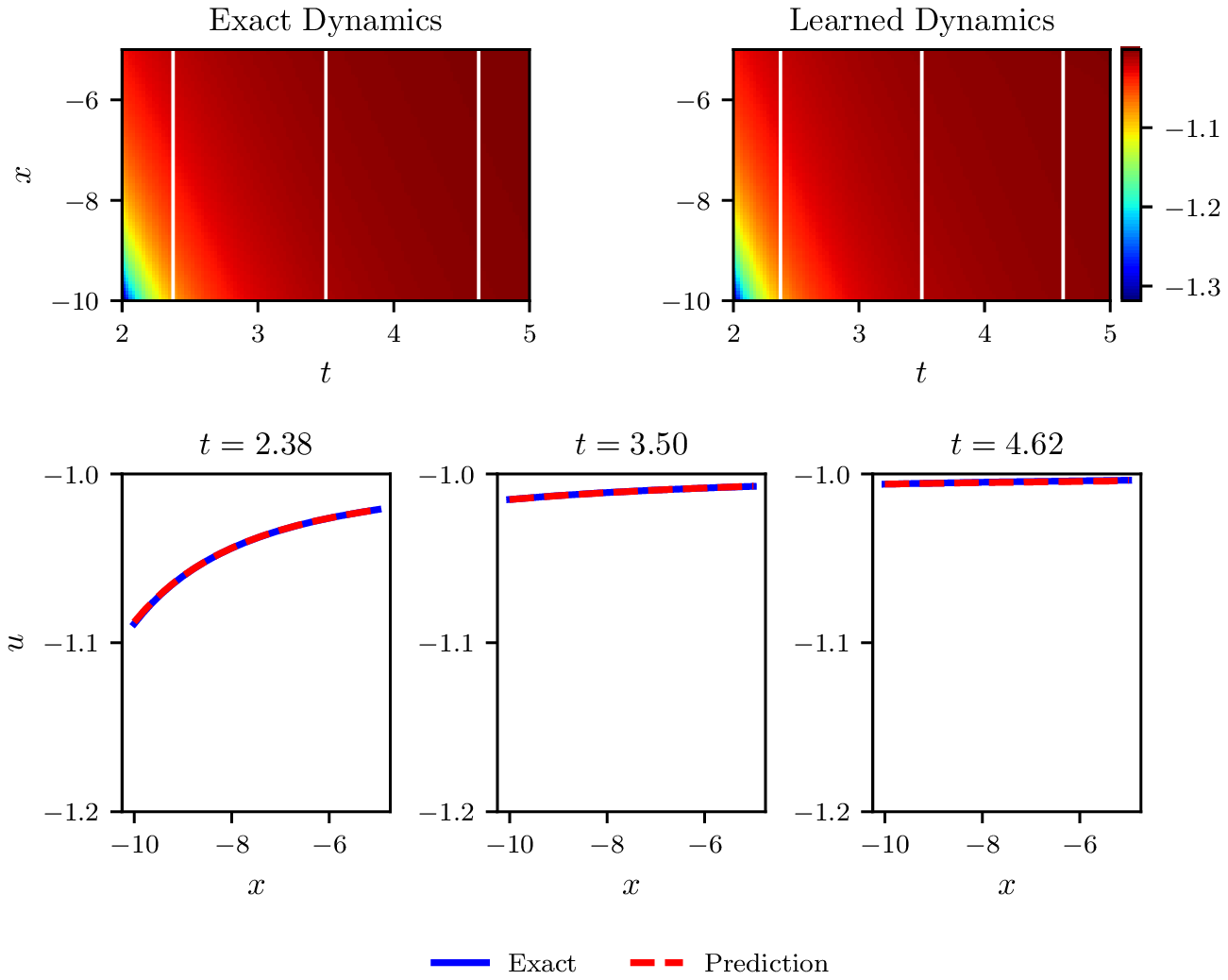}
$a$
\includegraphics[width=7cm,height=5cm]{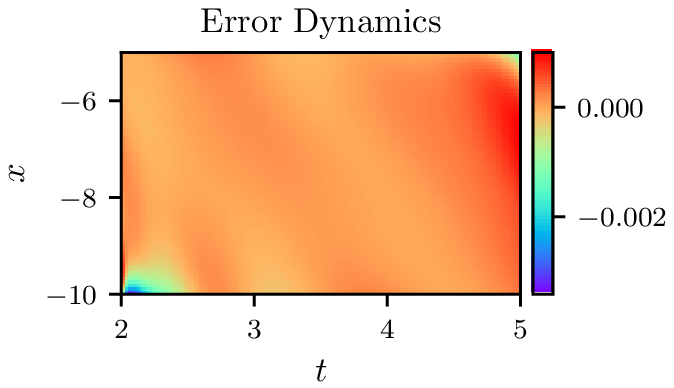}
$b$
\caption{(Color online) Data-driven solution $u(x,t)$ of the KdV equation by Scheme \uppercase\expandafter{\romannumeral 1}: (a) The density diagrams and comparison between the predicted solutions and exact solutions at the three temporal snapshots of $u(x,t)$; (b) The error density diagram of $u(x,t)$.}
\label{fig3-4}
\end{figure}

\begin{figure}[htbp]
\centering
\includegraphics[width=7cm,height=5cm]{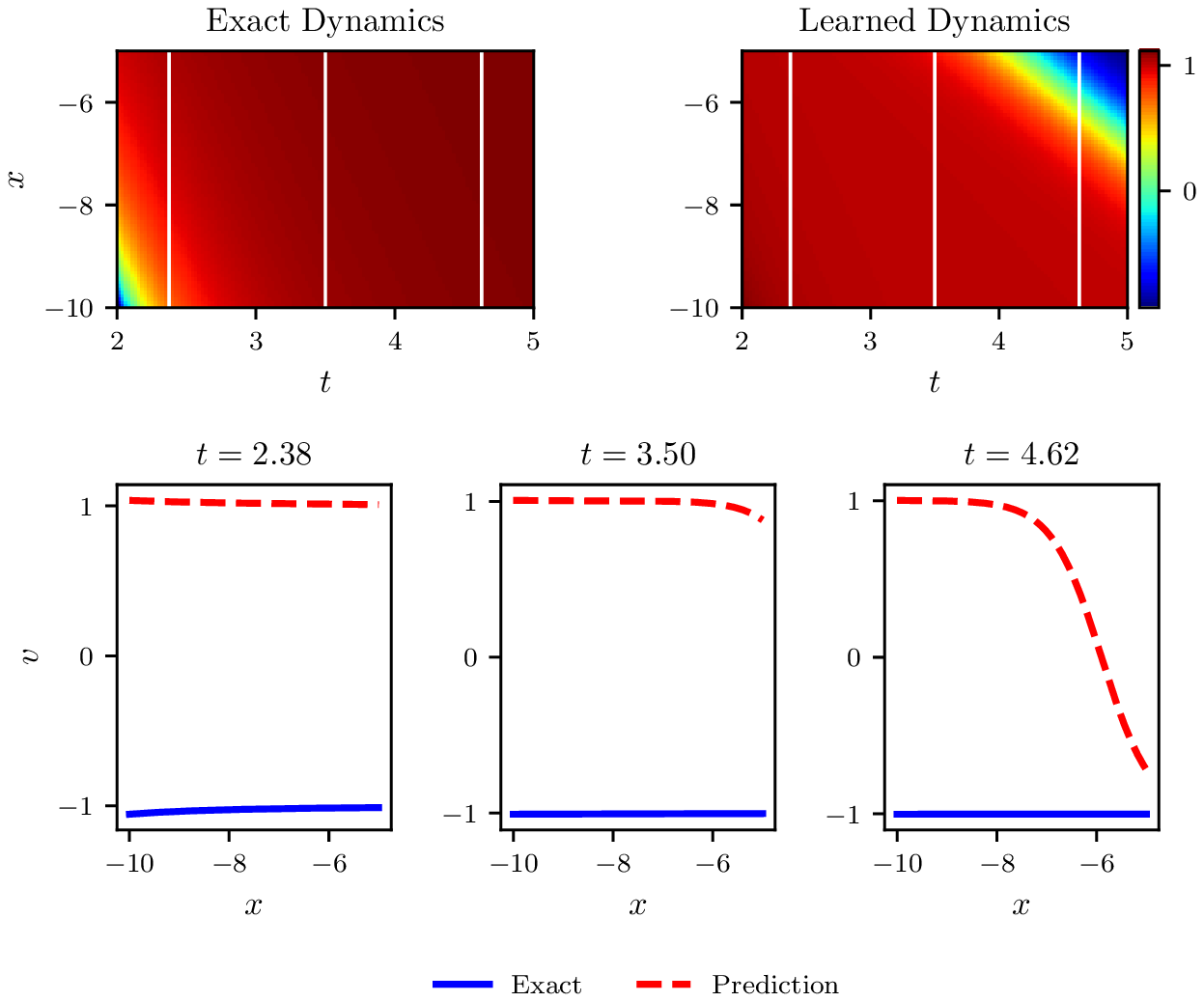}
$a$
\includegraphics[width=7cm,height=5cm]{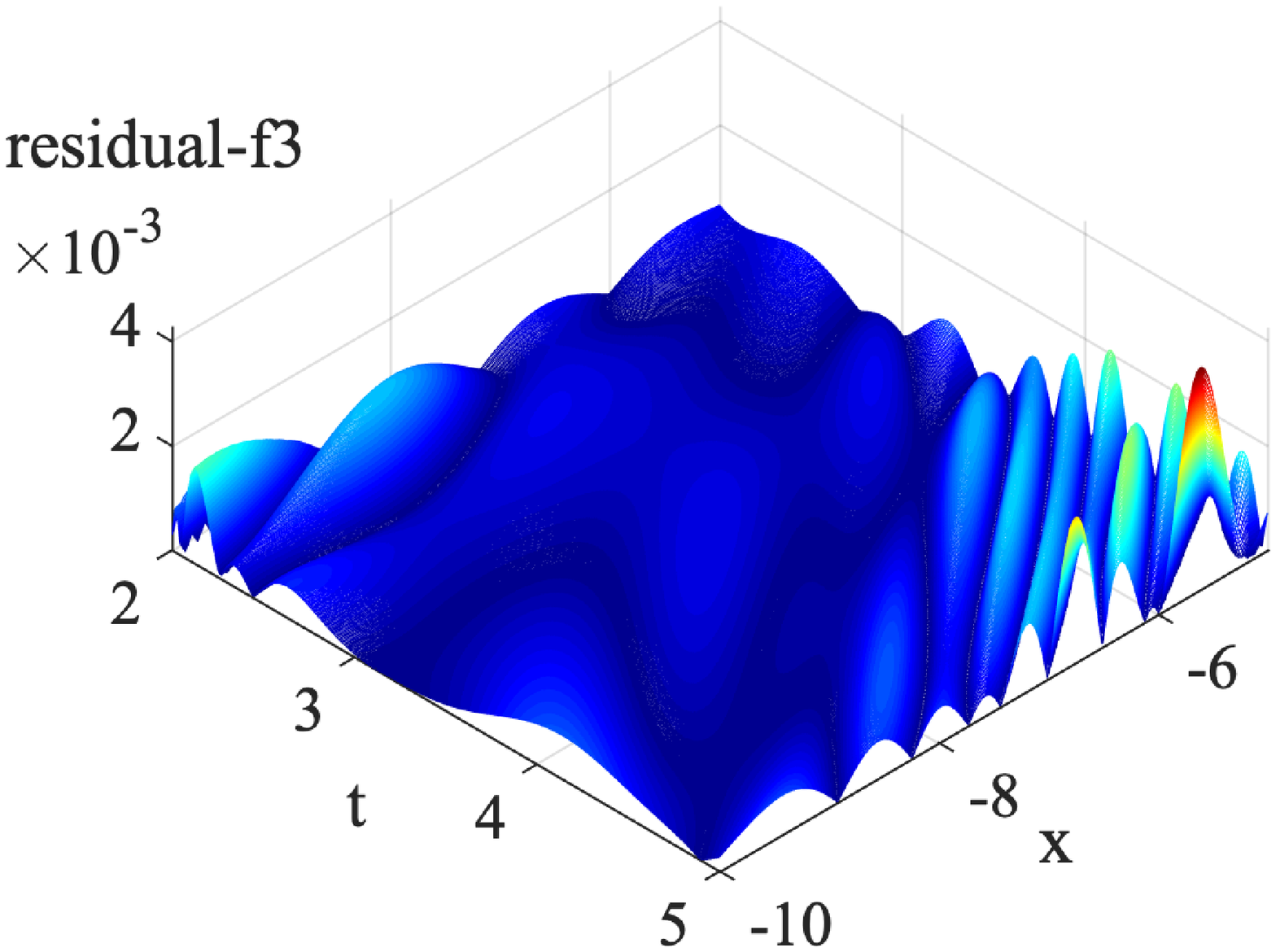}
$b$
\caption{(Color online) First-order rational solution $v(x,t)$ of the defocusing mKdV equation by Scheme \uppercase\expandafter{\romannumeral 1}: (a) The density diagrams and comparison between the predicted solutions and exact solutions at the three temporal snapshots of $v(x,t)$; (b) The three-dimensional plot of the residual corresponding to $f_3$.}
\label{fig3-5}
\end{figure}

\begin{figure}[htbp]
\centering
\includegraphics[width=7cm,height=5.5cm]{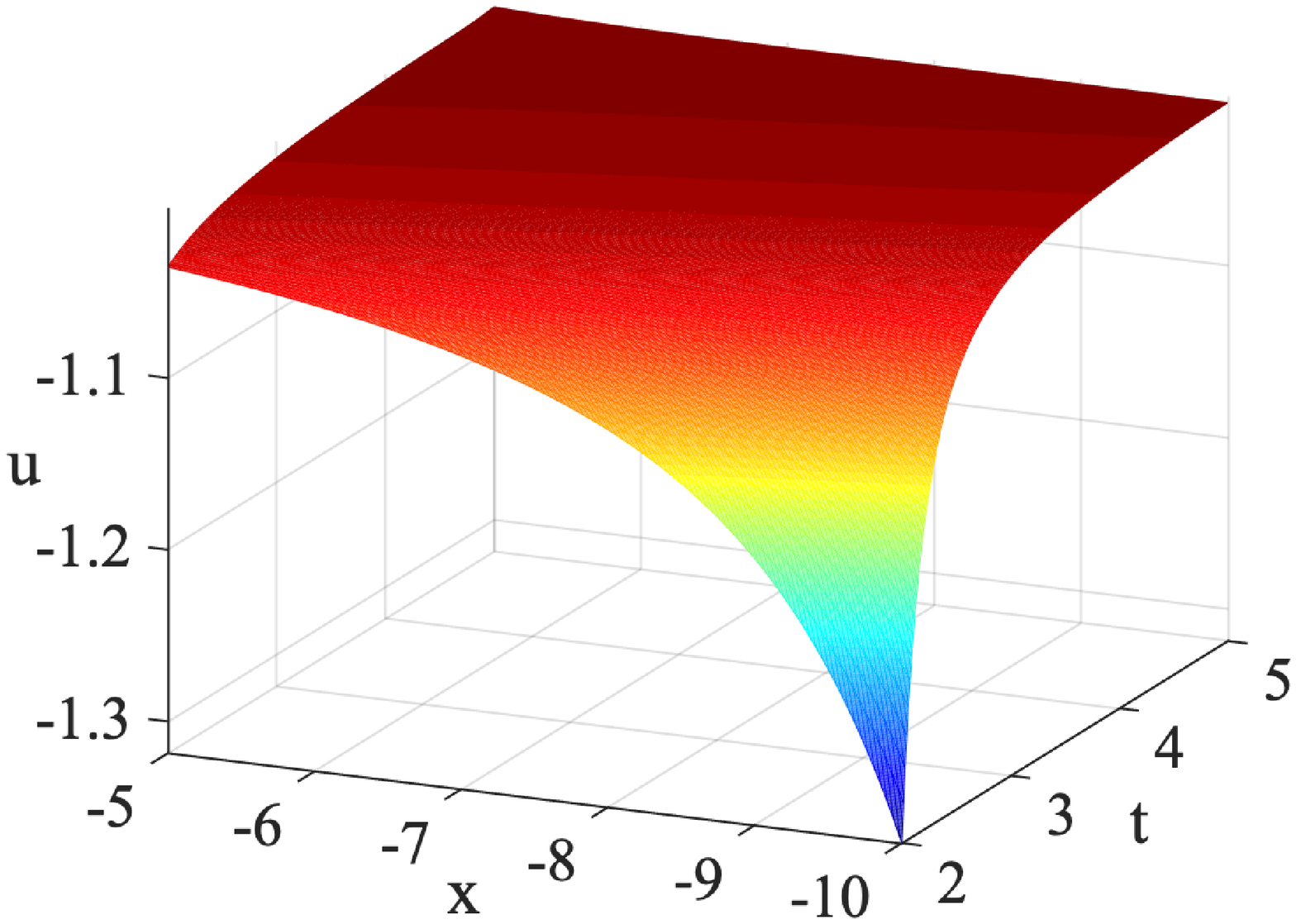}
$a$
\includegraphics[width=7cm,height=5.5cm]{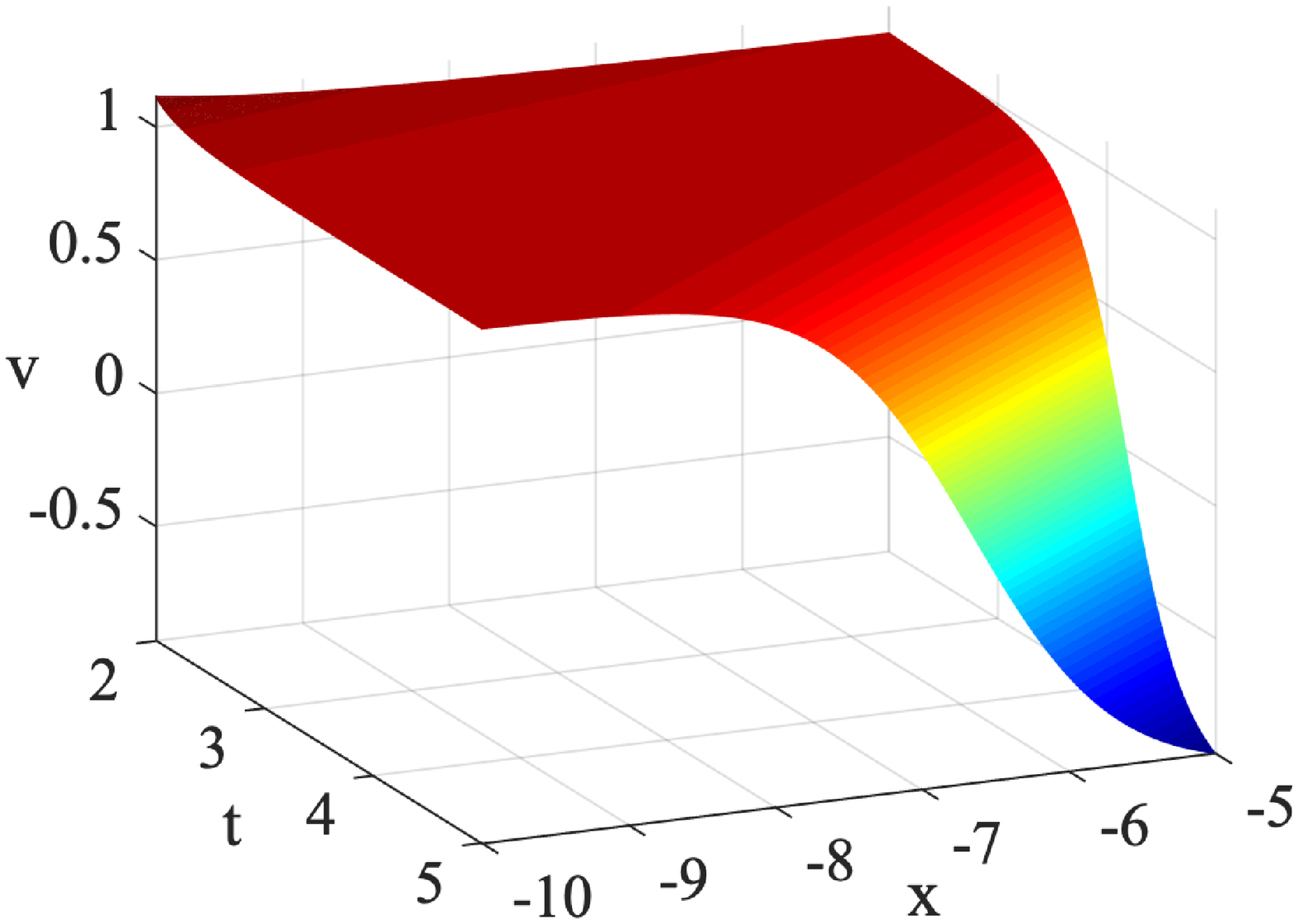}
$b$
\caption{(Color online) Data-driven solutions $u(x,t)$ of the KdV equation and $v(x,t)$ of the defocusing mKdV equation by Scheme \uppercase\expandafter{\romannumeral 1}: (a) The three-dimensional plot of $u(x,t)$; (b) The three-dimensional plot of $v(x,t)$.}
\label{fig3-6}
\end{figure}

With the advantage of the PINN method based on Miura transformations, we simulate precisely the data-driven solution of the KdV solution and it achieves the relative $\mathbb{L}_2$ error of 2.470634e-04 after 6544 times iterations in about 604.4544 seconds. The detailed image information is given in Fig. \ref{fig3-4}, including the density diagrams, comparison between the predicted solutions and exact solutions at the three temporal snapshots and the error density diagram. Noticeably, a new numerical solution of the defocusing mKdV solution is acquired by observing comparison between the predicted solution and exact solution in Fig. \ref{fig3-5} (a). It clearly reveals that the wave height of the exact solution is around -1 and sharply decreases near the point $(x,t)=(-10,2)$ while the wave height of the predicted one is around 1 and rapidly falls off near $(x,t)=(-5,5)$. To determine whether the predicted solution numerically satisfies the defocusing mKdV equation, the 3D plot of the residuals corresponding to $f_3$ ($|f_3(x^i,t^i)|$) is displayed in Fig. \ref{fig3-5} (b). Thus, we can conclude from the results that a new numerical solution of rational type has been obtained and this conclusion is reliable and there is enough evidence to back up since the residual is also kept small in the order of magnitude of $10^{-3}$ in the given spatiotemporal region and the mean squared error $MSE_{G}$ of $N_x \times N_t=513 \times 201$ grid points corresponding to $f_3$ is 2.854879e-07. In addition, the three-dimensional plots of data-driven solutions are shown in Fig. \ref{fig3-6}.

\section{Applications in the complex Miura transformation between the focusing mKdV equation and the KdV equation}\label{complex Miura}

Here, we mainly focus on the research of achieving the data-driven solutions by Scheme \uppercase\expandafter{\romannumeral 2} on the basis of the complex Miura transformation between the focusing mKdV equation and the KdV equation, since Scheme \uppercase\expandafter{\romannumeral 2} works better than Scheme \uppercase\expandafter{\romannumeral 1} in the following cases.

\subsection{Case 1}\label{4.1}
\quad

The well-known complex Miura transformation
\begin{equation}\label{E4-1}
u={\rm{i}}v_x+v^2,	
\end{equation}
is also presented by Miura\cite{Miura1968}, which is utilized to transform the solution of the focusing mKdV equation
\begin{equation}\label{E4-2}
v_t+6v^2 v_x+v_{xxx}=0,	
\end{equation}
into that of the KdV equation
\begin{equation}\label{E4-3}
u_t+6u u_x+u_{xxx}=0.	
\end{equation}

By the means of the Hirota bilinear method, we can derive one-soliton solution \cite{CDY2006} of the focusing mKdV equation
\begin{equation}\label{E4-4}
v=\frac{\mathrm{e}^{-k^{3} t+k x+p}}{1+\frac{\mathrm{e}^{-2 k^{3} t+2 k x+2 p}}{4 k^{2}}},
\end{equation}
and meanwhile obtain the solution of the KdV equation
\begin{equation}\label{E4-5}
u={\rm{i}}\left(\frac{k \mathrm{e}^{-k^{3} t+k x+p}}{1+\frac{\mathrm{e}^{-2 k^{3} t+2 k x+2 p}}{4 k^{2}}}-\frac{\mathrm{e}^{-k^{3} t+k x+p} \mathrm{e}^{-2 k^{3} t+2 k x+2 p}}{2\left(1+\frac{\mathrm{e}^{-2 k^{3} t+2 k x+2 p}}{4 k^{2}}\right)^{2} k}\right)+\left(\frac{\mathrm{e}^{-k^{3} t+k x+p}}{1+\frac{\mathrm{e}^{-2 k^{3} t+2 k x+2 p}}{4 k^{2}}}\right)^2,
\end{equation}
by complex Miura transformation \eqref{E4-1}. In this case, $u(x,t)$ is a complex-valued function and $v(x,t)$ is a real-valued one.

In step one of Scheme \uppercase\expandafter{\romannumeral 2}, we consider the KdV equation with the first kind of boundary condition \eqref{E3-6} as well. After setting $k=1, p=0, [x_0,x_1]=[-5,5], [t_0,t_1]=[-5,5]$, we have:
\begin{align}
&u_0(x)={\rm{i}}\left( \frac{\mathrm{e}^{5+x}}{1+\frac{\mathrm{e}^{10+2x}}{4}} -\frac{\mathrm{e}^{5+x} \mathrm{e}^{10+2x}}{2(1+\frac{\mathrm{e}^{10+2x}}{4})^2}\right)+\frac{(\mathrm{e}^{5+x})^2}{\left( 1+\frac{\mathrm{e}^{10+2x}}{4} \right)^2},\\
&u(-5,t)={\rm{i}}\left( \frac{\mathrm{e}^{-t-5}}{1+\frac{\mathrm{e}^{-2t-10}}{4}} -\frac{\mathrm{e}^{-t-5} \mathrm{e}^{-2t-10}}{2(1+\frac{\mathrm{e}^{-2t-10}}{4})^2}\right)+\frac{(\mathrm{e}^{-t-5})^2}{\left( 1+\frac{\mathrm{e}^{-2t-10}}{4} \right)^2},\\
&u(5,t)={\rm{i}}\left( \frac{\mathrm{e}^{-t+5}}{1+\frac{\mathrm{e}^{-2t+10}}{4}} -\frac{\mathrm{e}^{-t+5} \mathrm{e}^{-2t+10}}{2(1+\frac{\mathrm{e}^{-2t+10}}{4})^2}\right)+\frac{(\mathrm{e}^{-t+5})^2}{\left( 1+\frac{\mathrm{e}^{-2t+10}}{4} \right)^2}.	
\end{align}

We divide the spatial region $[x_0,x_1]=[-5,5]$ into $N_x =513$ discrete equidistance points and time region $[t_0,t_1]=[-5,5]$ into $N_t = 201$ discrete equidistance points separately. Then the solution $u$ is discretized into $513 \times 201$ data points in the given spatiotemporal domain. The initial-boundary data ($\{x^i_u,t^i_u,u^i\}^{N_u}_{i=1}, N_u=200$) is sampled randomly as the input to the neural network for training and the Latin hypercube sampling method is used to select $N_f =5000$ configuration points ($\{x_{f}^i,t_{f}^i\}^{N_{f}}_{i=1}$).

Given the complexity of the structure of complex-valued solution $u(x,t)$, we decompose it into into real part $u_r(x,t)$ and imaginary part $u_i(x,t)$, i.e. $u=u_r+{\rm{i}}u_i$. Therefore, the neural networks of these two parts can be learned by minimizing the mean squared error loss
\begin{equation}
MSE_1=MSE_u+MSE_{F_1}+MSE_{F_2},
\end{equation}
where
\begin{equation}
MSE_u=\frac{1}{N_u}\sum^{N_u}_{i=1}|\Widehat{u}(x_u^i,t_u^i)-u^i|^2,
\end{equation}
\begin{equation}
MSE_{F_1}=\frac{1}{N_f}\sum^{N_f}_{i=1}|F_1(x_{f}^i,t_{f}^i)|^2,
\end{equation}
\begin{equation}
MSE_{F_2}=\frac{1}{N_f}\sum^{N_f}_{i=1}|F_2(x_{f}^i,t_{f}^i)|^2,
\end{equation}
and $F_1$ and $F_2$ correspond to the following governing equations separately
\begin{align}
&f_1:=F_1=(u_r)_t+6u_r (u_r)_x-6u_i (u_i)_x+(u_r)_{xxx},\\
&f_2:=F_2=(u_i)_t+6u_i (u_r)_x+6u_r (u_i)_x+(u_i)_{xxx}.
\end{align}

\begin{figure}[htbp]
\centering
\includegraphics[width=7cm,height=5cm]{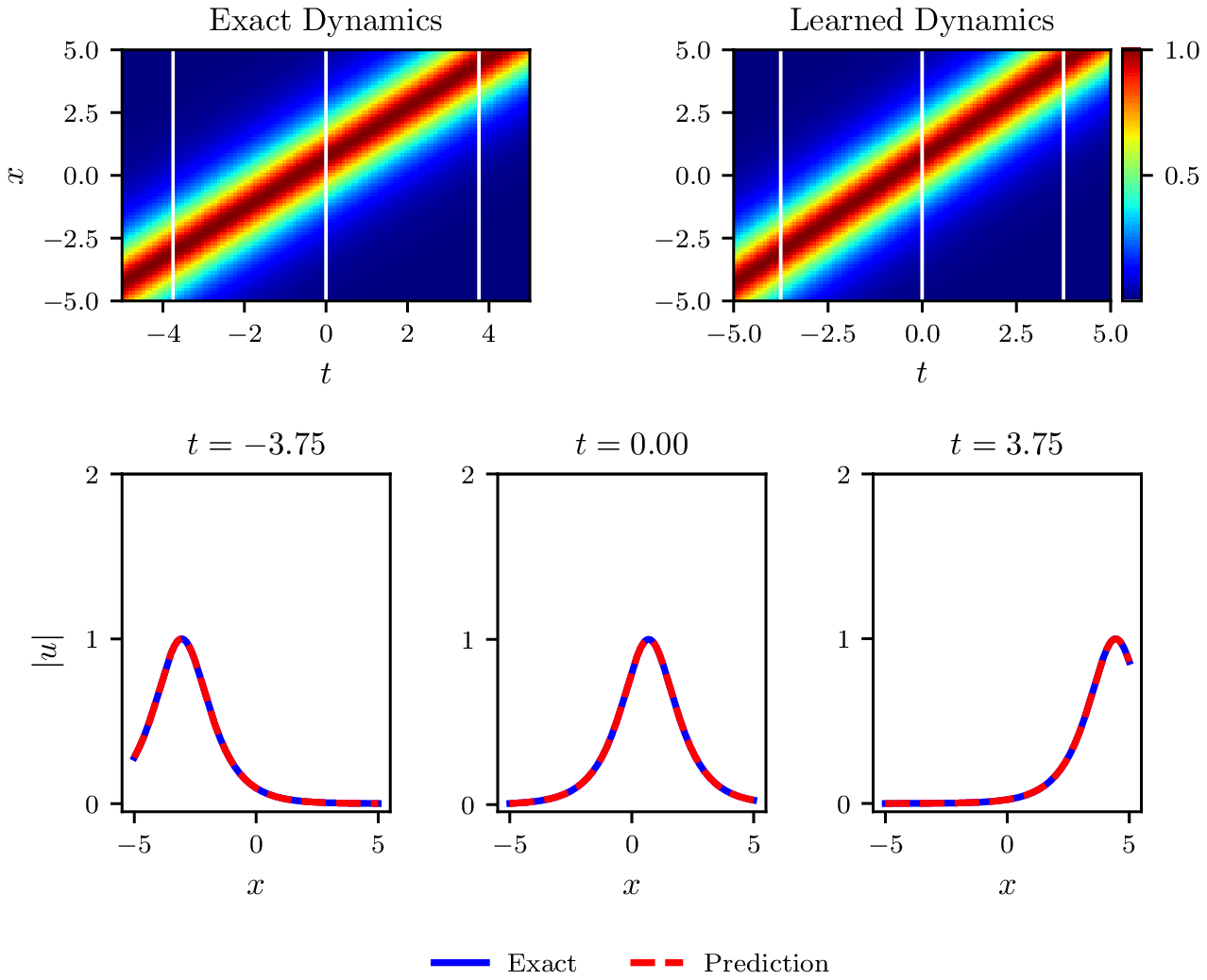}
$a$
\includegraphics[width=7cm,height=5cm]{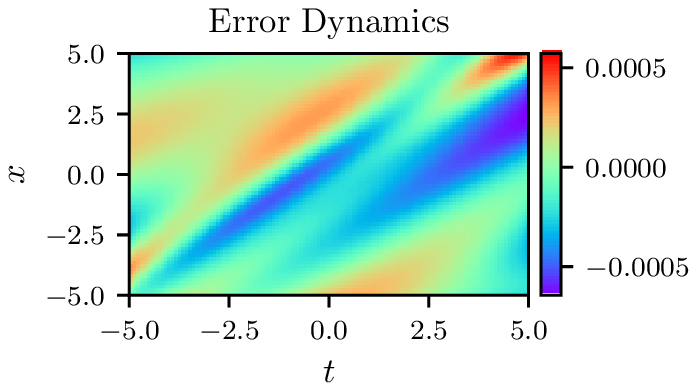}
$b$
\caption{(Color online) Data-driven solution $u(x,t)$ of the KdV equation by Scheme \uppercase\expandafter{\romannumeral 2}: (a) The density diagrams and comparison between the predicted solutions and exact solutions at the three temporal snapshots of $u(x,t)$; (b) The error density diagram of $u(x,t)$.}
\label{fig4-1}
\end{figure}

Then a 5-layer feedforward neural network with 50 neurons per hidden layer is constructed to learn the solution of the KdV equation. Here we also use the hyperbolic tangent ($tanh$) activation function and initialize weights of the neural network with the Xavier initialization. The derivatives of the network $u_r$ and $u_i$ with respect to time $t$ and space $x$ are derived by automatic differentiation and the above loss functions are optimized by L-BFGS algorithm. After 643 times iterations in about 40.6527 seconds, the relative $\mathbb{L}_2$ errors  of the real part $u_r$, the imaginary part $u_i$ and the modulus $|u|$ are 8.743754e-04, 9.836247e-04 and 5.424276e-04, respectively. Fig. \ref{fig4-1} shows the density diagrams, comparison between the predicted solutions and exact solutions as well as the error density diagram of the KdV equation. It can be observed that this data-driven solution propagates along the positive direction of the x-axis as time goes by according to the bottom panel of Fig. \ref{fig4-1} (a), which presents the comparisons between exact solutions and predicted solutions at three different time points $t=-3.75, 0, 3.75$.

Then in step two, a new physics-informed neural network is constructed to acquire the data-driven solution of the focusing mKdV equation based on the numerical solution $\Widehat{u}(x,t)$ obtained in the first step. Although the exact solution $v$ \eqref{E4-4} is real-valued, we should assume that it is a complex-valued function, i.e. $v=v_r+{\rm{i}}v_i$,  since the known initial-boundary data of $u$ is complex-valued. Thus the mean squared error loss is given by
\begin{equation}
MSE_2=MSE_{G_1}+MSE_{G_2}+MSE_{M_1}+MSE_{M_2},
\end{equation}
where
\begin{equation}
MSE_{G_1}=\frac{1}{N_g}\sum^{N_g}_{i=1}|G_1(x_{g}^i,t_{g}^i)|^2,
\end{equation}
\begin{equation}
MSE_{G_2}=\frac{1}{N_g}\sum^{N_g}_{i=1}|G_2(x_{g}^i,t_{g}^i)|^2,
\end{equation}
\begin{equation}
MSE_{M_1}=\frac{1}{N_g}\sum^{N_g}_{i=1}|M_1(x_g^i,t_g^i)|^2,
\end{equation}
\begin{equation}
MSE_{M_2}=\frac{1}{N_g}\sum^{N_g}_{i=1}|M_2(x_g^i,t_g^i)|^2,
\end{equation}
and $G_1$, $G_2$, $M_1$ and $M_2$ respectively correspond to the following governing equations
\begin{align}
&f_3:=G_1=(v_r)_t+6(v_r)_x (v_r^2-v_i^2)-12v_r v_i (v_i)_x +(v_r)_{xxx},\\
&f_4:=G_2=(v_i)_t+6(v_i)_x (v_r^2-v_i^2)+12(v_r)_x v_r v_i +(v_i)_{xxx},\\
&f_5:=M_1=u_r+(v_i)_x-v_r^2+v_i^2,\\
&f_6:=M_2=u_i-(v_r)_x-2v_r v_i.	
\end{align}
Here $N_g$ denotes the number of collocation points ($\{x_g^i,t_g^i,\Widehat{u}(x_g^i,t_g^i)\}^{N_{g}}_{i=1}$) in the region $[x_0,x_1]\times[t_0,t_1]=[-5,5]\times[-5,5]$, which are sampled randomly from $N_x \times N_t=513 \times 201$ grid points and we take $N_g=2000$ in this case.

\begin{figure}[htbp]
\centering
\includegraphics[width=7cm,height=5cm]{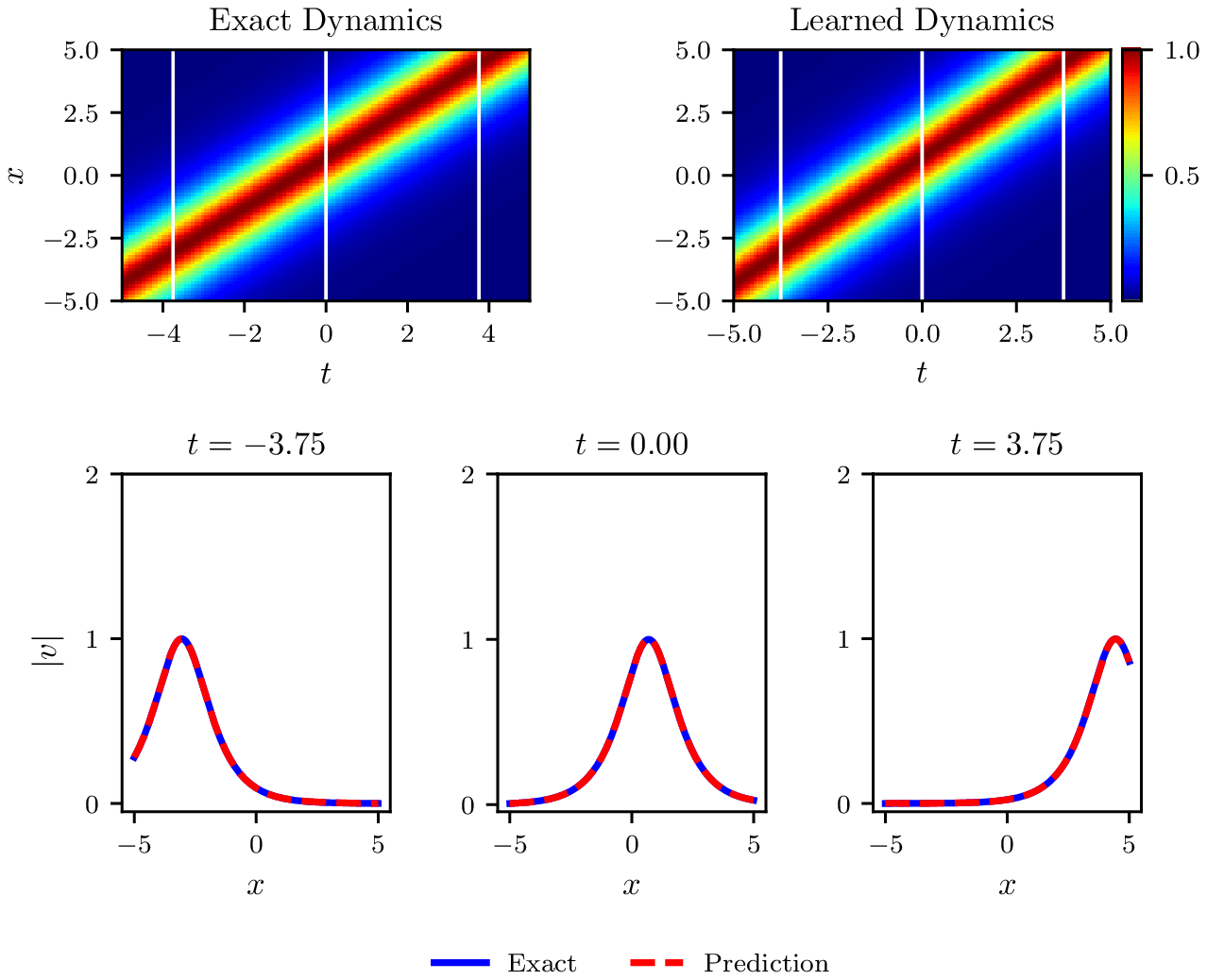}
$a$
\includegraphics[width=7cm,height=5cm]{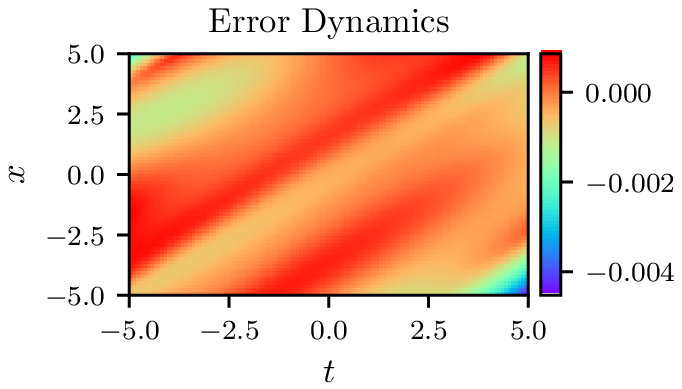}
$b$
\caption{(Color online) One-soliton solution $v(x,t)$ of the focusing mKdV equation by Scheme \uppercase\expandafter{\romannumeral 2}: (a) The density diagrams and comparison between the predicted solutions and exact solutions at the three temporal snapshots of $v(x,t)$; (b) The error density diagram of $v(x,t)$.}
\label{fig4-2}
\end{figure}

\begin{figure}[htbp]
\centering
\includegraphics[width=5.5cm,height=4.5cm]{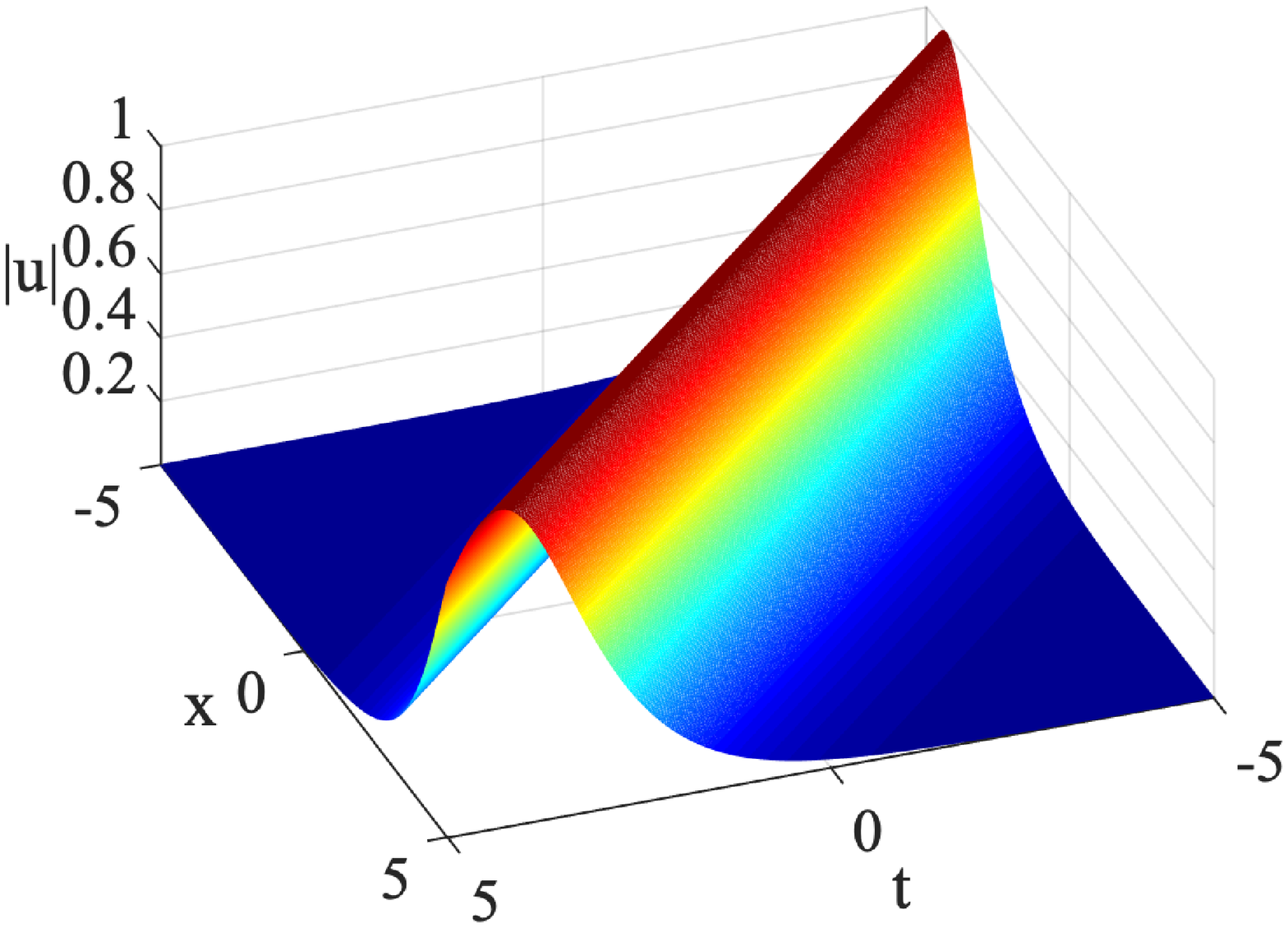}
$a$
\includegraphics[width=5.5cm,height=4.5cm]{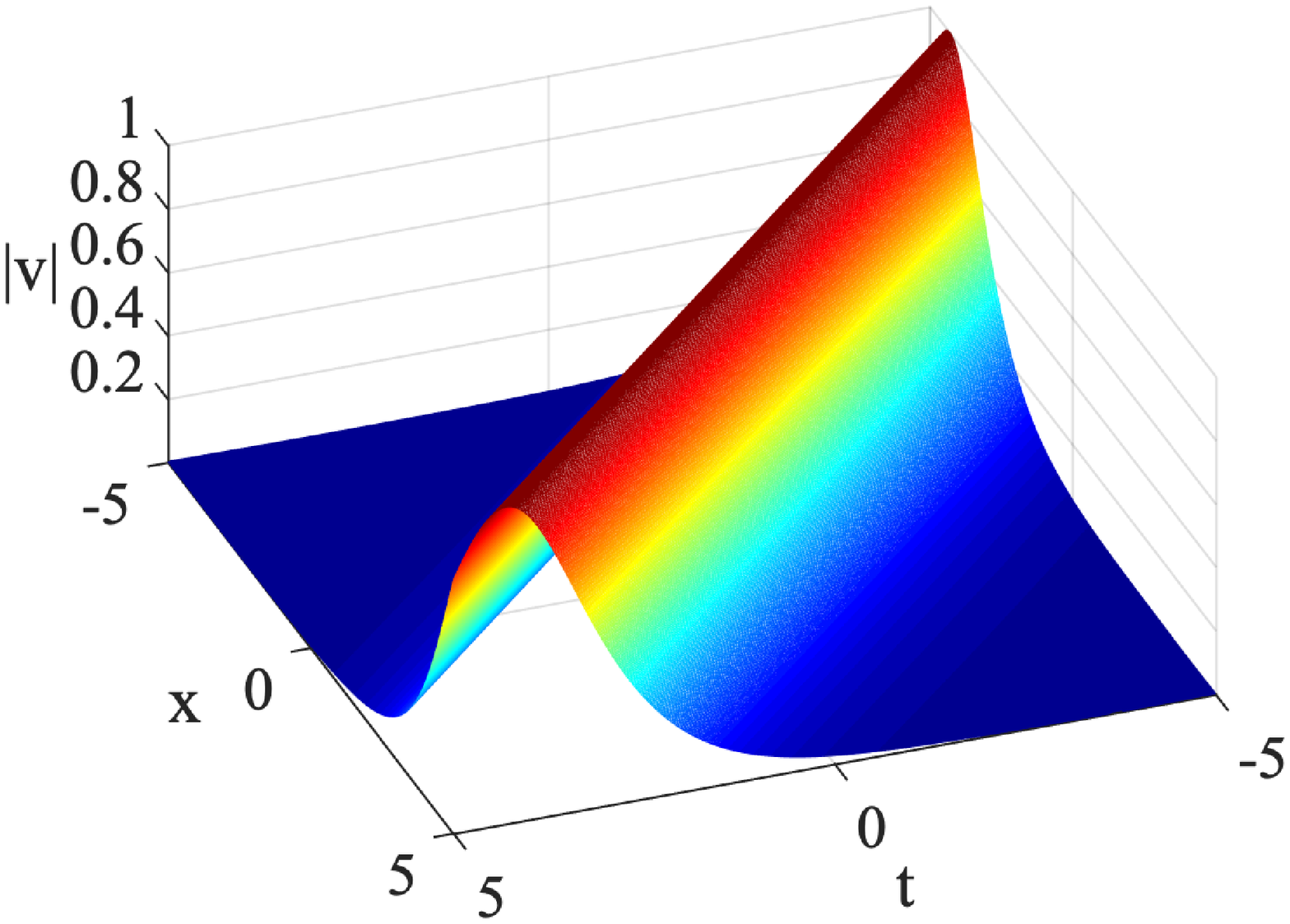}
$b$
\includegraphics[width=5.5cm,height=4.5cm]{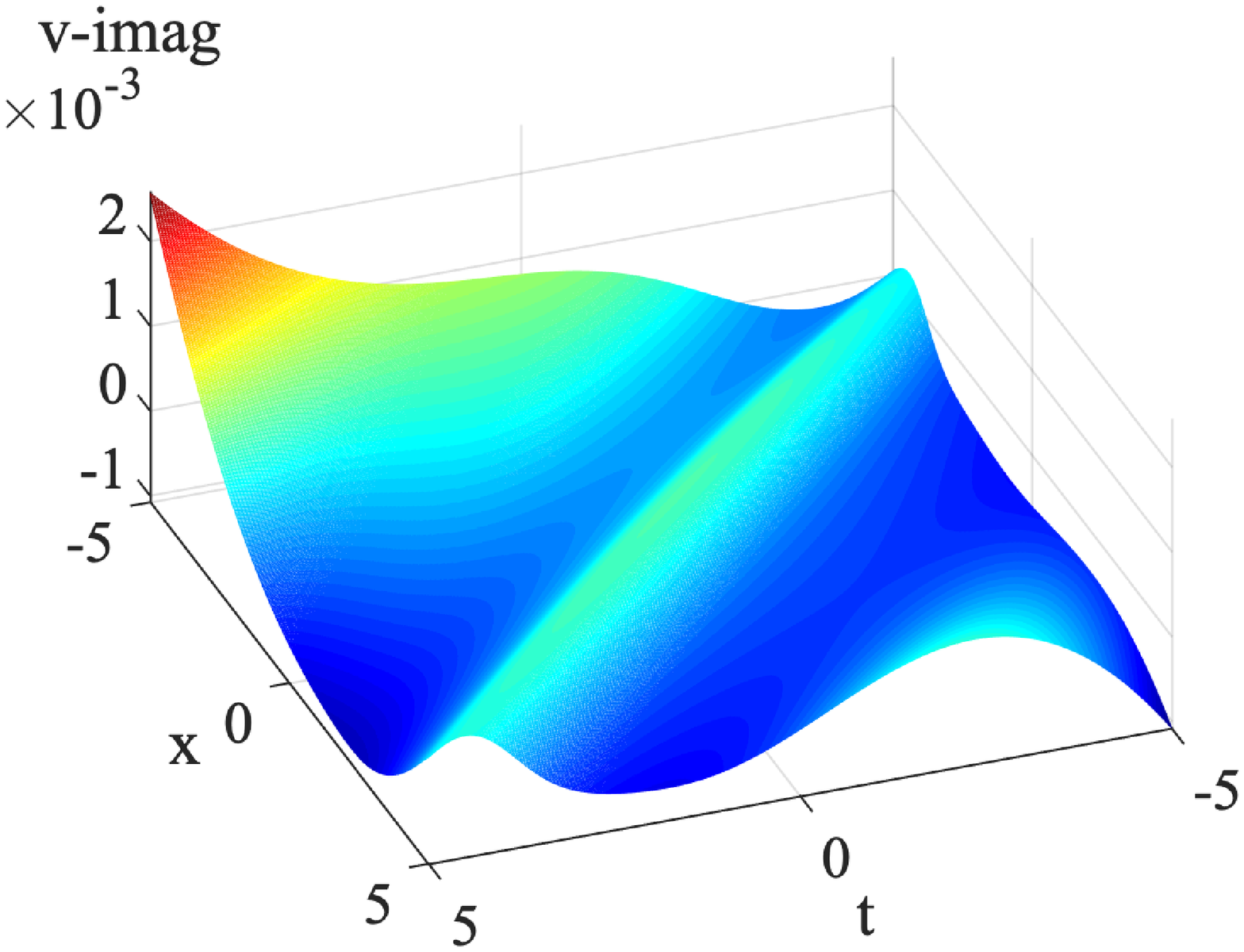}
$c$
\caption{(Color online) Data-driven solutions $u(x,t)$ of the KdV equation and $v(x,t)$ of the focusing mKdV equation by Scheme \uppercase\expandafter{\romannumeral 2}: (a) The three-dimensional plot of $|u|(x,t)$; (b) The three-dimensional plot of $|v|(x,t)$; (c) The three-dimensional plot of the imaginary part $v_i(x,t)$.}
\label{fig4-3}
\end{figure}

Eventually, the one-soliton solution of the focusing mKdV equation is successfully simulated with the aid of a feedforward neural network which contains 2 hidden layers and each layer has 40 neurons. After 457 times iterations in about 19.2721 seconds, the relative $\mathbb{L}_2$ errors  of the real part $v_r$ and the modulus $|v|$ are 1.217201e-03 and 1.285430e-03. Fig. \ref{fig4-2} clearly compares the exact solution with the predicted one and the three-dimensional plots of $|u|(x,t)$, $|v|(x,t)$ as well as the imaginary part $v_i(x,t)$ are displayed in Fig. \ref{fig4-3}. From Fig. \ref{fig4-3}, we can clearly observe that the wave patterns of these two data-driven solutions are very similar, and moreover, the predicted one-soliton solution of the focusing mKdV equation can be regarded as the real-valued one and is close to the exact real-valued solution $v$ \eqref{E4-4} since the imaginary part $v_i(x,t)$ is small enough in the order of magnitude of $10^{-3}$.

Details of data-driven solutions of the KdV equation and the focusing mKdV equation are listed in Table \ref{table-case1}.

\begin{table}[H]
\caption{Data-driven solutions in Case 4.1 by Scheme \uppercase\expandafter{\romannumeral 2}: iteration times, elapsed time, relative $\mathbb{L}_2$ errors and type of solution.}
\label{table-case1}
\centering
\begin{tabular}{cc|cc}
\bottomrule
\multicolumn{2}{c|}{\textbf{Step One}}            & \multicolumn{2}{c}{\textbf{Step Two}}             \\ \hline
\multicolumn{1}{c|}{Hidden layers-Neurons}  & 4-50 & \multicolumn{1}{c|}{Hidden layers-Neurons}  & 2-40 \\ \hline
\multicolumn{1}{c|}{Iteration times}  & 643 & \multicolumn{1}{c|}{Iteration times}  & 457 \\ \hline
\multicolumn{1}{c|}{Elapsed time (s)} & 40.6527 & \multicolumn{1}{c|}{Elapsed time (s)} & 19.2721 \\ \hline
\multicolumn{1}{c|}{$u_r(x,t)$}                 & 8.743754e-04 & \multicolumn{1}{c|}{$v_r(x,t)$}                 & 1.217201e-03 \\ \hline
\multicolumn{1}{c|}{$u_i(x,t)$}                 & 9.836247e-04 & \multicolumn{1}{c|}{$|v|(x,t)$}                 & 1.285430e-03 \\ \hline
\multicolumn{1}{c|}{$|u|(x,t)$}                 & 5.424276e-04 & \multicolumn{1}{c|}{Type of solution}      & One soliton \\ \toprule
\end{tabular}
\end{table}

\subsection{Case 2}
\quad

For the focusing mKdV equation, the two-soliton solution \cite{CDY2006}
\begin{equation}\label{E4-6}
v=2\left( {\rm{arctan}} \frac{r}{q} \right) _x,	
\end{equation}
where
\begin{align}
q&=1-\frac{1}{4k_1 k_2}\left( \frac{k_1-k_2}{k_1+k_2} \right)^2 \mathrm{e}^{\xi_1+\xi_2}, \\
r&=\frac{1}{2 k_1}\mathrm{e}^{\xi_1}+\frac{1}{2 k_2} \mathrm{e}^{\xi_2},\\
\xi_j&=k_j x-k_j^3 t+p_j (j=1,2),
\end{align}
can also be derived by Hirota bilinear method and then the complex Miura transformation transforms it into the following solution $u(x,t)$ of the KdV equation by choosing corresponding parameters as $k_1=2, k_2=1, p_1=0, p_2=1$
\begin{align}
u={\rm{i}}(A-\frac{B}{C})+D,
\end{align}
where
\begin{align}
&A=\frac{2 \mathrm{e}^{-8t+2x}+\mathrm{e}^{-t+x+1}+\frac{5\mathrm{e}^{-17t+5x+1}}{144}+\frac{\mathrm{e}^{-10t+4x+2}}{9}}{1+\frac{\mathrm{e}^{-16t+4x}}{16}+\frac{2\mathrm{e}^{-9t+3x+1}}{9}+\frac{\mathrm{e}^{-2t+2x+2}}{4}+\frac{\mathrm{e}^{-18t+6x+2}}{5184}}	,\\
&B=\left(\mathrm{e}^{-8t+2x}+\mathrm{e}^{-t+x+1}+\frac{\mathrm{e}^{-17t+5x+1}}{144}+\frac{\mathrm{e}^{-10t+4x+2}}{36}\right)\left(\frac{\mathrm{e}^{-16t+4x}}{4}+\frac{2\mathrm{e}^{-9t+3x+1}}{3}+\frac{\mathrm{e}^{-2t+2x+2}}{2}+\frac{\mathrm{e}^{-18t+6x+2}}{864}\right),\\
&C=\left( 1+\frac{\mathrm{e}^{-16t+4x}}{16} +\frac{2\mathrm{e}^{-9t+3x+1}}{9}+\frac{\mathrm{e}^{-2t+2x+2}}{4}+\frac{\mathrm{e}^{-18t+6x+2}}{5184} \right)^2,\\
&D=\frac{\left(\mathrm{e}^{-8t+2x}+\mathrm{e}^{-t+x+1}+\frac{\mathrm{e}^{-17t+5x+1}}{144}+\frac{\mathrm{e}^{-10t+4x+2}}{36} \right)^2}{\left( 1+\frac{\mathrm{e}^{-16t+4x}}{16}+\frac{2\mathrm{e}^{-9t+3x+1}}{9}+ \frac{\mathrm{e}^{-2t+2x+2}}{4}+\frac{\mathrm{e}^{-18t+6x+2}}{5184}\right)^2}.
\end{align}
After taking $[x_0,x_1]=[-1,2],[t_0,t_1]=[-5,5]$,
the Dirichlet boundary condition of $u(x,t)$ can be given, which is no longer presented here due to space limitation.

The training dataset ($N_u = 200$, $N_f =5000$) is sampled randomly by exploiting the same data discretization and sampling method in Section \ref{4.1}. In step one, we establish a 7-layer feedforward neural network with 60 neurons per hidden layer to solve the initial-boundary value problem of the KdV equation. After 7217 times iterations in about 675.7553 seconds, the relative $\mathbb{L}_2$ errors of the real part $u_r$, the imaginary part $u_i$ and the modulus $|u|$ are 4.337120e-03, 8.181105e-03 and 2.868770e-03 separately.

The structure of neural networks in step two is different from step one. Here we construct a 3-layer feedforward neural network with 60 neurons per hidden layer and set $N_g=5000$. The key constituents of these two steps are the same as the previous subsection, including parameter-setting, loss functions,  the method of initializing weights (Xavier initialization), activation function ($tanh$) as well as the optimization algorithm (L-BFGS). Ultimately, the two-soliton solution of the focusing mKdV equation has been successfully simulated. This model achieves that the relative $\mathbb{L}_2$ errors of the real part $v_r$ and the modulus $|v|$ are 5.026626e-03 and 5.020651e-03 in about 378.4642 seconds and the number of iterations is 7545.

\begin{figure}[htbp]
\centering
\includegraphics[width=7cm,height=5cm]{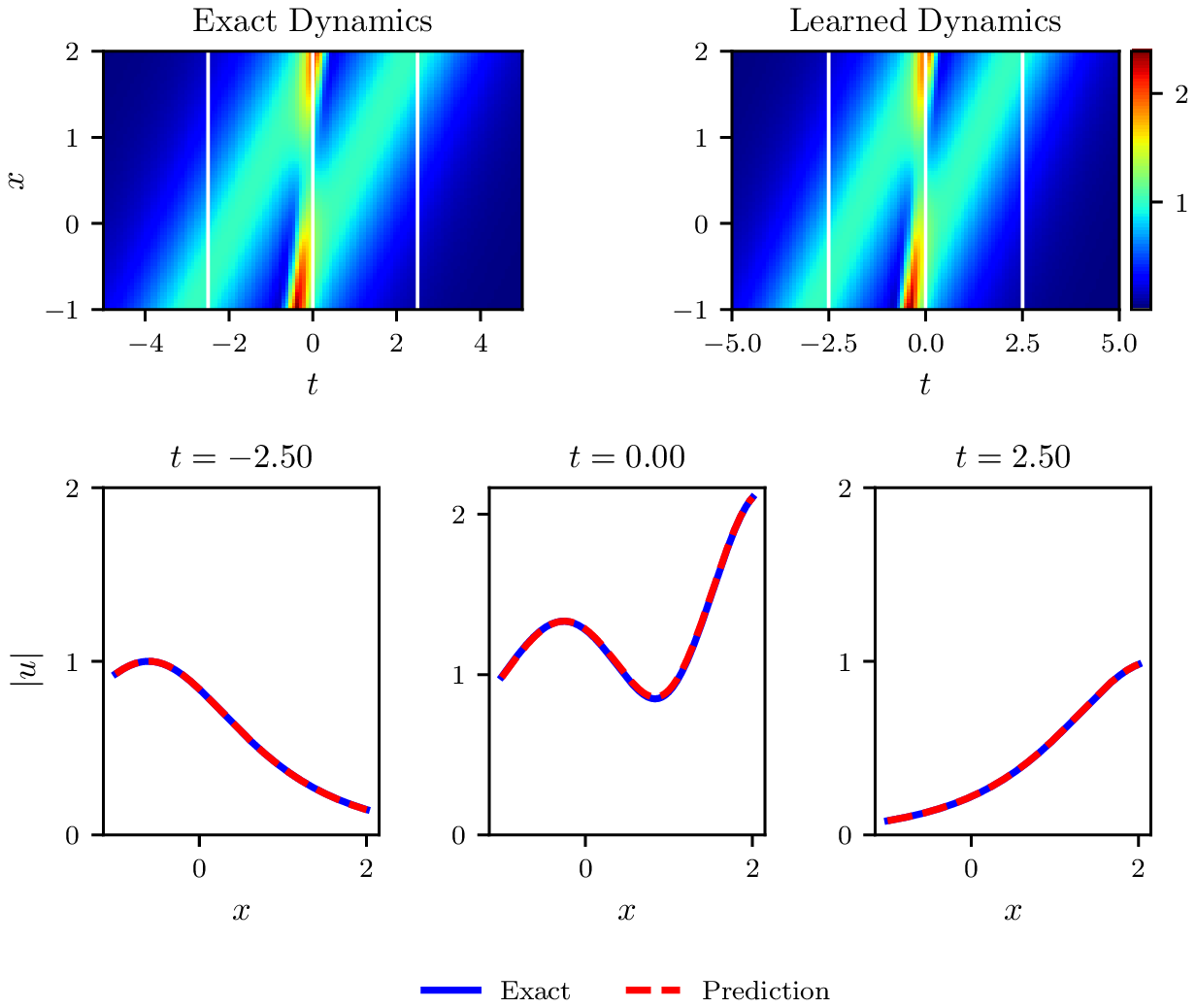}
$a$
\includegraphics[width=7cm,height=5cm]{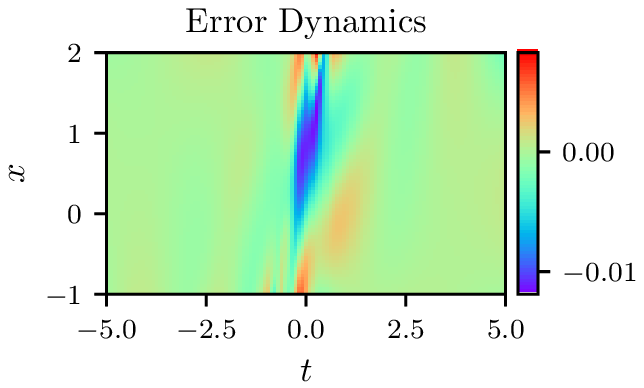}
$b$
\caption{(Color online) Data-driven solution $u(x,t)$ of the KdV equation by Scheme \uppercase\expandafter{\romannumeral 2}: (a) The density diagrams and comparison between the predicted solutions and exact solutions at the three temporal snapshots of $u(x,t)$; (b) The error density diagram of $u(x,t)$.}
\label{fig4-4}
\end{figure}

\begin{figure}[htbp]
\centering
\includegraphics[width=7cm,height=5cm]{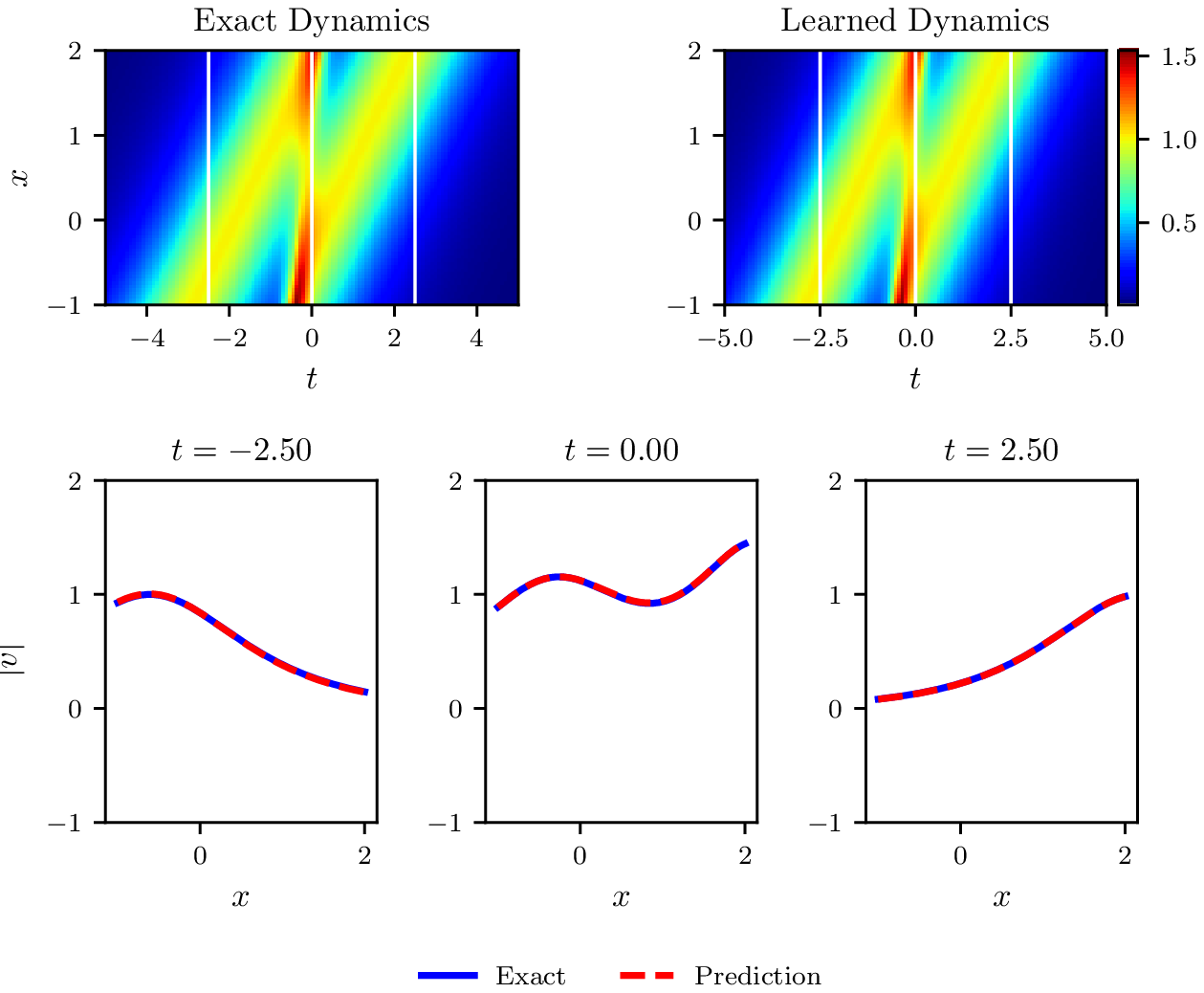}
$a$
\includegraphics[width=7cm,height=5cm]{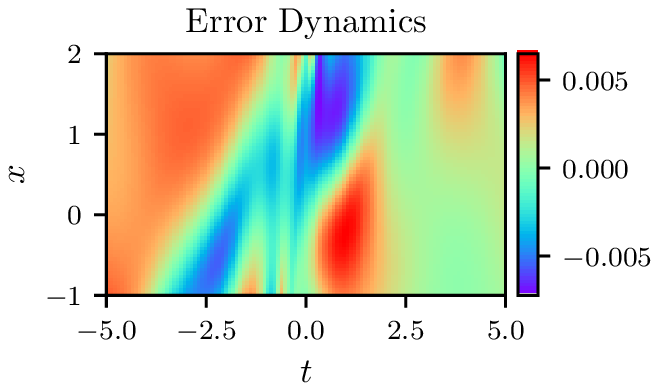}
$b$
\caption{(Color online) Two-soliton solution $v(x,t)$ of the focusing mKdV equation by Scheme \uppercase\expandafter{\romannumeral 2}: (a) The density diagrams and comparison between the predicted solutions and exact solutions at the three temporal snapshots of $v(x,t)$; (b) The error density diagram of $v(x,t)$.}
\label{fig4-5}
\end{figure}

\begin{figure}[htbp]
\centering
\includegraphics[width=5.5cm,height=4.5cm]{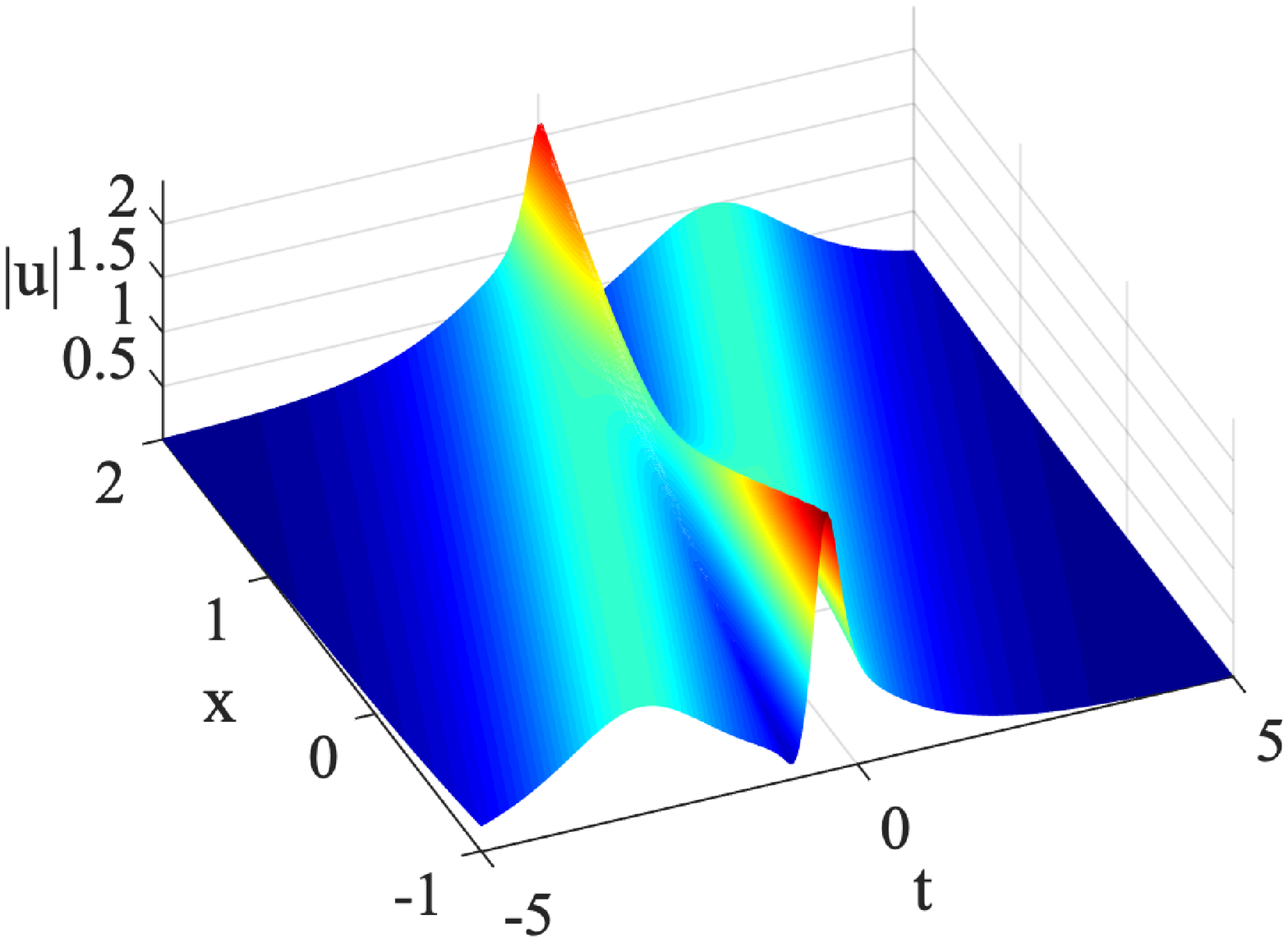}
$a$
\includegraphics[width=5.5cm,height=4.5cm]{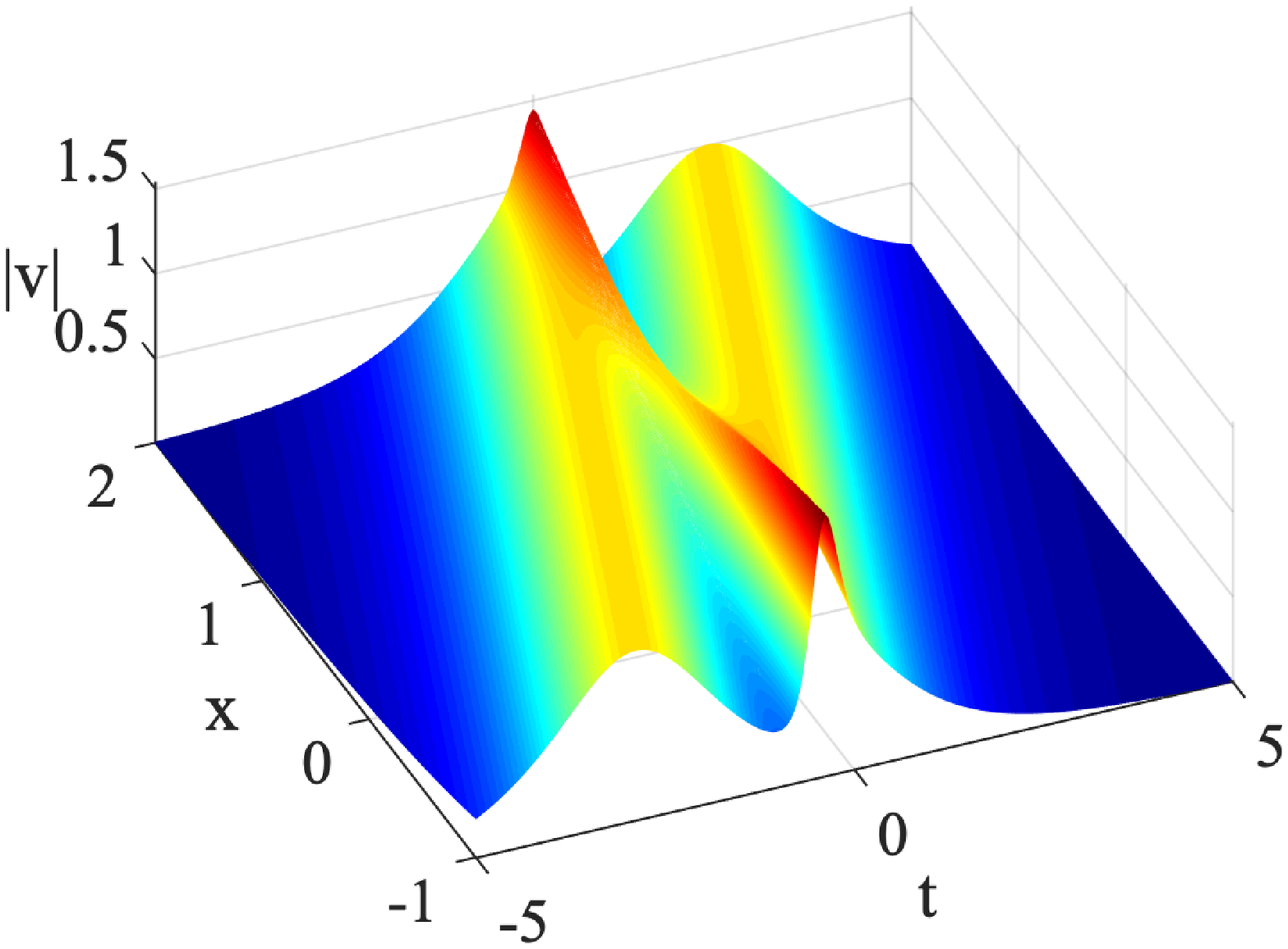}
$b$
\includegraphics[width=5.5cm,height=4.5cm]{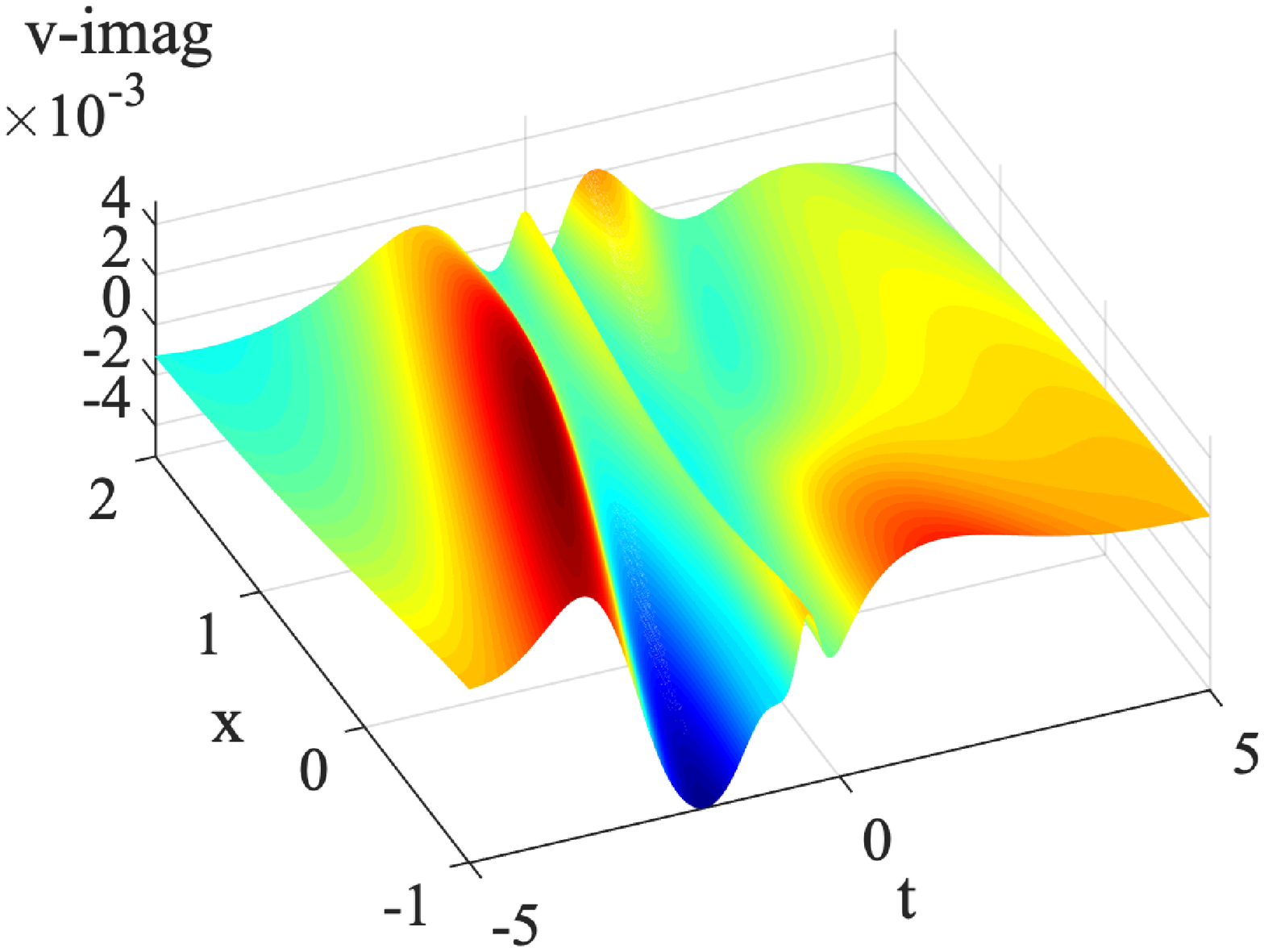}
$c$
\caption{(Color online) Data-driven solutions $u(x,t)$ of the KdV equation and $v(x,t)$ of the focusing mKdV equation by Scheme \uppercase\expandafter{\romannumeral 2}: (a) The three-dimensional plot of $|u|(x,t)$; (b) The three-dimensional plot of $|v|(x,t)$; (c) The three-dimensional plot of the imaginary part $v_i(x,t)$.}
\label{fig4-6}
\end{figure}

Fig. \ref{fig4-4} - Fig. \ref{fig4-6} present the density diagrams, comparison between the predicted solutions and exact solutions, error density diagrams and the predicted 3D plots. Similarly, the predicted one-soliton solution of the focusing mKdV equation can be regarded as the real-valued one and is close to the exact real-valued solution $v$ \eqref{E4-6} since the imaginary part $v_i(x,t)$ is small enough according to Fig. \ref{fig4-6}(c).

Details of data-driven solutions of the KdV equation and the focusing mKdV equation are shown in Table \ref{table-case2}.

\begin{table}[H]
\caption{Data-driven solutions in Case 4.2 by Scheme \uppercase\expandafter{\romannumeral 2}: iteration times, elapsed time, relative $\mathbb{L}_2$ errors and type of solution.}
\label{table-case2}
\centering
\begin{tabular}{cc|cc}
\bottomrule
\multicolumn{2}{c|}{\textbf{Step One}}            & \multicolumn{2}{c}{\textbf{Step Two}}             \\ \hline
\multicolumn{1}{c|}{Hidden layers-Neurons}  & 6-60 & \multicolumn{1}{c|}{Hidden layers-Neurons}  & 2-60 \\ \hline
\multicolumn{1}{c|}{Iteration times}  & 7217 & \multicolumn{1}{c|}{Iteration times}  & 7545 \\ \hline
\multicolumn{1}{c|}{Elapsed time (s)} & 675.7553 & \multicolumn{1}{c|}{Elapsed time (s)} & 378.4642 \\ \hline
\multicolumn{1}{c|}{$u_r(x,t)$}    & 4.337120e-03 & \multicolumn{1}{c|}{$v_r(x,t)$}  & 5.026626e-03 \\ \hline
\multicolumn{1}{c|}{$u_i(x,t)$}    & 8.181105e-03 & \multicolumn{1}{c|}{$|v|(x,t)$}  & 5.020651e-03 \\ \hline
\multicolumn{1}{c|}{$|u|(x,t)$}    & 2.868770e-03 & \multicolumn{1}{c|}{Type of solution}                 & Two soliton \\ \toprule
\end{tabular}
\end{table}

\subsection{Case 3}
\quad

Based on \eqref{rational}, we can obtain the first-order rational solution \cite{Ankiewicz2018} of the focusing mKdV equation
\begin{equation}\label{rationalfocusing}
v=-\frac{12}{-3-12(x-6t)^2}-1,	
\end{equation}
after taking $\beta=6, \gamma=-1$, and then derive the solution of the KdV equation
\begin{equation}
u=\frac{12 {\rm{i}}(-24 x+144 t)}{\left(-3-12(x-6 t)^{2}\right)^{2}}+\left(-\frac{12}{-3-12(x-6 t)^{2}}-1\right)^{2},
\end{equation}
by complex Miura transformation.

We choose $[x_0,x_1]=[-2,2], [t_0,t_1]=[-0.5,0.5]$,  which yields
\begin{align}
&u_0(x)=\frac{12 {\rm{i}} (-24x-72)}{\left( -3-12(x+3)^2 \right)^2}+\left(-\frac{12}{-3-12(x+3)^2}-1\right)^2,\\
&u(-2,t)=\frac{12 {\rm{i}} (48+144t)}{\left( -3-12(-2-6t)^2 \right)^2}+\left(-\frac{12}{-3-12(-2-6t)^2}-1\right)^2,\\
&u(2,t)=\frac{12 {\rm{i}} (-48+144t)}{\left( -3-12(2-6t)^2 \right)^2}+\left(-\frac{12}{-3-12(2-6t)^2}-1\right)^2.	
\end{align}
In this case, we aim to utilize the above initial-boundary conditions of the KdV equation to simulate numerically the solution of the focusing mKdV equation.

\begin{figure}[htbp]
\centering
\includegraphics[width=7cm,height=5cm]{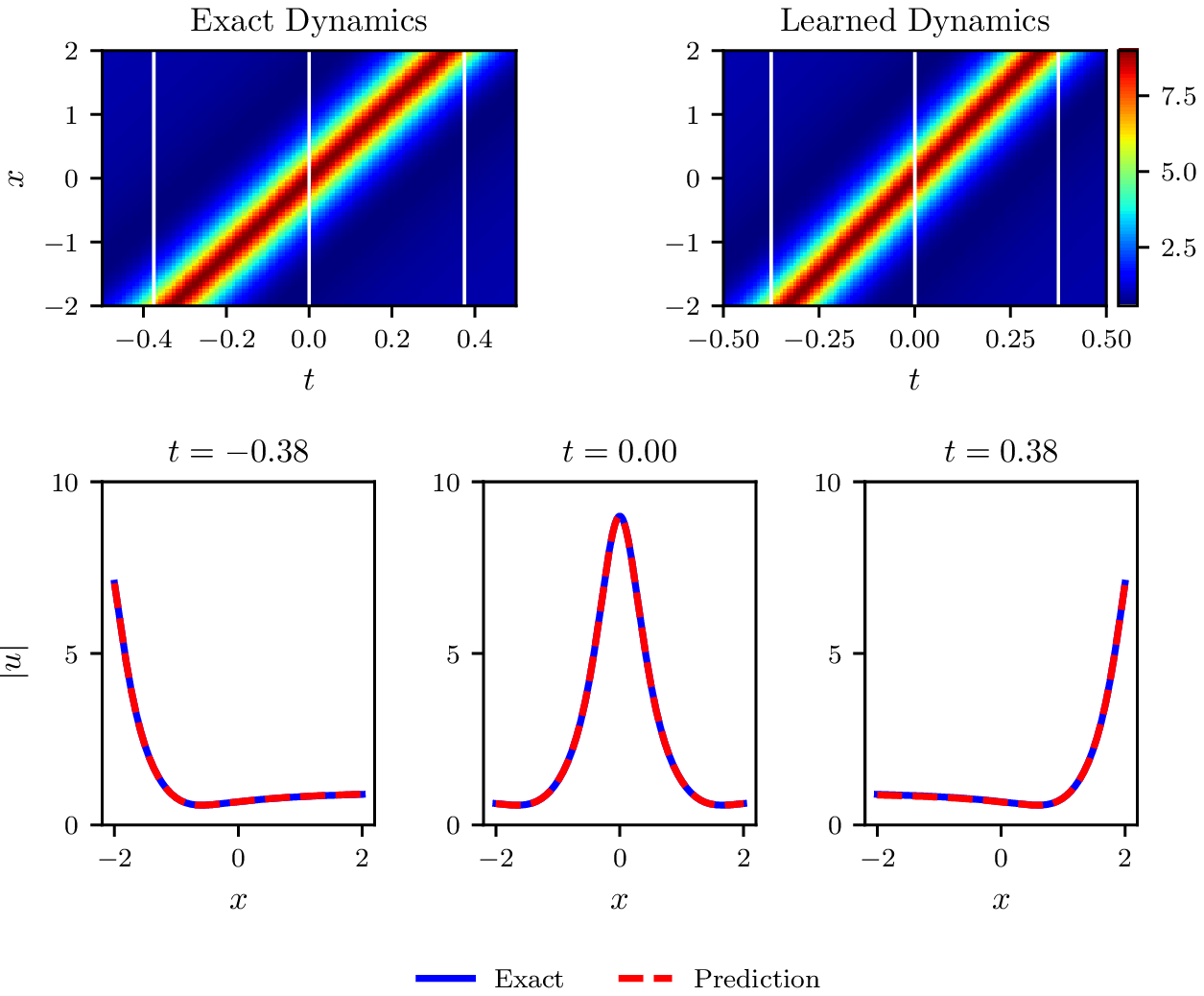}
$a$
\includegraphics[width=7cm,height=5cm]{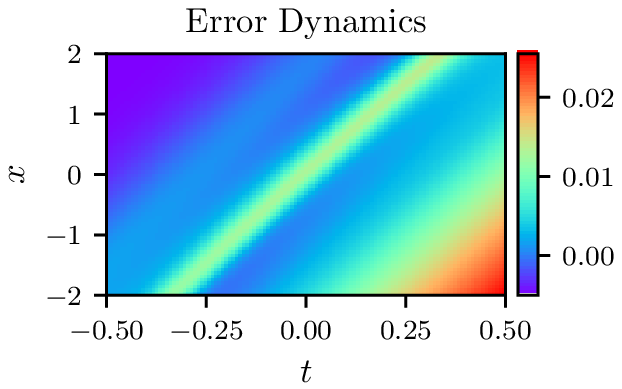}
$b$
\caption{(Color online) Data-driven solution $u(x,t)$ of the KdV equation by Scheme \uppercase\expandafter{\romannumeral 2}: (a) The density diagrams and comparison between the predicted solutions and exact solutions at the three temporal snapshots of $u(x,t)$; (b) The error density diagram of $u(x,t)$.}
\label{fig4-7}
\end{figure}

Likewise, the initial-boundary data is sampled randomly ($N_u=200$) as the input of the neural network for training after exploiting the same data discretization and sampling method in Section \ref{4.1}. In addition, $N_f=5000$ collocation points are extracted by Latin hypercube sampling method. In step one, a 9-layer feedforward neural network with 40 neurons per hidden layer is constructed to solve the initial-boundary value problems of the KdV equation. After 2617 times iterations in about 273.2680 seconds, the relative $\mathbb{L}_2$ errors  of the real part $u_r$, the imaginary part $u_i$ and the modulus $|u|$ are 2.772086e-03, 1.914695e-03 and 2.262575e-03, respectively. The detailed image information is given in Fig. \ref{fig4-7}, including the density diagrams, comparison between the predicted solutions and exact solutions at the three temporal snapshots as well as the error density diagram.

\begin{figure}[htbp]
\centering
\includegraphics[width=5.5cm,height=4.5cm]{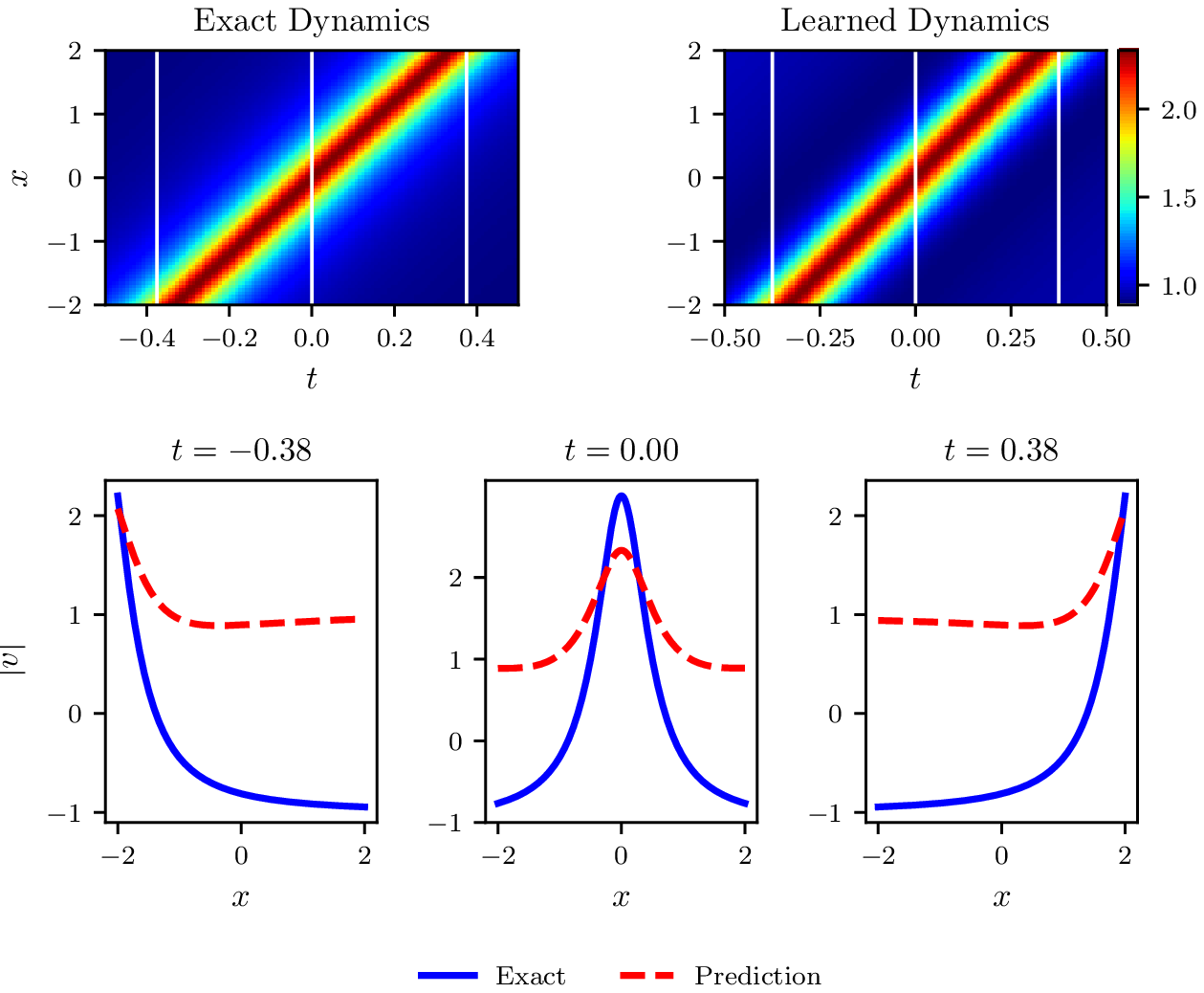}
$a$
\includegraphics[width=5.5cm,height=4.5cm]{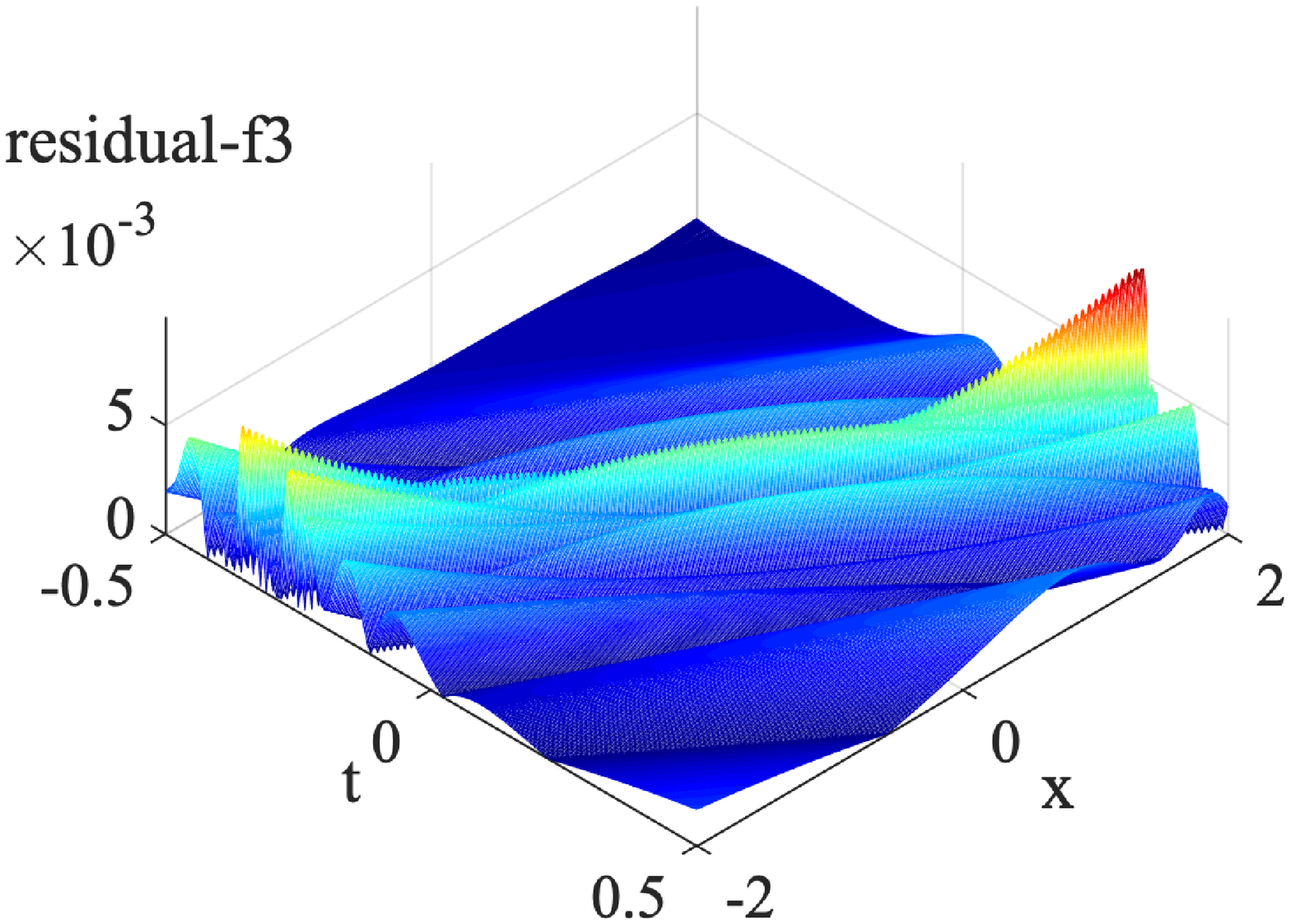}
$b$
\includegraphics[width=5.5cm,height=4.5cm]{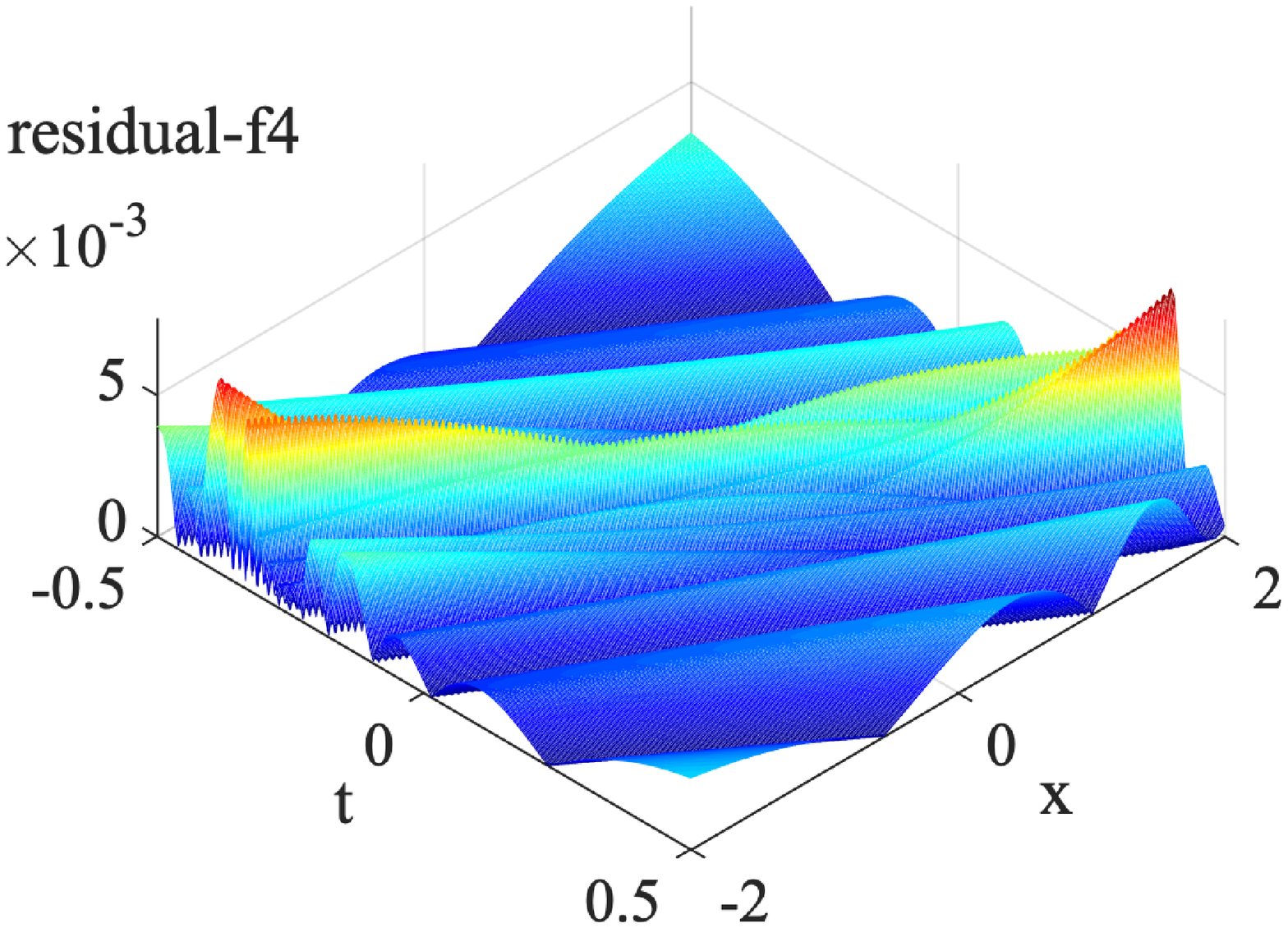}
$c$
\caption{(Color online) Data-driven solution $v(x,t)$ of the defocusing mKdV equation by Scheme \uppercase\expandafter{\romannumeral 2}: (a) The density diagrams and comparison between the predicted solutions and exact solutions at the three temporal snapshots of $v(x,t)$; (b) The three-dimensional plot of the residual corresponding to $f_3$; (c) The three-dimensional plot of the residual corresponding to $f_4$.}
\label{fig4-8}
\end{figure}

\begin{figure}[htbp]
\centering
\includegraphics[width=5.5cm,height=4.5cm]{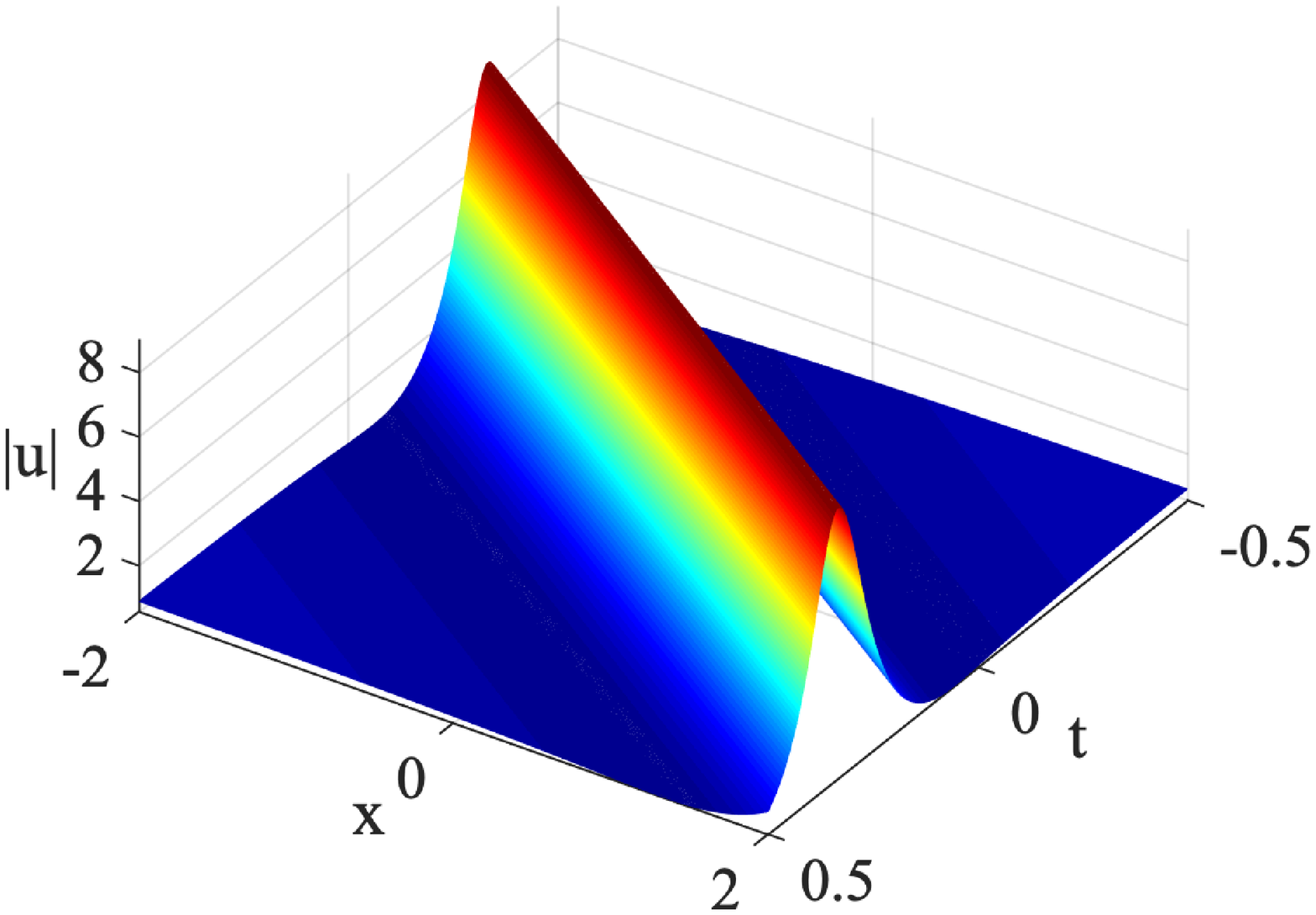}
$a$
\includegraphics[width=5.5cm,height=4.5cm]{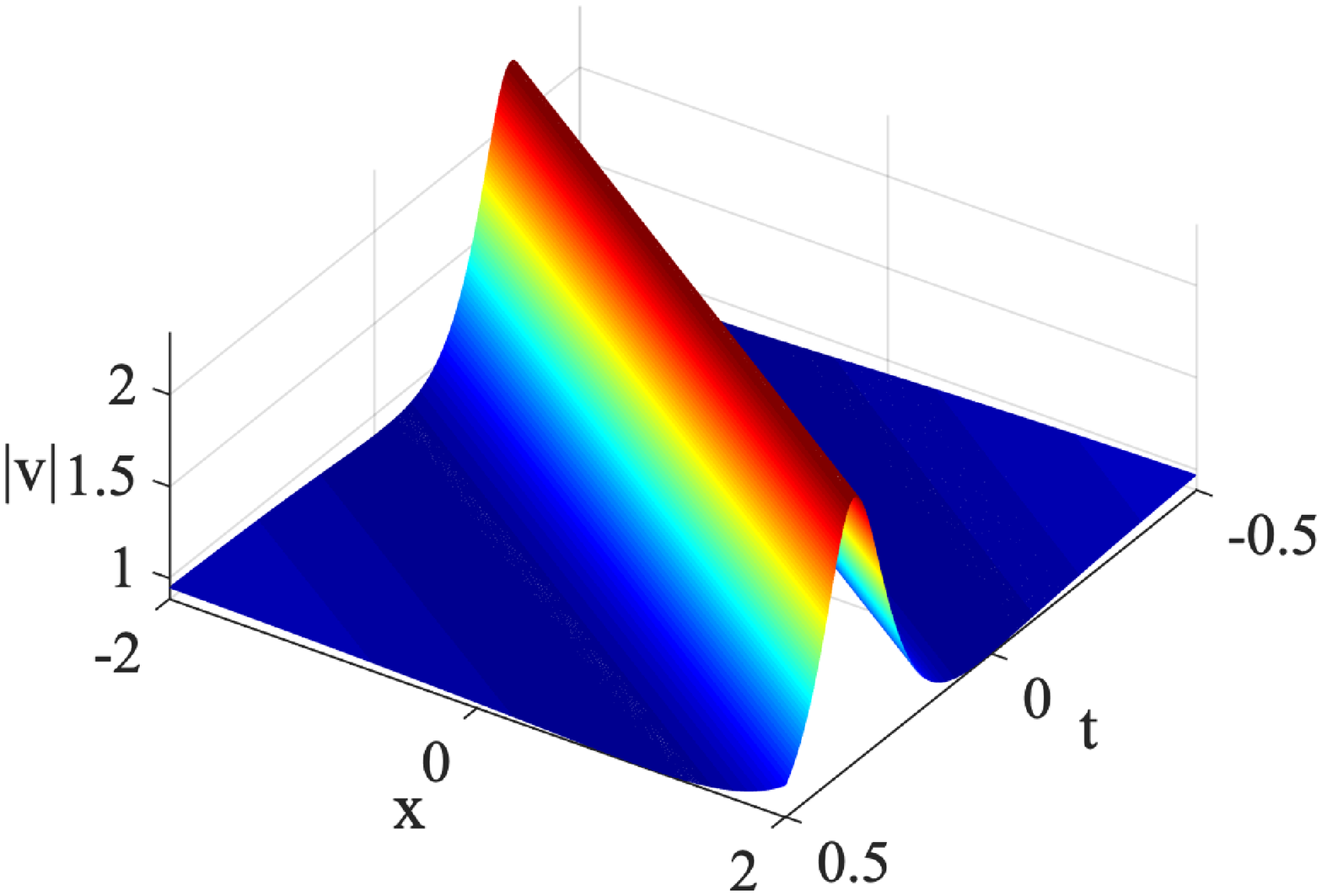}
$b$
\includegraphics[width=5.5cm,height=4.5cm]{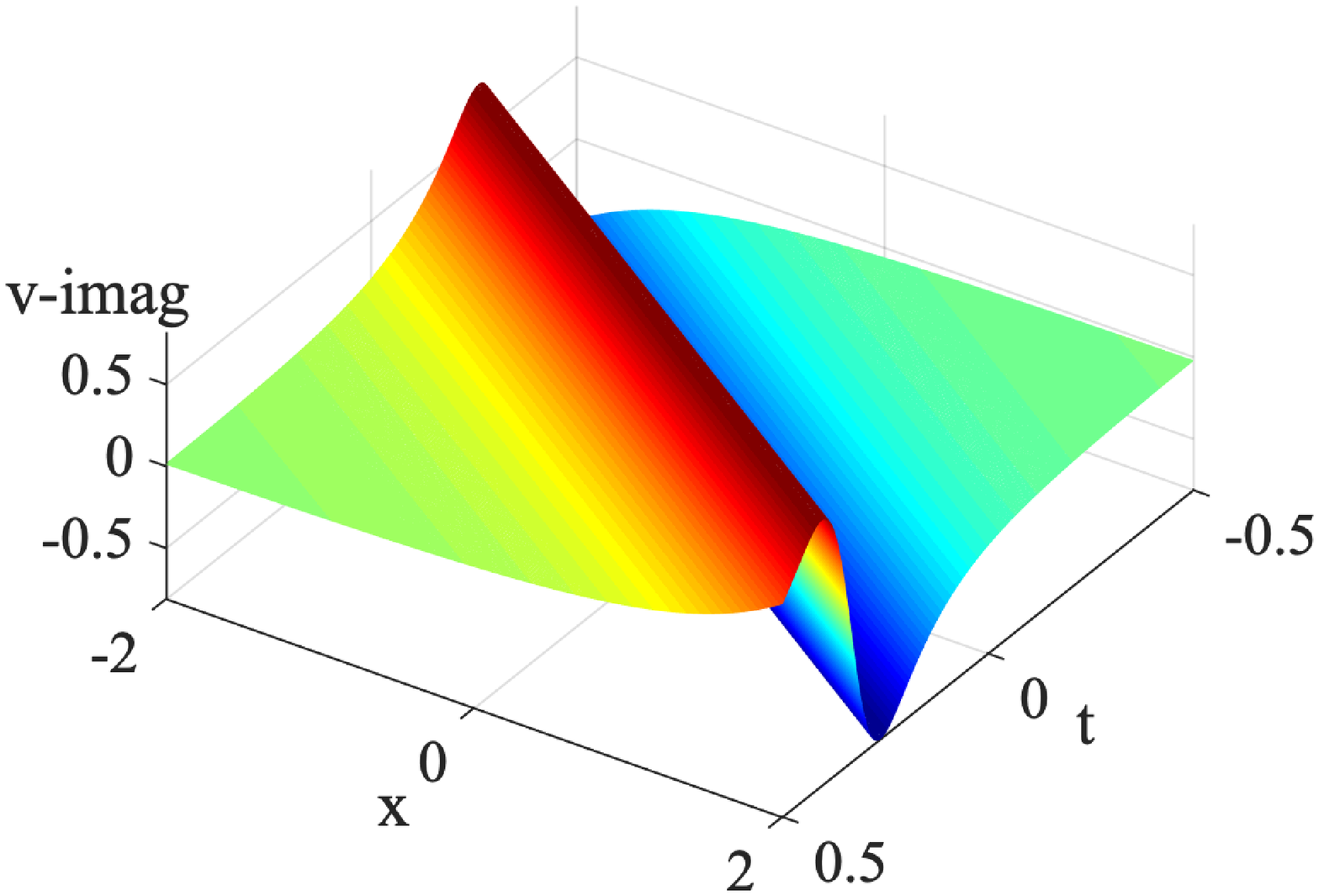}
$c$
\caption{(Color online) Data-driven solutions $u(x,t)$ of the KdV equation and $v(x,t)$ of the focusing mKdV equation by Scheme \uppercase\expandafter{\romannumeral 2}: (a) The three-dimensional plot of $|u|(x,t)$; (b) The three-dimensional plot of $|v|(x,t)$; (c) The three-dimensional plot of the imaginary part $v_i(x,t)$.}
\label{fig4-9}
\end{figure}

In step two, we take $N_g=5000$ and establish a new physics-informed neural network to obtain the data-driven solution of the focusing mKdV equation, which has 3 hidden layers with 40 neurons per hidden layer. Eventually, a new numerical solution different from \eqref{rationalfocusing} is simulated after  2194 times iterations in about 150.9557 seconds. Fig. \ref{fig4-8} (a) explicitly reveals the difference of wave patterns between the exact solution and the learned one. What's more, the three-dimensional plots of $|u|(x,t),|v|(x,t)$ and $v_i(x,t)$ are shown in Fig. \ref{fig4-9} and $v_i$ is not close to 0 where the wave height is large in Fig. \ref{fig4-9}(c), which means that this new numerical solution is not (or is not close to) the real-valued one. Since we wonder if the data-driven one is the solution of the focusing mKdV equation, the 3D plots of the residuals corresponding to $f_3$ ($|f_3(x^i,t^i)|$) and $f_4$ ($|f_4(x^i,t^i)|$) are displayed in Fig. \ref{fig4-8} (b)-(c). Evidently, the residuals are kept small below the order of magnitude of $10^{-3}$ in the given region and the mean squared error corresponding to $f_3$ ($MSE_{G_1}$) and $f_4$ ($MSE_{G_2}$) are 2.439725e-06 and 2.865712e-06. It provides a powerful evidence for the proof of the correctness of this new numerical solution.

Considering that we assume $v(x,t)$ is a complex-valued function since the known initial-boundary data of $u$ is complex-valued, which may be the reason why the above numerical solution of the focusing mKdV equation is different from \eqref{rationalfocusing}, we wonder whether the real-valued first-order rational solution \eqref{rationalfocusing} can be got if we assume $v(x,t)$ is real-valued in step two.

\begin{figure}[htbp]
\centering
\includegraphics[width=5.5cm,height=4.5cm]{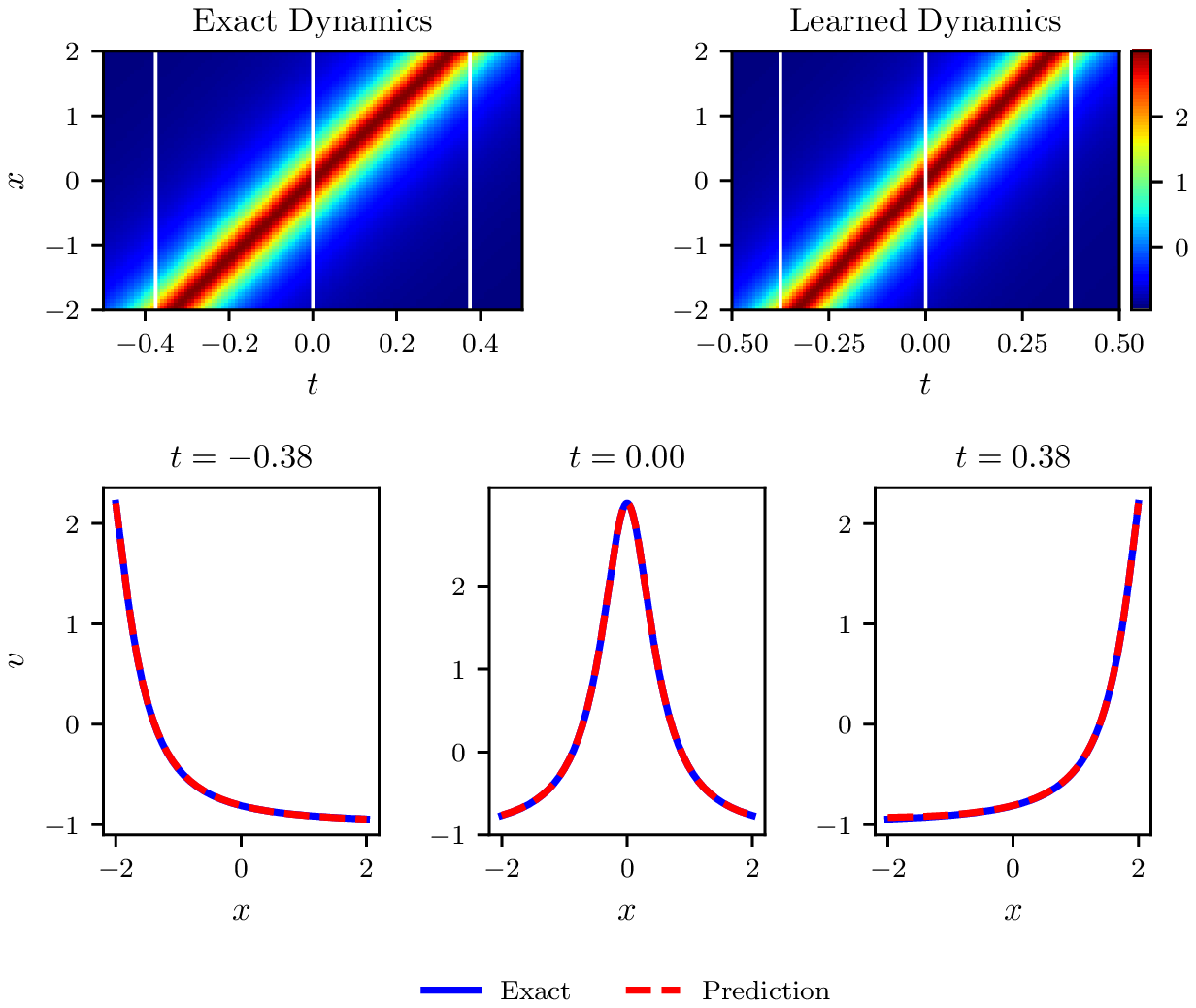}
$a$
\includegraphics[width=5.5cm,height=4.5cm]{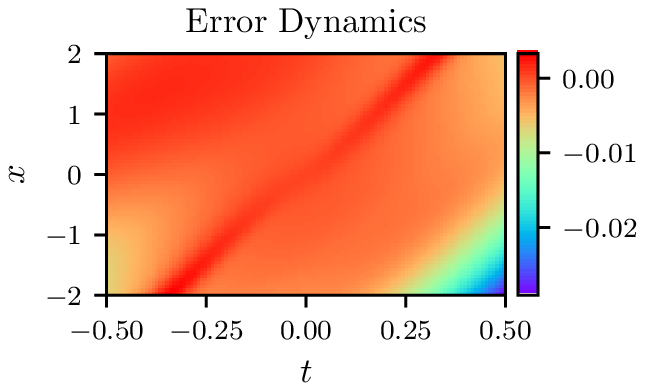}
$b$
\includegraphics[width=5.5cm,height=4.5cm]{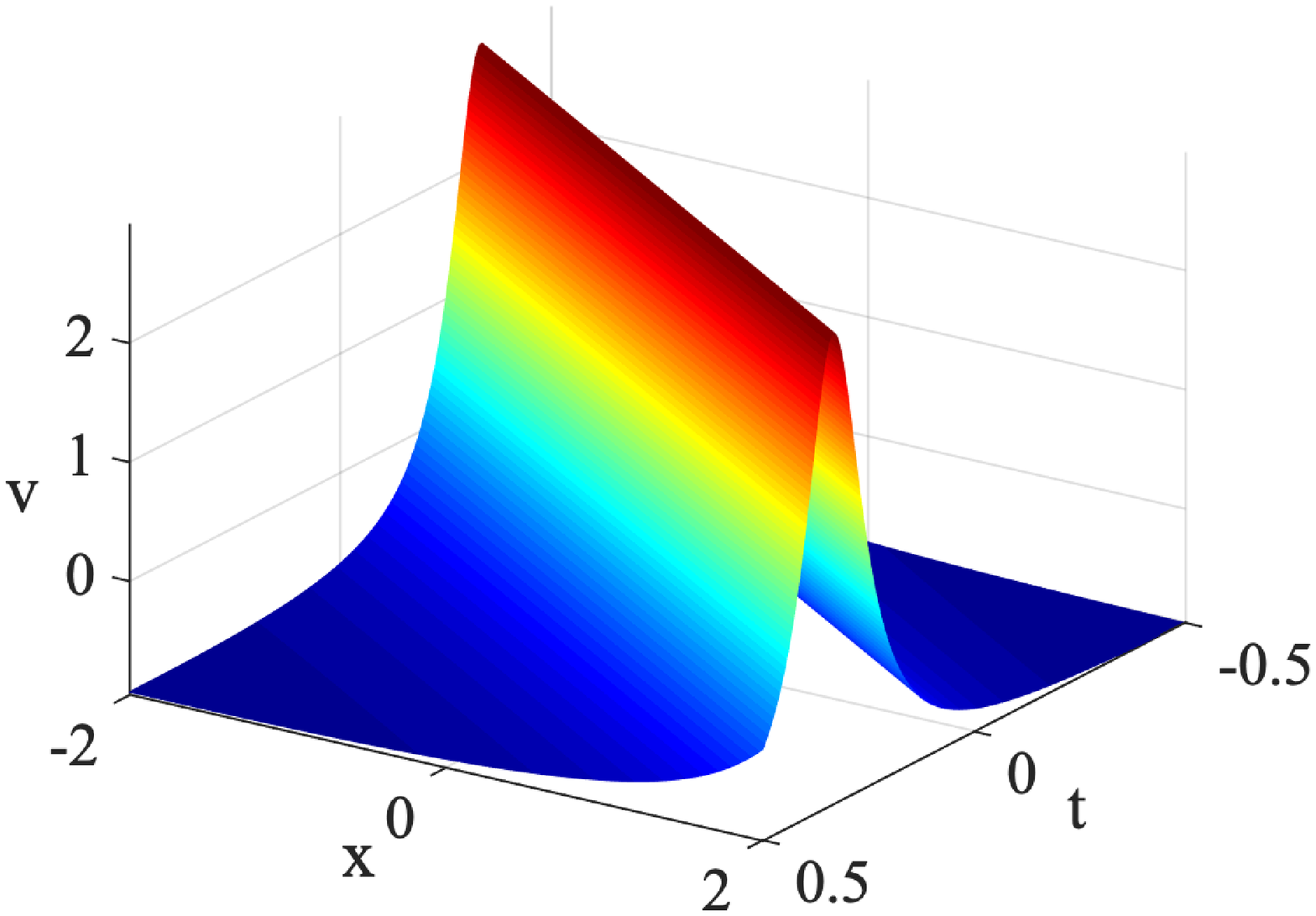}
$c$
\caption{(Color online) First-order rational solution $v(x,t)$ of the focusing mKdV equation by Scheme \uppercase\expandafter{\romannumeral 2}: (a) The density diagrams and comparison between the predicted solutions and exact solutions at the three temporal snapshots of $v(x,t)$; (b) The error density diagram of $v(x,t)$;(c) The three-dimensional plot of $v(x,t)$.}
\label{fig4-10}
\end{figure}

On the basis of step one mentioned above, the mean squared error loss of step two is changed into
\begin{equation}
MSE_2=MSE_{G}+MSE_{M_1}+MSE_{M_2},
\end{equation}
where
\begin{equation}
MSE_{G}=\frac{1}{N_g}\sum^{N_g}_{i=1}|G(x_{g}^i,t_{g}^i)|^2,
\end{equation}
\begin{equation}
MSE_{M_1}=\frac{1}{N_g}\sum^{N_g}_{i=1}|M_1(x_g^i,t_g^i)|^2,
\end{equation}
\begin{equation}
MSE_{M_2}=\frac{1}{N_g}\sum^{N_g}_{i=1}|M_2(x_g^i,t_g^i)|^2,
\end{equation}
and $G$, $M_1$ and $M_2$ respectively correspond to the following governing equations
\begin{align}
&f_3:=G=v_t+6v^2 v_x+v_{xxx},\\
&f_4:=M_1=u_r-v^2,\\
&f_5:=M_2=u_i-v_x.	
\end{align}
Then we also take $N_g=5000$ and establish a 3-layer feedforward neural network with 40 neurons per hidden layer. After 899 times iterations in about 30.6301 seconds, the first-order rational solution of the focusing mKdV equation is successfully simulated and the relative $\mathbb{L}_2$ error of $v$ is 3.500186e-03. Besides, the density diagrams, comparison between the predicted solutions and exact solutions, the error density diagram as well as the 3D plot are displayed in Fig. \ref{fig4-10}.

Finally, details of data-driven solutions of the KdV equation and the focusing mKdV equation are listed in Table \ref{table-case3}.

\begin{table}[H]
\caption{Data-driven solutions in Case 4.3 by Scheme \uppercase\expandafter{\romannumeral 2}: iteration times, elapsed time, relative $\mathbb{L}_2$ errors, mean squared errors.}
\label{table-case3}
\centering\begin{tabular}{cc|c|c}
\bottomrule
\multicolumn{2}{c|}{\multirow{6}{*}{\textbf{Step One}}}                   & Hidden layers-Neurons & 8-40 \\ \cline{3-4}
\multicolumn{2}{c|}{}                                              & Iteration times  &2617  \\ \cline{3-4}
\multicolumn{2}{c|}{}                                              & Elapsed time (s) & 273.2680  \\ \cline{3-4}
\multicolumn{2}{c|}{}                                              & $u_r(x,t)$ & 2.772086e-03 \\ \cline{3-4}
\multicolumn{2}{c|}{}                                              & $u_i(x,t)$ & 1.914695e-03  \\ \cline{3-4}
\multicolumn{2}{c|}{}                                              & $|u|(x,t)$ & 2.262575e-03 \\ \hline
\multicolumn{1}{c|}{\multirow{9}{*}{\textbf{Step Two}}} & \multirow{5}{*}{$v$ is complex-valued} & Hidden layers-Neurons & 3-40  \\ \cline{3-4}
\multicolumn{1}{c|}{}                          &                   & Iteration times & 2194 \\ \cline{3-4}
\multicolumn{1}{c|}{}                          &                   & Elapsed time (s) & 150.9557  \\ \cline{3-4}
\multicolumn{1}{c|}{}                          &                   & $MSE_{G_1}$ & 2.439725e-06 \\ \cline{3-4}
\multicolumn{1}{c|}{}                          &                   & $MSE_{G_2}$ & 2.865712e-06  \\ \cline{2-4}
\multicolumn{1}{c|}{}                          & \multirow{4}{*}{$v$ is real-valued} & Hidden layers-Neurons & 2-40 \\ \cline{3-4}
\multicolumn{1}{c|}{}                          &                   & Iteration times &899  \\ \cline{3-4}
\multicolumn{1}{c|}{}                          &                   & Elapsed time (s) & 30.6301  \\ \cline{3-4}
\multicolumn{1}{c|}{}                          &                   & $v(x,t)$ & 3.500186e-03 \\ \toprule
\end{tabular}
\end{table}

\section{Analysis and comparison concerning advantages and disadvantages of two schemes}\label{comparison}

\subsection{Performance comparison of two schemes}\label{5.1}
\quad

In this part, we carry out numerical experiments of different cases above (Case 3.1, Case 4.1 and Case 4.2) to investigate and compare the performances of Scheme \uppercase\expandafter{\romannumeral 1} and Scheme \uppercase\expandafter{\romannumeral 2}.

Given that the most noteworthy feature of our proposed method is that we can simply use small amounts of initial-boundary data of a solution $u$ of a certain nonlinear equation to obtain the data-driven solution $v$ of another evolution equation with the aid of PINNs and Miura transformations, the accuracy of the data-driven solution $v$ is undoubtedly what we focus on and is of much higher concern than the accuracy of $u$ here.

The effect of changes of the number of hidden layers and the number of neurons per hidden layer is considered here because they are main factors affecting the training results of PINNs. Meanwhile, the variable-controlling method is adopted and the structure of neural networks of Scheme \uppercase\expandafter{\romannumeral 1} should be kept the same as that in step one of Scheme \uppercase\expandafter{\romannumeral 2}, while the number of hidden layers and neurons in step two can be seen as additional hyper-parameters to be optimized and arbitrarily assigned since Scheme \uppercase\expandafter{\romannumeral 2} has two steps and the network structure is more flexible.

\textbf{(1) Data-driven solution of the defocusing mKdV solution (Case 3.1)}

Since the obtained numerical solution of Case 3.1 is different from the exact one, we conduct numerical experiments of this case and evaluate the performance of two schemes in terms of accuracy ($MSE_G$), efficiency (the elapsed time) and diversity of solutions.

$\bullet$ \textbf{Changes of the number of hidden layers}

The number of hidden layers changes from 2 to 20 with step size 2 and each hidden layer has 40 neurons. In all numerical tests, invariant hyper-parameters are: $N_u=200, N_f=5000, N_g=2000$.

\begin{figure}[htbp]
\centering
\includegraphics[width=7cm,height=5cm]{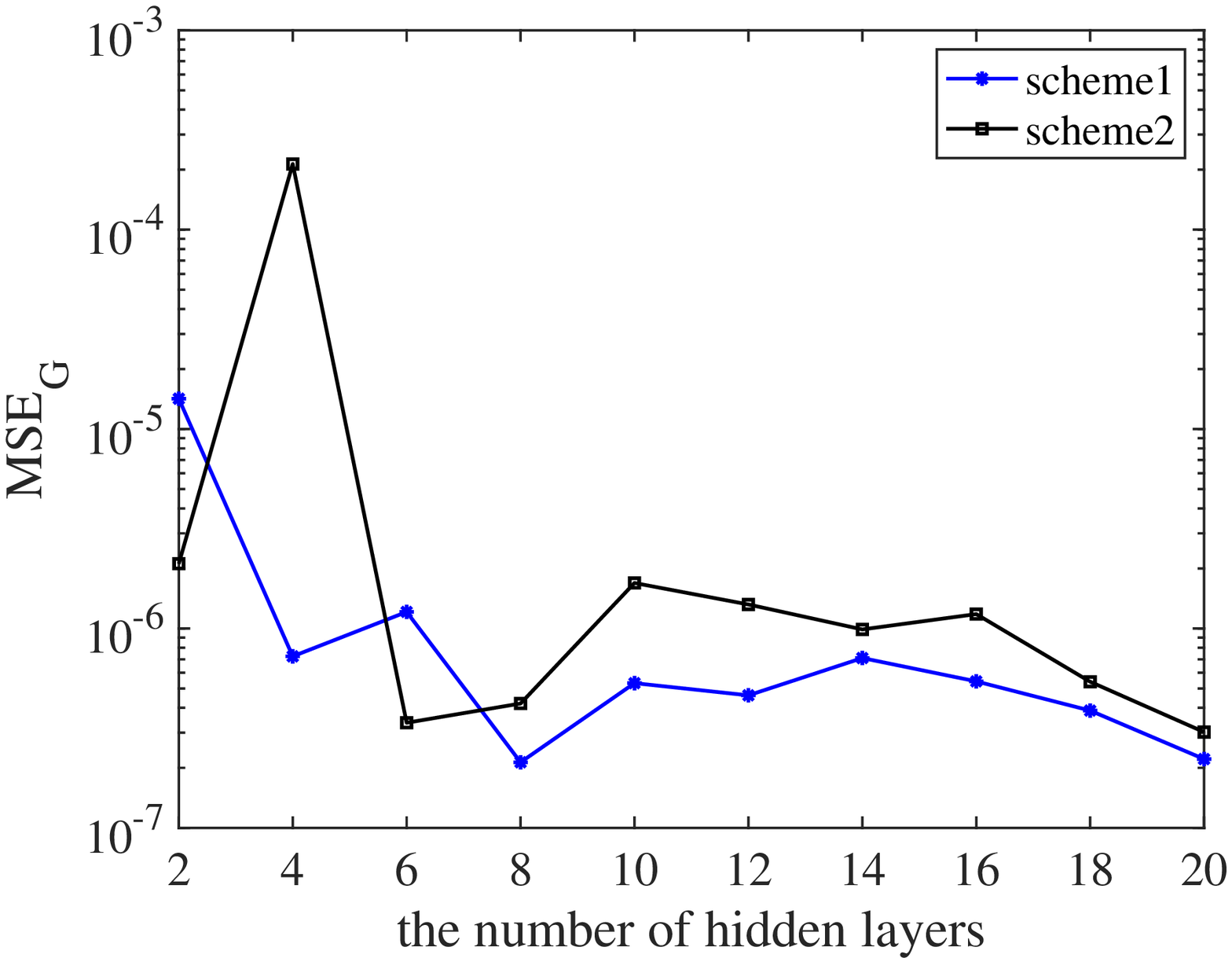}
$a$
\includegraphics[width=7cm,height=5cm]{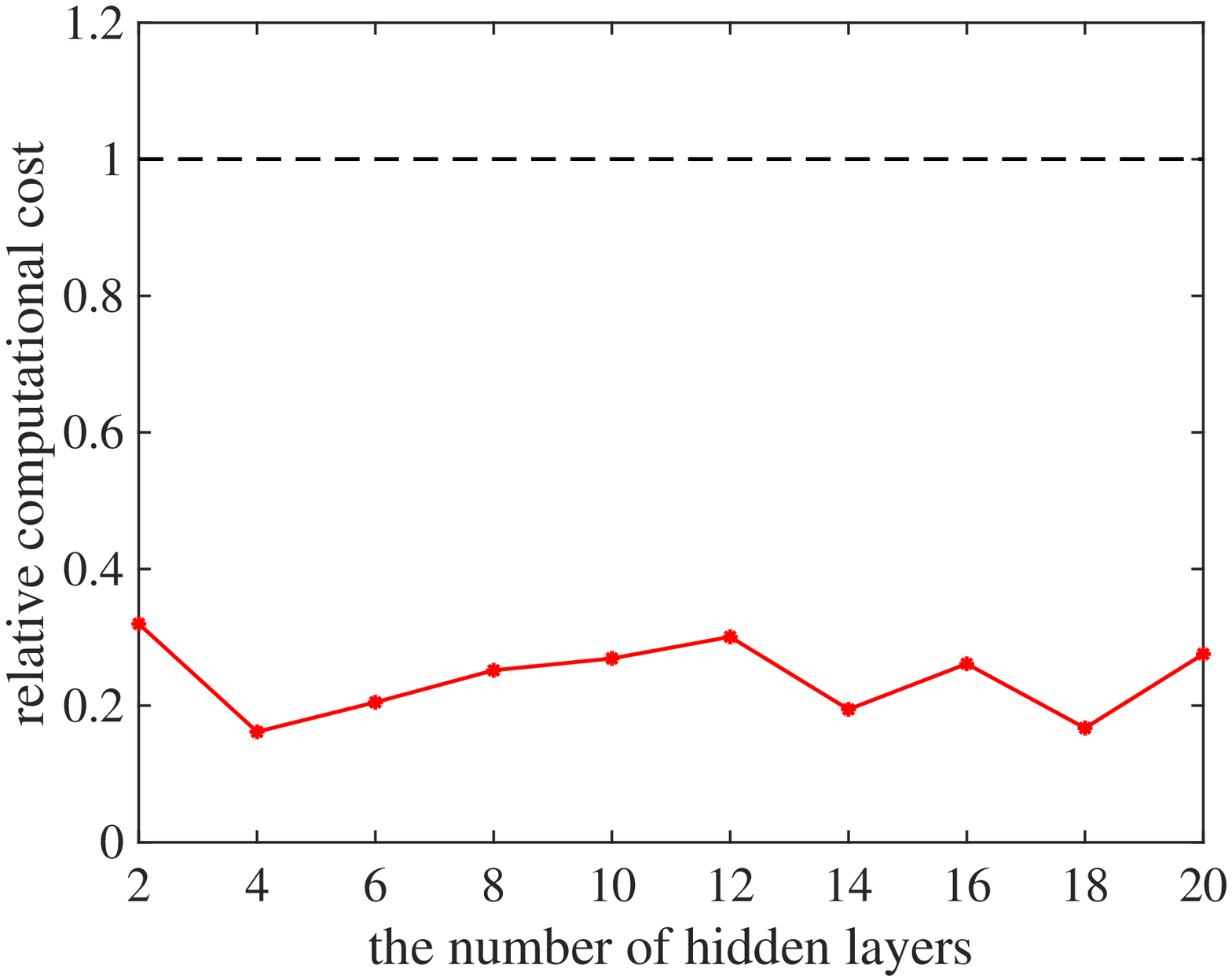}
$b$
\caption{(Color online) Data-driven solution of the defocusing mKdV equation: comparison of two schemes by using different numbers of hidden layers (a) The mean squared errors $MSE_G$; (b) The relative computational costs.}
\label{fig5-1}
\end{figure}

We perform two groups of ten independent numerical experiments corresponding to two schemes and the numerical results are shown in Table \ref{tableA-1-1}-Table \ref{tableA-1-2} in Appendix A. For the sake of comparison of accuracy, we use the mean squared error corresponding to $f_3$ ($MSE_G$)
\begin{align}
MSE_{G}&=\frac{1}{N}\sum^{N}_{i=1}|G(x^i,t^i)|^2, N=N_x \times N_t,\\
f_3:&=G=v_t-6v^2 v_x+v_{xxx},	
\end{align}
of $N_x \times N_t=513 \times 201$ grid points to determine whether the numerical solution satisfies the equation since the data-driven solution may be different from the solution \eqref{E3-4} of the defocusing mKdV equation. To compare the efficiency, the relative computational cost (RCC) is defined by the elapsed time of Scheme \uppercase\expandafter{\romannumeral 2} divided by that of Scheme \uppercase\expandafter{\romannumeral 1}
 \begin{align}
 RCC=\frac{t_{Scheme
 \uppercase\expandafter{\romannumeral 2}}}{t_{Scheme \uppercase\expandafter{\romannumeral 1}}}.	
 \end{align}
The mean squared error of $N_x \times N_t=513 \times 201$ grid points ($MSE_G$) and the relative computational costs are plotted in Fig. \ref{fig5-1}, and Table \ref{table5-1-1} presents average $MSE_G$ and elapsed time of ten experiments by using different numbers of hidden layers.

\begin{table}[htbp]
\caption{Data-driven solution of the defocusing mKdV equation: average mean squared errors and elapsed time by using different numbers of hidden layers.}
\label{table5-1-1}
\centering
\begin{tabular}{c|c|c}
\bottomrule
Method (neurons=40)                  & \textbf{Scheme I} & \textbf{Scheme II} \\ \hline
Average $MSE_G$      & 1.920152E-06         & 2.223562E-05          \\ \hline
Average elapsed time (s) & 727.75602         & 172.807473          \\ \toprule
\end{tabular}
\end{table}

As is shown above, the mean squared error $MSE_G$ of Scheme \uppercase\expandafter{\romannumeral 1} is less than that of Scheme \uppercase\expandafter{\romannumeral 2} in most cases and the average mean squared error of Scheme \uppercase\expandafter{\romannumeral 1} outperforms Scheme \uppercase\expandafter{\romannumeral 2} by an order of magnitude, which implies that the data-driven solution derived by Scheme \uppercase\expandafter{\romannumeral 1} satisfies the defocusing mKdV equation more ideally. From Table \ref{tableA-1-2} in Appendix A, it can also be seen that even the solution $\widehat{u}$ of the KdV equation cannot be simulated well in the second experiment and thus the numerical solution $\widehat{v}$ of the defocusing mKdV equation based on $\widehat{u}$ is not reliable. The failure of this experiment  is the main reason which leads to worse accuracy of Scheme \uppercase\expandafter{\romannumeral 2} and the $MSE_G$ values of two schemes are relatively similar in other experiments. However, Scheme \uppercase\expandafter{\romannumeral 2} has greater advantage on efficiency compared with Scheme \uppercase\expandafter{\romannumeral 1} according to the relative computational costs and the average elapsed time.

Besides, a variety of data-driven solutions of the defocusing mKdV equation are exhibited in Table \ref{table5-1-2}, where the contents in brackets of the first column 'Hidden layers-Neurons' denote the number of hidden layers and neurons per hidden layer in step two of Scheme \uppercase\expandafter{\romannumeral 2}. By using different numbers of hidden layers with 40 neurons per hidden layer, three types of data-driven solutions (bright soliton, dark soliton and kink-bell type solution) can be obtained by Scheme \uppercase\expandafter{\romannumeral 1} while two types (kink and kink-bell type solution) by Scheme \uppercase\expandafter{\romannumeral 2}. Both two schemes exhibit their good performance in diversity of solutions.

\begin{table}[htbp]
\caption{Data-driven solutions of the defocusing mKdV equation: types of solutions, density diagrams and three-dimensional plots by using different numbers of hidden layers.}
\label{table5-1-2}
\centering
\begin{tabular}{c|ccc}
\bottomrule
\multicolumn{1}{c|}{\begin{tabular}[c]{@{}c@{}}Hidden layers\\ -Neurons\end{tabular}} & Type of solution & Density diagram & 3D plot \\ \hline
\multicolumn{1}{c|}{\begin{tabular}[c]{@{}c@{}}Scheme \uppercase\expandafter{\romannumeral 1}\\ 2-40/6-40/10-40/12-40/\\ 14-40/16-40/20-40 \end{tabular}}      & bright soliton                 &  \multicolumn{1}{c}{\begin{tabular}[c]{@{}c@{}}\includegraphics[width=4cm,height=3cm]{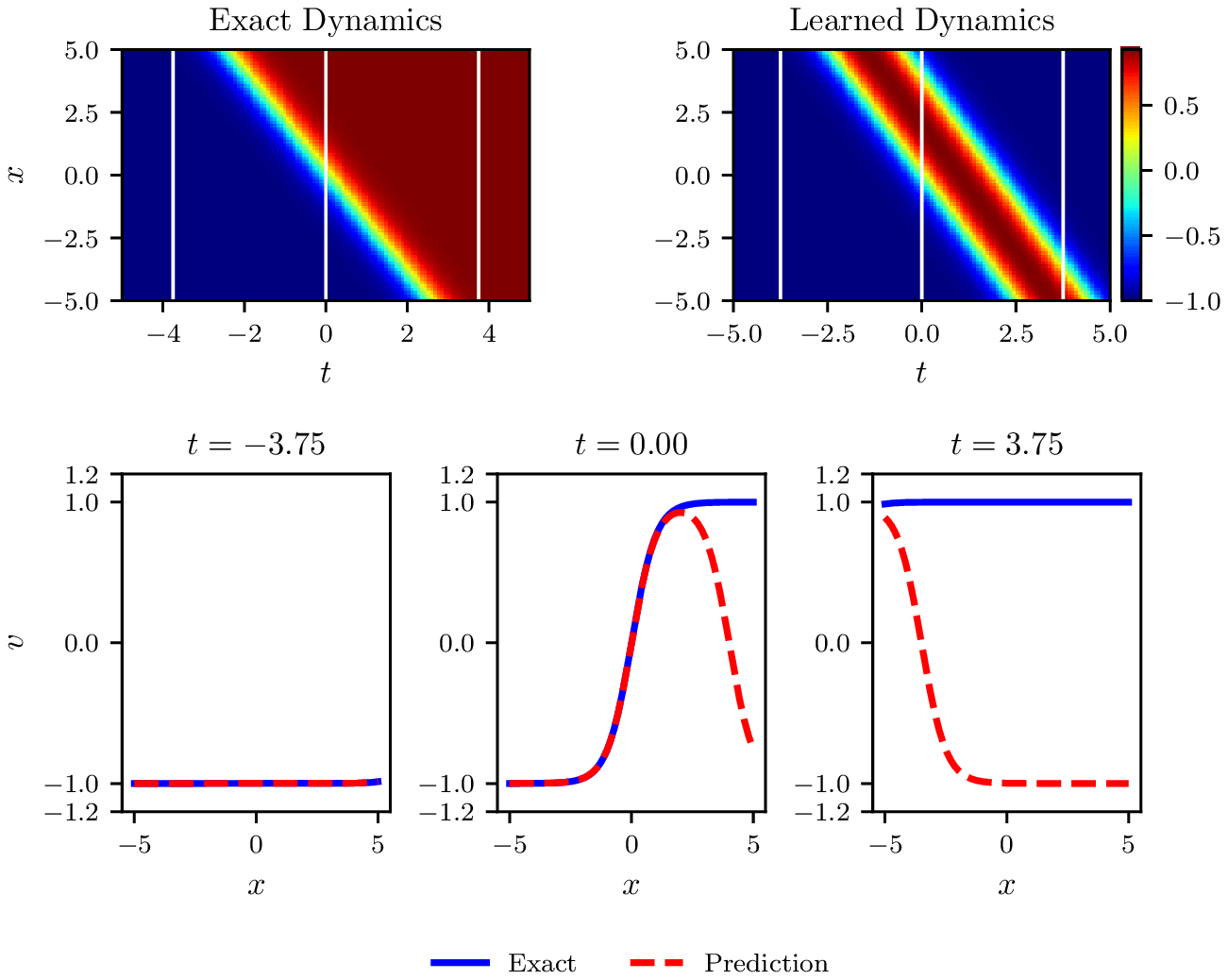} \end{tabular}}               &  \multicolumn{1}{c}{\begin{tabular}[c]{@{}c@{}}\includegraphics[width=4cm,height=3cm]{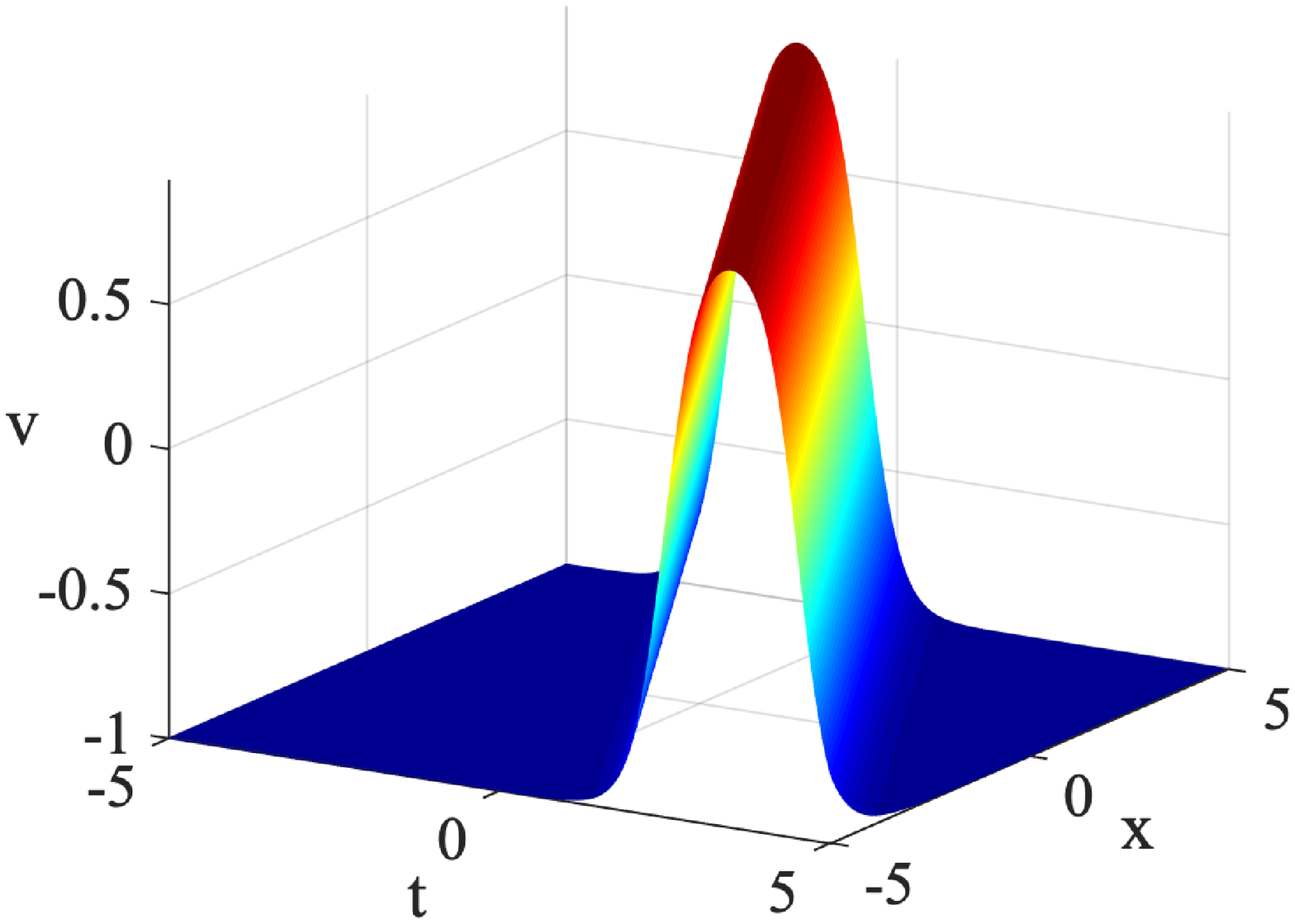} \end{tabular}}       \\
\multicolumn{1}{c|}{\begin{tabular}[c]{@{}c@{}}Scheme \uppercase\expandafter{\romannumeral 1}\\ 8-40/18-40 \end{tabular}}       & dark soliton                 &  \multicolumn{1}{c}{\begin{tabular}[c]{@{}c@{}}\includegraphics[width=4cm,height=3cm]{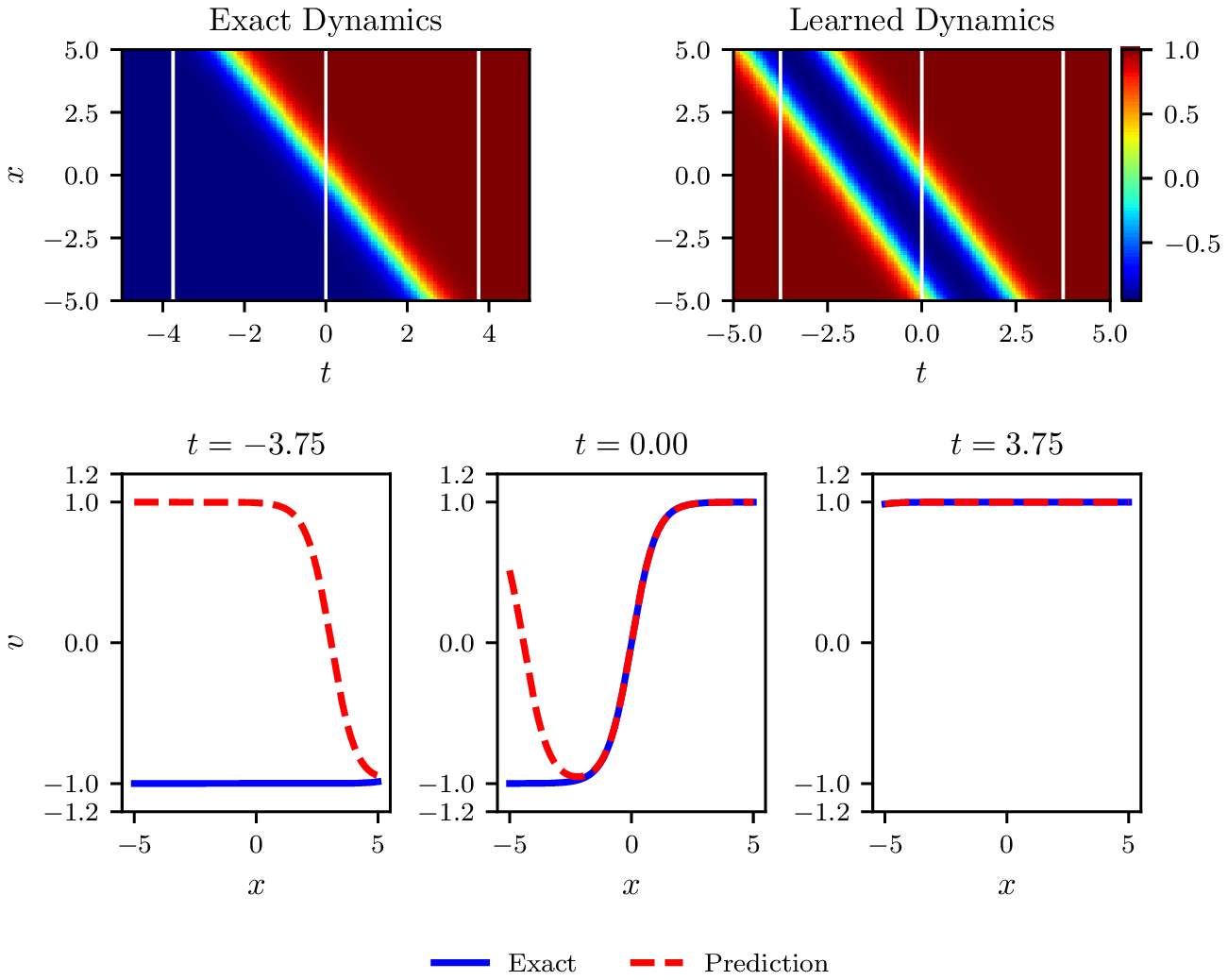} \end{tabular}}                              &  \multicolumn{1}{c}{\begin{tabular}[c]{@{}c@{}}\includegraphics[width=4cm,height=3cm]{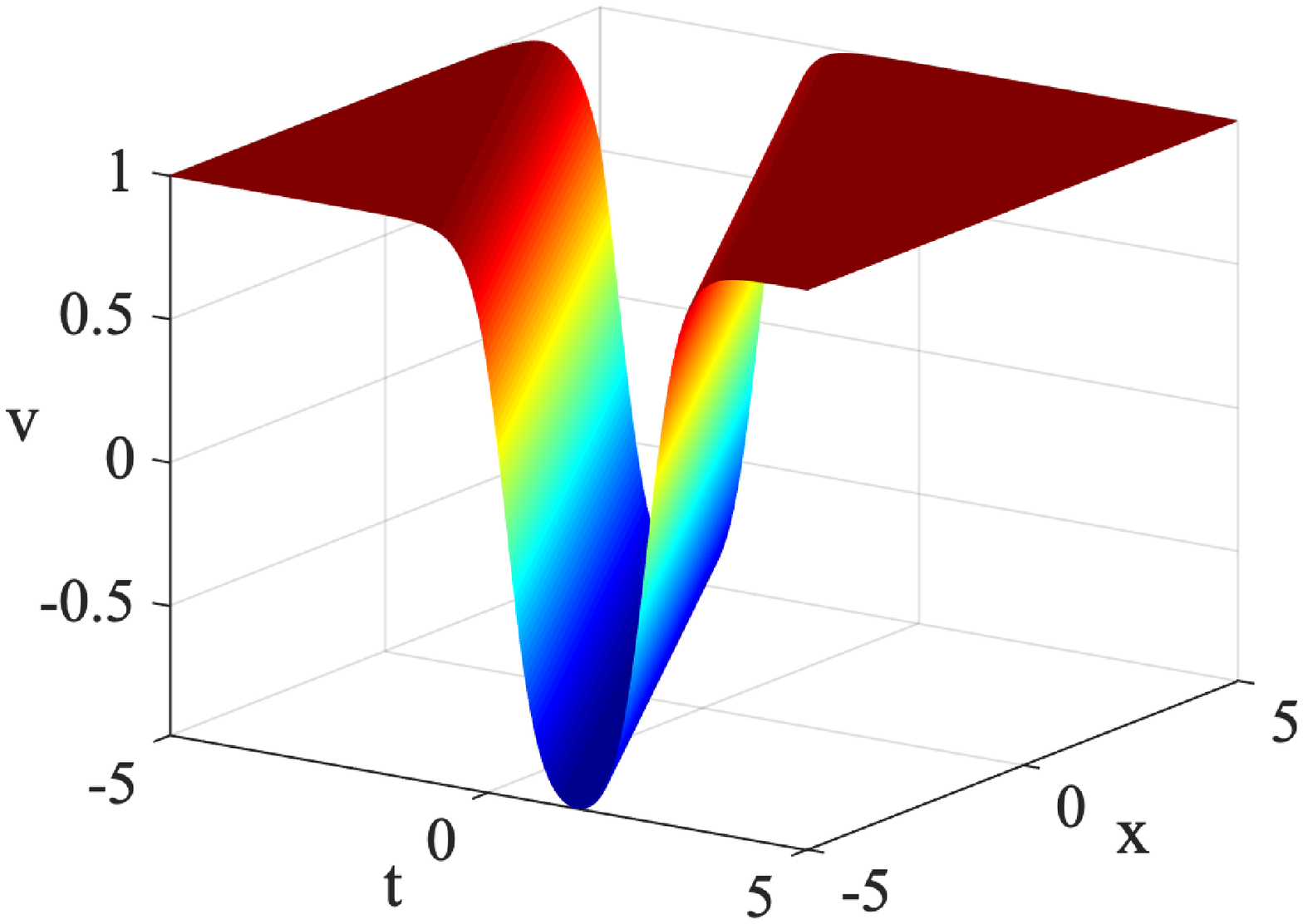} \end{tabular}}                      \\
\multicolumn{1}{c|}{\begin{tabular}[c]{@{}c@{}}Scheme \uppercase\expandafter{\romannumeral 1}\\ 4-40 \end{tabular}}       & kink-bell type solution                 & \multicolumn{1}{c}{\begin{tabular}[c]{@{}c@{}}\includegraphics[width=4cm,height=3cm]{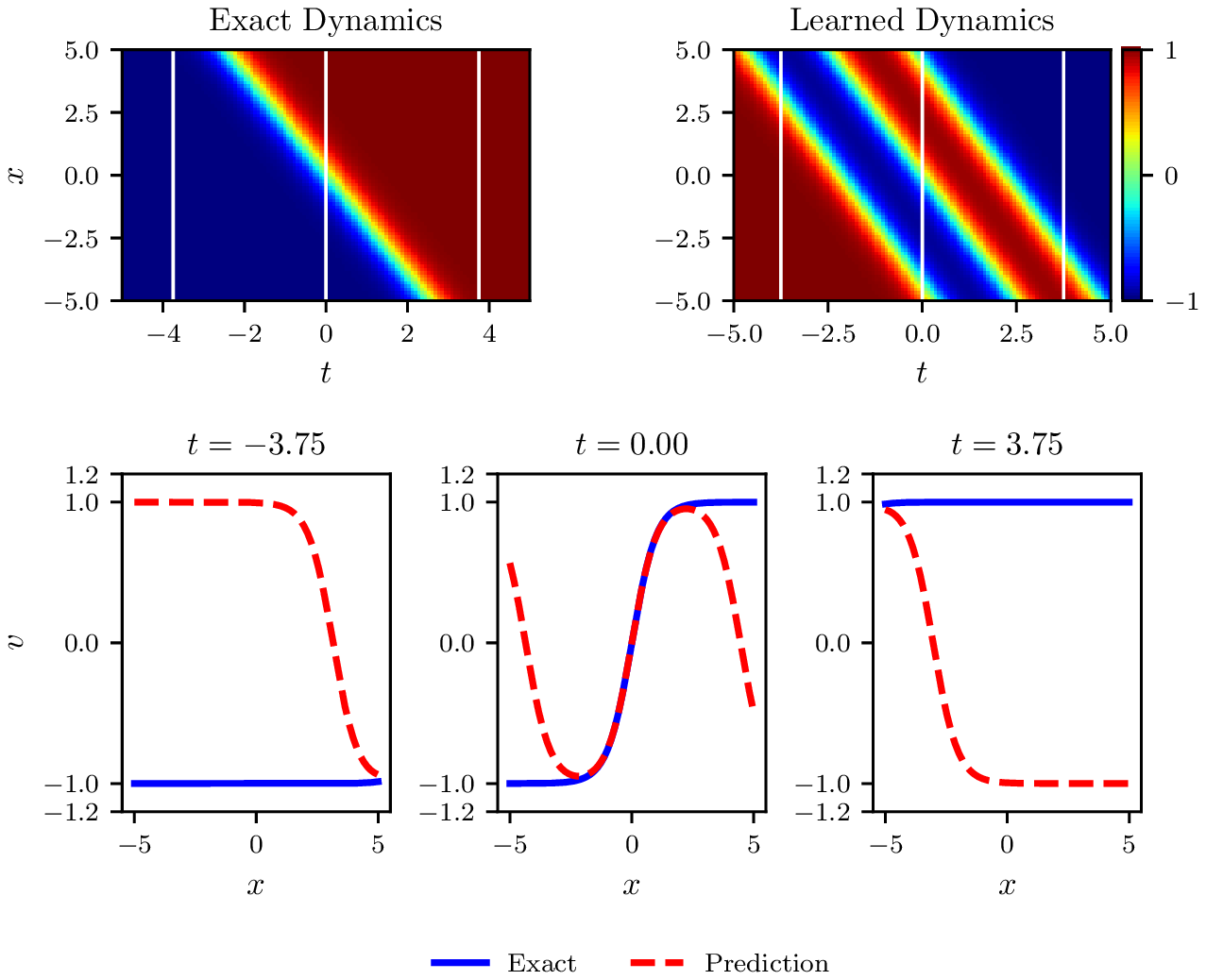} \end{tabular}}                               & \multicolumn{1}{c}{\begin{tabular}[c]{@{}c@{}}\includegraphics[width=4cm,height=3cm]{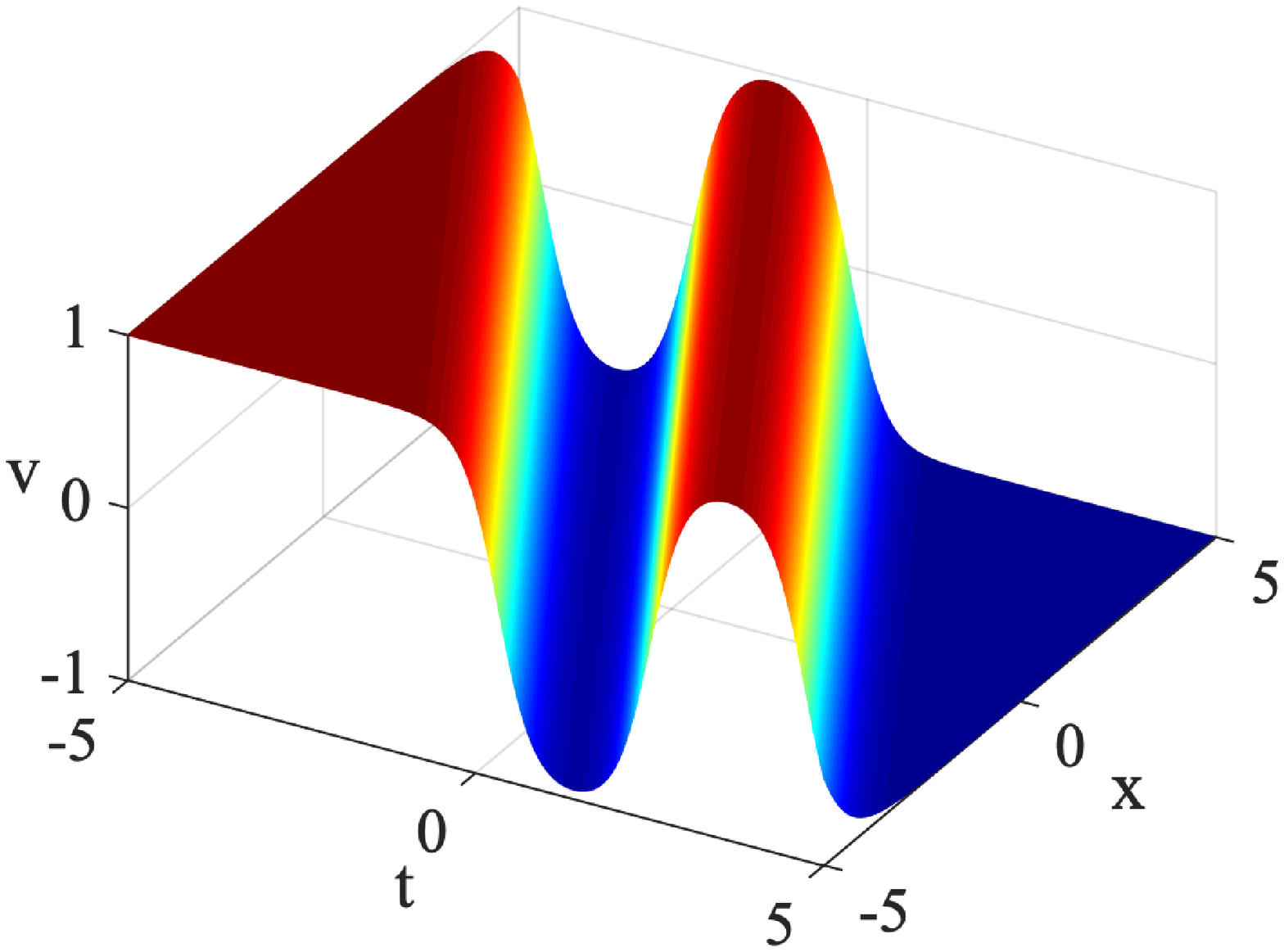} \end{tabular}}                       \\
\multicolumn{1}{c|}{\begin{tabular}[c]{@{}c@{}}Scheme \uppercase\expandafter{\romannumeral 2}\\ 16-40(2-40)/18-40(2-40)/\\ 20-40(2-40) \end{tabular}}       & kink                  &\multicolumn{1}{c}{\begin{tabular}[c]{@{}c@{}}\includegraphics[width=4cm,height=3cm]{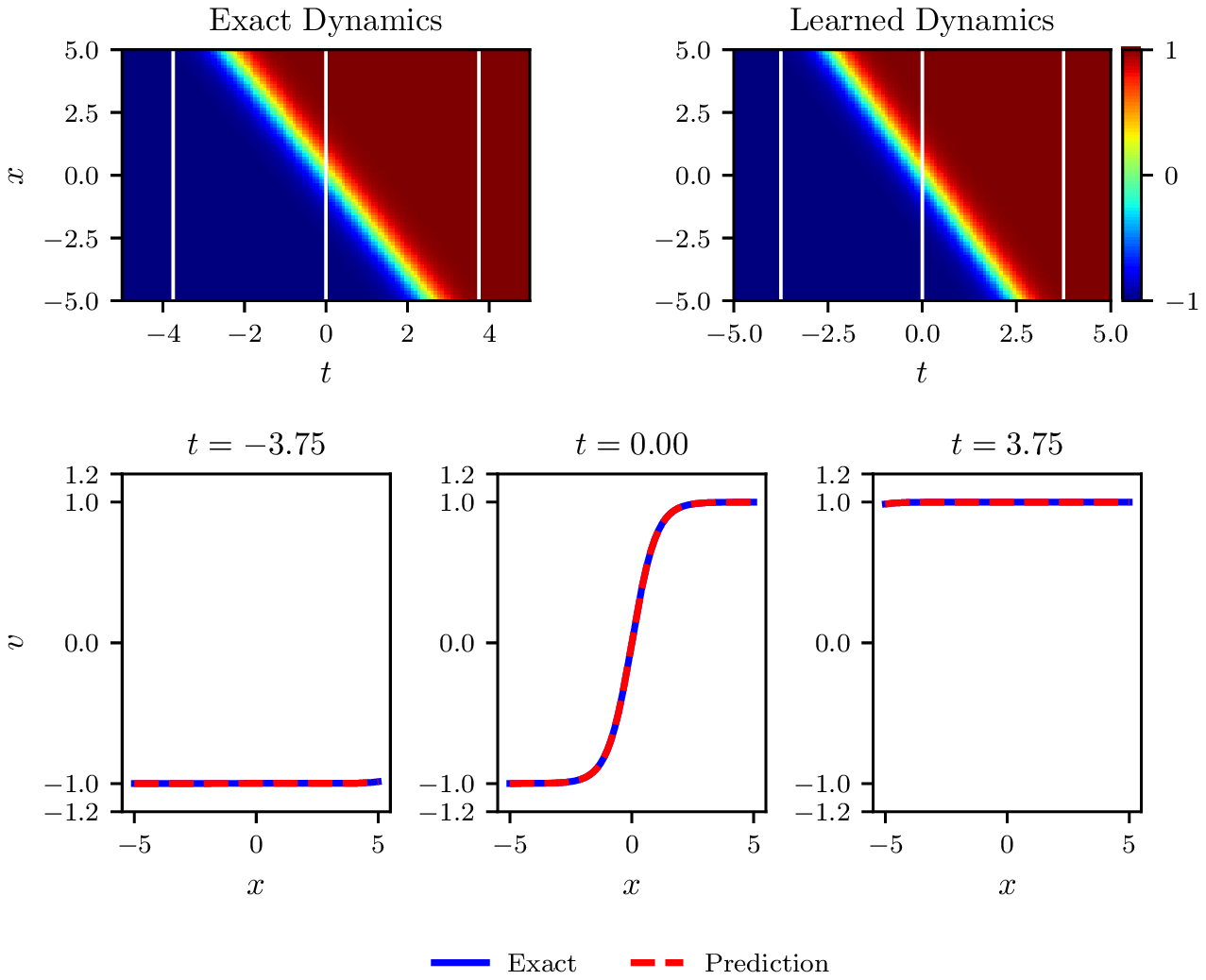} \end{tabular}}                                & \multicolumn{1}{c}{\begin{tabular}[c]{@{}c@{}}\includegraphics[width=4cm,height=3cm]{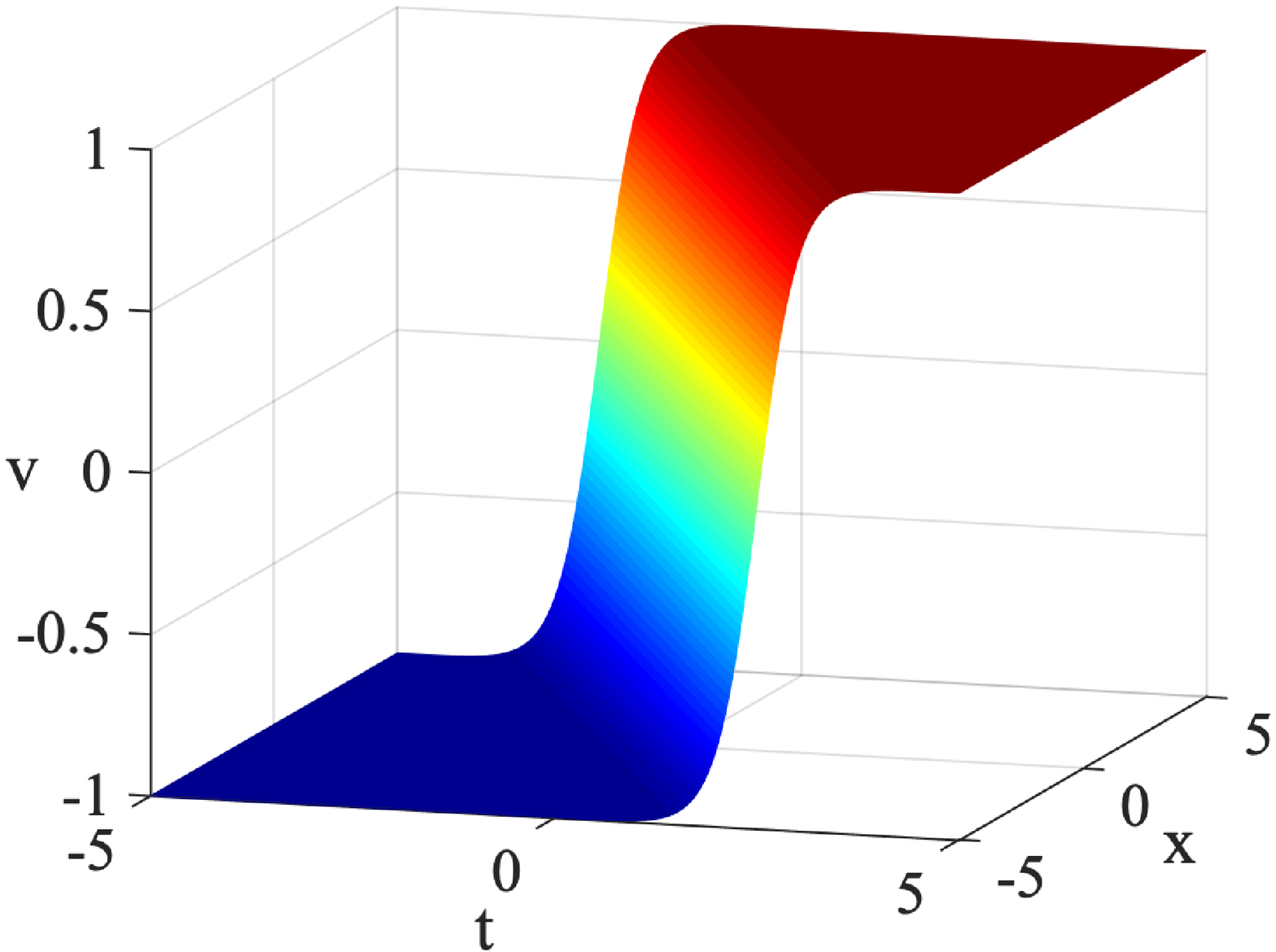} \end{tabular}}                       \\
\multicolumn{1}{c|}{\begin{tabular}[c]{@{}c@{}}Scheme \uppercase\expandafter{\romannumeral 2}\\ 2-40(2-40)/6-40(2-40)/\\ 8-40(2-40)/10-40(2-40)/\\ 12-40(2-40)/14-40(2-40) \end{tabular}}       & kink-bell type solution                 &\multicolumn{1}{c}{\begin{tabular}[c]{@{}c@{}}\includegraphics[width=4cm,height=3cm]{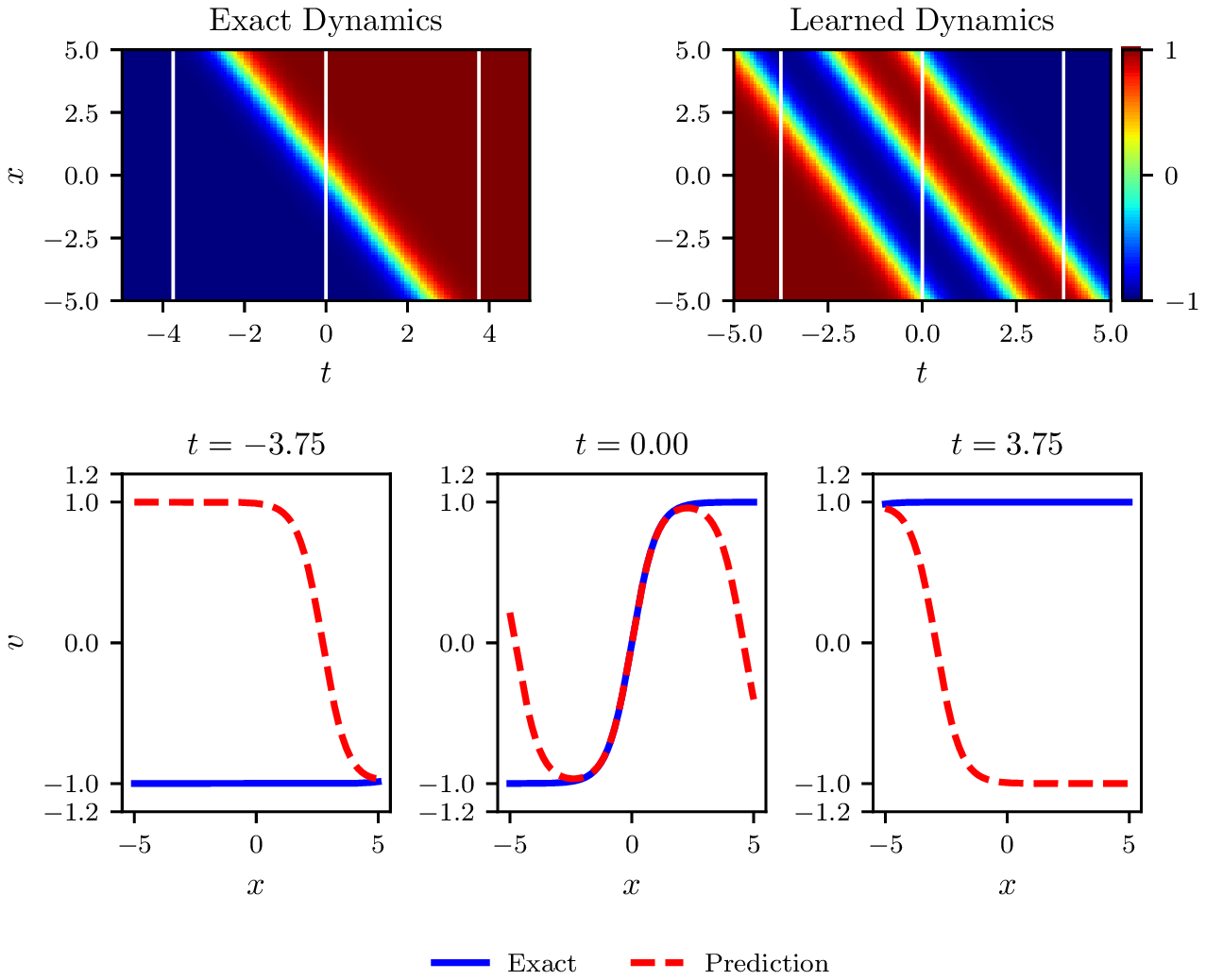} \end{tabular}}                                & \multicolumn{1}{c}{\begin{tabular}[c]{@{}c@{}}\includegraphics[width=4cm,height=3cm]{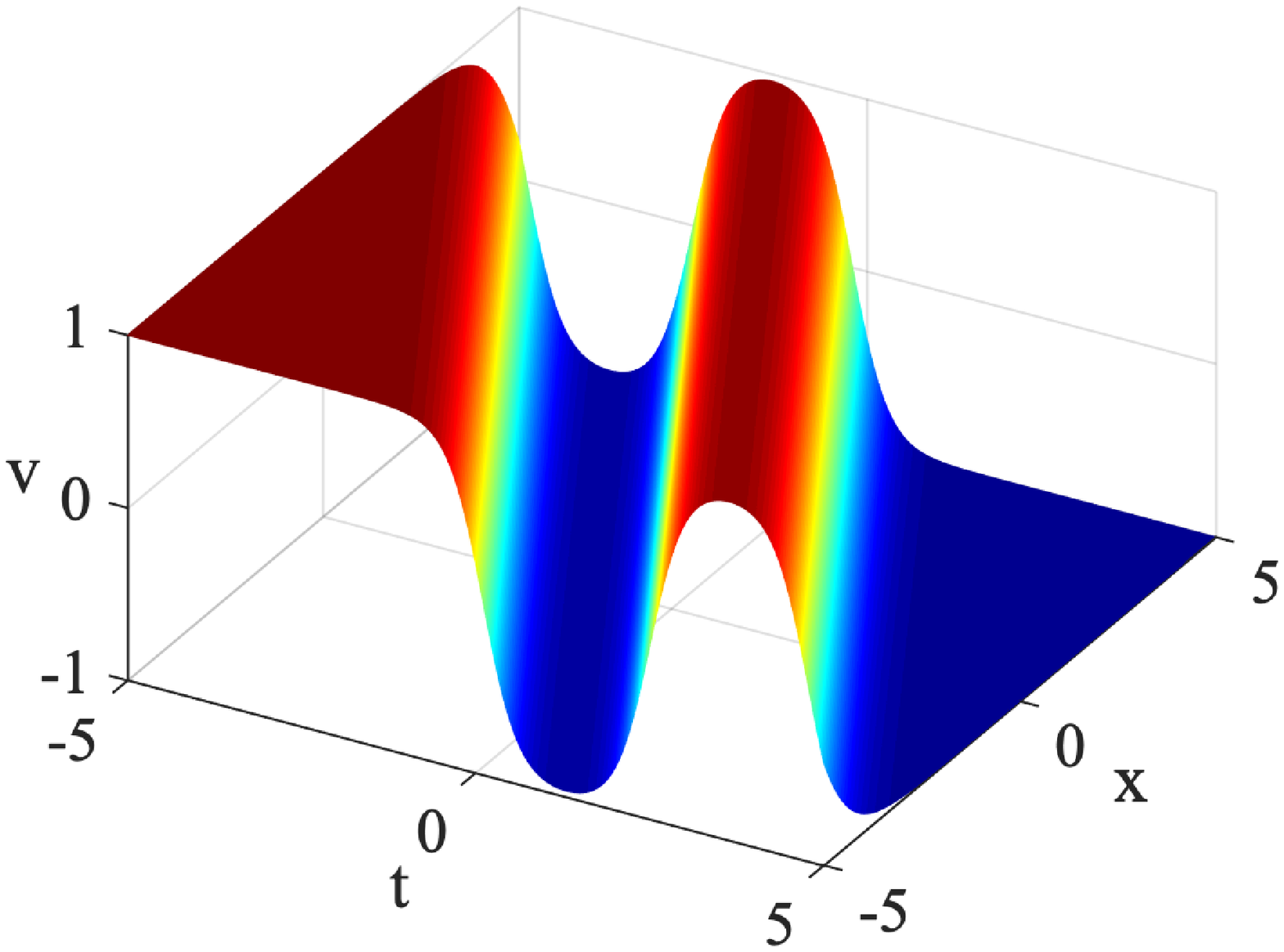} \end{tabular}}                       \\\toprule
\end{tabular}
\end{table}

$\bullet$ \textbf{Changes of the number of neurons per hidden layer}

The number of neurons in each hidden layer changes from 10 to 100 with step size 10 and and the number of hidden layers is 4. In all numerical tests, invariant hyper-parameters are: $N_u=200, N_f=5000, N_g=2000$.

\begin{figure}[htbp]
\centering
\includegraphics[width=7cm,height=5cm]{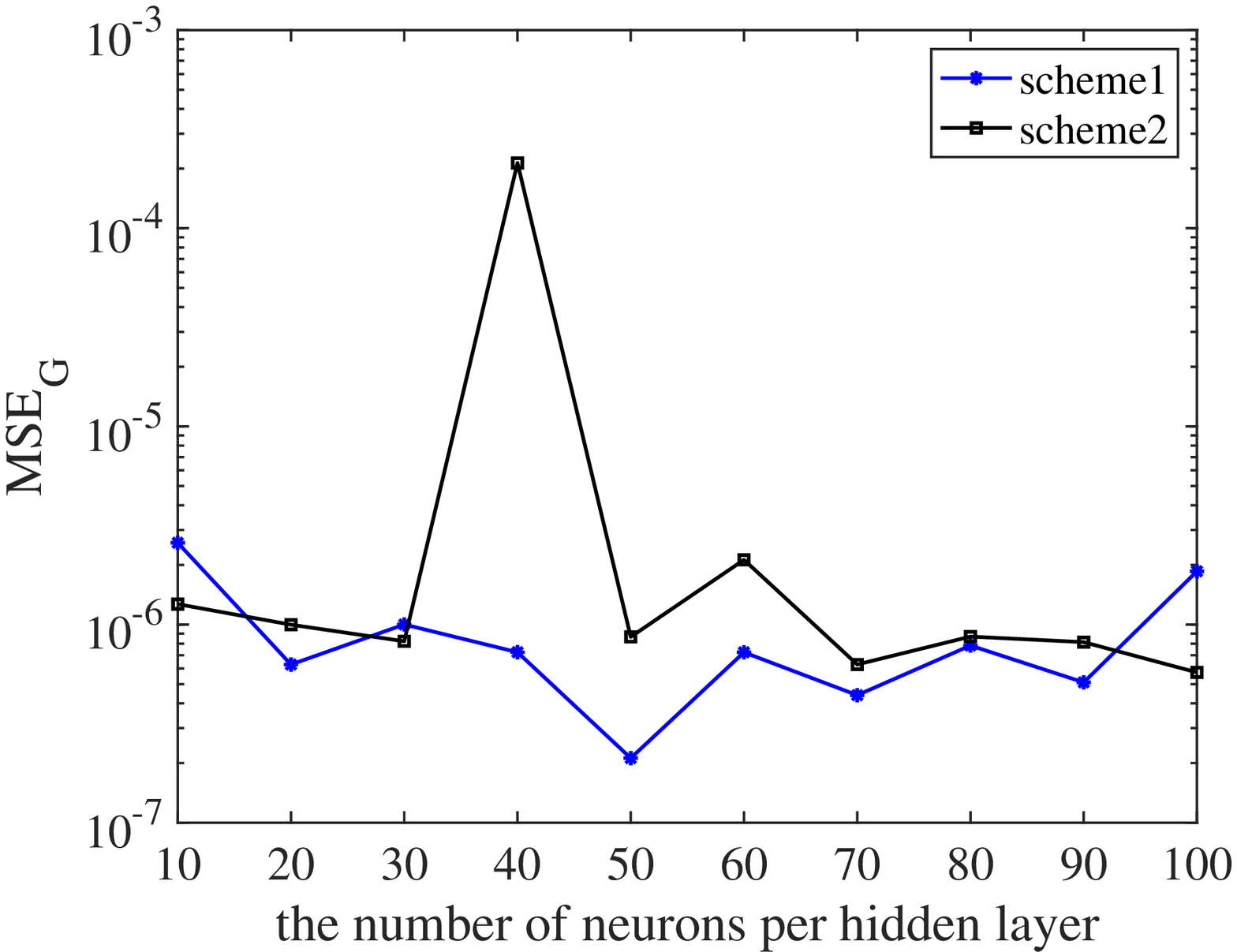}
$a$
\includegraphics[width=7cm,height=5cm]{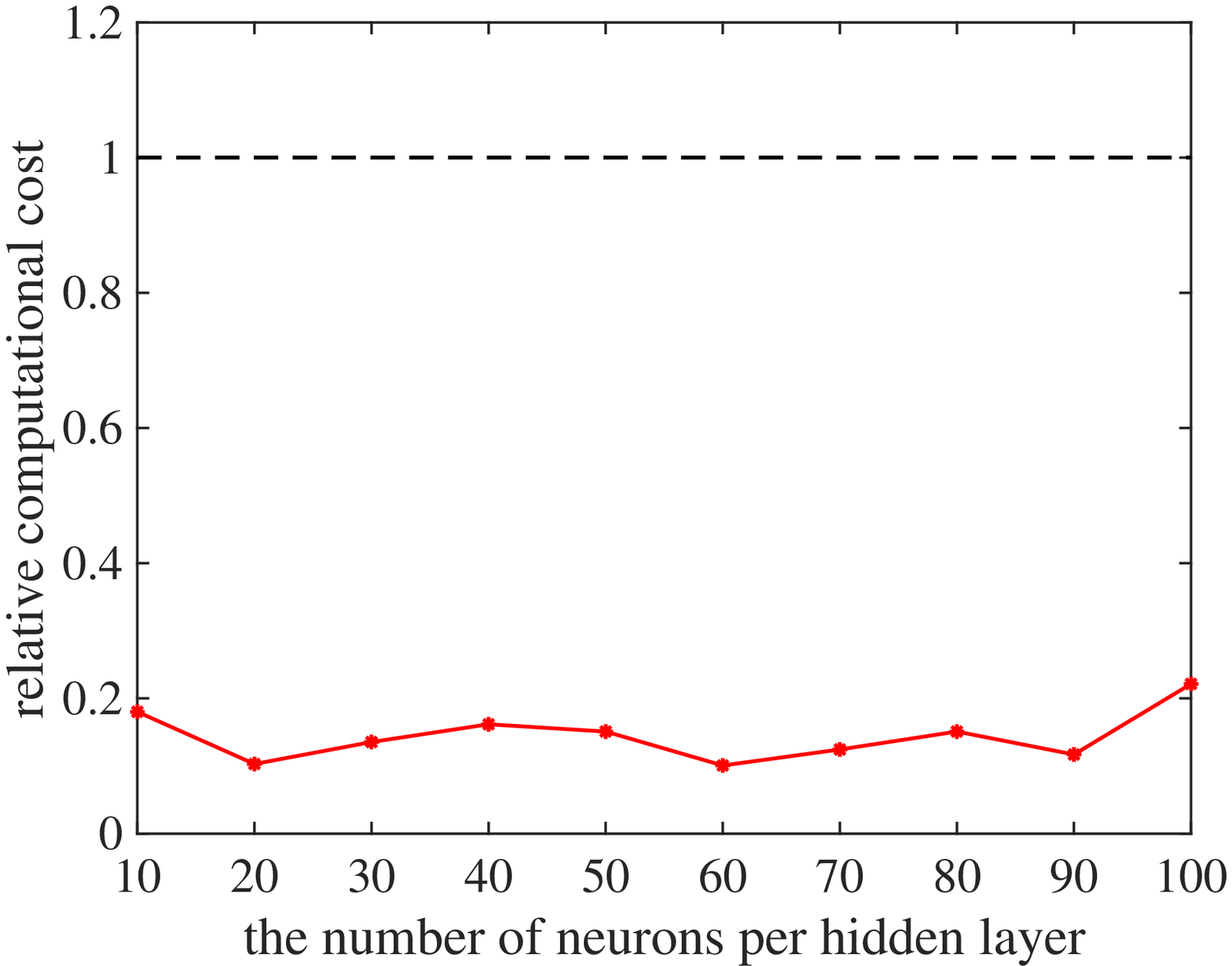}
$b$
\caption{(Color online) Data-driven solution of the defocusing mKdV equation: comparison of two schemes by using different numbers of neurons per hidden layer (a) The mean squared errors $MSE_G$; (b) The relative computational costs.}
\label{fig5-2}
\end{figure}

Numerical results of experiments are summarized in Table \ref{tableA-1-3}-Table \ref{tableA-1-4} in Appendix A. In addition, The mean squared errors $MSE_G$ and the relative computational costs are plotted in Fig. \ref{fig5-2}, and Table \ref{table5-1-3} presents average $MSE_G$ and elapsed time of ten experiments by using different numbers of neurons per hidden layer.

\begin{table}[htbp]
\caption{Data-driven solution of the defocusing mKdV equation: average mean squared errors and elapsed time by using different numbers of neurons per hidden layer.}
\label{table5-1-3}
\centering
\begin{tabular}{c|c|c}
\bottomrule
Method (hidden layers=4)                  & \textbf{Scheme I} & \textbf{Scheme II} \\ \hline
Average $MSE_G$     & 9.464863E-07         & 2.224271E-05          \\ \hline
Average elapsed time (s) & 977.14973         & 146.5134          \\ \toprule
\end{tabular}
\end{table}

The performance of two schemes under the changes of the number of neurons per hidden layer is similar to performance under the changes of the number of hidden layers. The simulation results show that Scheme \uppercase\expandafter{\romannumeral 1} is more precise for the most part but Scheme \uppercase\expandafter{\romannumeral 2} excels at efficiency markedly.

Table \ref{table5-1-4} shows the details of data-driven solutions of the defocusing mKdV equation, where the contents in brackets of the first column 'Hidden layers-Neurons' denote the number of hidden layers and neurons per hidden layer in step two of Scheme \uppercase\expandafter{\romannumeral 2}. It indicates that we can simulate three types of numerical solutions (bright soliton, dark soliton and kink-bell type solution) via Scheme \uppercase\expandafter{\romannumeral 1} and one type of numerical solutions (kink-bell type solution) via Scheme \uppercase\expandafter{\romannumeral 2} by using different numbers of neurons per hidden layer in the numerical experiments that have been carried out.

\begin{table}[htbp]
\caption{Data-driven solutions of the defocusing mKdV equation: types of solutions, density diagrams and three-dimensional plots by using different numbers of neurons per hidden layer.}
\label{table5-1-4}
\centering
\begin{tabular}{c|ccc}
\bottomrule
\multicolumn{1}{c|}{\begin{tabular}[c]{@{}c@{}}Hidden layers\\ -Neurons\end{tabular}} & Type of solution & Density diagram & 3D plot \\ \hline
\multicolumn{1}{c|}{\begin{tabular}[c]{@{}c@{}}Scheme \uppercase\expandafter{\romannumeral 1}\\ 4-50 \end{tabular}}      & bright soliton                 &  \multicolumn{1}{c}{\begin{tabular}[c]{@{}c@{}}\includegraphics[width=4cm,height=3cm]{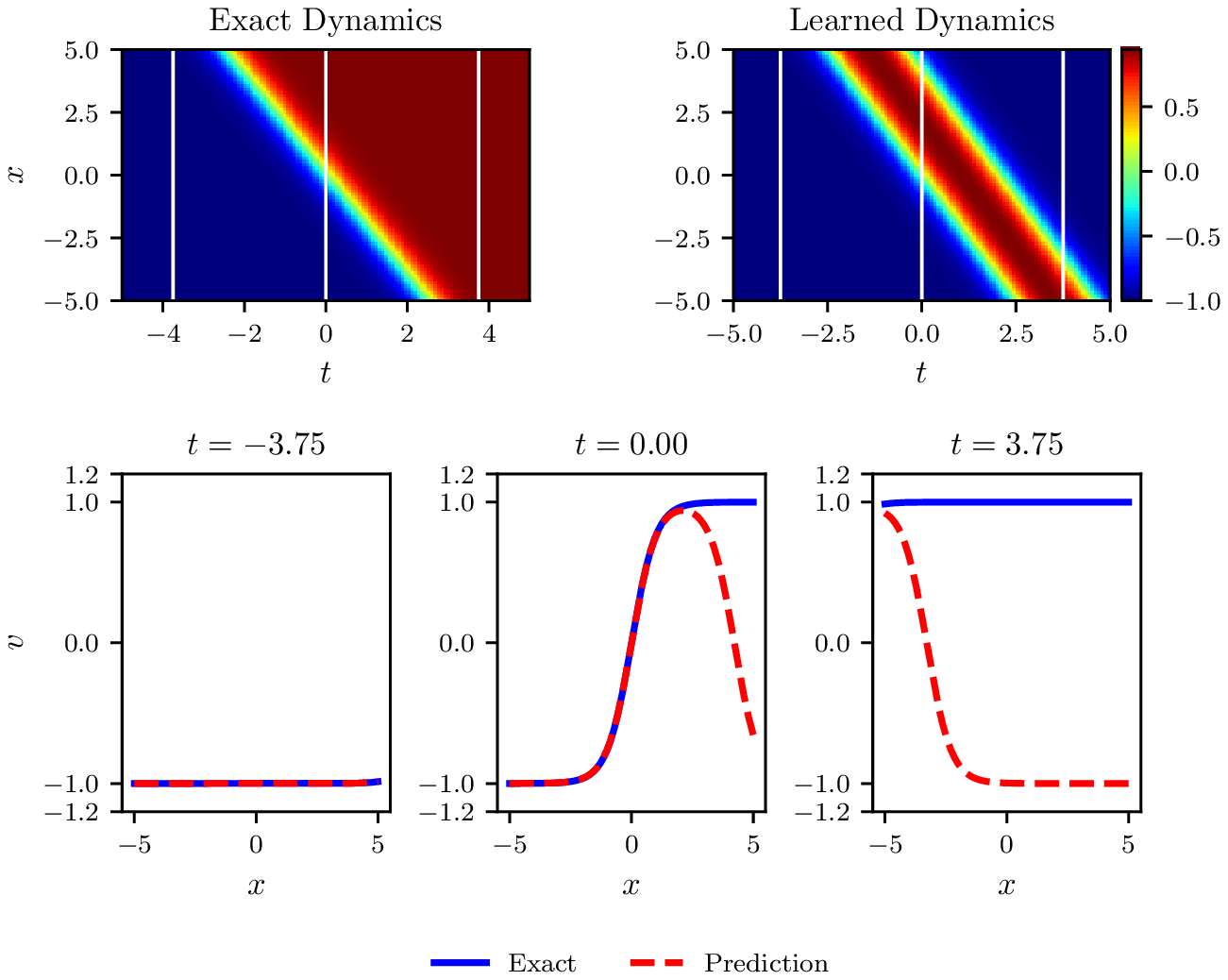} \end{tabular}}               &  \multicolumn{1}{c}{\begin{tabular}[c]{@{}c@{}}\includegraphics[width=4cm,height=3cm]{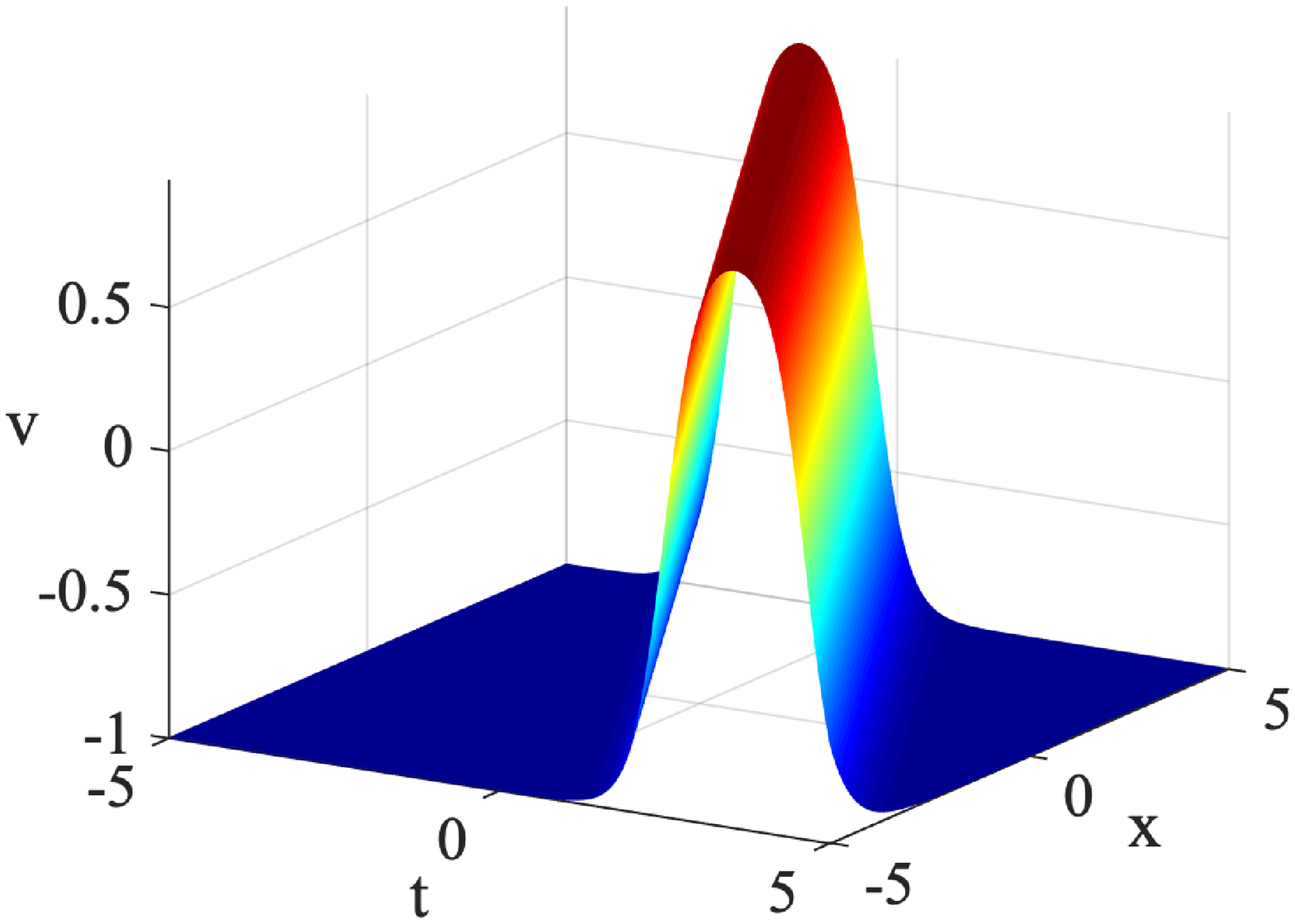} \end{tabular}}       \\
\multicolumn{1}{c|}{\begin{tabular}[c]{@{}c@{}}Scheme \uppercase\expandafter{\romannumeral 1}\\ 4-10/4-20/4-30/4-60\\ 4-70/4-80/4-90/4-100  \end{tabular}}       & dark soliton                 &  \multicolumn{1}{c}{\begin{tabular}[c]{@{}c@{}}\includegraphics[width=4cm,height=3cm]{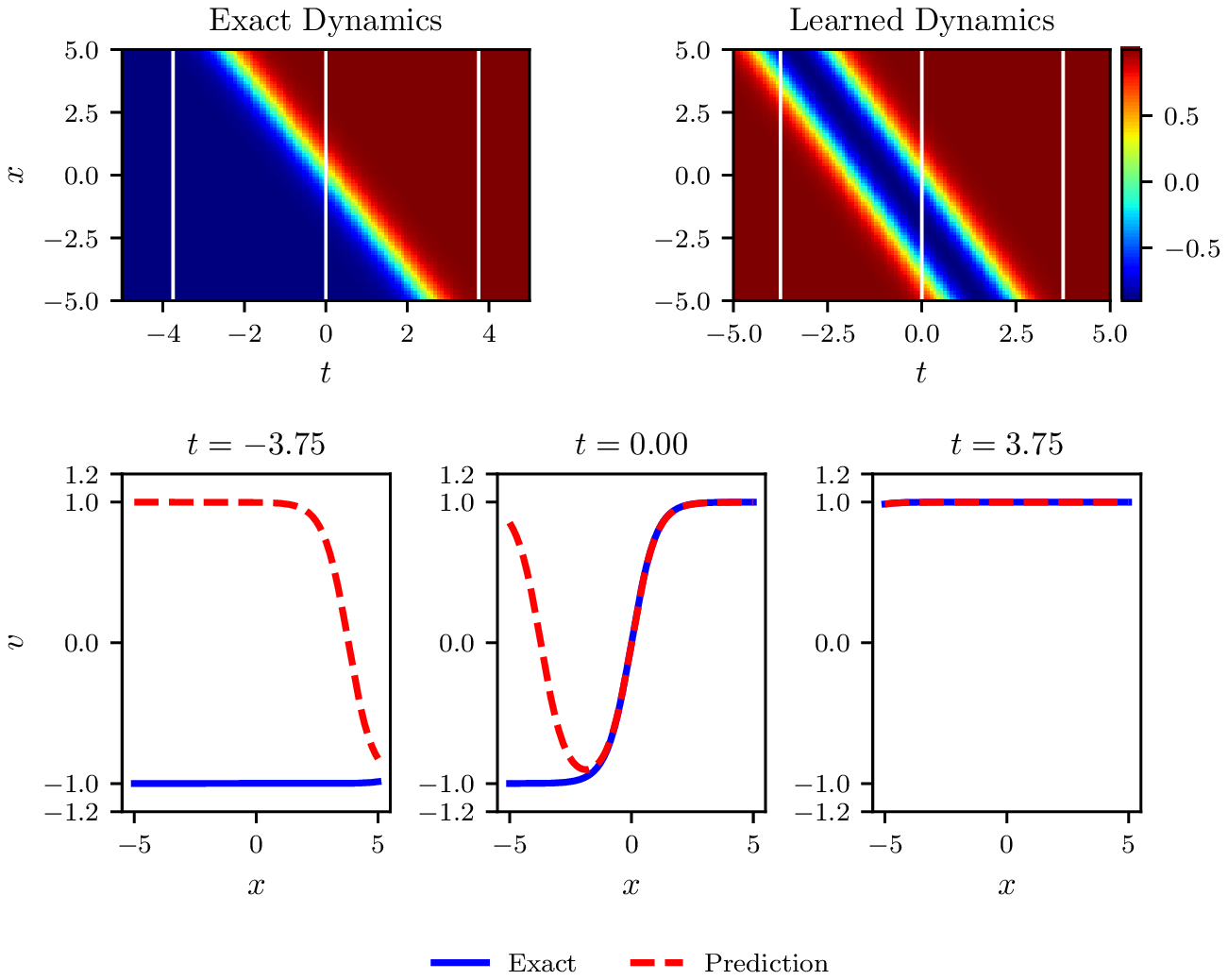} \end{tabular}}                              &  \multicolumn{1}{c}{\begin{tabular}[c]{@{}c@{}}\includegraphics[width=4cm,height=3cm]{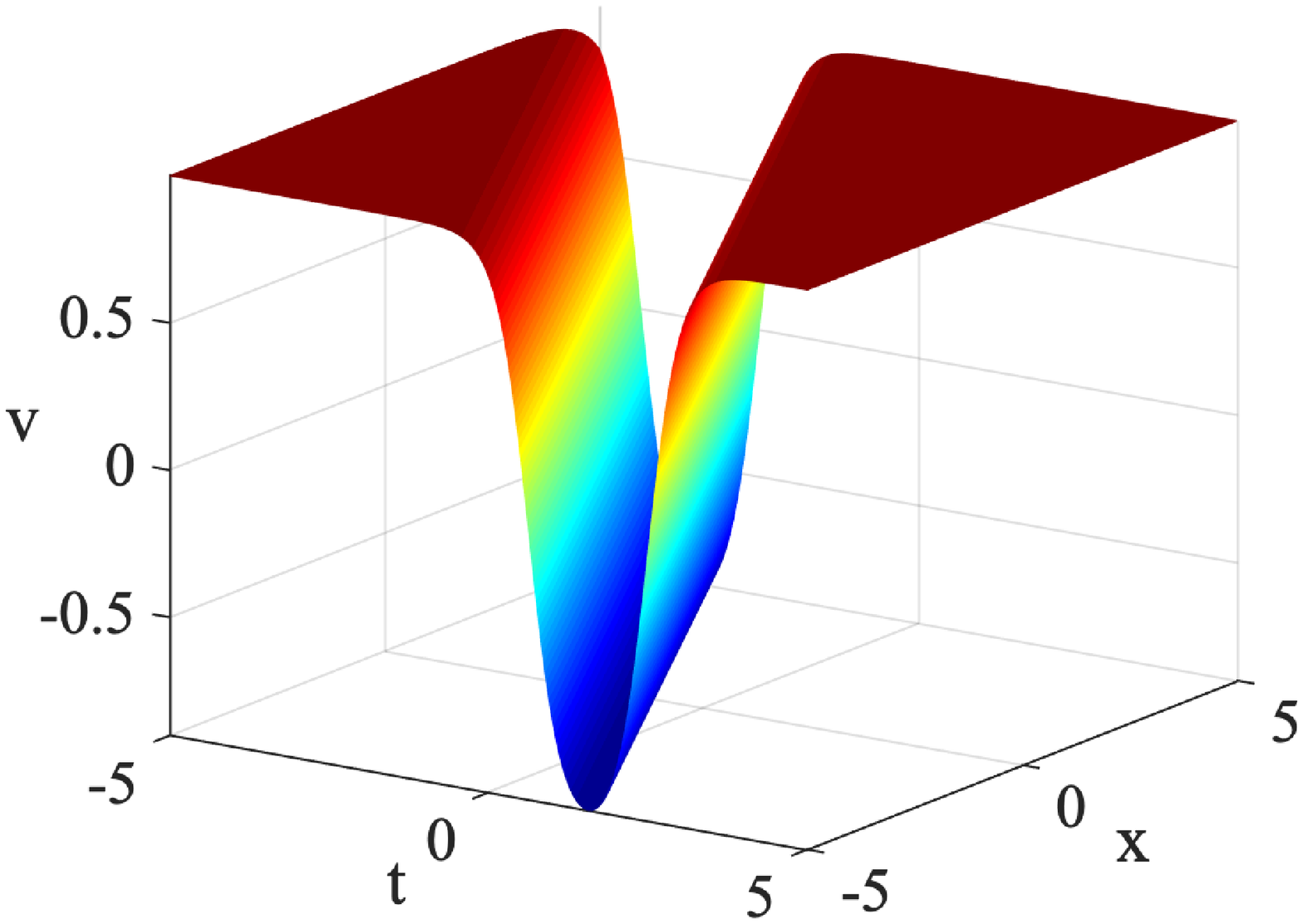} \end{tabular}}                      \\
\multicolumn{1}{c|}{\begin{tabular}[c]{@{}c@{}}Scheme \uppercase\expandafter{\romannumeral 1}\\ 4-40  \end{tabular}}       & kink-bell type solution                 & \multicolumn{1}{c}{\begin{tabular}[c]{@{}c@{}}\includegraphics[width=4cm,height=3cm]{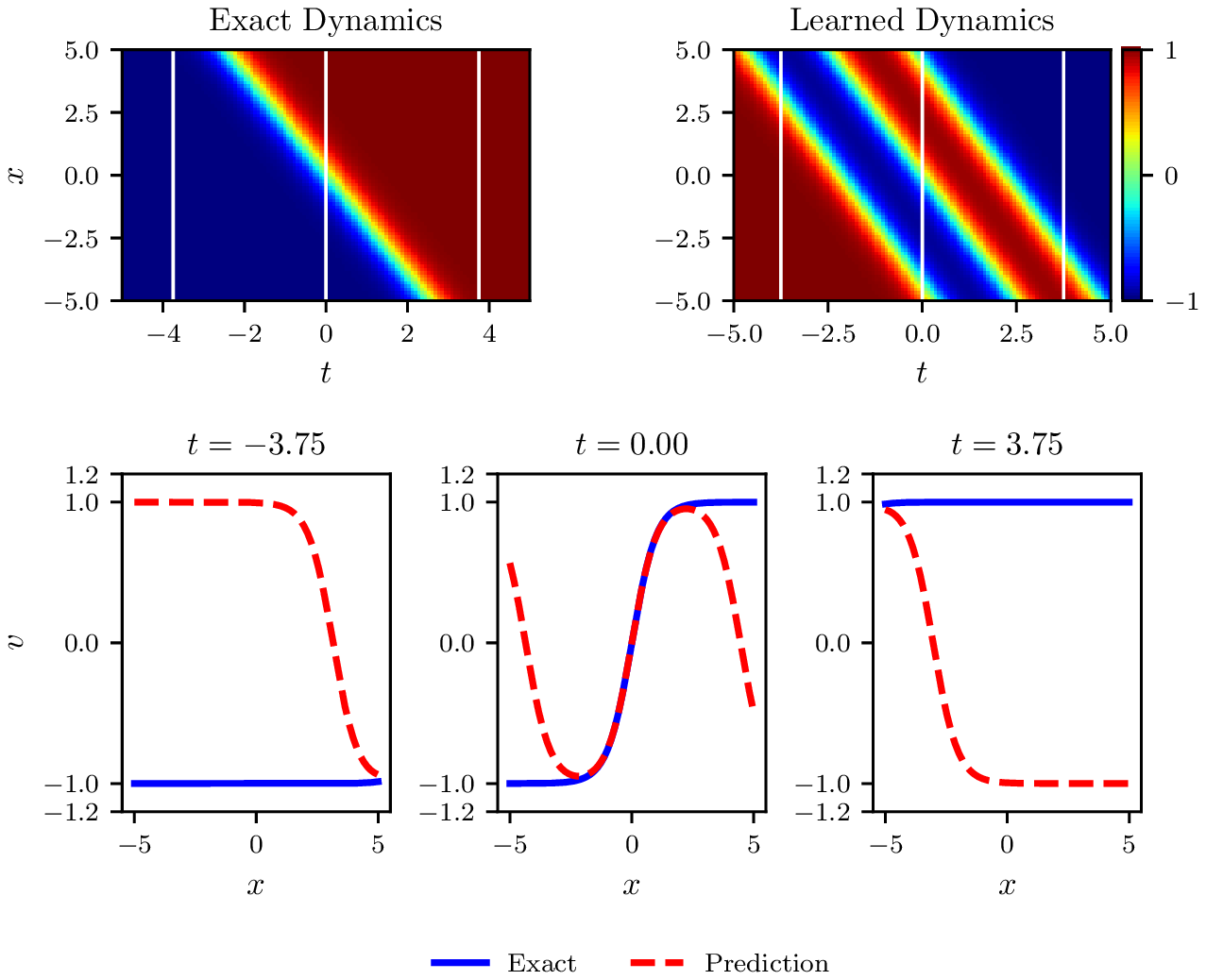} \end{tabular}}                               & \multicolumn{1}{c}{\begin{tabular}[c]{@{}c@{}}\includegraphics[width=4cm,height=3cm]{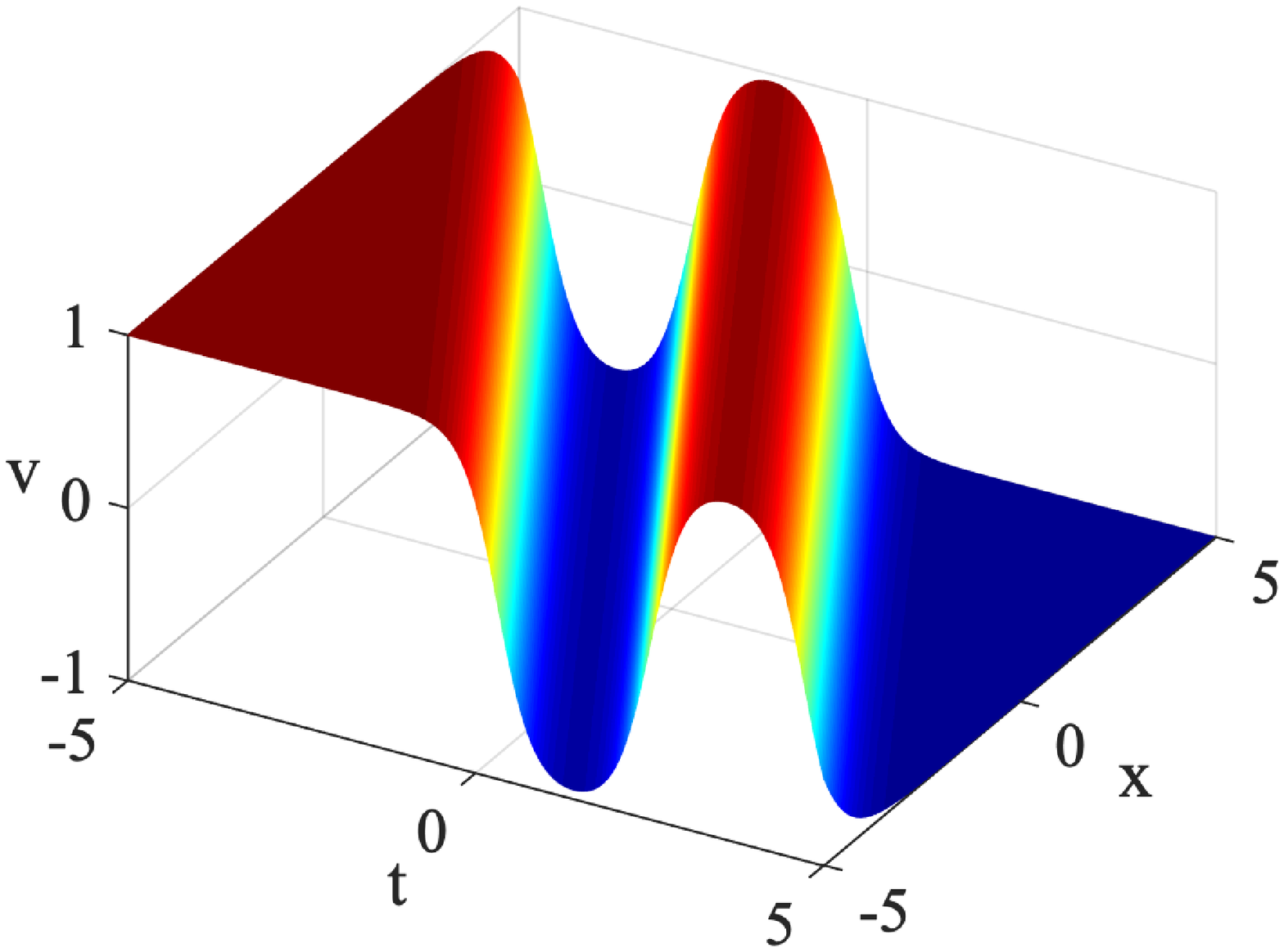} \end{tabular}}                       \\
\multicolumn{1}{c|}{\begin{tabular}[c]{@{}c@{}}Scheme \uppercase\expandafter{\romannumeral 2}\\ 4-10(2-40)/4-20(2-40)/\\ 4-30(2-40)/4-50(2-40)/\\ 4-60(2-40)/4-70(2-40)/\\ 4-80(2-40)/4-90(2-40)/\\ 4-100(2-40)  \end{tabular}}       & kink-bell type solution                 &\multicolumn{1}{c}{\begin{tabular}[c]{@{}c@{}}\includegraphics[width=4cm,height=3cm]{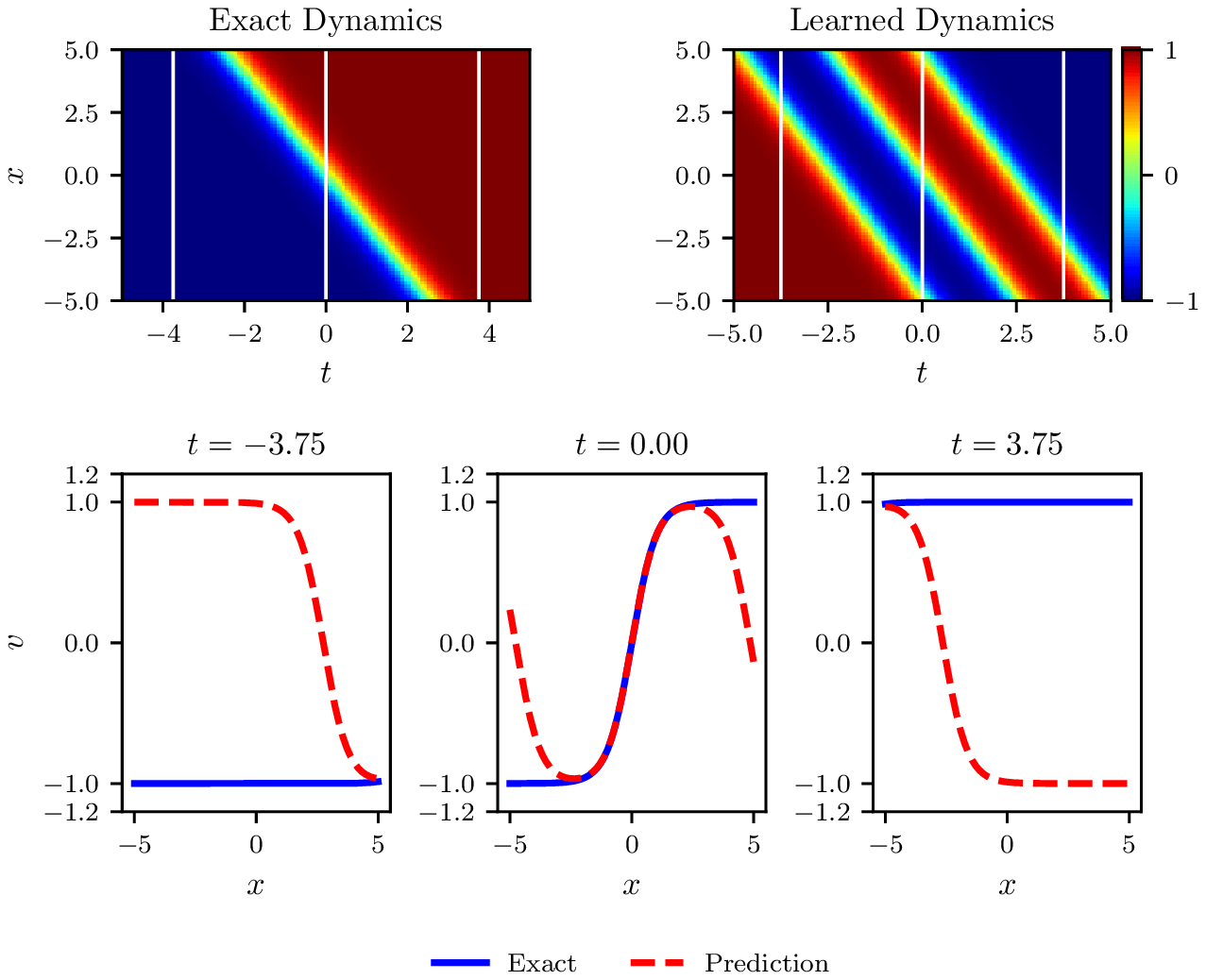} \end{tabular}}                                & \multicolumn{1}{c}{\begin{tabular}[c]{@{}c@{}}\includegraphics[width=4cm,height=3cm]{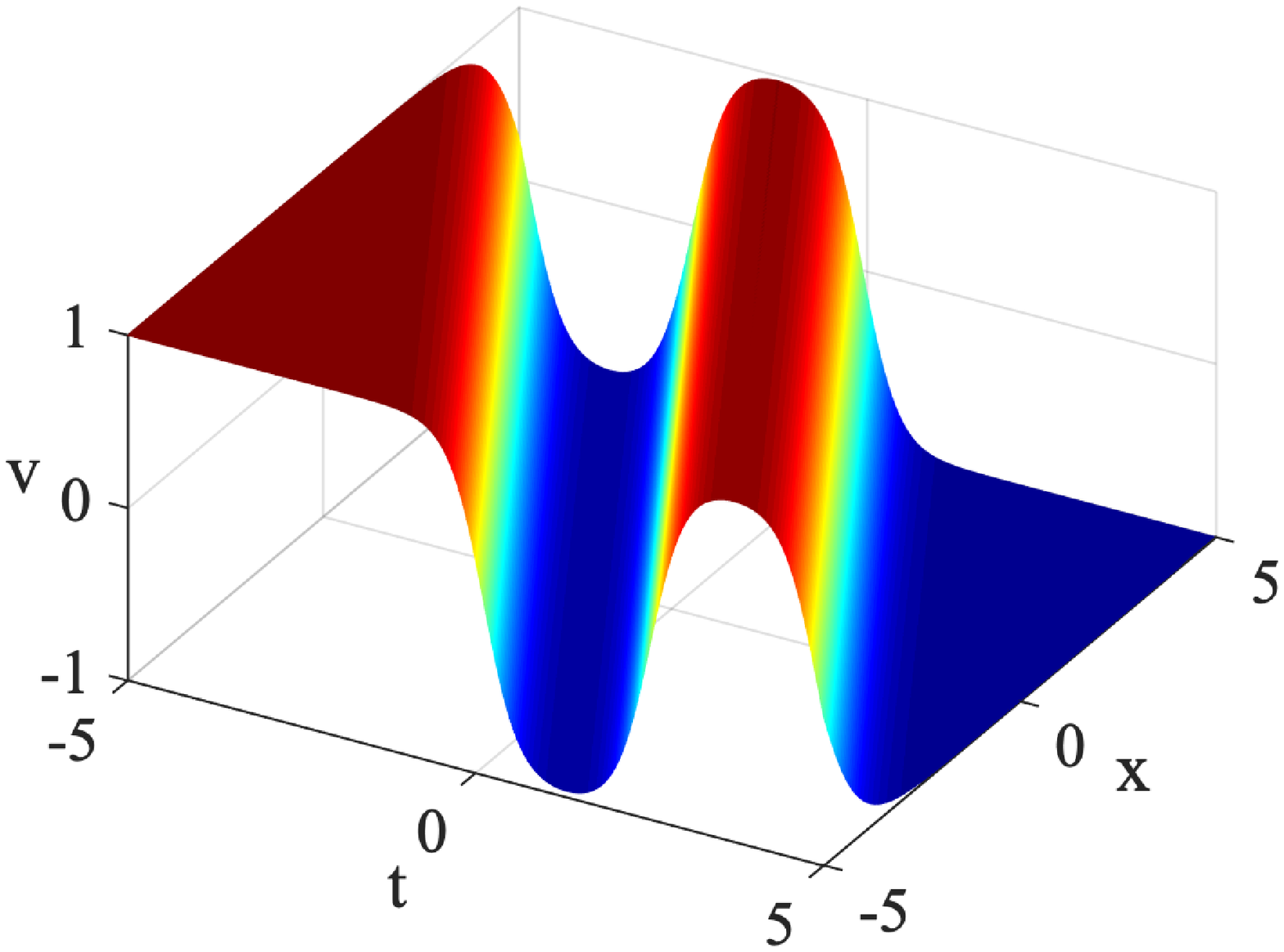} \end{tabular}}                       \\\toprule
\end{tabular}
\end{table}

\textbf{(2) Data-driven one-soliton solution of the focusing mKdV solution (Case 4.1)}

Considering that data-driven solutions of the first two cases in Section \ref{complex Miura} can be successfully simulated, which are close to the exact ones, while numerical solution of Case 4.3 is a new solution of the focusing mKdV equation different from \eqref{rationalfocusing},  numerical experiments of Case 4.1 (data-driven one-soliton solution of the focusing mKdV solution) and Case 4.2 (data-driven two-soliton solution of the focusing mKdV solution) are performed here to compare the results of two schemes in terms of accuracy and efficiency, more specifically, i.e. the relative $\mathbb{L}_2$ error of $v$ and elapsed time.

$\bullet$ \textbf{Changes of the number of hidden layers}

The number of hidden layers changes from 2 to 20 with step size 2 and each hidden layer has 40 neurons. In all numerical tests, invariant hyper-parameters are: $N_u=200, N_f=5000, N_g=2000$.

\begin{figure}[htbp]
\centering
\includegraphics[width=5.5cm,height=4.2cm]{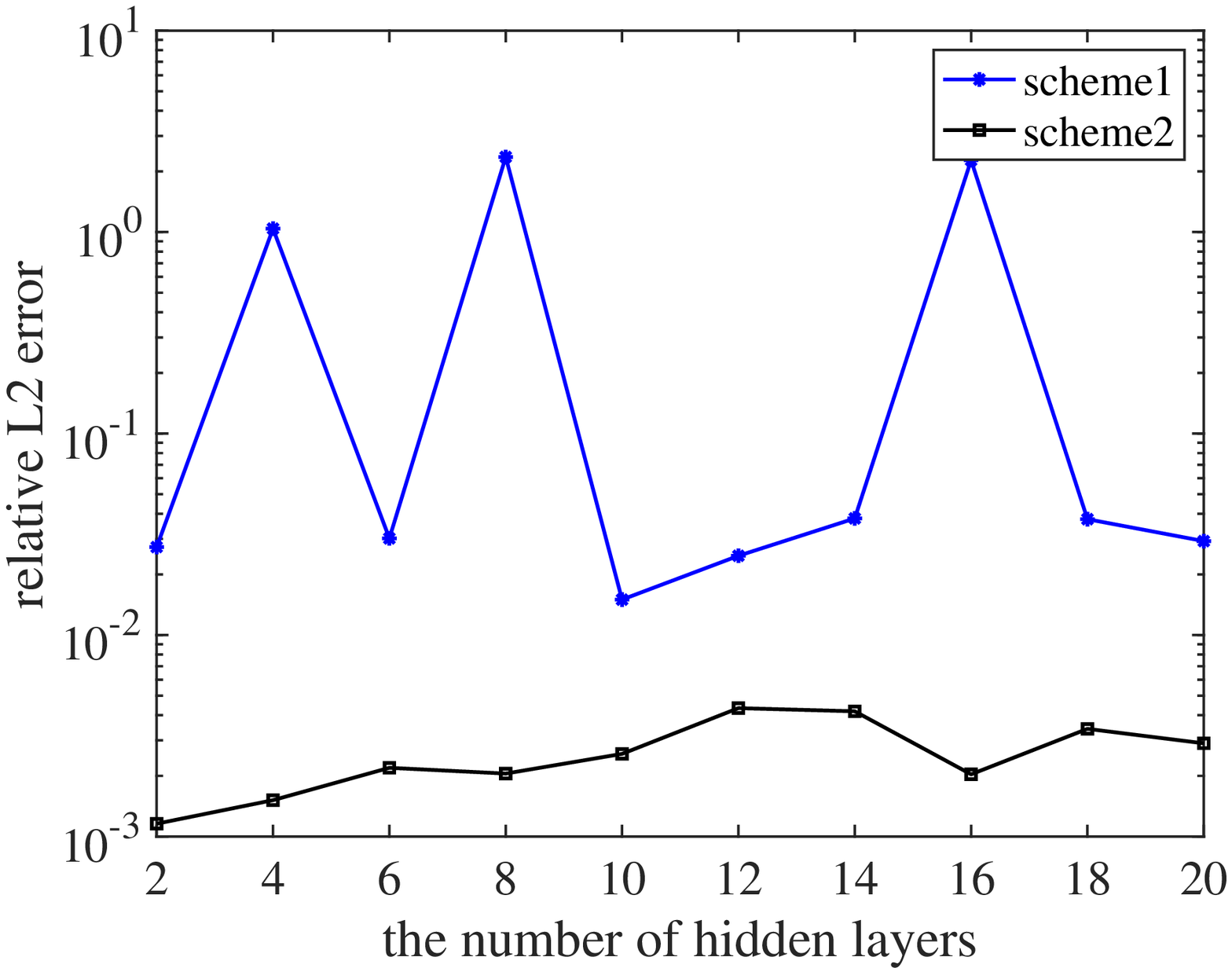}
$a$
\includegraphics[width=5.5cm,height=4.2cm]{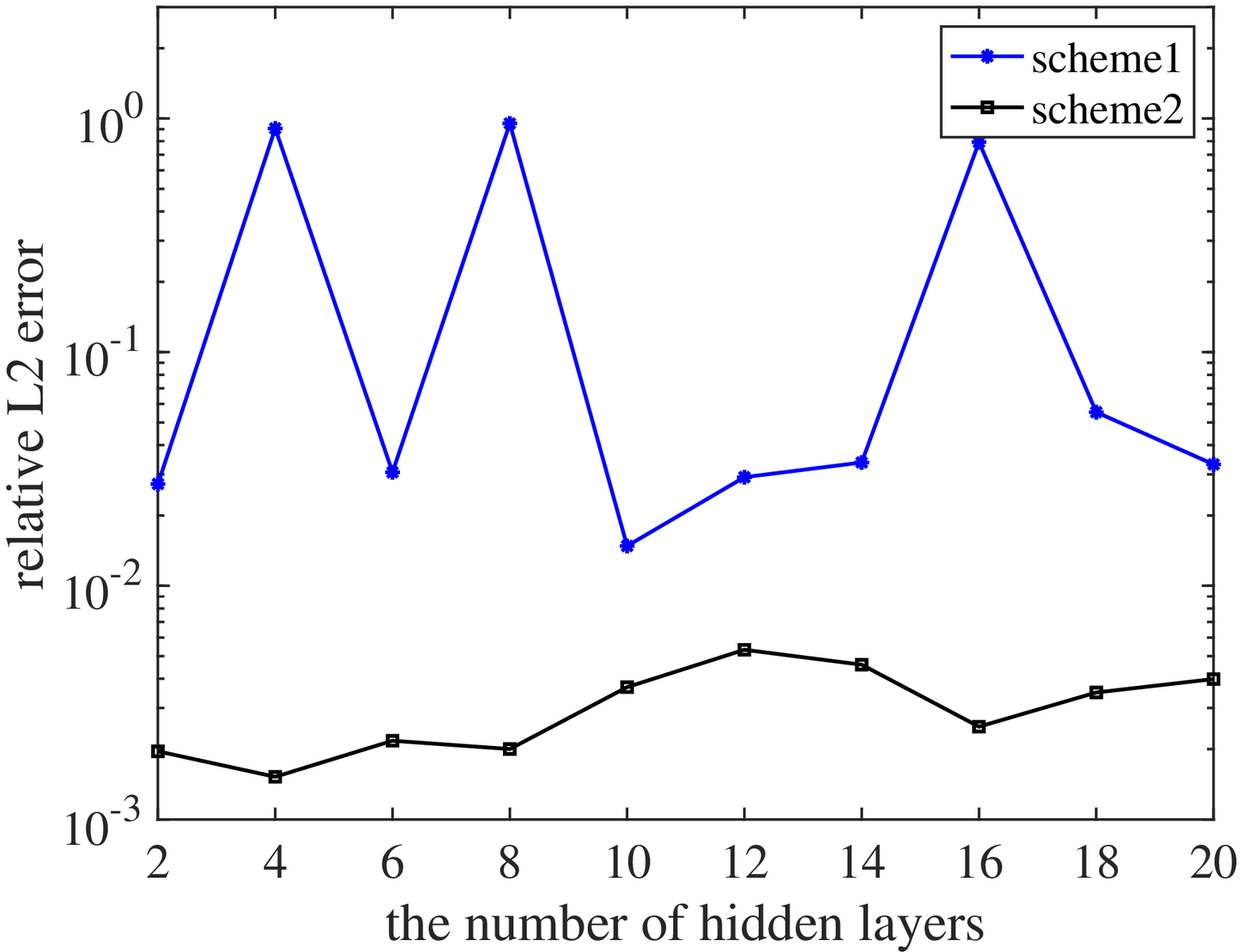}
$b$
\includegraphics[width=5.5cm,height=4.2cm]{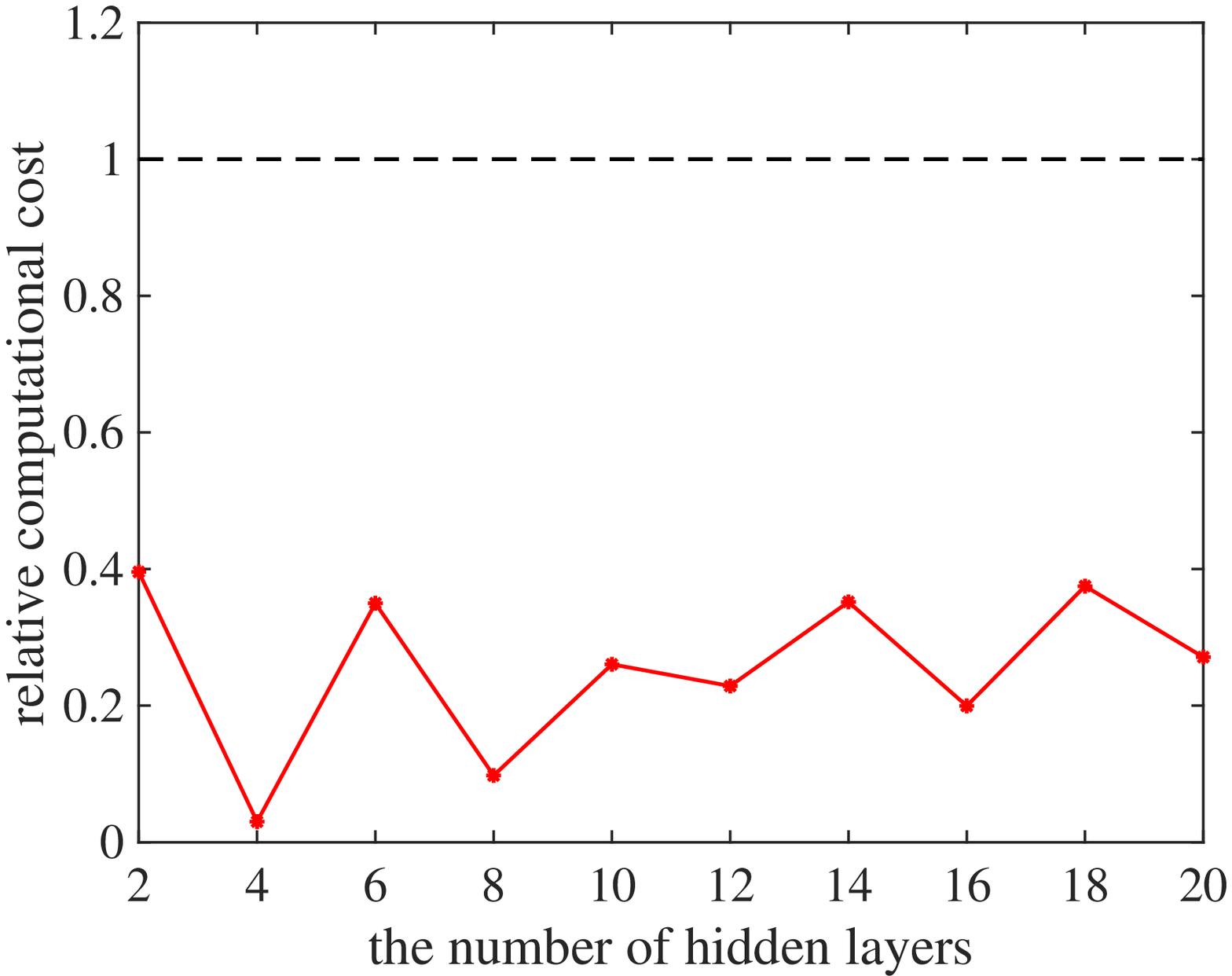}
$c$
\caption{(Color online) Data-driven one-soliton solution of the focusing mKdV equation: comparison of two schemes by using different numbers of hidden layers (a) The relative $\mathbb{L}_2$ errors of $v_r$; (b) The relative $\mathbb{L}_2$ errors of $|v|$; (c) The relative computational costs.}
\label{fig5-3}
\end{figure}

Two groups of ten independent numerical experiments corresponding to Scheme \uppercase\expandafter{\romannumeral 1} and Scheme \uppercase\expandafter{\romannumeral 2} are performed and the detailed results are shown in Table \ref{tableA-2-1}-Table \ref{tableA-2-2} in Appendix A. Then the relative $\mathbb{L}_2$ errors of the real part $v_r$ and the modulus $|v|$ as well as the relative computational costs are plotted in Fig. \ref{fig5-3}, and Table \ref{table5-2-1} presents average relative $\mathbb{L}_2$ errors and elapsed time of ten experiments by using different numbers of hidden layers.

\begin{table}[htbp]
\caption{Data-driven one-soliton solution of the focusing mKdV equation: average relative $\mathbb{L}_2$ errors and elapsed time by using different numbers of hidden layers.}
\label{table5-2-1}
\centering
\begin{tabular}{c|c|c}
\bottomrule
Method (neurons=40)                  & \textbf{Scheme I} & \textbf{Scheme II} \\ \hline
$v_r$      & 5.874070E-01         & 2.636346E-03          \\ \hline
$|v|$            & 2.875090E-01         & 3.123145E-03          \\ \hline
Average elapsed time (s) & 1263.96375         &  272.64193         \\ \toprule
\end{tabular}
\end{table}

In this case, Scheme \uppercase\expandafter{\romannumeral 2} results in less relative $\mathbb{L}_2$ errors of the real part $v_r$ and the modulus $|v|$ while Scheme \uppercase\expandafter{\romannumeral 1} is not suitable here and achieves unsatisfactory results. Also remarkably, the average mean squared errors of Scheme \uppercase\expandafter{\romannumeral 2} outperform Scheme \uppercase\expandafter{\romannumeral 1} by two orders of magnitude and meanwhile Scheme \uppercase\expandafter{\romannumeral 2} can speed up learning process and reduce elapsed time greatly. In short, Scheme \uppercase\expandafter{\romannumeral 2} has higher accuracy and faster computation speed compared with Scheme \uppercase\expandafter{\romannumeral 1}.

$\bullet$ \textbf{Changes of the number of neurons per hidden layer}

The number of neurons in each hidden layer changes from 10 to 100 with step size 10 and and the number of hidden layers is 4 and 6. In all numerical tests, invariant hyper-parameters are: $N_u=200, N_f=5000, N_g=2000$.

\begin{figure}[htbp]
\centering
\includegraphics[width=5.5cm,height=4.2cm]{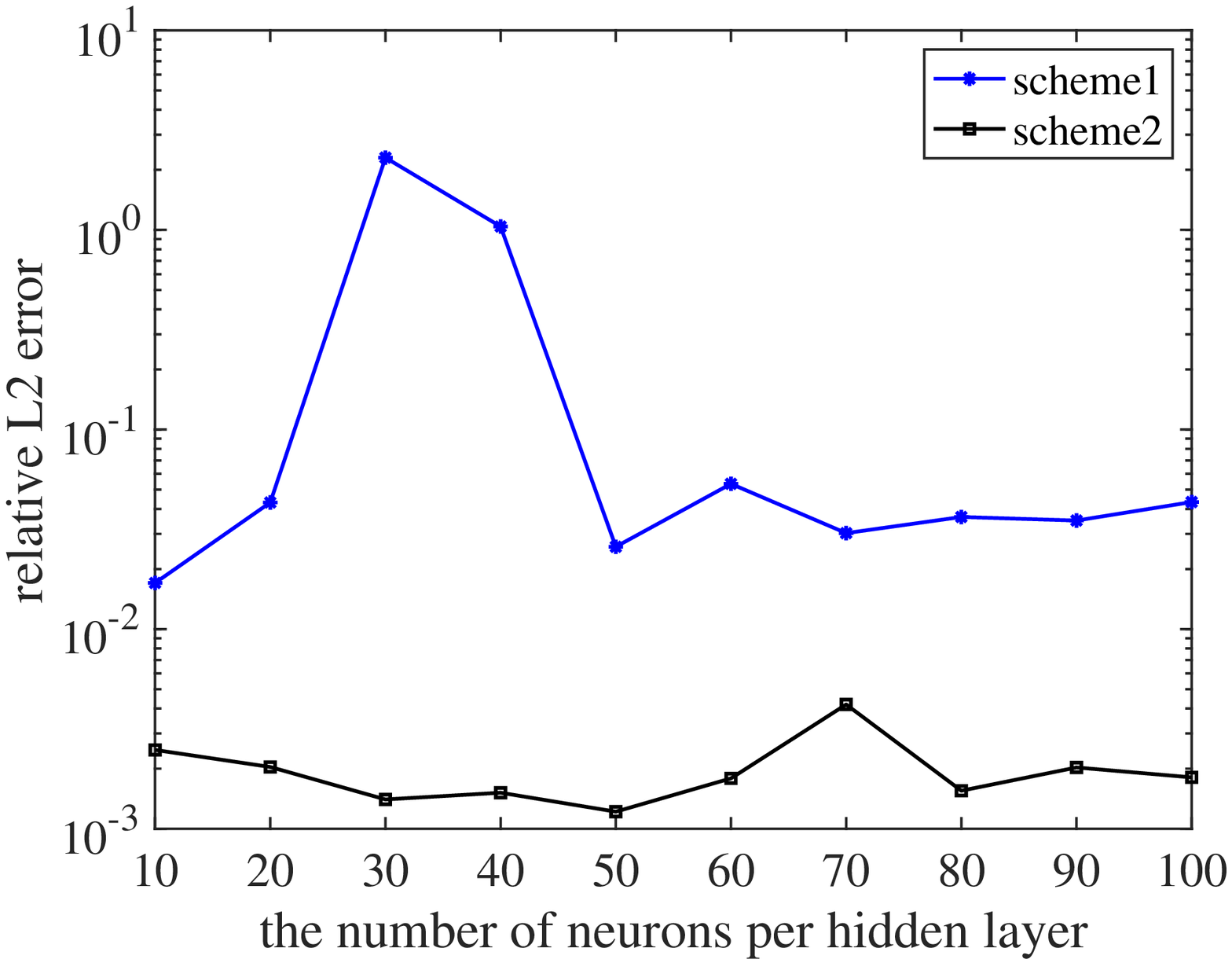}
$a$
\includegraphics[width=5.5cm,height=4.2cm]{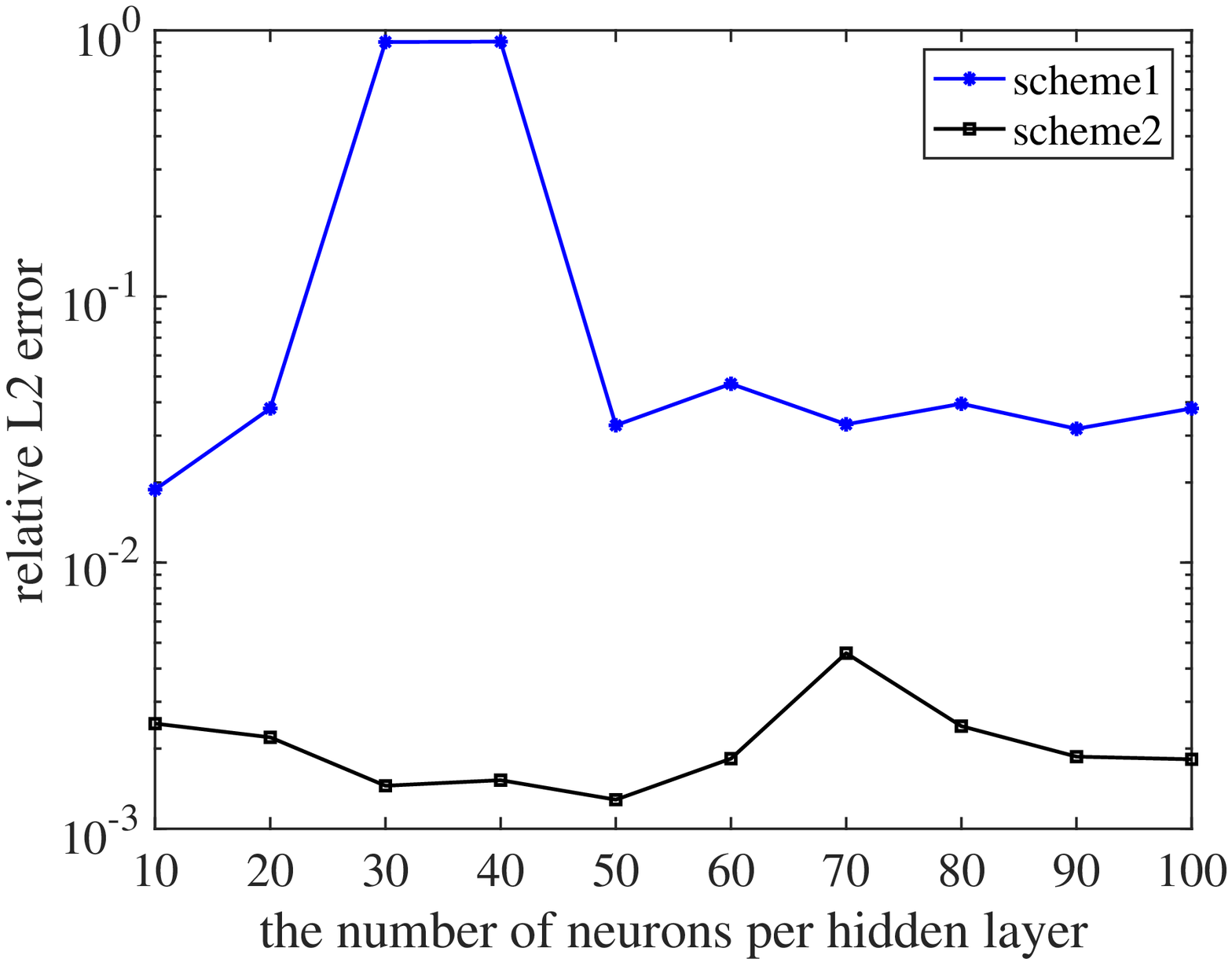}
$b$
\includegraphics[width=5.5cm,height=4.2cm]{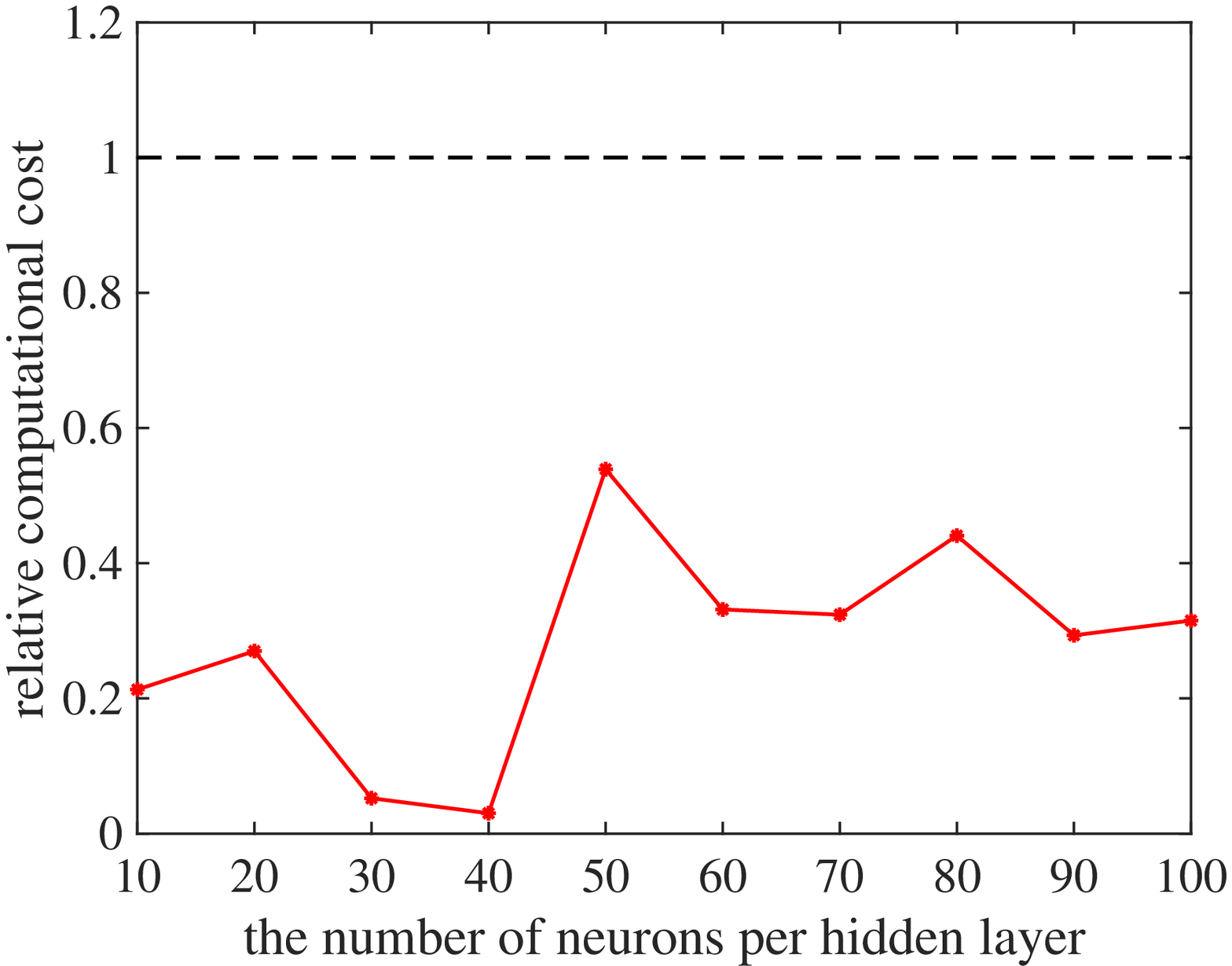}
$c$ \\
\includegraphics[width=5.5cm,height=4.2cm]{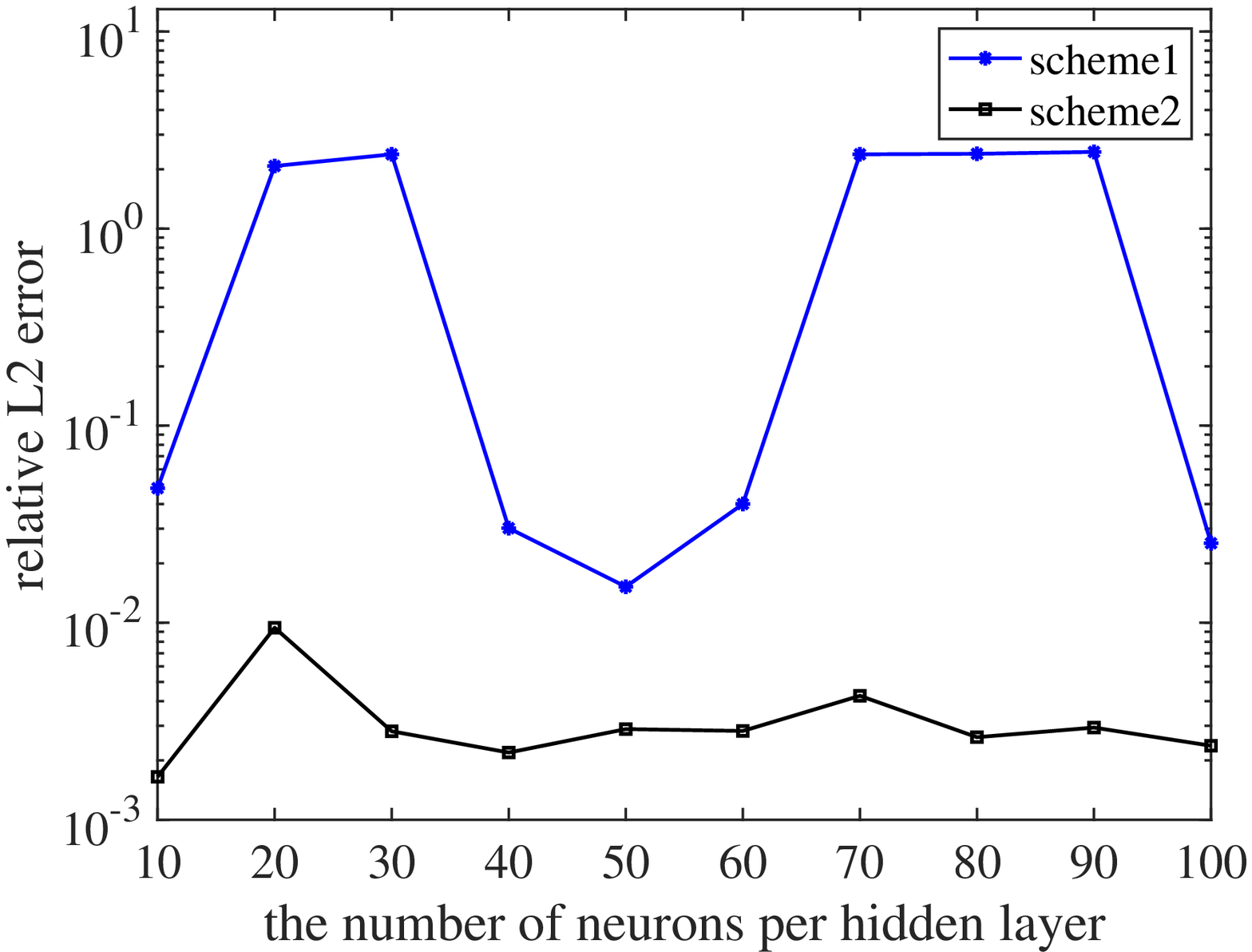}
$d$
\includegraphics[width=5.5cm,height=4.2cm]{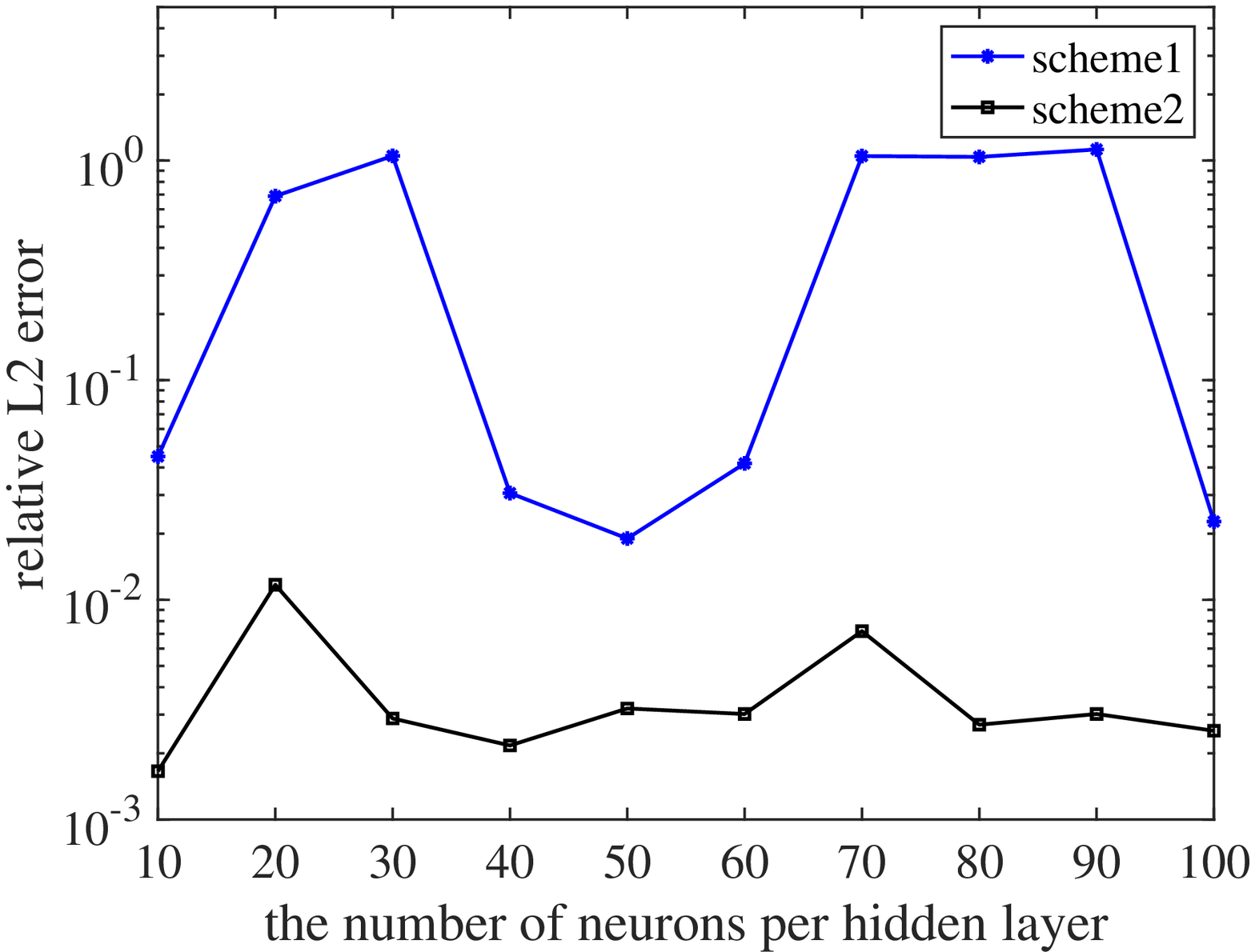}
$e$
\includegraphics[width=5.5cm,height=4.2cm]{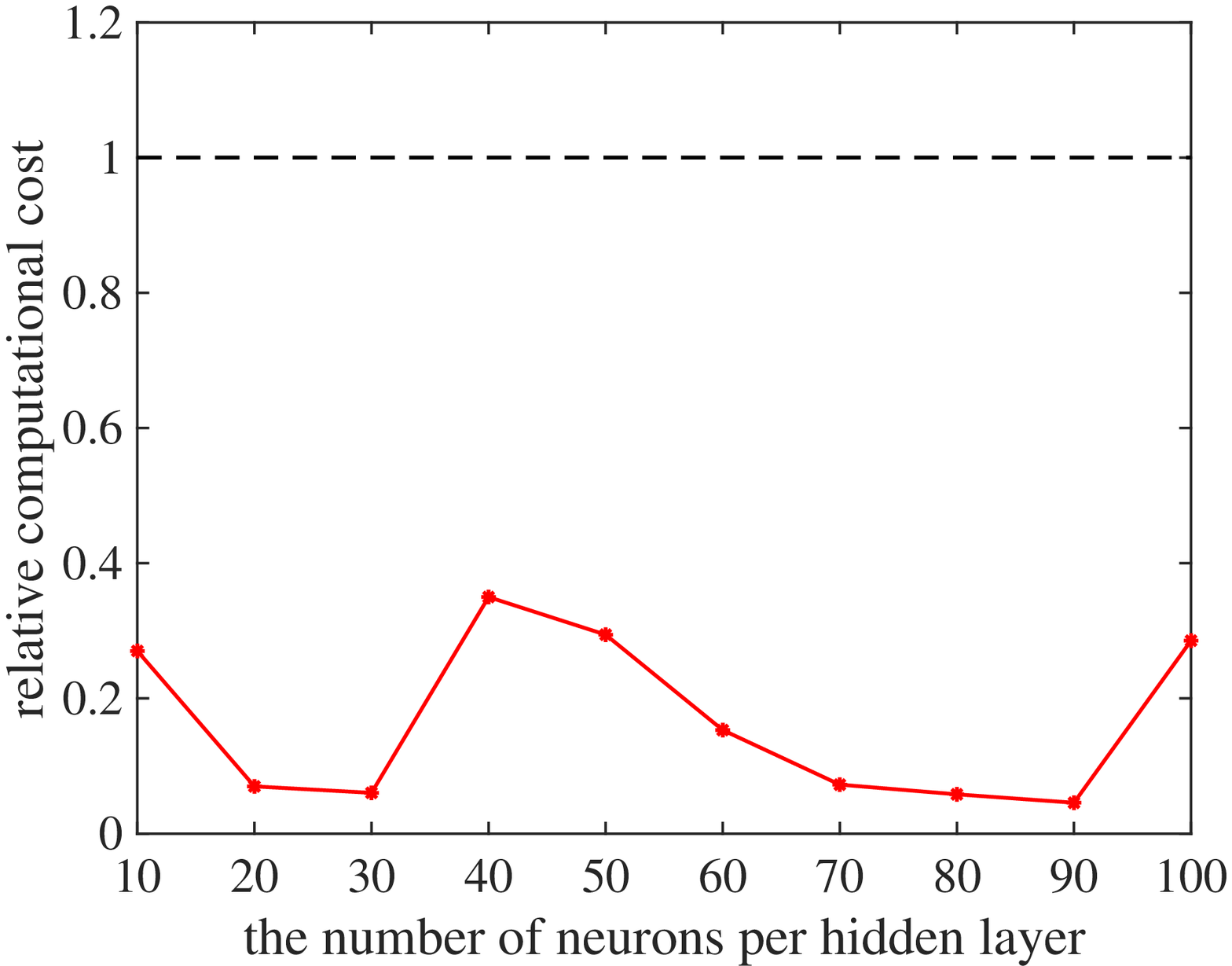}
$f$
\caption{(Color online) Data-driven one-soliton solution of the focusing mKdV equation: comparison of two schemes by using different numbers of neurons per hidden layer. Hidden layers = 4: (a) The relative $\mathbb{L}_2$ errors of $v_r$; (b) The relative $\mathbb{L}_2$ errors of $|v|$; (c) The relative computational costs; Hidden layers = 6: (d) The relative $\mathbb{L}_2$ errors of $v_r$; (e) The relative $\mathbb{L}_2$ errors of $|v|$; (f) The relative computational costs.}
\label{fig5-4}
\end{figure}

Numerical results of twenty experiments are summarized in in Table \ref{tableA-2-3}-Table \ref{tableA-2-6} in Appendix A. Besides, the relative $\mathbb{L}_2$ errors of the real part $v_r$ and the modulus $|v|$ as well as the relative computational costs are plotted in Fig. \ref{fig5-4}, and Table \ref{table5-2-2} presents average relative $\mathbb{L}_2$ errors and elapsed time of ten experiments by using different numbers of neurons per hidden layer.

\begin{table}[htbp]
\caption{Data-driven one-soliton solution of the focusing mKdV equation: average relative $\mathbb{L}_2$ errors and elapsed time by using different numbers of neurons per hidden layer.}
\label{table5-2-2}
\centering
\begin{tabular}{c|cc|cc}
\bottomrule
\multirow{2}{*}{Method} & \multicolumn{2}{c|}{hidden layers=4}      & \multicolumn{2}{c}{hidden layers=6}       \\ \cline{2-5}
                        & \multicolumn{1}{c|}{\textbf{Scheme I}} & \textbf{Scheme II} & \multicolumn{1}{c|}{\textbf{Scheme I}} & \textbf{Scheme II} \\ \hline
$v_r$                      & \multicolumn{1}{c|}{3.625614E-01}         &  2.001957E-03         & \multicolumn{1}{c|}{1.184843E+00}         & 3.397581E-03          \\ \hline
$|v|$                      & \multicolumn{1}{c|}{2.089086E-01}         & 2.146581E-03         & \multicolumn{1}{c|}{5.107828E-01}         & 4.009470E-03          \\ \hline
Average elapsed time (s)                       & \multicolumn{1}{c|}{439.35558}         & 64.26636          & \multicolumn{1}{c|}{1291.88992}         & 110.85687          \\ \toprule
\end{tabular}
\end{table}

As is shown in figures, the blue lines are always plotted above the black lines, which shows that Scheme \uppercase\expandafter{\romannumeral 2} is more accurate and can realize smaller relative $\mathbb{L}_2$ errors. Besides, the relative $\mathbb{L}_2$ errors of Scheme \uppercase\expandafter{\romannumeral 2} remain stable at a low level while the results of Scheme \uppercase\expandafter{\romannumeral 1} are always unstable with larger errors. No matter the number of hidden layers is 4 or 6, Scheme \uppercase\expandafter{\romannumeral 2} far outperforms Scheme \uppercase\expandafter{\romannumeral 1} in both accuracy and efficiency and the average relative $\mathbb{L}_2$ errors are significantly reduced by at least two orders of magnitude.

\textbf{(3) Data-driven two-soliton solution of the focusing mKdV solution (Case 4.2)}

$\bullet$ \textbf{Changes of the number of hidden layers}

The number of hidden layers changes from 2 to 20 with step size 2 and each hidden layer has 60 neurons. In all numerical tests, invariant hyper-parameters are: $N_u=200, N_f=5000, N_g=5000$.

Two groups of ten independent numerical experiments corresponding to Scheme \uppercase\expandafter{\romannumeral 1} and Scheme \uppercase\expandafter{\romannumeral 2} are carried out and the detailed results are shown in in Table \ref{tableA-3-1}-Table \ref{tableA-3-2} in Appendix A. Fig. \ref{fig5-5} shows the relative $\mathbb{L}_2$ errors of the real part $v_r$ and the modulus $|v|$ as well as the relative computational costs and Table \ref{table5-3-1} presents average relative $\mathbb{L}_2$ errors and elapsed time of ten experiments by using different numbers of hidden layers.

\begin{figure}[htbp]
\centering
\includegraphics[width=5.5cm,height=4.2cm]{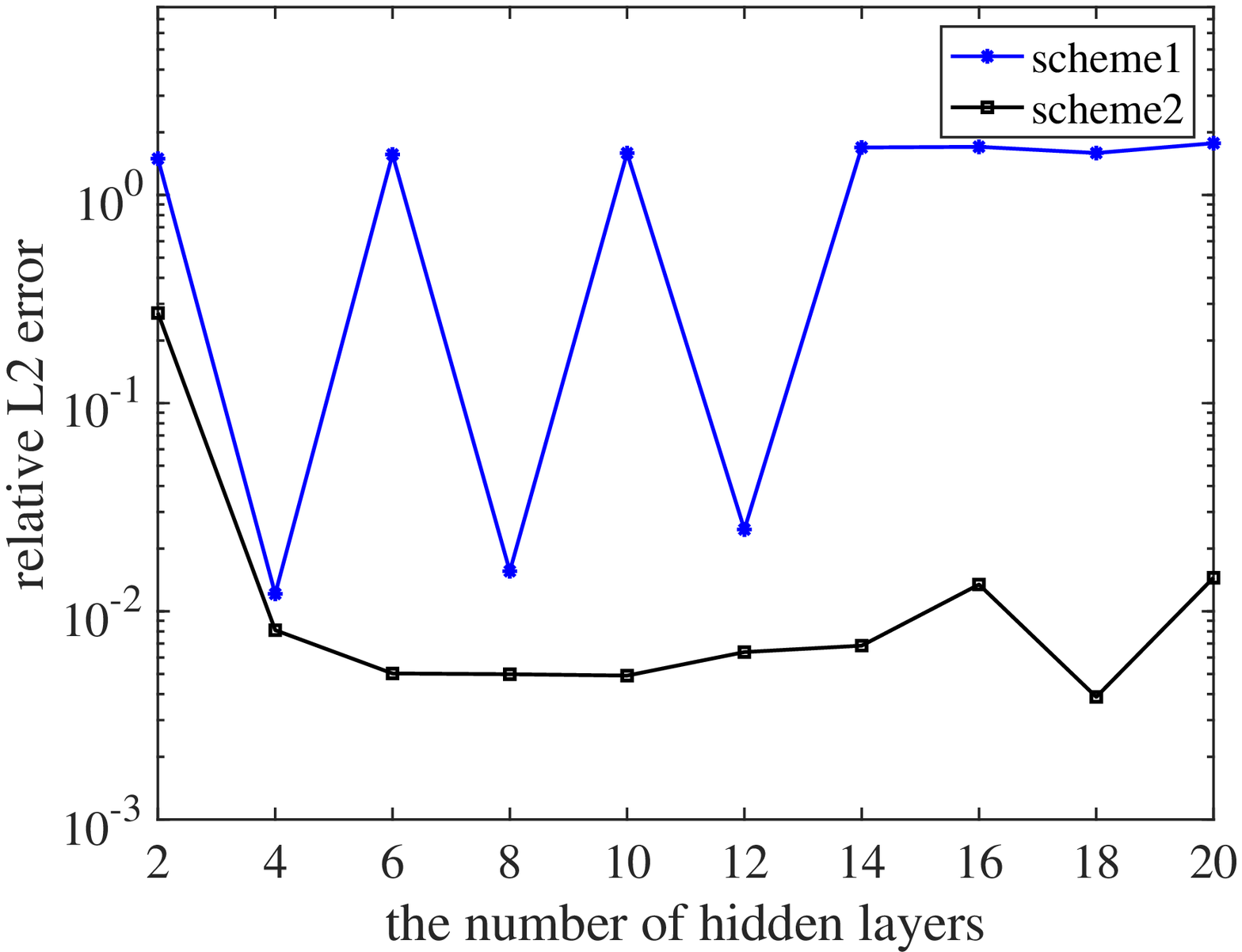}
$a$
\includegraphics[width=5.5cm,height=4.2cm]{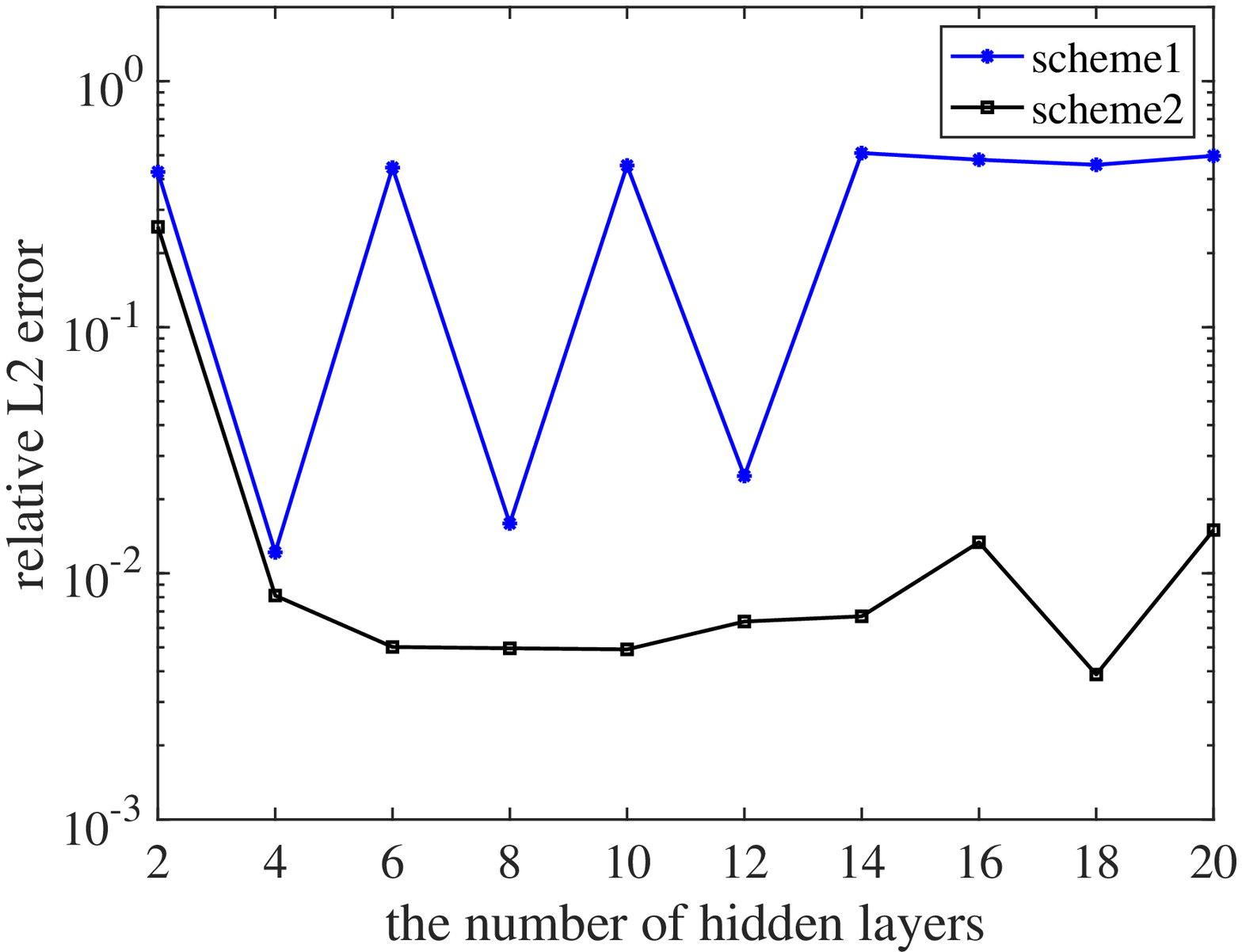}
$b$
\includegraphics[width=5.5cm,height=4.2cm]{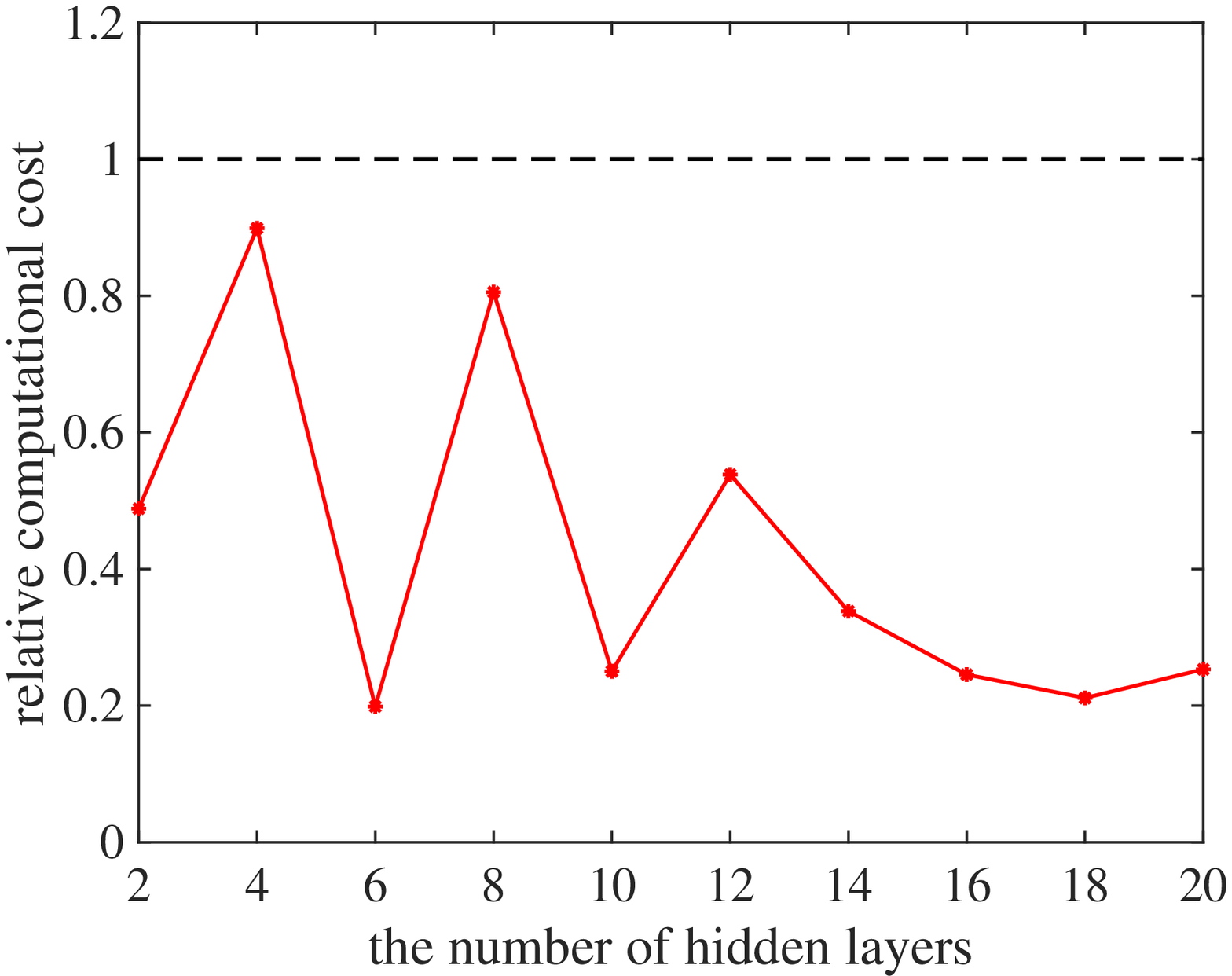}
$c$
\caption{(Color online) Data-driven two-soliton solution of the focusing mKdV equation: comparison of two schemes by using different numbers of hidden layers (a) The relative $\mathbb{L}_2$ errors of $v_r$; (b) The relative $\mathbb{L}_2$ errors of $|v|$; (c) The relative computational costs.}
\label{fig5-5}
\end{figure}

\begin{table}[htbp]
\caption{Data-driven two-soliton solution of the focusing mKdV equation: average relative $\mathbb{L}_2$ errors and elapsed time by using different numbers of hidden layers.}
\label{table5-3-1}
\centering
\begin{tabular}{c|c|c}
\bottomrule
Method (neurons=60)                   & \textbf{Scheme I} & \textbf{Scheme II} \\ \hline
$v_r$      & 1.146272E+00         & 3.391156E-02         \\ \hline
$|v|$           & 3.323545E-01         & 3.241187E-02          \\ \hline
Average elapsed time (s) & 5810.44675         & 1771.0284          \\ \toprule
\end{tabular}
\end{table}

Based on data in Table \ref{tableA-3-2} in Appendix A, it is obvious that the relative $\mathbb{L}_2$ errors of Scheme \uppercase\expandafter{\romannumeral 2} remain stable at a relatively low level except in the first experiment. In that experiment, the accuracy of the data-driven solution $\Widehat{u}$ obtained in step one is far from satisfactory presumably because the excessively shallow neural networks are not suitable for initial-boundary value problem of the KdV equation and may lead to remarkable errors. Then the failure of the first step further affects the precision of solution $\Widehat{v}$ of the focusing mKdV equation in step two. However, Scheme \uppercase\expandafter{\romannumeral 2} still has higher accuracy and much smaller errors than Scheme \uppercase\expandafter{\romannumeral 1}. Moreover, less elapsed time is required in Scheme \uppercase\expandafter{\romannumeral 2} compared with Scheme \uppercase\expandafter{\romannumeral 1} and the relative computational cost decreases in a fluctuation way and tends to be stable as the number of hidden layers increases according to Fig. \ref{fig5-5} (c).

$\bullet$ \textbf{Changes of the number of neurons per hidden layer}

The number of neurons in each hidden layer changes from 10 to 100 with step size 10 and and the number of hidden layers is 4 and 6. In all numerical tests, invariant hyper-parameters are: $N_u=200, N_f=5000, N_g=5000$.

\begin{figure}[htbp]
\centering
\includegraphics[width=5.5cm,height=4.2cm]{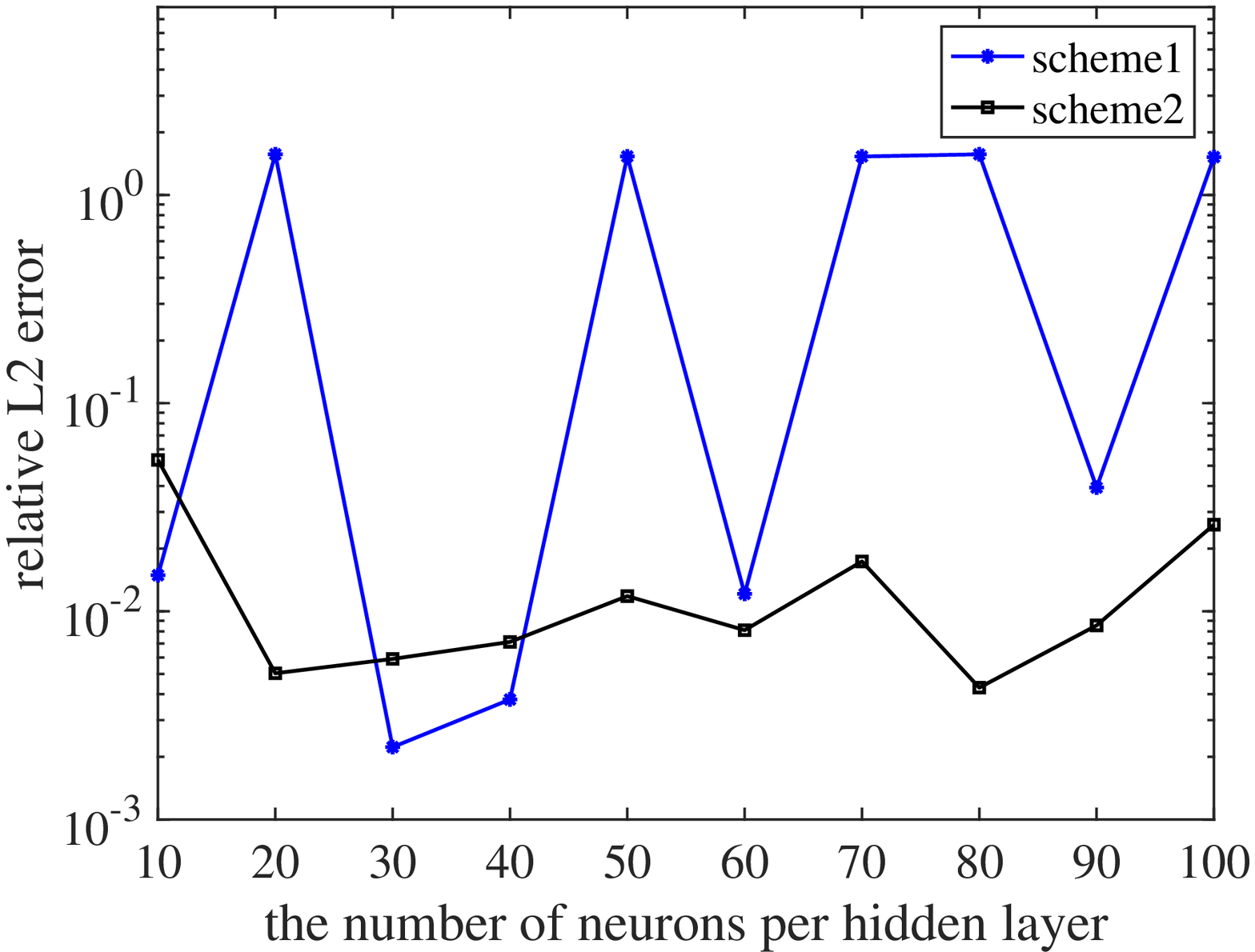}
$a$
\includegraphics[width=5.5cm,height=4.2cm]{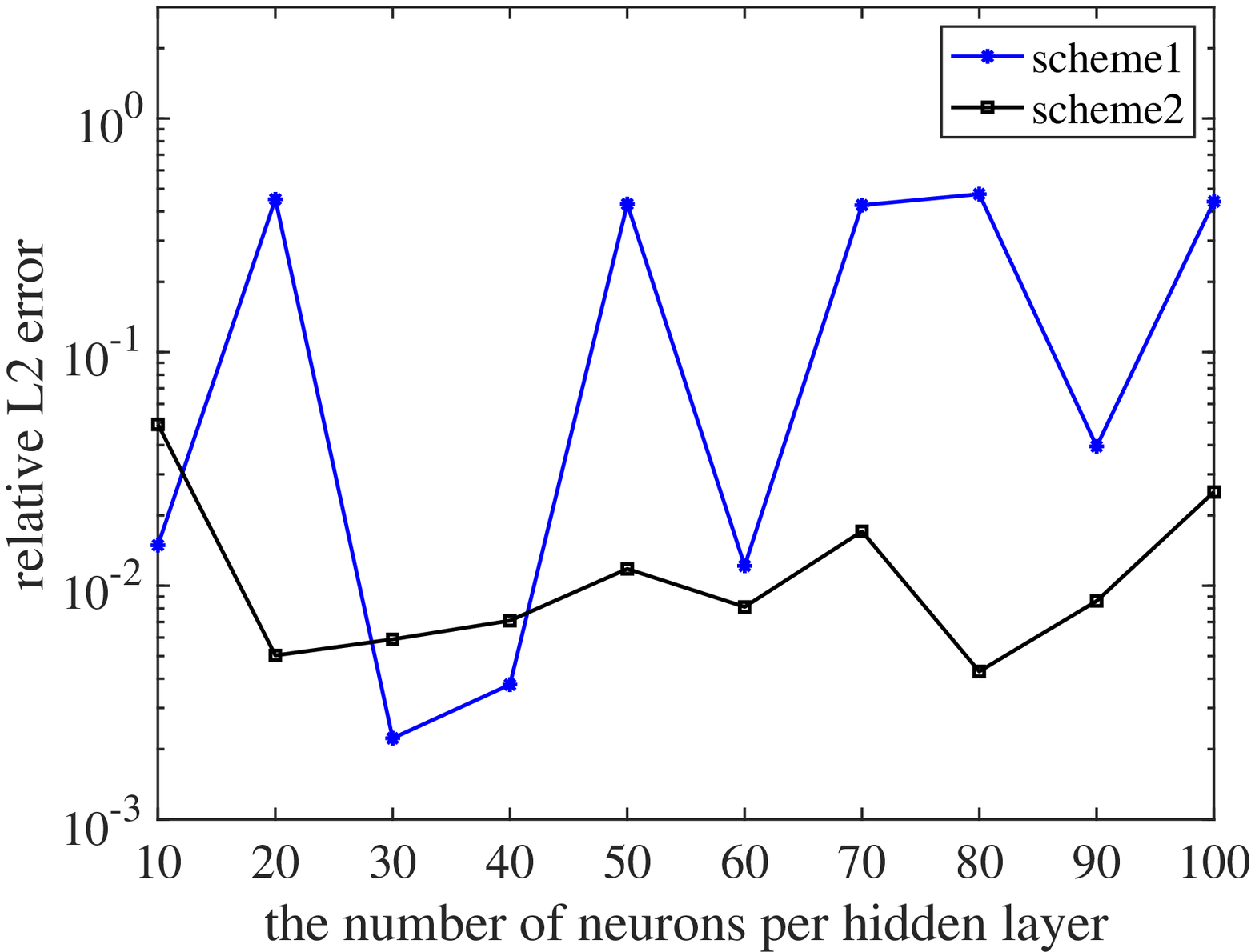}
$b$
\includegraphics[width=5.5cm,height=4.2cm]{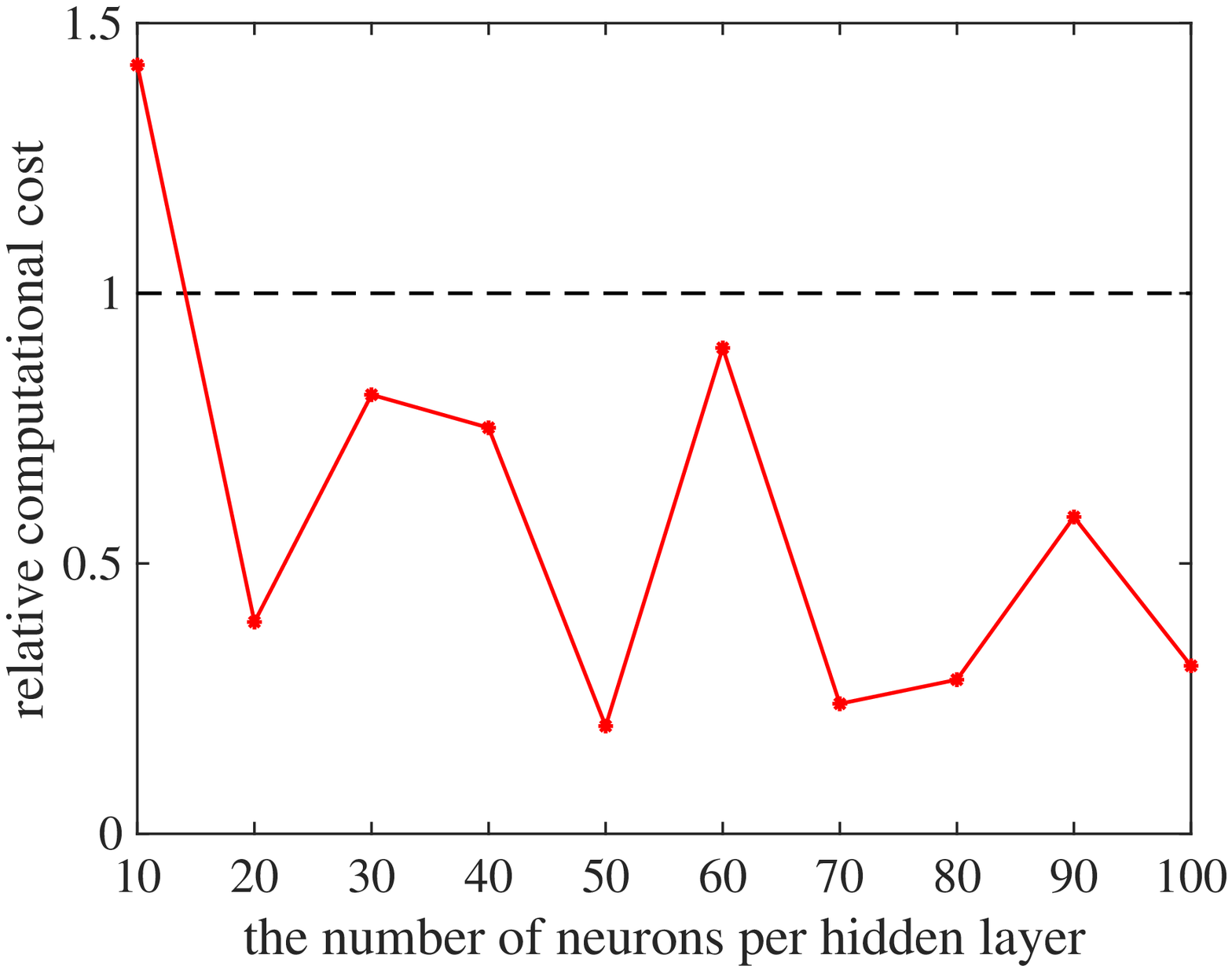}
$c$\\
\includegraphics[width=5.5cm,height=4.2cm]{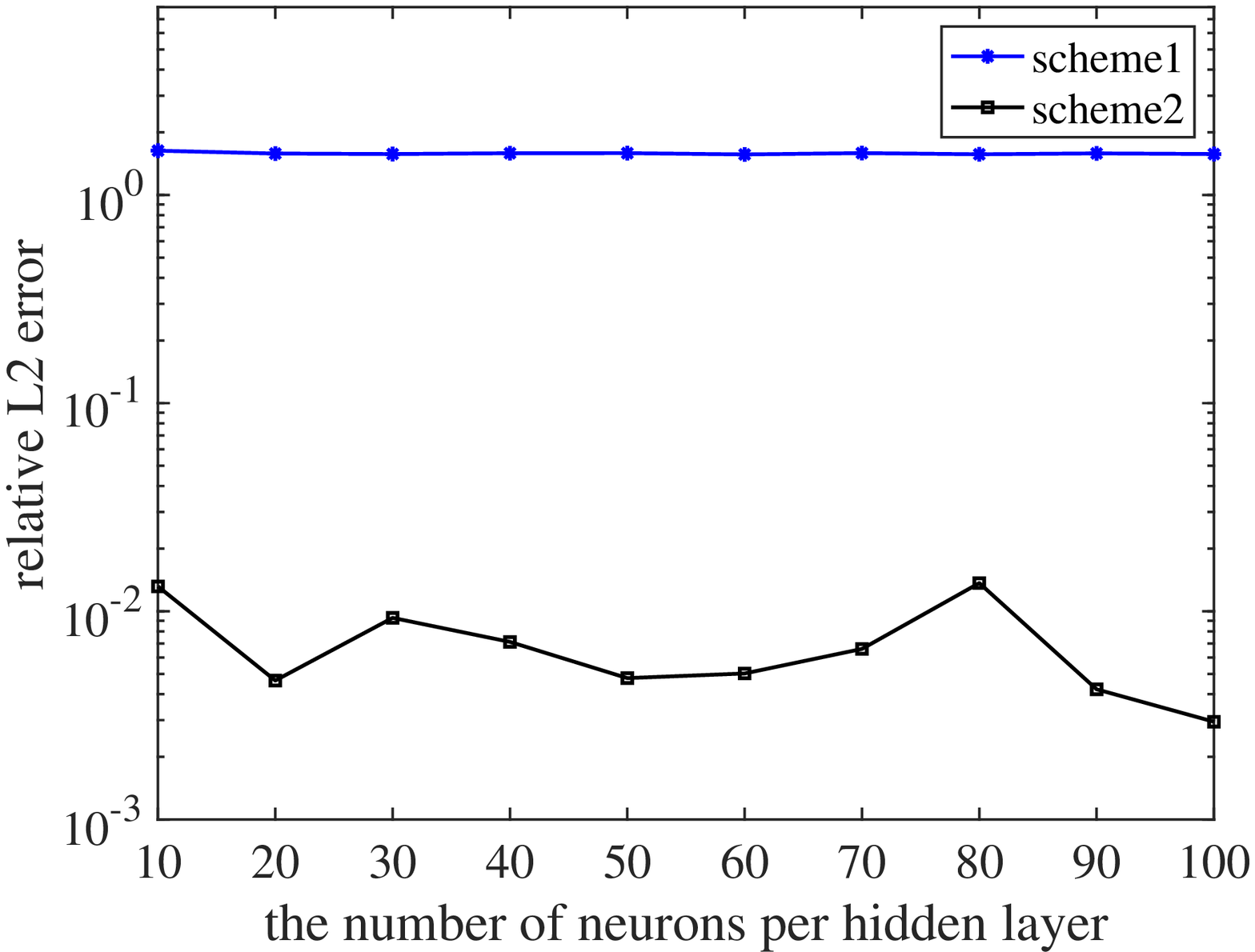}
$d$
\includegraphics[width=5.5cm,height=4.2cm]{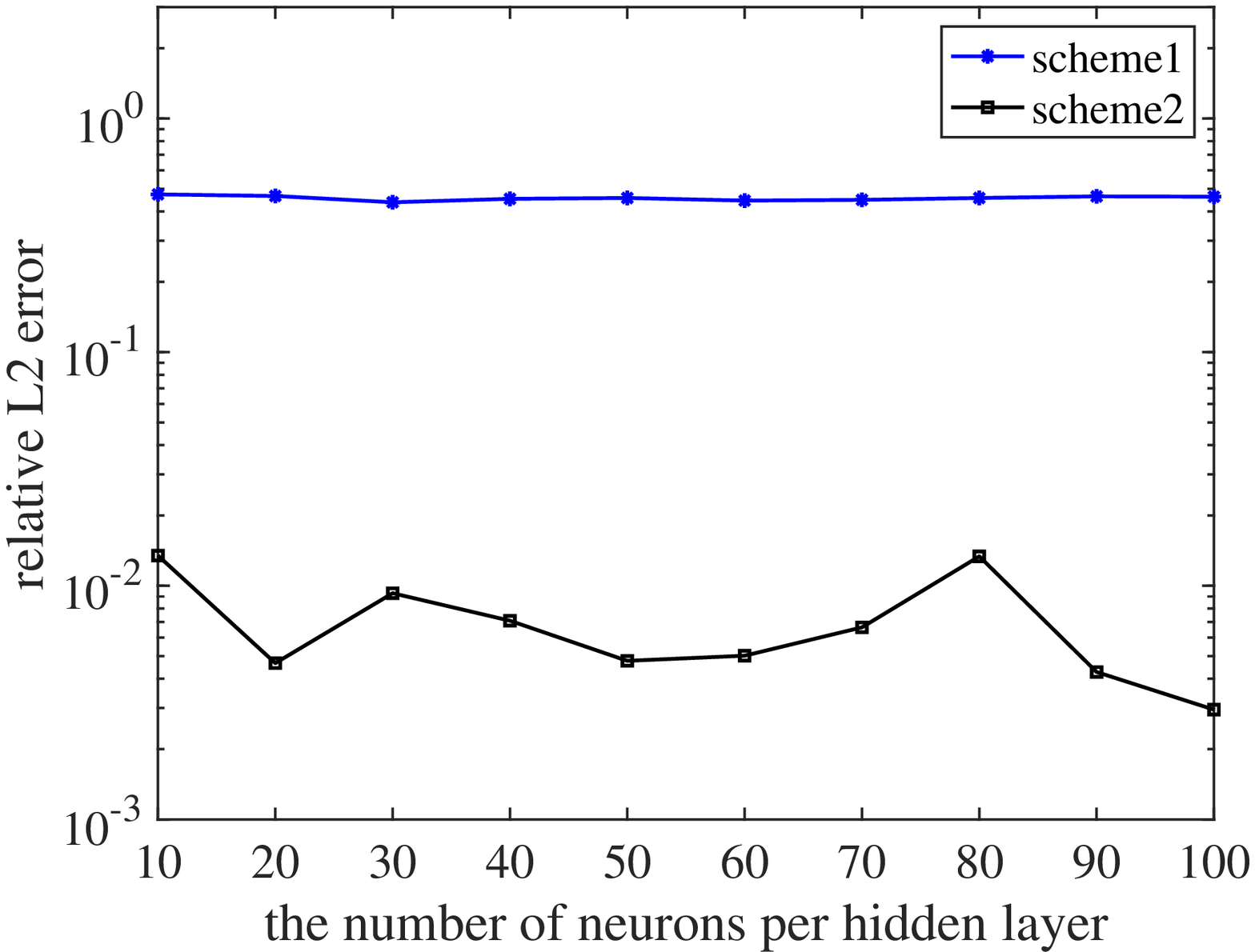}
$e$
\includegraphics[width=5.5cm,height=4.2cm]{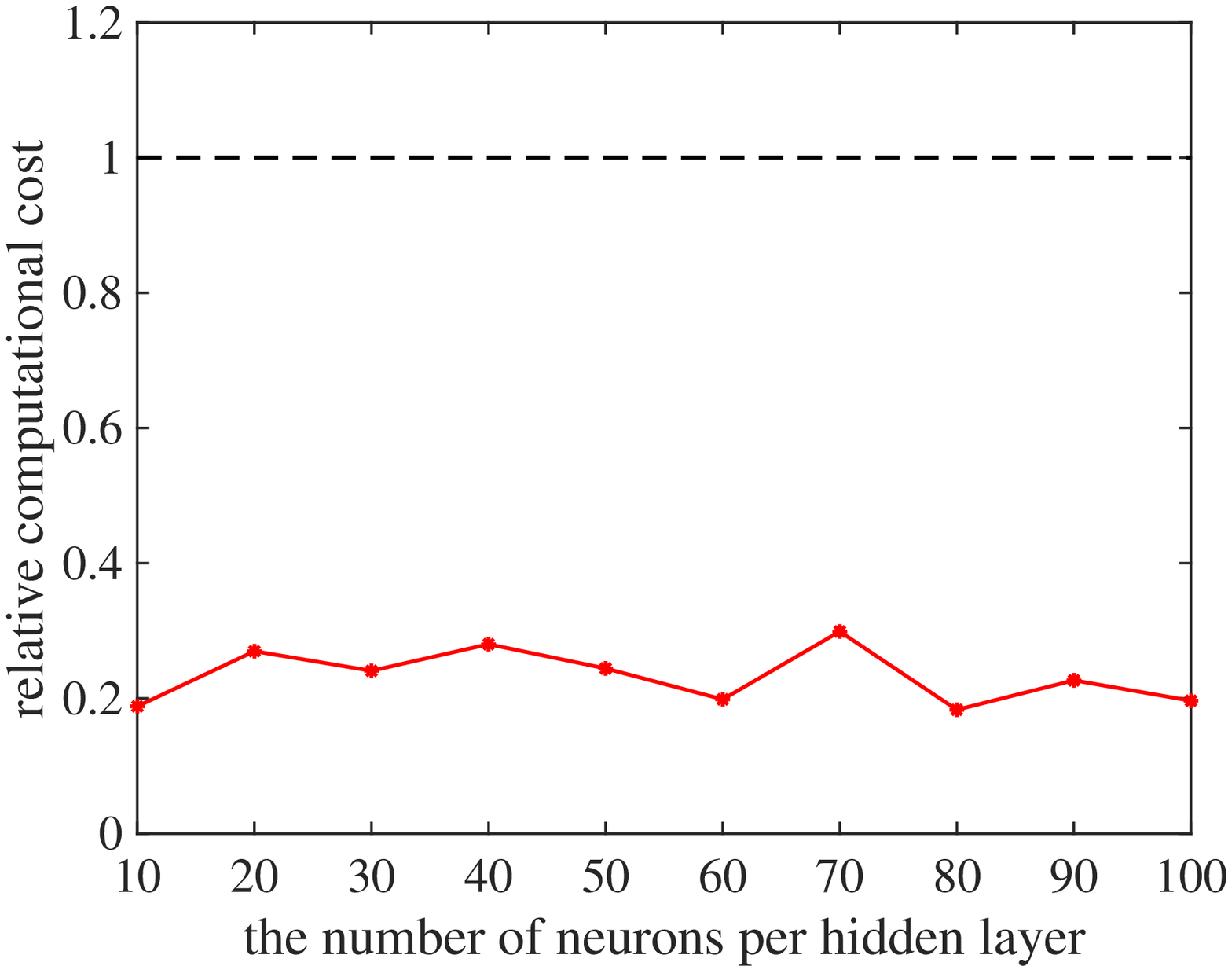}
$f$
\caption{(Color online) Data-driven two-soliton solution of the focusing mKdV solution: comparison of two schemes by using different numbers of neurons per hidden layer. Hidden layers = 4: (a) The relative $\mathbb{L}_2$ errors of $v_r$; (b) The relative $\mathbb{L}_2$ errors of $|v|$; (c) The relative computational costs; Hidden layers = 6: (d) The relative $\mathbb{L}_2$ errors of $v_r$; (e) The relative $\mathbb{L}_2$ errors of $|v|$; (f) The relative computational costs.}
\label{fig5-6}
\end{figure}

Numerical results of twenty experiments are summarized in in Table \ref{tableA-3-3}-Table \ref{tableA-3-6} in Appendix A. Besides, the relative $\mathbb{L}_2$ errors of the real part $v_r$ and the modulus $|v|$ as well as the relative computational costs are plotted in Fig. \ref{fig5-6}. Average relative $\mathbb{L}_2$ errors and elapsed time of ten experiments by using different numbers of neurons per hidden layer are presented in Table \ref{table5-3-2}.

\begin{table}[htbp]
\caption{Data-driven two-soliton solution of the focusing mKdV solution: average relative $\mathbb{L}_2$ errors and elapsed time by using different numbers of neurons per hidden layer.}
\label{table5-3-2}
\centering
\begin{tabular}{c|cc|cc}
\bottomrule
\multirow{2}{*}{Method} & \multicolumn{2}{c|}{hidden layers=4}      & \multicolumn{2}{c}{hidden layers=6}       \\ \cline{2-5}
                        & \multicolumn{1}{c|}{\textbf{Scheme I}} & \textbf{Scheme II} & \multicolumn{1}{c|}{\textbf{Scheme I}} & \textbf{Scheme II} \\ \hline
$v_r$                      & \multicolumn{1}{c|}{7.795786E-01}         & 1.476204E-02          & \multicolumn{1}{c|}{1.584703E+00}         & 7.146170E-03          \\ \hline
$|v|$                      & \multicolumn{1}{c|}{2.295433E-01}         & 1.422621E-02          & \multicolumn{1}{c|}{4.565248E-01}         & 7.153764E-03          \\ \hline
Average elapsed time (s)                       & \multicolumn{1}{c|}{2550.5411}         & 1025.71848          & \multicolumn{1}{c|}{4992.30987}         & 1129.29153          \\ \toprule
\end{tabular}
\end{table}

With a few exceptions, the relative $\mathbb{L}_2$ error of Scheme \uppercase\expandafter{\romannumeral 2} are smaller and it has higher efficiency overall while Scheme \uppercase\expandafter{\romannumeral 1} shows deficiencies in speed, precision as well as stability when the number of hidden layers is 4. The experimental results also indicate that the performance of Scheme \uppercase\expandafter{\romannumeral 2} is considerably better in both accuracy and efficiency when the number of hidden layers is 6. More specifically, the relative $\mathbb{L}_2$ errors of Scheme \uppercase\expandafter{\romannumeral 2} are at least two orders of magnitude smaller than that of Scheme \uppercase\expandafter{\romannumeral 1} and the relative computational costs remain below 0.4.

\subsection{Influence factors of types of data-driven solutions}
\quad

For a Miura transformation which transforms the solution $v$ of a certain partial differential equation into the solution $u$ of another one, the remarkable advantage of these two proposed schemes is that we can simply utilize the initial-boundary data of $u$ to simulate the data-driven solution $\widehat{v}$ of another equation. Most notably, various numerical solutions can be successfully derived with the aid of the many-to-one relationship between solutions before and after Miura transformations. For example, based on the kink solution \eqref{E3-4} of the defocusing mKdV equation, the soliton solution \eqref{E3-5} of the KdV equation can be obtained by real Miura transformation \eqref{E3-1} in Case 3.1, while conversely, not only the kink solution \eqref{E3-4} but also other three types of solutions (bright soliton, dark soliton and kink-bell type solution) can be simulated by using the initial-boundary data of the soliton solution \eqref{E3-5}, which shows that our method provides a possibility for new types of numerical solutions of nonlinear PDEs.

Inspired by the numerical results of Case 3.1 in Section \ref{5.1}, we discuss the following three influence factors of types of data-driven solutions in this part.

$\bullet$ \textbf{Type of the scheme}

Both Scheme \uppercase\expandafter{\romannumeral 1} and Scheme \uppercase\expandafter{\romannumeral 2} have their own merits in the diversity of solutions. In the numerical experiments of Case 3.1 that have been carried out in Section \ref{5.1}, three types of data-driven solutions (bright soliton, dark soliton and kink-bell type solution) can be obtained by Scheme \uppercase\expandafter{\romannumeral 1} while two types (kink and kink-bell type solution) by Scheme \uppercase\expandafter{\romannumeral 2}.

$\bullet$ \textbf{Structure of neural networks}

The structure of neural networks (the number of hidden layers and neurons per hidden layer) is also the main factor that affect the type of data-driven solution. When other conditions keep the same, it can be seen that multiple changes of types of solutions will occur by using different numbers of hidden layers or neurons per hidden layer from Table \ref{table5-1-2} and Table \ref{table5-1-4}.

$\bullet$ \textbf{Size of the spatiotemporal region}

\begin{figure}[htbp]
\centering
\includegraphics[width=5.5cm,height=4.5cm]{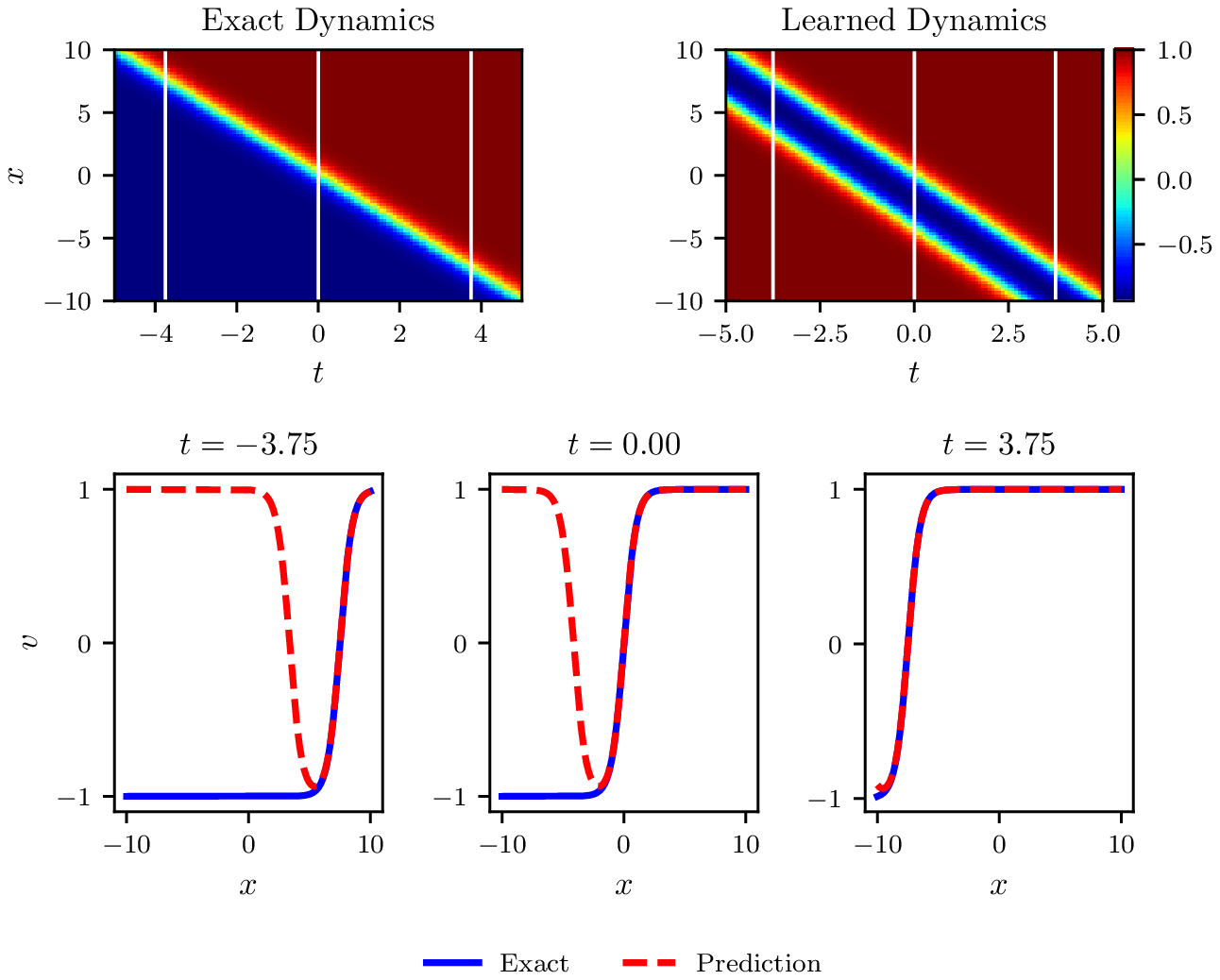}
$a$
\includegraphics[width=5.5cm,height=4.5cm]{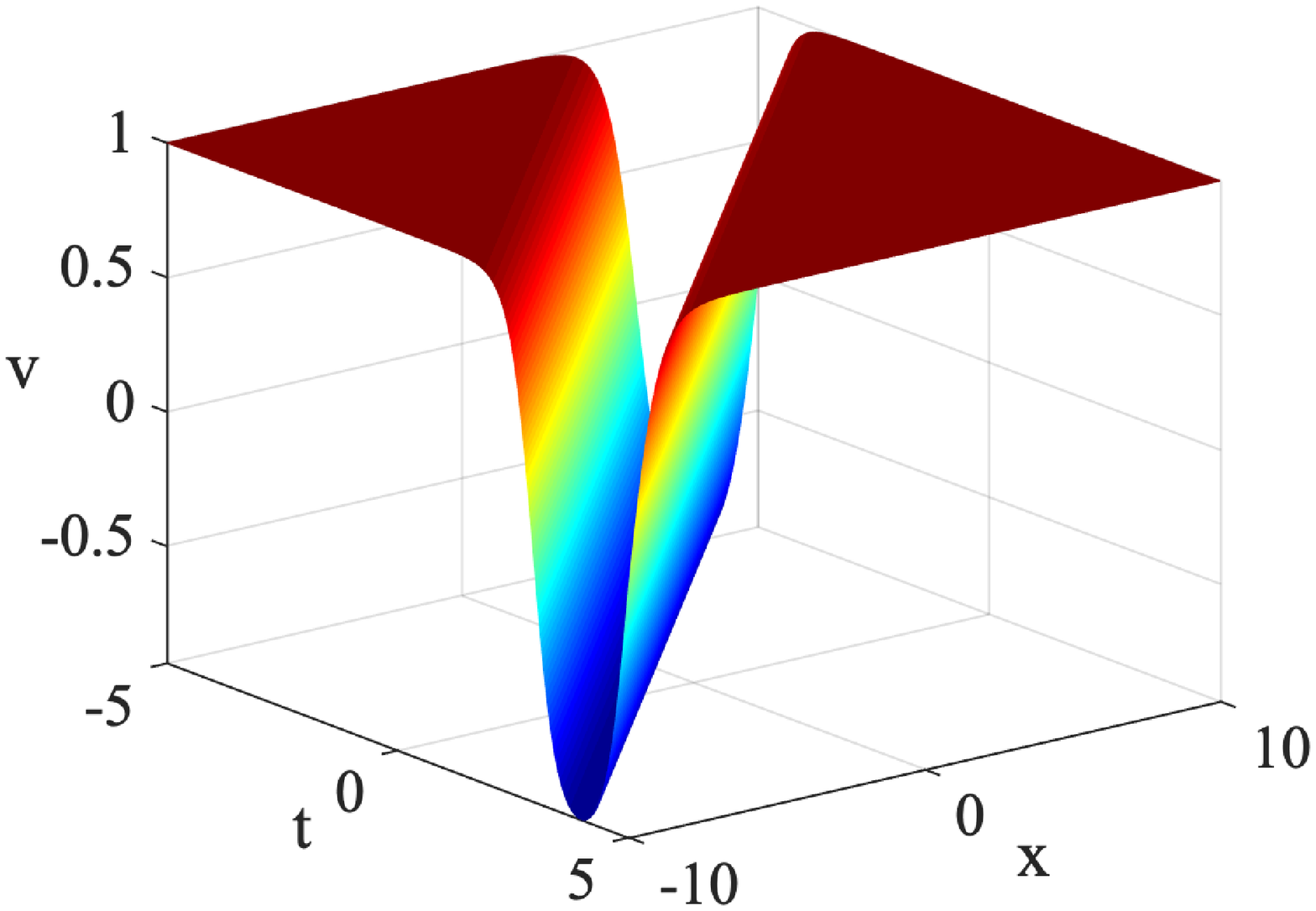}
$b$
\includegraphics[width=5.5cm,height=4.5cm]{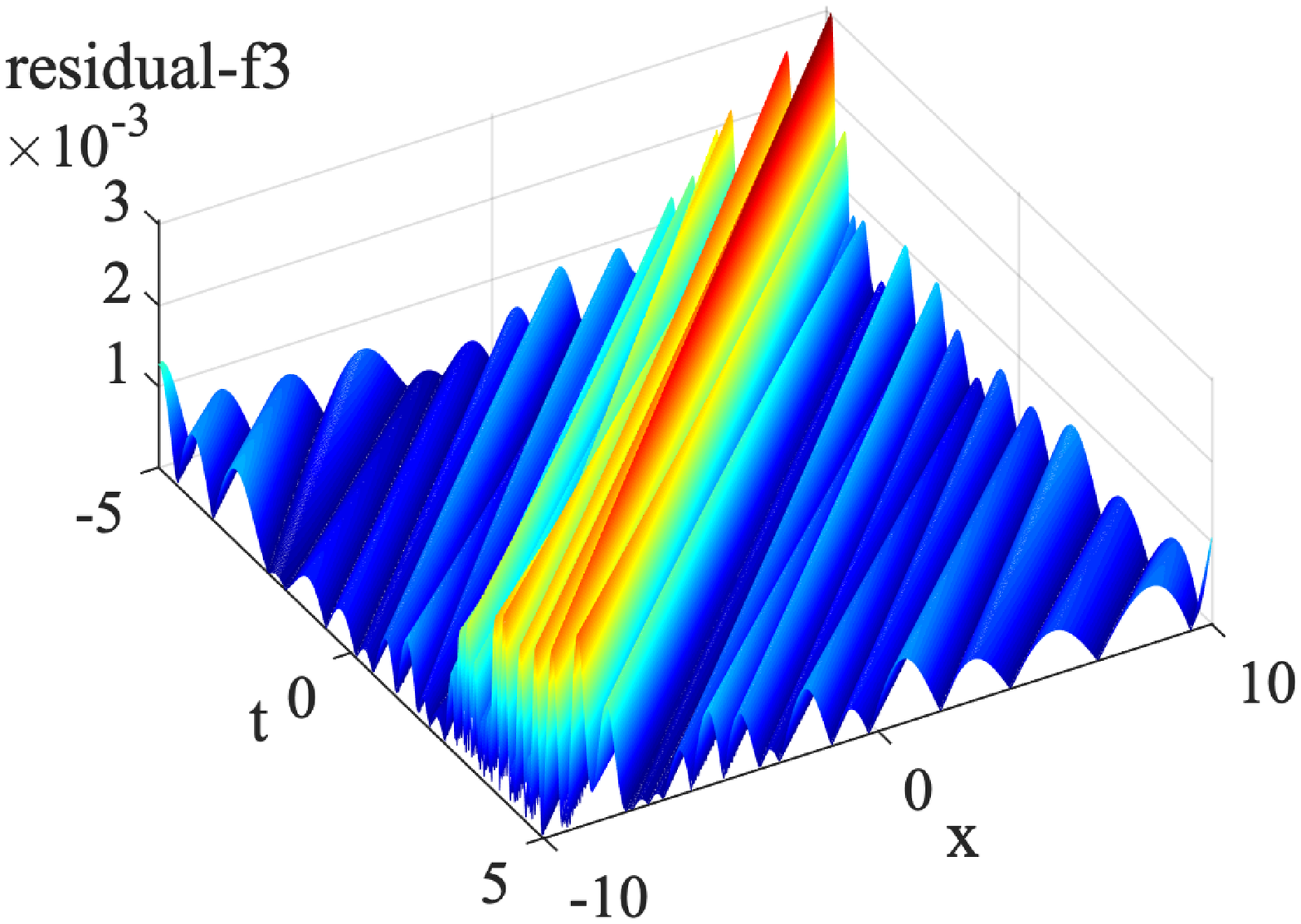}
$c$
\caption{(Color online) Data-driven solution $v(x,t)$ of the defocusing mKdV equation by Scheme \uppercase\expandafter{\romannumeral 1} in the spatiotemporal region $[x_0,x_1]\times[t_0,t_1]=[-10,10]\times[-5,5]$: (a) The density diagrams and comparison between the predicted solutions and exact solutions at the three temporal snapshots of $v(x,t)$; (b) The three-dimensional plot of $v(x,t)$; (c) The three-dimensional plot of the residual corresponding to $f_3$.}
\label{fig5-7}
\end{figure}

In Case 3.1, the size of spatiotemporal region is $[x_0,x_1]\times[t_0,t_1]=[-5,5]\times[-5,5]$ and then a 5-layer feedforward neural network with 40 neurons per hidden layer is constructed to successfully simulate the kink-bell type solution of the defocusing mKdV with the aid of Scheme \uppercase\expandafter{\romannumeral 1}. Under the same conditions, we choose $[x_0,x_1]\times[t_0,t_1]=[-10,10]\times[-5,5]$ as the training region and the dark soliton solution can be got after 6825 times iterations in about 641.2162 seconds, which is exhibited in Fig. \ref{fig5-7}. The mean squared error of $N_x \times N_t=513 \times 201$ grid points corresponding to $f_3$ is 5.880324e-07 and the PDE residual $|f_3(x^i,t^i)|$ is kept small in the order of magnitude of $10^{-3}$. This example illustrates that the size of the spatiotemporal region may affect the type of numerical solution.

\subsection{Advantages and disadvantages analysis of two schemes}
\quad

In this subsection, we discuss advantages and disadvantages of two schemes in terms of accuracy, efficiency, flexibility and diversity of solutions based on the numerical results in Section \ref{5.1}.

$\bullet$ \textbf{Accuracy}

According to the performance of two schemes in Case 3.1, it can be seen that the mean squared error $MSE_G$ of Scheme \uppercase\expandafter{\romannumeral 1} is less than that of Scheme \uppercase\expandafter{\romannumeral 2} in most cases but they are close on the whole except when the number of hidden layers and the number of neurons per hidden layers are 4 and 40, respectively. The accuracy of Scheme \uppercase\expandafter{\romannumeral 1} is slightly better than that of Scheme \uppercase\expandafter{\romannumeral 2} in Case 3.1, a relatively simpler example. Generally, we prefer to try Scheme \uppercase\expandafter{\romannumeral 1} first since only one feedforward neural network needs to be trained and it is more convenient. However, it may lead to unsatisfactory results under the circumstances of complicated equations, a large number of constraints and so on, which imply that there are too many objectives to be optimized. For example, there are more constraints in three cases in Section \ref{complex Miura} than in Section \ref{real Miura}. Thus Scheme \uppercase\expandafter{\romannumeral 1} no longer shows satisfactory performance in the repeated numerical tests and Scheme \uppercase\expandafter{\romannumeral 2} works better than Scheme \uppercase\expandafter{\romannumeral 1}. As for Scheme \uppercase\expandafter{\romannumeral 2}, we can acquire the numerical solution $\Widehat{u}(x,t)$ with satisfied accuracy in step one since the PINN method has been relatively mature and  effective in solving the initial-boundary value problems of partial differential equations. The success of the first step will contribute to the reliable numerical solution $\Widehat{v}(x,t)$ in step two. As it should be, the existing errors of $\Widehat{u}(x,t)$ in step one will further effect the accuracy of $\Widehat{v}(x,t)$ inevitably in step two.

$\bullet$ \textbf{Efficiency}

Theoretically, a major advantage of Scheme \uppercase\expandafter{\romannumeral 1} should be high efficiency and low training cost since we only need to train one physics-informed neural network to obtain the numerical solutions $\Widehat{u}(x,t)$ and $\Widehat{v}(x,t)$ but it turns out it doesn't. In almost all numerical experiments in Section \ref{5.1}, the relative computational costs remain below 1 with only one exception in Fig. \ref{fig5-6} (c). It demonstrates that the elapsed time of Scheme \uppercase\expandafter{\romannumeral 2} is always shorter and Scheme \uppercase\expandafter{\romannumeral 2} has the overwhelming superiority in efficiency, probably because the difficulty of this task is decomposed by its structure of two steps and then accelerates the training process, which follows the principle of gradual improvement.

$\bullet$ \textbf{Flexibility}

Compared with Scheme \uppercase\expandafter{\romannumeral 1}, the structure of neural networks of $\Widehat{u}(x,t)$ and $\Widehat{v}(x,t)$ in Scheme \uppercase\expandafter{\romannumeral 2} can be different and the number of hidden layers and neurons in step two of Scheme \uppercase\expandafter{\romannumeral 2} can be seen as additional hyper-parameters to be optimized, which is obviously more flexible and targeted. It should be noted that gratifying accuracy can be achieved by fairly shallow neural networks (the number of hidden layers is only 1 or 2) in step two based on experimental data presented in Appendix A, and it may work even better after optimization of the hyper-parameters in step two.

$\bullet$ \textbf{Diversity of solutions}

\begin{figure}[htbp]
\centering
\includegraphics[width=12cm,height=7.5cm]{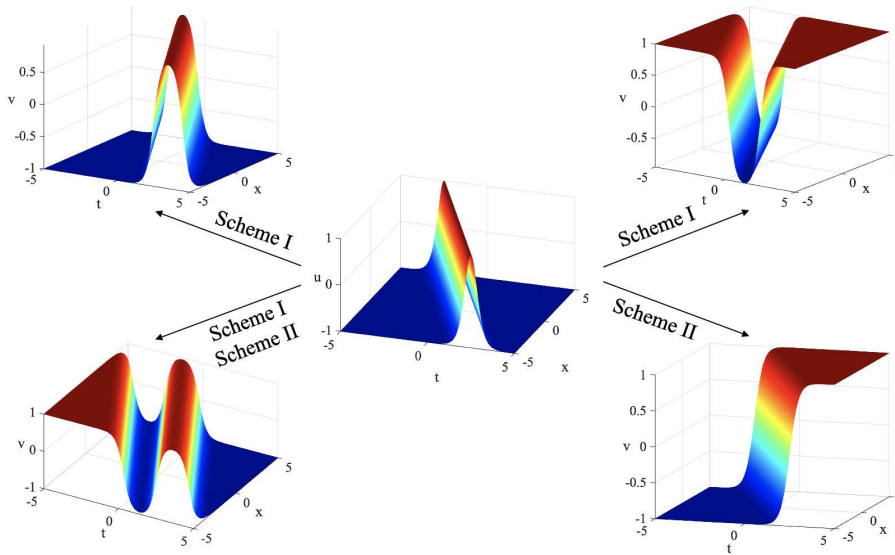}
\caption{(Color online) Various data-driven solutions of the defocusing mKdV equation based on the initial-boundary data of one-soliton solution of the KdV equation.}
\label{fig5-8}
\end{figure}

In Case 3.1, various numerical solutions of the defocusing mKdV equation are obtained by simply using the initial-boundary data of the one-soliton solution of the KdV equation. Based on the numerical experiments of Case 3.1 that have been carried out in Section \ref{5.1}, data-driven solutions and the corresponding schemes are shown in Fig. \ref{fig5-8}. Obviously, we can simulate three types of data-driven solutions (bright soliton, dark soliton and kink-bell type solution) and two types (kink and kink-bell type solution) by adopting Scheme \uppercase\expandafter{\romannumeral 1} and Scheme \uppercase\expandafter{\romannumeral 2}, respectively. Thus, these two schemes have their separate strengths in diversity of solutions.

In conclusion, Scheme \uppercase\expandafter{\romannumeral 2} has the overwhelming superiority in efficiency and its structure of neural networks is more flexible. Both two schemes have their own merits in the diversity of solutions and have the advantages of high accuracy in the relatively simple cases, such as Case 3.1. However, Scheme \uppercase\expandafter{\romannumeral 2} far outperforms Scheme \uppercase\expandafter{\romannumeral 1} in accuracy under the circumstances of complicated equations, a large number of constraints and so on, such as Case 4.1 and Case 4.2. Indeed, Scheme \uppercase\expandafter{\romannumeral 2} has better performance than Scheme \uppercase\expandafter{\romannumeral 1} in the numerical experiments of cases that have been carried out and discussed so far, but the effect of two schemes remains to be explored in other cases. In addition, no free lunch theorem \cite{Nofreelunch} indicates that no machine learning algorithm is always better than others in a sense. To sum up, two schemes of each has its own merits and thus more appropriate one should be chosen according to specific cases.

\section{Conclusion}

To our knowledge, there is rare correlative study that takes into account transformations in PINNs. The novelty of this research is the incorporation of Miura transformation constraints into neural networks to solve nonlinear PDEs. In the fields of integrable systems, the Miura transformation has great significance since it can transform the solution $v$ of a certain partial differential equation into the solution $u$ of another one, which provides convenience for solving the latter. Conversely, we wonder how to make use of Miura transformations to solve the former on the basis of initial and boundary conditions of the solution of the latter. In this paper, we devise two PINN schemes based on Miura transformations to acquire the numerical solution $\widehat{v}$ by leveraging the initial-boundary data of $u$.

The main difference of two schemes is that only one PINN is constructed to acquire the data-driven solutions (both $\Widehat{u}(x,t)$ and $\Widehat{v}(x,t)$) in Scheme \uppercase\expandafter{\romannumeral 1} while two PINNs are trained to derive $\Widehat{u}(x,t)$ and $\Widehat{v}(x,t)$ respectively in Scheme \uppercase\expandafter{\romannumeral 2}. Furthermore, we richly exemplify the use of these two schemes. Firstly, Scheme \uppercase\expandafter{\romannumeral 1} is adopted to carry out numerical experiments with respect to the real Miura transformation between the defocusing mKdV equation and the KdV equation. Meanwhile, the applications of Scheme \uppercase\expandafter{\romannumeral 2} in the complex Miura transformation between the focusing mKdV equation and the KdV equation are presented as well. Remarkably, various new data-driven solutions of the defocusing and focusing mKdV equations are simulated, which are different from exact ones utilized to derive the initial-boundary data of the solution for the KdV equation. Presumably it's because there is a many-to-one relationship between the solution of the defocusing/focusing mKdV equation and that of the KdV equation under the Miura transformation. One of the most important results is the discovery of a new localized wave solution: kink-bell type solution of the defocusing mKdV equation and it has not been previously observed and reported to our knowledge. Abundant numerical results show that the proposed methods can fulfill our purpose effectively and provides a possibility for new types of numerical solutions, which illustrates that the methodology possesses a great application merit. We also discuss the performance comparison in different cases and analyze advantages and disadvantages of two schemes  in detail. On the whole, Scheme \uppercase\expandafter{\romannumeral 2} has better performance than Scheme \uppercase\expandafter{\romannumeral 1} in the numerical experiments of cases that have been carried out and discussed above, and there is sufficient data to support this conclusion and our study. Ulteriorly, our proposed method can also be applied to the Miura transformations between other equations to observe dynamical behaviors of new numerical solutions in future studies and the effect of two schemes remains to be explored in other cases. Both schemes have their own advantages and no free lunch theorem indicates that in a sense, no machine learning algorithm is always better than others. Therefore, more appropriate one should be chosen according to specific cases.

The role of our proposed method is that the data-driven solution of another equation can be obtained if the initial-boundary data of the solution of a certain partial differential equation is known and there is a Miura transformation between these two equations. Meanwhile, our method is tailored to the inverse process of the Miura transformation since it is challenging to solve $v$ based on the Miura transformation \eqref{E2-3}, which is an implicit expression of $v$. Moreover, it provides a possibility for new numerical solutions of integrable equations attributed to the many-to-one relationship between solutions before and after Miura transformations, which may not be derived by classical methods or even have not the explicit expression. The combination of Miura transformations and PINNs may bring new inspirations to the field of integrable systems and it is of practical meaning and application foreground.

How to combine deep learning with integrable system theory more tightly and effectively to devise significant integrable-deep learning algorithms is an urgent problem to be solved in the future. Moreover, how to utilize the PINN method to pertinently solve problems arising in the field of integrable systems that cannot be solved by classical methods is also the goal we pursue. These new challenges will be considered in our future research.

\section*{Acknowledgments}
The project is supported by National Natural Science Foundation of China (No. 12175069 and No. 12235007) and Science and Technology Commission of Shanghai Municipality (No. 21JC1402500 and No. 22DZ2229014).

\section*{Appendix A.}

The detailed results of numerical experiments in Section \ref{5.1} are displayed in Table \ref{tableA-1-1}-Table \ref{tableA-3-6}, including the number of hidden layers and neurons per hidden layer, elapsed time, relative $\mathbb{L}_2$ errors, mean squared errors and type of solution. Note that the contents in brackets of the first column 'Hidden layers-Neurons' denote the number of hidden layers and neurons per hidden layer in step two of Scheme \uppercase\expandafter{\romannumeral 2}.

\begin{table}[H]
\caption{Data-driven solutions in Case 3.1 via Scheme \uppercase\expandafter{\romannumeral 1} by using different numbers of hidden layers.}
\label{tableA-1-1}
\centering
\begin{tabular}{c|ccccc}
\hline
\multicolumn{1}{c|}{\begin{tabular}[c]{@{}c@{}}Hidden layers\\ -Neurons\end{tabular}} & Elapsed time (s) & $u$ & $v$ & $MSE_G$ & Type of solution \\ \hline
 2-40& 567.483                                       & 4.133840E-03                       & 1.174788E+00                       & 1.419671E-05                            & bright soliton    \\
4-40 &  503.2289                                      & 7.385367E-04                       &  1.509559E+00                      & 7.247345E-07                            & kink-bell type    \\
 6-40&  422.4909                                      &8.504728E-04                        &1.107687E+00                        & 1.210860E-06                            &bright soliton     \\
 8-40& 453.2294                                       & 4.722505E-04                       & 1.065068E+00                       & 2.132264E-07                            &dark soliton     \\
 10-40& 582.3161                                       &  6.720320E-04                      & 1.084498E+00                       & 5.318142E-07                            &bright soliton     \\
12-40 & 579.0338                                      &3.706387E-04                        & 1.071584E+00                       &4.619493E-07                             &bright soliton     \\
14-40 & 856.6929                                       & 3.941899E-04                       & 1.064003E+00                       &  7.107039E-07                           &bright soliton     \\
16-40& 623.9675                                &6.276342E-04                        & 1.076917E+00                       & 5.428547E-07                            &bright soliton     \\
18-40& 1264.8702                                       & 7.273173E-04                       &  1.054744E+00                      &  3.873833E-07                           &dark solito     \\
20-40&   1424.2475                                     & 7.500523E-04                       &   1.068134E+00                     &  2.212841E-07                           &bright soliton    \\ \hline
\end{tabular}
\end{table}

\begin{table}[H]
\caption{Data-driven solutions in Case 3.1 via Scheme \uppercase\expandafter{\romannumeral 2} by using different numbers of hidden layers.}
\label{tableA-1-2}
\centering
\begin{tabular}{c|ccccc}
\hline
\multirow{2}{*}{\begin{tabular}[c]{@{}c@{}}Hidden layers\\ -Neurons\end{tabular}} & \multicolumn{1}{c|}{\multirow{2}{*}{\begin{tabular}[c]{@{}c@{}}Total \\ elapsed time (s)\end{tabular}}} & \multicolumn{1}{c|}{Step One} & \multicolumn{3}{c}{Step Two} \\ \cline{3-6} & \multicolumn{1}{c|}{}                     & \multicolumn{1}{c|}{$u$}        & $v$ & $MSE_G$ & Type of solution \\ \hline
 2-40(2-40)& 181.4906                                       & 3.099765E-04                       & 1.453627E+00                       &  2.112585E-06                           & kink-bell type    \\
4-40(2-40) &81.3834                                        & 4.984651E-01                       & 1.754573E+00                       & 2.134646E-04                            & Failed    \\
 6-40(2-40)& 86.5824                                       & 2.487746E-04                       & 1.436638E+00                       & 3.365896E-07                            & kink-bell type    \\
 8-40(2-40)& 114.09963                                       &  3.736209E-04                      & 1.441530E+00                       & 4.203775E-07                            &  kink-bell type   \\
 10-40(2-40)& 156.7211                                       &  3.942623E-04                      &  1.470885E+00                      & 1.690848E-06                            &  kink-bell type   \\
12-40(2-40) & 174.1295                                       &  4.323423E-04                      & 1.453320E+00                       & 1.319101E-06                            & kink-bell type    \\
14-40(2-40) & 166.6431                                       & 6.593621E-04                       &  1.377086E+00                      & 9.889965E-07                            & kink-bell type    \\
16-40(2-40)& 163.0173                                       &  5.464281E-04                      & 3.576556E-04                       &  1.180394E-06                           & kink    \\
18-40(2-40)&211.5204                                        &  8.079709E-04                      &  3.289345E-04                      & 5.404663E-07                            & kink    \\
20-40(2-40)& 392.4873                                       & 1.172494E-03                       & 6.173271E-04                       &  3.022606E-07                           & kink    \\ \hline
\end{tabular}
\end{table}

\begin{table}[H]
\caption{Data-driven solutions in Case 3.1 via Scheme \uppercase\expandafter{\romannumeral 1} by using different numbers of neurons.}
\label{tableA-1-3}
\centering
\begin{tabular}{c|ccccc}
\hline
\multicolumn{1}{c|}{\begin{tabular}[c]{@{}c@{}}Hidden layers\\ -Neurons\end{tabular}} & Elapsed time (s) & $u$ & $v$ & $MSE_G$ & Type of solution \\ \hline
 4-10& 188.5116                                       & 1.666771E-03                       & 1.134741E+00                       & 2.584902E-06                            & dark soliton    \\
 4-20& 294.2148                                       & 9.223519E-04                       & 1.107674E+00                       & 6.282552E-07                            & dark soliton    \\
 4-30& 417.4834                                       & 9.485671E-04                       & 1.112802E+00                       &1.000728E-06                             & dark soliton    \\
 4-40&  503.2289                                      & 7.385367E-04                       & 1.509559E+00                       & 7.247345E-07                            & kink-bell type    \\
 4-50& 619.6686                                       & 6.244513E-04                       & 1.086986E+00                       & 2.116692E-07                            &  bright soliton   \\
 4-60 & 781.9227                                       & 1.013358E-03                       & 1.115339E+00                       & 7.240079E-07                            &  dark soliton   \\
 4-70& 1246.9586                                       &  6.974275E-04                      &  1.094667E+00                      &  4.398950E-07                           & dark soliton    \\
 4-80&  1757.1154                                      &  9.608244E-04                      &  1.107847E+00                      & 7.834202E-07                            & dark soliton    \\
 4-90&  1987.1193                                      &   7.294770E-04                     &1.100771E+00                        & 5.105729E-07                            & dark soliton    \\
 4-100& 1975.274                                       & 1.452055E-03                       & 1.129787E+00                       & 1.856679E-06                            & dark soliton    \\ \hline
\end{tabular}
\end{table}

\begin{table}[H]
\caption{Data-driven solutions in Case 3.1 via Scheme \uppercase\expandafter{\romannumeral 2} by using different numbers of neurons.}
\label{tableA-1-4}
\centering
\begin{tabular}{c|ccccc}
\hline
\multirow{2}{*}{\begin{tabular}[c]{@{}c@{}}Hidden layers\\ -Neurons\end{tabular}} & \multicolumn{1}{c|}{\multirow{2}{*}{\begin{tabular}[c]{@{}c@{}}Total \\ elapsed time (s)\end{tabular}}} & \multicolumn{1}{c|}{Step One} & \multicolumn{3}{c}{Step Two} \\ \cline{3-6} & \multicolumn{1}{c|}{}                     & \multicolumn{1}{c|}{$u$}        & $v$ & $MSE_G$ & Type of solution \\ \hline
 4-10(2-40)& 33.962                                       &  1.004896E-04                      & 1.460607E+00                       &  1.267523E-06                           & kink-bell type    \\
 4-20(2-40) & 30.3064                                       & 2.639955E-04                       & 1.395289E+00                       & 9.979505E-07                            & kink-bell type    \\
 4-30(2-40)&56.5717                                        & 4.885976E-04                       & 1.438988E+00                       & 8.231744E-07                            & kink-bell type    \\
 4-40(2-40)&81.3834                                        &  4.984651E-01                      & 1.754573E+00                       & 2.134646E-04                            &  Failed   \\
 4-50(2-40)& 93.5626                                       & 2.615815E-04                       &  1.437260E+00                      &  8.659781E-07                           &  kink-bell type   \\
 4-60(2-40) &78.8367                                        &  1.016026E-03                      & 1.451707E+00                       & 2.121951E-06                            &  kink-bell type   \\
 4-70(2-40) &155.1951                                        & 5.887764E-04                       & 1.417840E+00                       & 6.285435E-07                            &  kink-bell type   \\
 4-80(2-40)&265.2676                                        & 9.530057E-04                       & 1.373539E+00                       & 8.692436E-07                            &  kink-bell type   \\
 4-90(2-40)&232.6271                                        & 1.438540E-03                       &  1.464685E+00                      & 8.149107E-07                            & kink-bell type    \\
 4-100(2-40)& 437.4214                                       & 2.184186E-03                       & 1.326573E+00                       & 5.732672E-07                            &  kink-bell type   \\ \hline
\end{tabular}
\end{table}

\begin{table}[H]
\caption{Data-driven solutions in Case 4.1 via Scheme \uppercase\expandafter{\romannumeral 1} by using different numbers of hidden layers.}
\label{tableA-2-1}
\centering
\begin{tabular}{c|cccccc}
\hline
\multicolumn{1}{c|}{\begin{tabular}[c]{@{}c@{}}Hidden layers\\ -Neurons\end{tabular}} & Elapsed time (s) & $u_r$ & $u_i$ & $|u|$ & $v_r$ & $|v|$ \\ \hline
2-40             &   204.3242                                    &    5.183421E-04                          &    6.065814E-04                          &    3.709793E-04
    &    2.731884E-02
    &    2.727806E-02\\
4-40             &   1419.7727                                    &  8.910368E-01                            &  1.293331E+00                            &  7.420655E-01                      & 1.039803E+00                             &  9.066031E-01 \\
6-40             &  260.1525                                     &  9.211512E-04                            &  1.174782E-03                            &  7.847132E-04                      &   3.018471E-02                           &  3.061304E-02 \\
8-40             & 1724.9562                                      &   5.879466E-01                           &  7.403070E-01                            &   5.522145E-01                     &     2.358843E+00                         &  9.519682E-01 \\
10-40            &  504.3492                                     &  1.193635E-03                            &   1.731766E-03                           &   8.661178E-04                     &   1.498732E-02                           &  1.480922E-02 \\
12-40            &  756.1334                                     &  3.505310E-03                            &  4.573557E-03                            &  2.264153E-03                      &   2.473040E-02                           &  2.916471E-02 \\
14-40            &  879.3537                                     &  1.767510E-03                            &   2.614627E-03                           &  1.216132E-03                      &   3.791111E-02                           &  3.370965E-02 \\
16-40            & 4169.47                                      &                              6.066911E-01&    7.653934E-01                          & 5.659317E-01                       & 2.273458E+00                             &  7.924997E-01 \\
18-40            & 1536.4741                                      &   2.293036E-03                           &  2.433681E-03                            &  1.333688E-03                      &  3.757963E-02                            &  5.536005E-02 \\
20-40            &   1184.6515                                    &   1.473440E-03                           &    2.016878E-03                          &  1.032837E-03                      &    2.925396E-02                          & 3.308443E-02  \\ \hline
\end{tabular}
\end{table}

\begin{table}[H]
\caption{Data-driven solutions in Case 4.1 via Scheme \uppercase\expandafter{\romannumeral 2} by using different numbers of hidden layers.}
\label{tableA-2-2}
\centering
\begin{tabular}{c|cccccc}
\hline
\multirow{2}{*}{\begin{tabular}[c]{@{}c@{}}Hidden layers\\ -Neurons \end{tabular}} & \multicolumn{1}{c|}{\multirow{2}{*}{\begin{tabular}[c]{@{}c@{}}Total \\ elapsed time (s)\end{tabular}}} & \multicolumn{3}{c|}{Step One}              & \multicolumn{2}{c}{Step Two} \\ \cline{3-7}
\multicolumn{1}{l|}{}                               & \multicolumn{1}{c|}{}                                                                                   & $u_r$ & $u_i$ & \multicolumn{1}{c|}{$|u|$} & $v_r$          & $|v|$         \\ \hline
2-40(1-40)             &  80.8637                                     &   1.403033E-03                           & 1.400772E-03                             &   8.318377E-04
    &  1.157315E-03
    &  1.957077E-03  \\
4-40(2-40)             &  42.7879                                     &  2.437708E-03                            & 3.227638E-03                             &  1.622590E-03                      &   1.514716E-03                           &  1.522125E-03 \\
6-40(2-40)             & 91.0395                                      &   1.046530E-03                           &  2.251383E-03                            & 1.307515E-03                       &    2.192511E-03                          & 2.171804E-03  \\
8-40(1-40)             &  168.5585                                     &   1.770486E-03                           &   2.149758E-03                           & 1.335227E-03                       &   2.052790E-03                           &  2.001277E-03 \\
10-40(2-40)             &  131.4029                                     & 2.124429E-03                             &  2.657149E-03                            &  1.408388E-03                      &   2.570683E-03                           &  3.683095E-03 \\
12-40(2-40)             &  172.9466                                     &   1.025497E-03                           &  1.359100E-03                            &  7.375549E-04                      &   4.336539E-03                           & 5.323622E-03  \\
14-40(2-40)             &  309.4216                                     &   1.439781E-03                           &   2.380187E-03                           &  1.127538E-03                      &   4.181549E-03                           &  4.594574E-03 \\
16-40(2-40)             &  832.1773                                     &  3.270299E-03                            &  4.133512E-03                            &   2.103457E-03                     &   2.035153E-03                           & 2.494883E-03  \\
18-40(2-40)             &  576.1047                                     &  2.335908E-03                            & 3.679665E-03                             &   2.207488E-03                     &  3.422012E-03                            & 3.494185E-03  \\
20-40(2-40)             &  321.1166                                     &  1.890436E-03                            &  2.021988E-03                            &  1.023345E-03                      &   2.900189E-03                           & 3.988804E-03  \\ \hline
\end{tabular}
\end{table}

\begin{table}[H]
\caption{Data-driven solutions in Case 4.1 via Scheme \uppercase\expandafter{\romannumeral 1} by using different numbers of neurons.}
\label{tableA-2-3}
\centering
\begin{tabular}{c|cccccc}
\hline
\multicolumn{1}{c|}{\begin{tabular}[c]{@{}c@{}}Hidden layers\\ -Neurons\end{tabular}} & Elapsed time (s) & $u_r$ & $u_i$ & $|u|$ & $v_r$ & $|v|$ \\ \hline
4-10             & 244.3881                                     &  2.571718E-04                            & 4.251052E-04                             &  2.451293E-04
    & 1.703381E-02
    & 1.880503E-02   \\
4-20             & 205.6731                                      & 4.076871E-04                             & 5.689918E-04                             &  3.265203E-04                      & 4.309674E-02                             & 3.795284E-02  \\
4-30             & 1329.9447                                      & 6.280789E-01                             & 7.854374E-01                             & 5.817879E-01                       & 2.301349E+00                             & 9.035948E-01  \\
4-40             & 1419.7727                                      & 8.910368E-01                             & 1.293331E+00                             & 7.420655E-01                       &  1.039803E+00                            & 9.066031E-01  \\
4-50            & 111.1782                                      & 5.512751E-04                             &  5.300699E-04                            &  3.493309E-04                      &  2.585492E-02                            & 3.279927E-02  \\
4-60            & 181.4277                                      &  9.474704E-04                            &  8.780018E-04                            & 4.887054E-04                       & 5.344326E-02                             & 4.695949E-02  \\
4-70            & 182.2225                                      & 6.684623E-04                             &  8.432546E-04                            &  5.397178E-04                      &  3.026379E-02                            & 3.306955E-02  \\
4-80            &  182.0109                                     &  3.978733E-04                            &  6.890699E-04                            & 4.173083E-04                       &  3.645947E-02                            &  3.952860E-02 \\
4-90            & 272.8169                                      &  9.510325E-04                            &  1.708237E-03                            &  9.834717E-04                      &  3.499474E-02                            & 3.181558E-02  \\
4-100            & 264.121                                      &  1.129744E-03                            & 1.558647E-03                             & 9.390967E-04                       &  4.331539E-02                            & 3.795801E-02  \\\hline
\end{tabular}
\end{table}

\begin{table}[H]
\caption{Data-driven solutions in Case 4.1 via Scheme \uppercase\expandafter{\romannumeral 2} by using different numbers of neurons.}
\label{tableA-2-4}
\centering
\begin{tabular}{c|cccccc}
\hline
\multirow{2}{*}{\begin{tabular}[c]{@{}c@{}}Hidden layers\\ -Neurons \end{tabular}} & \multicolumn{1}{c|}{\multirow{2}{*}{\begin{tabular}[c]{@{}c@{}}Total \\ elapsed time (s)\end{tabular}}} & \multicolumn{3}{c|}{Step One}              & \multicolumn{2}{c}{Step Two} \\ \cline{3-7}
\multicolumn{1}{l|}{}                               & \multicolumn{1}{c|}{}                                                                                   & $u_r$ & $u_i$ & \multicolumn{1}{c|}{$|u|$} & $v_r$          & $|v|$         \\ \hline
4-10(2-40)             & 52.0881                                      & 2.364126E-03                             & 2.867761E-03                             & 1.478256E-03
    & 2.481454E-03
    & 2.483139E-03   \\
4-20(2-40)             & 55.5413                                      &  6.387688E-04                            & 1.215582E-03                             & 7.247901E-04                       & 2.038422E-03                             &  2.205301E-03 \\
4-30(2-40)             & 69.6366                                      & 8.425673E-04                             & 1.340774E-03                             & 7.585947E-04                       & 1.401890E-03                             & 1.450353E-03  \\
4-40(2-40)             & 42.7879                                      & 2.437708E-03                             & 3.227638E-03                             & 1.622590E-03                       & 1.514716E-03                             &  1.522125E-03 \\
4-50(2-40)            & 59.9248                                      & 8.743754E-04                             &  9.836247E-04                            & 5.424276E-04                       & 1.217201E-03                             & 1.285430E-03  \\
4-60(2-40)            &60.1517                                       & 1.326843E-03                             & 1.411608E-03                             & 9.555646E-04                       &  1.787923E-03                            & 1.833829E-03  \\
4-70(2-40)            & 59.0152                                      &  1.250724E-03                            & 1.859204E-03                             & 1.101629E-03                       & 4.192471E-03                             & 4.567781E-03  \\
4-80(2-40)            &80.2356                                       &  2.490671E-03                            &  2.492770E-03                            &  1.264181E-03                      &  1.550268E-03                            & 2.429151E-03  \\
4-90(2-40)            & 80.0251                                      &  1.495744E-03                            &  2.866518E-03                            & 1.472167E-03                       & 2.027065E-03                             & 1.865521E-03  \\
4-100(2-40)            & 83.2573                                      & 1.445773E-03                             &  1.805627E-03                            &  8.738585E-04                      & 1.808155E-03                             & 1.823184E-03  \\ \hline
\end{tabular}
\end{table}

\begin{table}[H]
\caption{Data-driven solutions in Case 4.1 via Scheme \uppercase\expandafter{\romannumeral 1} by using different numbers of neurons.}
\label{tableA-2-5}
\centering
\begin{tabular}{c|cccccc}
\hline
\multicolumn{1}{c|}{\begin{tabular}[c]{@{}c@{}}Hidden layers\\ -Neurons\end{tabular}} & Elapsed time (s) & $u_r$ & $u_i$ & $|u|$ & $v_r$ & $|v|$ \\ \hline
6-10             &  284.9192                                     &  3.947548E-04                            &  9.096204E-04                            &  5.181661E-04
    & 4.814786E-02
    &  4.493648E-02  \\
6-20             &  1619.0398                                     &   7.124622E-01                           &  8.649321E-01                            &  6.589914E-01                      &  2.077955E+00                            &  6.888107E-01 \\
6-30             & 1484.5121                                      &   6.029013E-01                           &   7.644108E-01                           &  5.579018E-01                      &   2.382381E+00                           & 1.049778E+00  \\
6-40             &   260.1525                                    &  9.211512E-04                            &   1.174782E-03                           &  7.847132E-04                      &  3.018471E-02                            & 3.061304E-02  \\
6-50            &  281.6868                                     &  5.835207E-04                            &  7.107911E-04                            & 4.526346E-04                       &  1.522079E-02                            & 1.901190E-02  \\
6-60            &  645.3964                                     &  2.409360E-03                            &  3.012503E-03                            & 1.677761E-03                       &   4.002522E-02                           & 4.177472E-02  \\
6-70            &  1726.8101                                     &   6.121047E-01                           &  7.670547E-01                            &  5.629746E-01                      &    2.382579E+00                          & 1.048035E+00  \\
6-80            &  2256.5185                                     &   5.935319E-01                           &  7.440128E-01                            &  5.529531E-01                      &   2.396681E+00                           &  1.038981E+00 \\
6-90            &  3940.0338                                     &    5.484460E-01                          &  7.307185E-01                            & 5.142581E-01                       &   2.449943E+00                           & 1.123169E+00  \\
6-100            & 419.83                                      &  1.410361E-03                            &  1.844572E-03                            & 1.182368E-03                       &  2.531610E-02                            & 2.271839E-02  \\\hline
\end{tabular}
\end{table}

\begin{table}[H]
\caption{Data-driven solutions in Case 4.1 via Scheme \uppercase\expandafter{\romannumeral 2} by using different numbers of neurons.}
\label{tableA-2-6}
\centering
\begin{tabular}{c|cccccc}
\hline
\multirow{2}{*}{\begin{tabular}[c]{@{}c@{}}Hidden layers\\ -Neurons \end{tabular}} & \multicolumn{1}{c|}{\multirow{2}{*}{\begin{tabular}[c]{@{}c@{}}Total \\ elapsed time (s)\end{tabular}}} & \multicolumn{3}{c|}{Step One}              & \multicolumn{2}{c}{Step Two} \\ \cline{3-7}
\multicolumn{1}{l|}{}                               & \multicolumn{1}{c|}{}                                                                                   & $u_r$ & $u_i$ & \multicolumn{1}{c|}{$|u|$} & $v_r$          & $|v|$         \\ \hline
6-10(2-40)             & 76.9773                                      &   1.746472E-03                           &  2.413960E-03                            &  1.203520E-03
    &  1.651415E-03
    &  1.658844E-03  \\
6-20(2-40)             &  113.1193                                     &  7.435879E-04                            &  1.730013E-03                            &  7.934987E-04                      &  9.427861E-03                            &  1.170924E-02 \\
6-30(1-40)             &  89.5559                                     &  9.044041E-04                            &  2.475596E-03                            &  1.342138E-03                      & 2.811600E-03                             & 2.881950E-03  \\
6-40(2-40)             & 91.0395                                      & 1.046530E-03                             &  2.251383E-03                            & 1.307515E-03                       &   2.192511E-03                           & 2.171804E-03  \\
6-50(1-40)            & 82.8861                                      &  1.462668E-03                            &  2.539336E-03                            & 1.588425E-03                       &  2.883125E-03                            & 3.200710E-03  \\
6-60(2-40)           & 98.872                             &  1.231441E-03                            &  1.596648E-03                            &  1.052587E-03                      & 2.821355E-03                             & 3.019951E-03  \\
6-70(1-40)            &  125.0997                                     &  3.774040E-03                            &  5.174027E-03                            &  2.913835E-03                      &   4.255301E-03                           &  7.198678E-03 \\
6-80(1-40)            & 131.1341                                      &  3.635587E-03                            & 4.884044E-03                             &  2.659861E-03                      &   2.623133E-03                           &  2.699753E-03 \\
6-90(1-40)            &  180.0378                                     &   3.655364E-03                           &   4.090889E-03                           &  2.634729E-03                      &   2.934969E-03                           & 3.020450E-03  \\
6-100(1-40)            &  119.847                                     &   1.256882E-03                           &  1.844825E-03                            &  1.232265E-03                      &  2.374537E-03                            & 2.533324E-03  \\\hline
\end{tabular}
\end{table}

\begin{table}[H]
\caption{Data-driven solutions in Case 4.2 via Scheme \uppercase\expandafter{\romannumeral 1} by using different numbers of hidden layers.}
\label{tableA-3-1}
\centering
\begin{tabular}{c|cccccc}
\hline
\multicolumn{1}{c|}{\begin{tabular}[c]{@{}c@{}}Hidden layers\\ -Neurons\end{tabular}} & Elapsed time (s) & $u_r$ & $u_i$ & $|u|$ & $v_r$ & $|v|$ \\ \hline
2-60              & 2685.3878                                      & 7.629899E-01                             & 1.203793E+00                             &  3.654520E-01
    &  1.494403E+00
    & 4.274435E-01   \\
4-60             &  1007.1702                                     & 3.660134E-03                             & 7.797806E-03                             & 2.348684E-03                       & 1.211958E-02                             &  1.216880E-02 \\
6-60             &  5307.8999                                     & 7.752140E-01                             & 1.135267E+00                             & 4.085241E-01                       &  1.566295E+00                            & 4.453264E-01  \\
8-60             & 1845.9423                                      &  2.707277E-03                            & 5.681117E-03                             &  2.037050E-03                      &  1.558546E-02                            &  1.593920E-02 \\
10-60            &  5667.074                                     &  7.098843E-01                            &  8.580395E-01                            & 4.473454E-01                       &  1.589747E+00                            & 4.541926E-01  \\
12-60            & 3456.5594                                      &  1.683386E-03                            & 2.728771E-03                             & 1.657521E-03                       &   2.471447E-02                           & 2.484533E-02  \\
14-60            &   6370.3389                                    &  7.143915E-01                            & 7.714555E-01                             & 5.546292E-01                       &  1.691123E+00                            & 5.103338E-01  \\
16-60            & 8513.8075                                      &  6.576882E-01                            & 7.801949E-01                             & 5.323424E-01                       &  1.703944E+00                            & 4.792175E-01  \\
18-60            &  10915.8336                                     &   6.754926E-01                           & 7.803961E-01                             & 4.674965E-01                       &  1.591437E+00                            & 4.570077E-01  \\
20-60            &  12334.4539                                     &  7.146354E-01                            & 1.093869E+00                             & 5.372962E-01                       & 1.773351E+00                             & 4.970701E-01  \\ \hline
\end{tabular}
\end{table}

\begin{table}[H]
\caption{Data-driven solutions in Case 4.2 via Scheme \uppercase\expandafter{\romannumeral 2} by using different numbers of hidden layers.}
\label{tableA-3-2}
\centering
\begin{tabular}{c|cccccc}
\hline
\multirow{2}{*}{\begin{tabular}[c]{@{}c@{}}Hidden layers\\ -Neurons \end{tabular}} & \multicolumn{1}{c|}{\multirow{2}{*}{\begin{tabular}[c]{@{}c@{}}Total \\ elapsed time (s)\end{tabular}}} & \multicolumn{3}{c|}{Step One}              & \multicolumn{2}{c}{Step Two} \\ \cline{3-7}
\multicolumn{1}{l|}{}                               & \multicolumn{1}{c|}{}                                                                                   & $u_r$ & $u_i$ & \multicolumn{1}{c|}{$|u|$} & $v_r$          & $|v|$         \\ \hline
2-60(2-60)             & 1311.4765                                      &  2.248782E-01                            & 4.386757E-01                             & 1.507925E-01
    & 2.710537E-01
    & 2.558087E-01   \\
4-60(2-60)             & 905.4247                                      &  1.463081E-02                            & 2.462456E-02                             & 1.168516E-02                       & 8.116157E-03                             & 8.122335E-03  \\
6-60(2-60)             &  1054.2195                                     & 4.337120E-03                             & 8.181105E-03                             &  2.868770E-03                      & 5.026626E-03                             & 5.020651E-03  \\
8-60(2-60)             & 1486.7184                                      & 7.380453E-03                             & 1.540571E-02                             & 5.178559E-03                       & 4.984289E-03                             & 4.959158E-03  \\
10-60(2-60)            & 1418.3833                                      &  7.363330E-03                            & 1.478813E-02                             & 5.055910E-03                       & 4.908539E-03                             & 4.909086E-03  \\
12-60(2-60)            & 1860.4499                                      &  1.012838E-02                            & 2.337092E-02                             &  7.396188E-03                      &  6.368222E-03                            & 6.369645E-03  \\
14-60(2-60)            &  2155.1747                                     &  1.138666E-02                            & 2.711851E-02                             & 6.292783E-03                       &  6.837799E-03                            & 6.688996E-03  \\
16-60(2-60)            &  2089.5672                                     &  2.292037E-02                            & 5.465400E-02                             & 1.228248E-02                       & 1.344688E-02                             & 1.337199E-02  \\
18-60(2-60)            &  2306.5059                                     &  6.351039E-03                            & 1.478646E-02                             & 5.380513E-03                       &  3.877444E-03                            & 3.879812E-03   \\
20-60(2-60)            &  3122.3639                                     & 4.626637E-03                             & 1.065731E-02                             & 3.726550E-03                       & 1.449598E-02                             & 1.498831E-02  \\ \hline
\end{tabular}
\end{table}

\begin{table}[H]
\caption{Data-driven solutions in Case 4.2 via Scheme \uppercase\expandafter{\romannumeral 1} by using different numbers of neurons.}
\label{tableA-3-3}
\centering
\begin{tabular}{c|cccccc}
\hline
\multicolumn{1}{c|}{\begin{tabular}[c]{@{}c@{}}Hidden layers\\ -Neurons\end{tabular}} & Elapsed time (s) & $u_r$ & $u_i$ & $|u|$ & $v_r$ & $|v|$ \\ \hline
4-10             & 907.4656                                      & 1.221807E-02                             & 2.250158E-02                             & 9.766643E-03
    &  1.490645E-02
    & 1.492254E-02   \\
4-20             &  2125.2918                                     &  6.701543E-01                            &  7.408951E-01                            & 4.498599E-01                       & 1.570014E+00                             & 4.511622E-01  \\
4-30             & 1047.0144                                      & 3.460088E-03                             & 5.603351E-03                             & 3.263600E-03                       & 2.225615E-03                             & 2.225629E-03  \\
4-40             &  1172.531                                     &  5.812026E-03                            & 9.996955E-03                             & 4.921312E-03                       & 3.768596E-03                             & 3.775202E-03  \\
4-50            &  4292.577                                     & 7.762469E-01                            & 1.193470E+00                             & 3.811327E-01                       & 1.532484E+00                             & 4.302207E-01  \\
4-60            & 1007.1702                                      & 3.660134E-03                             & 7.797806E-03                             & 2.348684E-03                       & 1.211958E-02                             & 1.216880E-02  \\
4-70            & 4655.0913                                      & 7.873885E-01                             & 1.224162E+00                             & 4.022622E-01                       & 1.531313E+00                             & 4.257601E-01  \\
4-80            & 4349.5791                                      &  8.048643E-01                            & 1.209494E+00                             & 3.995942E-01                       & 1.568635E+00                             & 4.746459E-01  \\
4-90            &  1578.8459                                     &  4.904254E-03                            & 1.113737E-02                             &  3.049600E-03                      & 3.930203E-02                             & 3.951005E-02  \\
4-100            &  4369.8447                                     & 8.015748E-01                             & 1.277866E+00                             & 3.971405E-01                        & 1.521018E+00                             & 4.410420E-01  \\ \hline
\end{tabular}
\end{table}

\begin{table}[H]
\caption{Data-driven solutions in Case 4.2 via Scheme \uppercase\expandafter{\romannumeral 2} by using different numbers of neurons.}
\label{tableA-3-4}
\centering
\begin{tabular}{c|cccccc}
\hline
\multirow{2}{*}{\begin{tabular}[c]{@{}c@{}}Hidden layers\\ -Neurons \end{tabular}} & \multicolumn{1}{c|}{\multirow{2}{*}{\begin{tabular}[c]{@{}c@{}}Total \\ elapsed time (s)\end{tabular}}} & \multicolumn{3}{c|}{Step One}              & \multicolumn{2}{c}{Step Two} \\ \cline{3-7}
\multicolumn{1}{l|}{}                               & \multicolumn{1}{c|}{}                                                                                   & $u_r$ & $u_i$ & \multicolumn{1}{c|}{$|u|$} & $v_r$          & $|v|$         \\ \hline
4-10(2-60)             &  1290.7964                                     & 2.616543E-02                             & 5.585284E-02                             & 1.769006E-02
    & 5.333548E-02
    & 4.908704E-02   \\
4-20(2-60)             & 832.6408                                      & 6.077372E-03                             & 1.365479E-02                             & 4.099856E-03                       &  5.036887E-03                            & 5.034149E-03  \\
4-30(2-60)             & 850.5143                                      &  1.073505E-02                            & 1.700585E-02                             &  9.656258E-03                      & 5.904716E-03                             & 5.900611E-03  \\
4-40(2-60)             & 880.8518                                      &  1.026303E-02                            & 2.000605E-02                              & 7.917382E-03                       & 7.124518E-03                             & 7.094308E-03  \\
4-50(2-60)            & 854.8571                                      & 1.913489E-02                             & 4.220646E-02                             & 1.184004E-02                       & 1.182761E-02                             & 1.180926E-02  \\
4-60(2-60)            & 905.4247                                      &  1.463081E-02                            & 2.462456E-02                             & 1.168516E-02                       &  8.116157E-03                            & 8.122335E-03  \\
4-70(2-60)            &  1119.4328                                     & 1.735684E-02                             & 3.952177E-02                             & 1.123457E-02                       &  1.735642E-02                            & 1.709770E-02  \\
4-80(2-60)            &  1239.8304                                     & 6.172442E-03                             & 1.183725E-02                             & 4.910934E-03                       &  4.296357E-03                            & 4.298721E-03  \\
4-90(2-60)            &  925.3174                                     &  1.562113E-02                            & 3.411789E-02                             & 1.047664E-02                       &  8.554187E-03                            & 8.609257E-03  \\
4-100(2-60)            &  1357.5191                                     &  3.072888E-02                            & 6.587611E-02                             & 2.324392E-02                       & 2.606810E-02                             & 2.520870E-02  \\\hline
\end{tabular}
\end{table}

\begin{table}[H]
\caption{Data-driven solutions in Case 4.2 via Scheme \uppercase\expandafter{\romannumeral 1} by using different numbers of neurons.}
\label{tableA-3-5}
\centering
\begin{tabular}{c|cccccc}
\hline
\multicolumn{1}{c|}{\begin{tabular}[c]{@{}c@{}}Hidden layers\\ -Neurons\end{tabular}} & Elapsed time (s) & $u_r$ & $u_i$ & $|u|$ & $v_r$ & $|v|$ \\ \hline
6-10             & 4974.3366                                      & 6.441210E-01                             & 6.744598E-01                             & 4.789423E-01
    & 1.632854E+00
    & 4.734274E-01   \\
6-20             &  3651.9961                                     & 7.044630E-01                             & 7.666145E-01                             & 4.773265E-01                       &  1.581794E+00                            & 4.660028E-01  \\
6-30             &  4205.3262                                     &   7.197409E-01                           & 9.390197E-01                             & 4.064034E-01                       &  1.573120E+00                            & 4.380364E-01  \\
6-40             &  3426.2302                                     & 7.307312E-01                             &  8.700952E-01                            &  4.232221E-01                      & 1.587748E+00                             & 4.530368E-01  \\
6-50            &  4181.0131                                     & 7.458029E-01                             & 9.825304E-01                             &  3.955711E-01                      & 1.588834E+00                             & 4.569033E-01  \\
6-60            &  5307.8999                                     &  7.752140E-01                            & 1.135267E+00                             & 4.085241E-01                       & 1.566295E+00                             & 4.453264E-01  \\
6-70            & 4527.804                                      &  7.516539E-01                            & 9.763344E-01                             & 4.053878E-01                       & 1.589717E+00                             & 4.485103E-01  \\
6-80            &  6182.1128                                     &  7.807511E-01                            & 1.141780E+00                             & 3.932770E-01                       & 1.568960E+00                             & 4.569339E-01  \\
6-90            &  6266.6571                                     & 7.724391E-01                             & 1.096447E+00                             & 3.957096E-01                       &  1.584519E+00                            & 4.642414E-01  \\
6-100            & 7199.7227                                      & 7.808487E-01                             & 1.122319E+00                             & 3.921231E-01                       & 1.573187E+00                             &  4.628293E-01 \\ \hline
\end{tabular}
\end{table}

\begin{table}[H]
\caption{Data-driven solutions in Case 4.2 via Scheme \uppercase\expandafter{\romannumeral 2} by using different numbers of neurons.}
\label{tableA-3-6}
\centering
\begin{tabular}{c|cccccc}
\hline
\multirow{2}{*}{\begin{tabular}[c]{@{}c@{}}Hidden layers\\ -Neurons \end{tabular}} & \multicolumn{1}{c|}{\multirow{2}{*}{\begin{tabular}[c]{@{}c@{}}Total \\ elapsed time (s)\end{tabular}}} & \multicolumn{3}{c|}{Step One}              & \multicolumn{2}{c}{Step Two} \\ \cline{3-7}
\multicolumn{1}{l|}{}                               & \multicolumn{1}{c|}{}                                                                                   & $u_r$ & $u_i$ & \multicolumn{1}{c|}{$|u|$} & $v_r$          & $|v|$         \\ \hline
6-10(2-60)             & 936.411                                      & 1.032021E-02                             & 1.980933E-02                             & 8.457905E-03
    & 1.316230E-02
    & 1.346547E-02   \\
6-20(2-60)             & 985.4583                                      & 7.977485E-03                             & 1.052673E-02                             & 7.635404E-03                       & 4.649289E-03                             & 4.665464E-03  \\
6-30(2-60)             & 1012.7725                                      & 1.228665E-02                             & 2.643722E-02                             & 8.926824E-03                       & 9.298379E-03                             & 9.295697E-03  \\
6-40(2-60)             & 960.1164                                      &  3.621781E-03                            & 8.834826E-03                             &  1.961022E-03                      &  7.126951E-03                            & 7.087902E-03  \\
6-50(2-60)            & 1021.5179                                      & 5.270301E-03                             & 1.050430E-02                             & 3.949706E-03                       & 4.774229E-03                             & 4.772726E-03  \\
6-60(2-60)            &  1054.2195                                     & 4.337120E-03                             & 8.181105E-03                             & 2.868770E-03                       & 5.026626E-03                             & 5.020651E-03  \\
6-70(2-60)            &  1353.93                                     & 4.842935E-03                             & 8.821810E-03                             & 3.568311E-03                       & 6.594403E-03                             & 6.639293E-03  \\
6-80(2-60)            & 1132.469                                      &  7.717399E-03                            & 1.619077E-02                             & 5.810217E-03                       &  1.366428E-02                            & 1.336457E-02   \\
6-90(2-60)            &  1421.0988                                     & 5.089299E-03                             & 1.100261E-02                             & 3.527166E-03                       & 4.218376E-03                             & 4.275350E-03  \\
6-100(2-60)            &  1414.9219                                     &  4.101928E-03                            &  7.880932E-03                            & 2.820574E-03                       & 2.946871E-03                             & 2.950513E-03  \\ \hline
\end{tabular}
\end{table}

\end{document}